\documentclass[a4paper,12pt,times,print,index,custombib]{Classes/PhDThesisPSnPDF}

\input{Preamble/preamble}

\title{Hamiltonian Particle Dynamics in Fusion Plasmas: }

\subtitle{Orbital Tomography and Spectrum Analysis for Energy and Momentum Transport under Resonant Non-Axisymmetric Perturbations}

\author{Yiannis Antonenas}

\dept{School of Applied Mathematical and Physical Sciences
}

\university{National Technical University of Athens}
\crest{\includegraphics[width=0.3\textwidth]{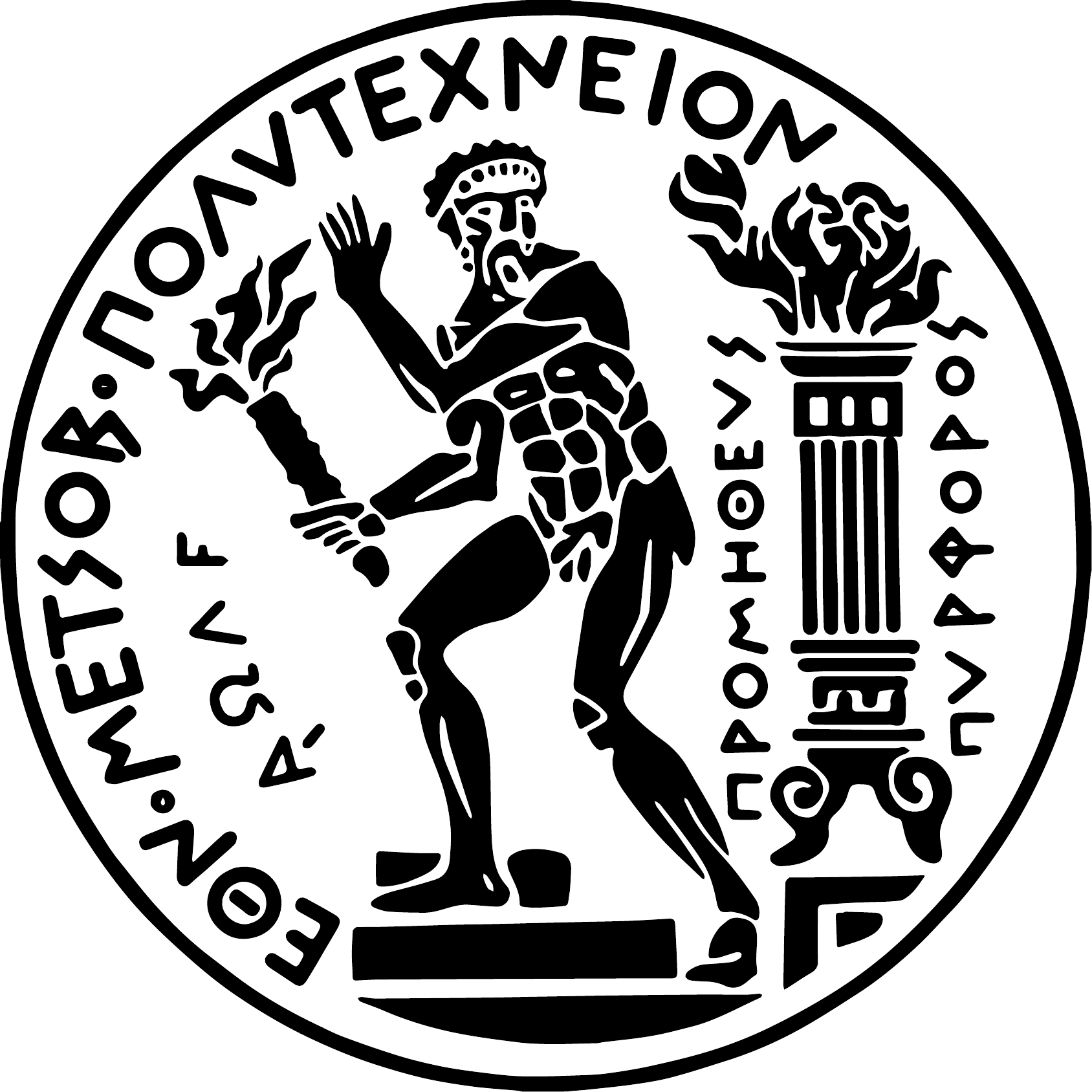}}

\degreetitle{Doctor of Philosophy}

\degreedate{April 2025} 

\subject{LaTeX} \keywords{{LaTeX} {PhD Thesis} {Engineering} {National Technical University of Athens}}

\ifdefineAbstract
\fi

\begin{document}

\frontmatter

\maketitle


\begin{declaration}

I hereby declare that except where specific reference is made to the work of 
others, the contents of this dissertation are original and have not been 
submitted in whole or in part for consideration for any other degree or 
qualification in this, or any other university. This dissertation is my own 
work and contains nothing which is the outcome of work done in collaboration 
with others, except as specified in the text and Acknowledgements. 

\end{declaration}

\begin{acknowledgements}

This work has been carried out within the framework of the EUROfusion Consortium, funded by the European Union via the Euratom Research and Training Programme (Grant Agreement No 101052200 — EUROfusion). Views and opinions expressed are however those of the author(s) only and do not necessarily reflect those of the European Union or the European Commission. Neither the European Union nor the European Commission can be held responsible for them. The work has also been partially supported by the National Fusion Programme of the Hellenic Republic – General Secretariat for Research and Innovation.

I would like to thank my close collaborators and friends. First and foremost, I am grateful to Giorgos Anastassiou, who supported me greatly— especially during my first steps in the field of plasma physics, but also throughout my journey ever since. I also thank Panos Zestanakis, with whom every discussion posed a new challenge in our work, continuously pushing us to improve.

During my PhD, I had the good fortune of meeting three great friends, Georgia Himona and Theodoros Bournelis, the other two PhD students under our shared supervision and Stratos Koukoutsis. They were always there to discuss our problems—problems that, as is well known, are rarely simple for PhD students.

A particularly fruitful and enjoyable collaboration was with Henok Moges and Professor Haris Skokos from the University of Cape Town. Together, we completed a project that led to a publication and laid the foundation for future work. But what made this collaboration truly special was my three-month stay in Cape Town, and even more so, the friendship I formed with Henok. Many thanks to him, to Professor Haris, and to his whole team—Malcolm, Arnold, and the rest—for their warm hospitality. I also gratefully acknowledge the Erasmus+ program for funding this collaboration.

A special thank you goes to my partner, Dafni. She has supported me from the very beginning of my PhD studies. During the most difficult period, while I was writing this thesis, she kept me going and motivated. Through all my worries along this journey, she was always there. A PhD may improve you, but the right person in your life has the power to truly shape and elevate your personality. I consider myself incredibly lucky to have her in my life.

Finally, I want to express my deepest gratitude to my supervisor and mentor, Professor Yiannis Kominis. I met Yiannis during my second undergraduate year in the School of Electrical and Computer Engineering at NTUA. From the very first lecture, I was intrigued. Once the course ended, I found myself wanting more. In the following years, I shaped my academic path so that I could attend more of the courses he taught. That’s when I realized I wanted to work with him. He supervised my diploma thesis, and naturally, I continued into a PhD under his guidance.

Yiannis is a true teacher—not just for me, but for many others as well. He inspires students to love the subject. Without ever being demanding or severe, he helps you grow, both scientifically and personally. I am deeply thankful for this journey with him—a journey that, I believe, is far from over.

\end{acknowledgements}


\begin{abstract}
This thesis investigates the impact of resonant mode-particle interactions on the transport and confinement properties of toroidal fusion plasmas. Using both analytical and numerical computational approaches, we analyze how magnetic perturbations, both intrinsic and externally applied, influence the behavior of plasma particles with different kinetic characteristics. The primary focus is on understanding the resonant interactions between perturbative modes and the guiding center motion of plasma particles.

We present a novel, low-computational-cost method, based on Action-Angle variables, to identify resonance locations and other key characteristics of resonances, such as the number of islands in a resonance island chain, as well as the formation of transport barriers, within phase space of the particle's guiding center motion. We introduce the drift center (DC) approximation to derive analytical expressions for the orbital frequencies and the kinetic $q$ factor in the case of a large aspect ratio (LAR) equilibrium. These expressions are essential for identifying the conditions under which mode-particle resonances occur, which significantly affect particle, momentum, and energy transport, and, consequently, the overall plasma confinement. We validate these results through comparisons with numerical simulations, providing an efficient and accurate tool for predicting transport barriers and the behavior of energetic particles in the presence of non-axisymmetric perturbations.

We extend the applicability of our orbital frequencies analysis methodology from LAR equilibria to numerically reconstructed, realistic equilibria. For this, we employ a computationally efficient, semi-analytical geometrical method that can be applied to any given unperturbed equilibrium.

Furthermore, for the LAR equilibrium, we extend our analysis to time-varying perturbations, demonstrating the emergence of Arnold diffusion in the guiding center phase space and highlighting the role of the Arnold web in particle transport. This work enhances our understanding of how mode-particle resonant interactions influence plasma behavior, paving the way for more accurate predictions of fusion device performance and offering computational tools applicable to realistic experimental configurations.
\clearpage

\end{abstract}

\begin{center}
   \textbf{\Large Dissemination of the results of this PhD research}
\end{center}

International peer-reviewed journals:
   \begin{itemize}
    \item Y.Antonenas, G.Anastasiou, and Y.Kominis, “Analytical calculation of the orbital spectrum of the guiding center motion in axisymmetric magnetic fields,” J. Plasma Phys. 87, 855870101 (2021). 
    \item G. Anastassiou, P. Zestanakis, Y. Antonenas, E. Viezzer, and Y. Kominis, “Role of the edge electric field in the resonant mode-particle interactions and the formation of transport barriers in toroidal plasmas,” J. Plasma Phys. 90, 905900110 (2024).
    \item H. T. Moges, Y. Antonenas, G. Anastassiou, C. Skokos, and Y. Kominis, “Kinetic vs. magnetic chaos in toroidal plasmas: A systematic quantitative comparison,” Phys. Plasmas 31, 012302 (2024).
    \item Y.Antonenas, G.Anastasiou, and Y.Kominis, “Analytical calculation of the kinetic q factor and resonant response of toroidally confined plasmas,” Phys. Plasmas 31,  102302 (2024).
 \end{itemize} 

 Indicative presentations in conferences:
\begin{itemize}
      \item Y.Antonenas, G.Anastassiou and Y. Kominis, "Systematic dissection of the guiding center phase space based on orbital spectrum analysis" FuseNet PHD Event, Padova, Italy, 04-06 July 2022, poster presentation.
      \item  H.T. Moges, Y. Antonenas, G. Anastassiou, Ch. Skokos and Y. Kominis, “Kinetic versus magnetic chaos in tokamak plasmas" 49th  EPS Conference on Plasma Physics , Bordeaux, France, 03-07 July 2023, poster presentation.
      \item G. Anastassiou, P. Zestanakis, Y. Antonenas, E. Viezzer, and Y. Kominis, “Resonant Mode-Particle Interactions and Transport Barriers in toroidal plasmas under the presence of an Edge Radial Electric field” 20th European Fusion Theory Conference (EFTC), Padova, Italy, 2-5 October 2023.
      \item H.T. Moges, Y. Antonenas, G. Anastassiou, Ch. Skokos, and Y. Kominis, “Quantification and Comparison of Magnetic and Kinetic Chaos in Toroidal Plasmas” 20th European Fusion Theory  Conference (EFTC), Padova, Italy, 2-5 October 2023.
      \item  Y.Antonenas, G.Anastassiou and Y. Kominis, "Particle energy and momentum transport on the Arnold web of the Guiding Center motion in toroidal fusion plasmas" 50th EPS Conference on Plasma Physics, Salamanca, Spain, 08-12 July 2024, poster presentation.
\end{itemize}



\tableofcontents

\listoffigures

\listoftables


\printnomenclature

\mainmatter


\chapter{Fusion}  

\ifpdf
    \graphicspath{{Chapter1/Figs/Raster/}{Chapter1/Figs/PDF/}{Chapter1/Figs/}}
\else
    \graphicspath{{Chapter1/Figs/Vector/}{Chapter1/Figs/}}
\fi

\section{The Power of the Sun on Earth} \label{section 1.1}

In our sun, at temperatures of 15 million degrees, hydrogen takes on a plasma form, which is characterized as the fourth state of matter. The gravitational forces, coupled with high temperatures, facilitate the overcoming of the nuclear electrostatic repulsion between the positively-charged hydrogen nuclei, leading to the combine of four protons into helium. This is a fusion reaction. Due to the slightly lower mass of helium, energy is produced.
Fusion is a fundamental process wherein nuclei of lighter atoms fuse to generate heavier atoms and vast amounts of energy during the reaction. Since the 1920s, when British astrophysicist Arthur Eddington  suggested that fusion is the primary mechanism driving energy production in stars, scientists have begun efforts to turn the dream of unlimited, clean and safe energy into reality by creating the so-called artificial sun.

At Earth, the immense gravitational force needed for the confinement of hydrogen nuclei does not exist. Thus, a different approach is employed to make fusion reactions feasible on Earth. This approach demands plasma at temperatures around 150 million degrees Celsius, which is ten times higher than the temperatures at the Sun's core. As a consequence, there is no material that can contain the plasma at this temperature. Therefore, the main idea is to confine the plasma in the air. To achieve this, scientists aim to create a container made of strong magnetic fields that prevent the plasma particles from escaping. Freidberg in Ref. \cite{Freidberg2007} analyses the environmental impact of fusion energy production, analyses the technical specifications of plasma necessary for generating fusion power, discusses the design of fusion reactors and offers an introduction to plasma physics along with the challenges encountered by researchers in the field.

In regard to the fuel required for fusion reactions, scientists have identified two potential scenarios. The first involves the deuterium and tritium fusion reaction (D-T reaction), which is the easiest to initiate, thus current fusion research is focusing on it. This reaction releases large amounts of energy, equivalent to $338 \times 10^6$ MJ/kg of fuel. Additionally, deuterium, one component of this reaction, is plentifully available as it can be extracted from ocean water at a low cost. However, the D-T reaction presents two significant challenges. Firstly, it generates high-energy neutrons, which cause material activation and radiation damage. Secondly, there is no natural tritium on Earth, and it must be produced by bombarding a lithium blanket in the wall of fusion reactors with neutrons. Despite adequate lithium supplies for the next thousands of years, tritium is also radioactive with a half-life of 12.26 years. Nowadays, nuclear engineers can deal with the aforementioned problems. An analysis of fusion wastes, as well as the requirements that future fusion power plants must meet, can be found in Ref. \cite{Sehila}. The second scenario is the deuterium and deuterium reaction (D-D), which also produces $78 \times 10^6$ or $96 \times 10^6$ MJ/kg depending on the final product. This scenario is more desirable from the point of view of the unlimited supplies of deuterium and its direct use. Additionally, most of the energy is carried by charged particles. However, the D-D reaction is more difficult to initiate.

In contrast to the fission reaction, fusion wastes do not contain many long-lived radioactive products, which, after careful management and disposal, will be safe for the environment and future generations. Additionally, while a chain reaction occurs in fission, allowing for self-sustaining operation, such a chain reaction does not happen in fusion. Thus, new fuel must be continually added to maintain the operation. This allows for the immediate stopping of the fusion reaction by halting the fuel flow. As a result, the risk of extended severe radioactive accidents in fusion reactors is markedly lower compared to fission reactors. Moreover, the energy released from nuclear reactions is one million times greater than that released from chemical reactions, and it requires a very small amount of fuel, which in the case of fusion does not involve mining. The aforementioned facts, combined with fusion's lack of greenhouse gas emissions, render it  the solution for climate change mitigation and for a sustainable future. In their article, H. Cabal et al. \cite{Cabal2017} analyze different scenarios of a future world and demonstrate that in a world with strong environmental responsibility and a strict target to minimize carbon emissions, fusion technologies make a very high contribution to the global electricity system. Additionally, in their study, Keii Gi et al. \cite{Gi2020} having analyzed the global energy system up to 2100, they concluded that a drastic decarbonization of energy systems is required to achieve the 2$^{\circ}C$ target of the Paris Agreement. For this purpose, fusion power plays a crucial role.

\section{Magnetic Confinement Fusion Experiments} \label{section 1.2}

Presently, two topologically toroidal devices are employed to achieve magnetic confinement fusion: tokamaks and stellarators. Tokamaks have exhibited greater efficiency in experiments thus far. The concept of the tokamak was developed in the Soviet Union by scientists Andrei Sakharov and Igor Tamm in 1950, and the Soviet scientist Artsimovich \cite{Artsimovich1972} demonstrated its greater efficiency compared to the stellarator, invented by Lyman Spitzer in 1951 \cite{Spitzeb1958}. In his book \cite{Wesson2005}, John Wesson outlines the contributions of current and past experiments conducted on small and medium-sized tokamaks, as well as evaluates the outcomes of experiments conducted on large tokamak devices. 

The world's largest tokamak device, which is expected to pave the way from experimental fusion to the production of self-sustained fusion energy for commercial use, is the international joint experiment in fusion ITER (International Thermonuclear Experimental Reactor). As ITER's successor, the Demonstration Power Plant (DEMO) will be designed to produce electricity for the grid. Technical specifications and physical requirements for ITER's operation, as well as potential problems of interest, are identified in the article by R. J. Hawryluk et al. \cite{Hawryluk2009}. In current experimental devices, plasma is heated externally to achieve fusion reactions and produce fusion power. Up to now, the Joint European Torus (JET) has achieved the world record for fusion power with an energy balance of $Q = 0.67$, i.e., $16$ MW of fusion power with an input heating of $24$ MW, while ITER is anticipated to produce $500$ MW of fusion power from its plasma with an input of $50$ MW heating power, i.e., $Q = 10$. JET, operating scenarios for ITER in a deuterium-tritium environment \cite{Mailloux2022}, reached another milestone on October 3, 2023, by releasing the most fusion energy in a single pulse: 69 MJ in a 6-second long pulse burning 0.2 milligrams of fusion fuel. The proximity of JET's technical specifications to those of ITER renders JET a strong foundation for ITER's operation. Details for the operation, experiments and contribution of JET machine in plasma physics can be found in the article by Fernanda G. Rimini and JET Contributors \cite{Rimini2024}. 

In addition, to support ITER's operation, the joint Japanese-European device JT-60SA,  superconducting tokamak, is being developed in Japan and is set to become the most powerful fusion device to date \cite{Tomarchio2017}. Furthermore, medium-sized Tokamaks such as the ASDEX Upgrade aim to develop a physics base for ITER and DEMO by studying essential plasma properties, primarily plasma density, plasma pressure, and wall load under future power plant conditions \cite{Stroth2022}. The Mega Amp Spherical Tokamak (MAST) Upgrade is a spherical tokamak primarily used for the study and development of the divertor, a system designed to handle the exhaust of particles and heat from the plasma, for future tokamak devices \cite{Sykes2001}. TCV is a device focused on the study of different plasma shapes \cite{Coda2017}, while WEST (W Environment in Steady-state Tokamak) is dedicated to the study of the divertor for ITER \cite{Bourdelle2015}.

\section{Fusion Plasma}  \label{section1.3}

In a magnetic fusion reactor, plasma must meet specific properties, including extremely high temperatures ($T\sim15$ keV) and pressures ($p\sim7$ atm), while also being confined for an adequate duration in a toroidal-shaped geometry to facilitate nuclear fusion. Achieving these conditions simultaneously is a complex task, requiring a detailed investigation into the behavior of plasma under specific circumstances, which is a main topic of the field of plasma physics. The development of fusion plasma physics has also promoted studies of space and astrophysical plasma. Comprehensive introductions to plasma physics can be found in the textbooks authored by Francis F. Chen (2016) \cite{Chen2016}, Robert Goldston and Paul Rutherford (1995) \cite{Goldston1995}, and Alexander Piel (2010) \cite{Piel2010}.

To be considered as plasma, an ionized gas must fulfill specific criteria summarized in the following expressions.
\begin{eqnarray}
    \lambda_{D_{e}} \ll L \label{DL}\\
    \omega_{p_{e}} \gg \omega_{T_{e}} \label{PF} \\
    \Lambda_{D} = \frac{4\pi}{3}n_e\lambda^{3}_{De} \gg 1 \label{SD}
\end{eqnarray}
The first two criteria (Eqs. \eqref{DL},\eqref{PF}) ensure the quasi-neutral behaviour of plasma and is linked to the Debye length ($\lambda_{D_{e}}$) and the electron plasma frequency ($\omega_{p_{e}}$) given by \cite{Freidberg2007}.
\begin{eqnarray}
    \lambda_{D_{e}} &=& \sqrt{\frac{\epsilon_{0}T_{e}}{e^2 n_{0}}}\label{Debye length}\\
    \omega_{p_{e}} &=& \frac{n_{0}e^2}{m_{e}\epsilon_{0}}
\end{eqnarray}
where $T_{e}$ is the electron temperature in the plasma, $e$ and $m_{e}$ are the charge and the mass of the electron, respectively, and $n_{0}$ is the electron density at a reference position. When $\lambda_{D_{e}}\ll L$, where $L$ represents the geometric dimensions of the plasma, the plasma effectively shields out DC electric fields. Additionally, when $\omega_{p_{e}}\gg\omega_{T_{e}}$, where $\omega_{T_{e}}$ denotes the shortest inverse thermal transit time for an electron across the plasma, the plasma screens out AC electric fields. This screening also leads to the high electrical conductivity of plasma.

The third criterion, Eq. \eqref{SD}, involves low collisionality and collective behavior, which leads to the dominance of long-range electric and magnetic fields over short-range Coulomb collisions in plasma behavior. When $\Lambda_{D}\gg 1$, where $\Lambda_{D}$ represents the number of particles in the sphere with a radius equal to the Debye length $\lambda_{D_{e}}$, low collisionality is ensured (Ref. \cite{Freidberg2007}, Sec. 7.4).

In the context of fusion applications, additional criteria must be met for the fusion plasma. These include the requirement of a very small Larmor radius ($\rho_{L_{i}}$) and a large gyro-frequency ($\omega_{c_{i}}$) for ions, which are associated with the perpendicular circular orbit of a particle around a magnetic field line:
\begin{eqnarray}
    \rho_{L_{i}}\ll L \label{Larmor radius con} \\
    \omega_{c_{i}}\gg \omega_{T_{i}} \label{gyro-f con}
\end{eqnarray}
For the electron Larmor radius and gyro-frequency it holds that $\rho_{L_{e}}\ll \rho_{L_{i}}$ and $\omega_{c_{e}} \gg \omega_{c_{i}}$ respectively thus the conditions \eqref{Larmor radius con} and \eqref{gyro-f con} are fulfilled by many orders of magnitude \cite{Freidberg2007}.

\section{Two-fluid model}
From the discussion above, it follows that plasma, as a gas, exhibits fluid behavior governed by fluid dynamics, such as the conservation laws of mass, energy, and momentum. However, the macroscopic electromagnetic fields governed by Maxwell's equations, which dominate in plasma, cause long-range collective effects. Moreover, for simplicity it is assumed that, plasma consists of two distinct species of particles: ions and electrons. Thus, a macroscopic two-fluid model (one for electrons and one for ions) along with Maxwell's equations can provide a self-consistent description of plasma behavior. This description takes into account not only the applied electromagnetic fields on plasma but also how these fields are affected by the particles' motion. The two-fluid model for ions (i) and electrons (e) can be described by the following sets of fluid differential equations \cite{Freidberg2007}, \cite{Chen2016}:

\begin{itemize}
    \item \emph{Conservation of mass}
        \begin{equation} \label{mass conserv}
            \frac{\partial n_{j}}{\partial t} + \nabla\cdot(n_{j}\mathbf{u}_{j}) = 0 \qquad j=i,e
        \end{equation}
    \item \emph{Conservation of momentum}
        \begin{equation} \label{moment conserv}
            m_{j} n_{j}\left(\frac{\partial}{\partial t} + \textbf{u}_{j}\cdot\nabla\right)\textbf{u}_{j} = -e n_{j}\left(\mathbf{E} + \mathbf{u}_{j}\times\mathbf{B}\right) - \mathbf{\nabla}p_{j} \pm m_{j}n_{j}\bar{\nu}_{ei}(\mathbf{u}_{e}-\mathbf{u}_{i}) \qquad j = i,e
        \end{equation}
        
        In the case of an isotropic fluid, such as when the plasma distribution function is Maxwellian, the pressure $p_{j}$ is a scalar quantity. The term $\pm m_{j} n_{j} \bar{\nu}_{ei} (\mathbf{u}_{e} - \mathbf{u}_{i})$ represents the collisional friction force (with $-$ for electrons and $+$ for ions), accounting for the net momentum exchange, where $\bar{\nu}_{ei}$ is the net momentum exchange collision frequency (Ref.~\cite{Freidberg2007}, Sec. 9.7.2).  

    \item \emph{Maxwell's equations}
        \begin{eqnarray}
            \mathbf{\nabla}\times\mathbf{E} &=& -\frac{\partial \mathbf{B}}{\partial t} \label{M1}\\
            \mathbf{\nabla}\times\mathbf{B} &=& \mu_{0}e\left(n_{i}\mathbf{u}_{i} - n_{e}\mathbf{u}_{e}\right) + \mu_{0}\epsilon_{0}\frac{\partial\mathbf{E}}{\partial t} \label{M2}\\
            \epsilon_{0}\nabla\cdot\mathbf{E} &=& e\left(n_{i}-n_{e}\right)\label{M3}\\
            \nabla\cdot\mathbf{B} &=& 0 \label{M4}
        \end{eqnarray}
\end{itemize}

To determine the pressure $p_{i,e}$, generally given by $p_{i,e} = n_{i,e}T_{i,e}$, one can either formulate a conservation of energy equation analogous to the conservation of momentum equation \eqref{moment conserv}, which also provides information on the thermal energy of the plasma \cite{Freidberg2007}, or, after particular simplifications in the energy conservation equation, utilize the thermodynamic equation of state that relates pressure to density, as presented in the following \cite{Chen2016}.
\begin{itemize}
    \item \emph{Equation of state}
        \begin{eqnarray}
            p_{j} = Cn_{j}^{\gamma}, \qquad j=i,e
        \end{eqnarray}
\end{itemize}

The above set of equations involves six basic unknowns: the average macroscopic velocities of fluid elements $\mathbf{u}_{i}$ and $\mathbf{u}_{e}$, the densities $n_{i}$ and $n_{e}$ for ions and electrons, respectively, as well as the electric field $\mathbf{E}$ and the magnetic field $\mathbf{B}$. These variables are determined by a set of six independent equations, comprising Eqs. \eqref{mass conserv} - \eqref{M2}. (In the time dependent case Maxwell's equations \eqref{M3} and \eqref{M4} can be derived from Eqs. \eqref{M1} and \eqref{M2} when Eqs. \eqref{M3} and \eqref{M4} are imposed as initial conditions). 

\section{Single-fluid model and Force Balance Equation} \label{single fluid model}

In order for fusion reactions to occur in continuous and steady operation, fusion plasma must be held stable over a sufficiently long period within a macroscopic magnetic field. The magnetic field configuration that produces the forces required for holding plasma is referred to as the equilibrium magnetic field. In order to be stable the equilibrium should exhibit magnetohydrodynamic stability. The derivation of a stable equilibrium can be based on the two-fluid model, which can be further simplified into a single-fluid model known as the MagnetoHydroDynamic (MHD) model. Under these simplifications, the plasma is treated as a single fluid without distinguishing between particle species. By introducing single-fluid variables and neglecting small terms (such as those associated with the mass difference between ions and electrons), this description focuses on macroscopic quantities and time/space scales (Ref. \cite{Freidberg2007}, Sec. 11).  

Specifically, the appropriate macroscopic length scale $L$ for the MHD model is the plasma radius, $L \sim \alpha$. The time scale is the ion thermal transit time across the plasma, $\tau \sim \alpha/u_{T_{i}}$, and as a result, the characteristic velocity is $u_{i}\sim u_{T_{i}}$ \cite{Freidberg2007}. Additionally, by the definition of the Debye length \eqref{Debye length}, in the limit $\epsilon_{0} \longrightarrow 0$, the condition \eqref{DL}, $\lambda_{D_{e}} \ll \alpha$, is fulfilled. Respectively, by letting $m_{e} \longrightarrow 0$, the condition \eqref{PF} is satisfied. As a result, at these limits, the right-hand side of the electron conservation of momentum equation, Eq. \eqref{moment conserv}, the second term on the left-hand side in Eq. \eqref{M2} as well as the left-hand side of Eq. \eqref{M3} can be neglected.

Either the very small Debye length or, equivalently, the elimination of the term $\epsilon_{0}\nabla\cdot\mathbf{E}$ in Eq. \eqref{M3} results in the quasi-neutrality property, i.e., $n_{e}\approx n_{i} = n$. This quasi-neutrality, combined with the electron mass tending to zero, leads to the conclusion that the mass density of the single-fluid MHD model will be that of the ions, i.e., $\rho = m_{i}n$ \cite{Freidberg2007}. Thus, by multiplying Eq. \eqref{mass conserv} for ions by $m_{i}$, we obtain:
\begin{equation} \label{mass consrv single}
    \frac{d\rho}{dt} + \rho\nabla\cdot\mathbf{v} = 0
\end{equation}
This equation represents the mass conservation equation for the single-fluid MHD model.

The dominant macroscopic fluid velocity in plasma is the $\mathbf{E}\times\mathbf{B}$ drift velocity (Ref. \cite{Freidberg2007}, Sec. 11.2.1). It can be proved that both electrons and ions move with the same $\mathbf{E}\times\mathbf{B}$ velocity. Since $m_{e}\ll m_{i}$, the momentum of the fluid is primarily carried by the ions, allowing us to consider the macroscopic fluid element velocity as $\mathbf{v} = \mathbf{u}_{i}$ in the ion conservation of momentum equation. Incorporating the simplifications mentioned above and adding the conservation of momentum equations for ions and electrons, we have
\begin{equation} \label{moment conserv single}
    \rho\frac{d \mathbf{v}}{d t} = \mathbf{J}\times\mathbf{B} - \nabla p
\end{equation}
where $\mathbf{J} = en(\mathbf{v}-\mathbf{u_{e}})$ represents the current density and $p = p_{i} + p_{e}$.

Substituting $\mathbf{u}_e = \mathbf{v} - \mathbf{J}/en$ and $m_e = 0$ into the electron conservation of momentum equation Eq. \eqref{moment conserv}, it is obtained:
\begin{equation}\label{gen Ohms law}
    \mathbf{E} + \mathbf{v}\times\mathbf{B} = \left(\mathbf{J}\times\mathbf{B} - \nabla p_{e}\right)/en + \eta\mathbf{J}
\end{equation}
where $\eta = m_e\bar{\nu}_{ei}/e^2n_e$ is the resistivity and $\bar{\nu}_{ei}$ is the net momentum exchange collision frequency (Ref. \cite{Freidberg2007}, Sec. 9.7.2). Neglecting collisions, $\bar{\nu}_{ei} = 0$, the resistivity becomes zero. Moreover, it can be shown that the magnitudes of $\mathbf{J}\times\mathbf{B}$ and $\nabla p$ are much smaller compared to the magnitude of $\mathbf{v}\times \mathbf{B}$, and hence they can be neglected in Eq. \eqref{gen Ohms law}, leading to the 'ideal' Ohm's law 
\begin{equation} \label{ideal Ohm's law}
    \mathbf{E} + \mathbf{v}\times\mathbf{B} = 0
\end{equation}

In summary the single fluid MHD model which is necessary to analyze the macroscopic equilibrium and stability of plasma is described by the following set of equations.

\begin{itemize}
    \item \emph{Conservation of mass}
        \begin{equation} \label{MHD mass}
             \frac{d\rho}{dt} + \rho\nabla\cdot\mathbf{v} = 0
        \end{equation}
    \item \emph{Conservation of momentum}
        \begin{equation} \label{MHD momnet}
            \rho\frac{d \mathbf{v}}{d t} = \mathbf{J}\times\mathbf{B} - \nabla p
        \end{equation}
    \item \emph{Equation of state}
        \begin{eqnarray}
            p = C\rho^{\gamma}
        \end{eqnarray}
    \item \emph{Maxwell's equations}
        \begin{eqnarray}
            \mathbf{\nabla}\times\mathbf{E} &=& -\frac{\partial \mathbf{B}}{\partial t} \label{MHD M1}\\
            \mathbf{\nabla}\times\mathbf{B} &=&  \mu_{0}\mathbf{J}\label{MHD M2}\\
            \nabla\cdot\mathbf{B} &=& 0 \label{MHD M4}
        \end{eqnarray}
    \item \emph{"Ideal" Ohm's law}
        \begin{equation} \label{Ideal Ohm's law}
            \mathbf{E} + \mathbf{v}\times\mathbf{B} = 0
        \end{equation}
\end{itemize}
Equations \eqref{MHD mass}-\eqref{MHD M2} and Eq. \eqref{Ideal Ohm's law} form a set of five nonlinear first-order partial differential equations with five unknown quantities: $\rho$, $\mathbf{v}$, $\mathbf{B}$, $\mathbf{J}$, and $\mathbf{E}$.

When the plasma reaches a steady-state situation, i.e., $\partial/\partial t = 0$, and assuming that $\bm{v} = 0$\footnote{In general there are steady states with $\bm{v} \ne 0$. One can define static, quasi-static and stationary steady states.}, which is identical with $d/dt = 0$ (Ref. \cite{Freidberg2007}, Sec. 11.5.1), the left-hand side of Eq. \eqref{MHD momnet} becomes zero, resulting in
\begin{equation} \label{force balance eq}
    \mathbf{J}\times\mathbf{B} - \nabla p = 0
\end{equation}
This equation is the force balance equation, which, along with Eq. \eqref{MHD M2}, is utilized to determine the geometry of various equilibrium magnetic fields. By employing these equations, the Grad-Shafranov equation, for two-dimensional equilibria, i.e for translationally symmetric, axisymmetric or helically symmetric equilibria, is derived (see Sec. \ref{Sec: Grad-Shafranov Equation}), which provides the relationship between the current, the magnetic field, and the pressure. As it will be discussed in next chapter, the axisymmetric toroidal and large aspect ratio magnetic field equilibrium used in this work stems from this equation.

\section{Linear Magnetohydrodynamic (MHD) modes}\label{MHD Modes}

After determining an equilibrium magnetic field that satisfies the single-fluid MHD model equations (Eqs. \eqref{MHD mass}-\eqref{Ideal Ohm's law}) with $\partial/\partial t = 0$ and $\bm{v} = 0$, the question arises: is this equilibrium stable? In other words, what happens if plasma is slightly perturbed  from its equilibrium state? If it is stable, it will either return to its initial equilibrium state or oscillate around it. However, if it is unstable, the perturbation will grow, moving the plasma away from its equilibrium condition.

Newcomb (1958) \cite{Newcomb1958} proved that it is permissible to picture the magnetic field lines as moving with a velocity $\mathbf{u}$, if $\mathbf{u}$ is flux-preserving. In addition, for every electromagnetic field there exists a flux-preserving velocity which is perpendicular to the magnetic field $\mathbf{B}$, thus, $\mathbf{u}$ will be denoted as $\mathbf{u}_{\perp}$. Flux-preserving velocity is not necessarily unique; nevertheless, any flux-preserving velocity could be ascribed as a velocity for the magnetic field lines without contradicting the general picture of the electromagnetic field \cite{Newcomb1958}.

Let's consider an arbitrary surface that moves with a velocity $\mathbf{u}_{\perp}$, perpendicular to the magnetic field $\mathbf{B}$. The total time derivative of the magnetic flux through this surface is given by:
\begin{equation} \label{general magnetic flux}
    \frac{d\psi}{dt} = \int{\frac{\partial \mathbf{B}}{\partial t}}\cdot\mathbf{\hat{n}}dS + \oint{\mathbf{B}\times\mathbf{u}_{\perp}d\mathbf{l}}
\end{equation}
In the case of ideal MHD, by making use of Eqs. \eqref{MHD M1} and \eqref{Ideal Ohm's law} and applying Stokes' theorem, the previous equation can be expressed as (Ref. \cite{Freidberg2007}, Sec. 12.2.2):
\begin{equation}
    \frac{d\psi}{dt} = \oint{\left[\left(\mathbf{v}_{\perp} - \mathbf{u}_{\perp}\right)\times \mathbf{B}\right]\cdot d\mathbf{l}}
\end{equation}
where, $\mathbf{v}_{\perp}$ represents the fluid element velocity perpendicular to the magnetic field, which is analogous to the $\mathbf{E}\times\mathbf{B}$ drift macroscopic velocity that dominates in plasma. If $\mathbf{u}_{\perp} = \mathbf{v}_{\perp}$, the equation above demonstrates that the fluid element velocity $\mathbf{v}_{\perp}$ is a flux-preserving velocity in ideal MHD. As a result, $\mathbf{v}_{\perp}$ can be chosen as the magnetic field lines' velocity, allowing one to say that within the ideal MHD the magnetic field lines move with the plasma's fluid elements (frozen-in condition). For this reason, the magnetic field line topology is preserved during the plasma motion \cite{Freidberg2007}. Consequently, in ideal MHD (Eqs. \eqref{MHD mass}-\eqref{MHD M2} and Eq. \eqref{Ideal Ohm's law}), the electromagnetic and fluid perturbations are not independent and both can be described by the ideal displacement of the fluid element denoted with $\bm{\xi}$. 

The ideal MHD instabilities are relatively robust, making ideal MHD modes dangerous and thus necessary to avoid in fusion reactors \cite{Freidberg2007}. These modes can be analyzed through a linear stability procedure, where the quantities in the ideal MHD model are expressed as the sum of an equilibrium solution and a perturbation, i.e., $Q = Q_{0} + Q_{1}$. Here, $Q_{0}$ represents the equilibrium quantity, while $Q_{1}$ is the perturbative term. Non-linear perturbation terms (i.e., terms of order $Q_1^2$, $Q_1\bm{\xi}$, and higher) are neglected. The perturbed quantities are expressed using a normal mode expansion, $Q_{1}(\mathbf{r},t) = Q_{1}(\mathbf{r})\exp(-j\omega t)$, and the fluid displacement $\bm{\xi}$ is related to the perturbed velocity $\bm{v}_1$ through:
\begin{eqnarray}
    \bm{v}_1 = \frac{d\bm{\xi}}{dt} = \frac{\partial \bm{\xi}}{\partial t} + (\bm{v}_1\cdot\nabla)\bm{\xi},
\end{eqnarray}
where the second term on the right-hand side, $(\bm{v}_1 \cdot \nabla)\bm{\xi}$, is second-order with respect to the perturbation, as it involves the product $\bm{v}_1\bm{\xi}$. Neglecting this term, the expression simplifies to
\begin{equation}\label{xi_t}
    \mathbf{v}_{1} = \frac{\partial \bm{\xi}(\mathbf{r},t)}{\partial t} = -j\omega\bm{\xi}(\mathbf{r})\exp(-j\omega t).
\end{equation}
Substituting these forms into Eqs. \eqref{MHD M1} and \eqref{Ideal Ohm's law}, while neglecting non-linear perturbation terms, the perturbed magnetic field as a function of the ideal displacement is expressed as follows:
\begin{equation} \label{delta B ideal}
    \mathbf{B} = \mathbf{B_{0}} + \nabla\times(\bm{\xi}\times\mathbf{B_{0}})
\end{equation}
Using the equation above, as well as linearizing the rest of the equations of the single fluid ideal MHD model (Eqs. \eqref{MHD mass}-\eqref{Ideal Ohm's law}), an eigenvalue problem is obtained. The determination of the complex eigenvalue $\omega$ characterizes the equilibrium as stable or unstable. Whether through an energy principle method \cite{White2013}, it is proven that $\omega^2$ is real, and the mode is either purely growing or stable. 

It is important to note that when expressing the magnetic field through the linear stability procedure, i.e., Eq. \eqref{delta B ideal}, the condition that the magnetic field lines are frozen in the plasma is violated because only the linear perturbation terms of the fluid displacement are considered in the single-fluid model MHD equations \cite{White2013a}. In other words, the full plasma response to the magnetic perturbation and the subsequent magnetic field alteration due to this response, in a self-consistent way according to the ideal MHD model equations, has not been taken into account. As a result, the magnetic flux is not conserved, and the topology of the magnetic field lines changes due to the perturbation. This is evident from the formation of extraneous magnetic islands due to the resonant interactions between the equilibrium magnetic field and the perturbation (Eq. \eqref{delta B ideal}) \cite{White2013b}. Under certain resonance conditions, this can even lead to the stochasticization of the magnetic field lines \cite{White2013}.

To satisfy the frozen-in condition—ensuring that the magnetic field is expressed such that $\mathbf{u}_{\perp} = \mathbf{v}_{\perp}$ and $d\psi/dt = 0$ even after the fluid displacement—a complete solution of the single-fluid MHD equations with the displacement $\bm{\xi}$ is required. This solution must retain all perturbation terms, making the analysis inherently nonlinear \cite{White2013a}. Alternatively, a form of the magnetic field that preserves the flux after a fluid perturbation was derived by S. Lundquist in 1951 for an incompressible fluid (i.e., $\nabla \cdot \bm{v} = 0$, which is equivalent to setting $d\rho/dt = 0$ in Eq. \eqref{MHD mass}) \cite{Lundquist1951}, expressed as:
\begin{equation} \label{Lundquist B}
\mathbf{B} = \mathbf{B_{0}} + (\mathbf{B_{0}} \cdot \nabla) \bm{\xi}
\end{equation}
Although the above expression preserves the topology of the magnetic field lines, it is not as convenient as the linearized perturbation expression \eqref{delta B ideal} of the magnetic field for the purposes of this work. As R. B. White demonstrates in \cite{White2013b}, regardless of which representation of the magnetic field is used, particles' motion will not be significantly affected.

In toroidal plasma geometry, the symmetry with respect to poloidal ($\theta$) and toroidal ($\zeta$) coordinates allows for the Fourier series expansion of the perturbation $\bm{\xi}(\mathbf{r})$ in space as:
\begin{equation} \label{xi_r fourier}
    \bm{\xi}(\mathbf{r}) = \sum_{m,n}{\bm{\xi}(r) e^{j(m\theta + n\zeta)}}
\end{equation}
In this way, the ideal displacement $\bm{\xi}(\mathbf{r},t)$ as well as all perturbation quantities $Q_{1}(\mathbf{r},t)$ are written as waves with $m$ and $n$ being the poloidal and toroidal wave numbers, respectively. Substituting Eq. \eqref{xi_r fourier} into Eq. \eqref{xi_t} and then incorporating the resulting expression into Eqs. \eqref{MHD mass}-\eqref{Ideal Ohm's law} and keeping only the linear perturbation terms, the resulting MHD stability equations are reduced from 3-D to 1-D, involving differentiation only with respect to the radial coordinate $r$. The solution of the 1-D MHD differential equations allows the determination of the eigenvalue $\omega^2$ as a function of the mode numbers $(m,n)$ and the radial coordinate $r$. Consequently, the choice of mode numbers $(m,n)$ can render the system either stable or unstable depending on the sign of $\omega^2$, the continuous dependence of which on $r$ forms the continuous spectrum of the wave \cite{Freidberg2007}.

A characteristic category of MHD waves, which are often driven unstable by energetic particles affecting the performance of fusion devices, comprises the Shear Alfven Waves (SAWs). In the case of the simplest MHD equilibrium consisting of a constant magnetic field $\mathbf{B} = B_{0}\mathbf{e_{z}}$, constant pressure and density, $p = p_{0}$ and $\rho = \rho_{0}$, and zero current and velocity, $\mathbf{J} = 0$ and $\mathbf{v} = 0$, an eigenvalue that arises is given by $\omega^{2} = k_{\parallel}^{2} u_{a}^{2}$, which is the shear Alfven eigenvalue that describes a stable oscillatory wave. Here $k_{\parallel}$ is the wave mode number in the direction of the magnetic field, and $u_{a} = \left(B_{0}^{2}/\mu_{0}\rho_{0}\right)^{1/2}$ is the Alfven velocity. The displacement and the perturbed magnetic field that arise are perpendicular to the original magnetic field \cite{Freidberg2007}. A general formulation for solving the linear MHD eigenmode equations in the case of a general axisymmetric toroidal magnetic field equilibrium is provided by C. Z. Cheng and M. S. Chance (1986) \cite{Cheng1986}. They show that the breakups of the Alfvén continuous spectrum are caused by the coupling of different poloidal modes $m$ due to the non-uniform toroidal magnetic field. Inside the gap of the continuum the toroidal coupling effects can induce discrete ($\omega$ independent on $r$), global (the magnitude of the perpendicular to $\mathbf{B}_{0}$ component of $\bm{\xi}$ is non-localized around a particular $r$ for low-$n$ modes) Toroidal Alfvén Eigenmodes (TAEs). In addition to toroidicity-induced gaps, other types of gaps can also exist in the continuous spectrum of SAWs and discrete Alfvén Eigenmodes (AEs) can be localized in these gaps. SAWs with frequencies in the continuum are damped; however, those with discrete frequencies within the gaps experience weaker damping and are thus more easily excited by particles whose velocities resonate with the wave. Studying this interaction is crucial for fusion plasma research. A comprehensive theoretical overview of Alfvén waves and their resonant interactions with energetic particles is provided by L. Chen and F. Zonca (2007) and the references therein.



\chapter{Research Motivation and Scope}\label{Sec: Chapter 2}

\ifpdf
    \graphicspath{{Chapter2/Figs/Raster/}{Chapter2/Figs/PDF/}{Chapter2/Figs/}}
\else
    \graphicspath{{Chapter2/Figs/Vector/}{Chapter2/Figs/}}
\fi

In Section \ref{MHD Modes}, we demonstrated that magnetohydrodynamic (MHD) instabilities perturb the equilibrium magnetic field, as described by Eq. \eqref{delta B ideal}, leading to modifications of the topology of the magnetic field lines. These topological changes are linked to wave-wave resonant interactions between the perturbations and the equilibrium magnetic field, which can result in the formation of magnetic islands or even chaotic magnetic field lines behavior \cite{Evans2008, Abdullaev2008}.

The interaction of MHD waves with particles involves resonances that facilitate the transport of particles, energy, and momentum. The effects of these resonances are most important for high-energy particles, including fusion-born alpha particles and particles generated by external power sources for plasma control, such as neutral beam injection and radiofrequency (RF) heating. Confining these particles is crucial for magnetic confinement fusion, as they are essential for heating the plasma \cite{Heidbrink1994}. Their interactions with MHD modes, in addition to destabilizing these modes and jeopardizing macroscopic plasma confinement, can result in modifications to the particles' distribution or even significant particle losses.

To suppress MHD instabilities, such as edge localized modes (ELMs) occurring at the edge of high-confinement (H-mode) plasma \cite{Zohm1996} or TAEs, or to mitigate their effects without appreciable reduction in
performance of fusion device, external static magnetic perturbations with zero frequency ($\omega = 0$) are applied to the plasma through external coils to control the instabilities \cite{Munoz2019}. These perturbations, known as Resonant Magnetic Perturbations (RMPs), also interact with the plasma's energetic particles via resonances, leading to particle losses despite their benefits in controlling MHD instabilities.

While extensive research has been conducted on the field of resonant mode-particle interactions, this chapter reviews a selection of the most relevant works, focusing on those that are crucial to the objectives of this thesis. The review is organized into two main sections: experimental observations and theoretical studies, with the works in each subsection presented chronologically.

\section{Experimental Studies} \label{Experimental Studies}
Experimental research has been pivotal in revealing the complexities and consequences of resonant mode-particle interactions on magnetic confinement fusion.
Heidbrink et al. (1991) \cite{Heidbrink1991} simulated fusion born alpha particles with neutral beam injection to create an energetic particle population in DIII-D experiments. They observed TAE modes near the frequency gap, which interacted with the energetic particles, leading to instabilities that caused large fast ion losses, potentially preventing a reactor from igniting. In further DIII-D experiments, Duong, Heidbrink, et al. (1993) \cite{Duong1993} observed that in the presence of violent TAE activity, losses can be as large as 70$\%$ of the injected beam power. Their experimental data suggest resonant transport as the main loss mechanism, with resonance overlap leading to stochastic particle motion as a secondary possible mechanism. 

In the field of energetic particle driven MHD waves, Fasoli et al. (1997) \cite{Fasoli1997} reviewed the experimental results obtained using an active diagnostic system developed at JET \cite{Fasoli1995} to investigate different classes of Alfvén eigenmodes (AEs) and their interactions with fast particles, comparing these results with theoretical model predictions for mode stability. While, Fu et al. (1998) \cite{Fu1998} provided a detailed report on observations of alpha particle-driven TAEs in Tokamak Fusion Test Reactor deuterium–tritium (TFTR) experiments, comparing them with theoretical predictions.

Fiksel, Hudson et al. (2005) \cite{Fiksel2005} observed that fast ions were well confined in the Madison Symmetric Torus reversed field pinch experiments, despite the presence of a stochastic magnetic field due to resonant MHD modes. They explained these observations using numerical simulations that tracked ions' guiding center orbits in the magnetic field. By comparing the radial location and widths of magnetic field and mode-particle resonances through the magnetic safety factor and the "ion guiding center (IGC) safety factor" they concluded that at certain radii, mode-particle resonances do not overlap, allowing ions to remain confined, while magnetic resonances overlap, leading to a stochastic magnetic field.

In experiments with RMPs, Evans et al. (2006) \cite{Evans2006} reported low-collisionality, ITER-relevant experiments on the DIII-D tokamak, where edge localized modes (ELMs) were controlled using RMPs with $n=3$ at the plasma edge. They demonstrated that RMPs completely eliminated ELMs while maintaining steady-state, high-confinement (H-mode) plasma. The elimination of ELMs was attributed to the chaotic behavior of magnetic field lines, which occurs when the RMPs resonate with the equilibrium magnetic field. While, Liang et al. (2007) \cite{Liang2007} employed an external coil system to generate $n=1$ RMPs for the active control of ELMs in experiments on JET. They observed that both the frequency and amplitude of type-I ELMs could be actively controlled with an acceptable reduction in plasma confinement. Additionally, they noted a decrease in energetic particles and energy losses during the ELMs mitigation phase. These results are linked to the stochasticity of the magnetic field lines near the plasma edge, induced by the RMPs.

In further experiments with resonant magnetic perturbations, Suttrop et al. (2011) \cite{Suttrop2011}, in the ASDEX Upgrade tokamak with toroidal mode number $n=2$ RMPs, observed a significant reduction in plasma energy loss associated with type-I edge localized modes (ELMs) in high-confinement mode plasmas, with no performance penalty for ELMs mitigation. They also observed that magnetic field resonances occurring over a wide range of magnetic safety factor values near the plasma edge led to ELMs mitigation. Later, Bortolon, Heidbrink, et al. (2013) \cite{Bortolon2013} presented observations of the mitigation of fast-ion driven Alfvén modes following the application of RMPs in experiments at the NSTX tokamak. They demonstrated that the RMPs modify the nonlinear evolution of fast-particle driven instability. Based on the experimental parameters, they simulated the full orbits of test particles, showing that RMPs alter the particle distribution function in velocity space, leading to the loss of some fast ions that drive the instability, thereby reducing the fast-ion drive for the instability. They suggest that the ability to modify the fast-ion distribution function in velocity space through the application of properly tailored static fields or propagating waves could be seen as a method of 'phase-space engineering' to control fast-ion instabilities.

Collins, Heidbrink, et al. (2016) \cite{Collins2016} presented measurements in equilibria with a minimum in the magnetic safety factor profile, showing that beyond a critical threshold— when the region of particles' phase space measured by the diagnostic becomes stochastic due to many overlapping small amplitude TAE wave-particle resonances—the equilibrium fast-ion density profile remains unchanged despite intense fast-ion transport. They propose that these measurements can be used for the qualitative validation of AE transport models that are not fully self-consistent, as they do not account for the evolution of AE structures, amplitudes, and frequencies with the fast-ion distribution function and equilibrium plasma profiles, as including such effects would be computationally demanding. Thus, these models could be adapted for application in ITER, where the alpha particles population will be characterized by an isotropic distribution. More recently, Kim et al. (2022) \cite{Kim2022} in experiments with advanced operation scenarios on the KSTAR tokamak, showed that the electron cyclotron current drive (ECCD) AE control tool is able to suppress beam-ion driven TAEs. They also observed that in the non-suppression TAE phase, enhanced fast-ion losses originate from the interaction between several TAEs and the beam ions. Notably, only specific TAEs had a significant impact on fast-ion transport. Consequently, the prior detection of these specific TAEs could enhance the efficiency of the control tool by allowing it to target these particular modes. 

\section{Theoretical Studies} \label{Theoretical Studies}
The experiments were both guided by and provided a foundation for the theoretical studies on resonant wave-particle interactions in fusion plasma, including the development of theoretical models and simulation codes. In this review, we primarily focus on studies examining the dynamics of single particle guiding center during its interaction with waves, as it pertains to the objectives of this thesis. Unlike thermal particles, whose motion is determined by collisions—allowing their distribution function to be described by a Maxwellian and their motion by fluid dynamics—the collisions in energetic particle (EP) populations, characterized by low densities and high velocities, are rare. As a result, EP distribution functions have complex dependencies on particle energy and velocity. Because different EP velocities behave quite differently, a single-particle picture is the appropriate starting point for EP transport theory \cite{Heidbrink2020}. Kaufman (1972) \cite{Kaufman1972} utilizes the Lagrangian that describes single particle guiding center motion in electromagnetic fields, proposed by Taylor (1964) \cite{Taylor1964}, to develop a quasilinear diffusion theory for an axisymmetric toroidal system. From the guiding center Lagrangian, he derives the canonical guiding center Hamiltonian and three invariants of single particle motion in axisymmetric field configuration: the magnetic moment $\mu$, the canonical angular momentum $P_{\zeta}$ and the energy $E$ of the system. He then transforms the system into Action-Angle variables. To study particle diffusion due to their interaction with waves, he expresses the evolution of the particle distribution function in action-angle variables and derives the diffusion tensor in terms of actions, showing that it is non-zero when the normal perturbative mode-single particle resonance condition is fulfilled in action-angle space. Thus, under static perturbations where particle energy is conserved, and assuming $\mu$ is conserved even in the presence of perturbations, net particle diffusion occurs when resonance widths overlap on constant energy surfaces. While, in the case of static perturbations with non-conserved $\mu$, Arnold diffusion becomes possible.

Littlejohn (1983) \cite{Littlejohn1983}, based on non-canonical Hamiltonian formulation, provided a rigorous derivation of the phase space Lagrangian that describes single particle guiding center drift motion in electromagnetic fields, using a variational principle and thus paved the way for its canonical Hamiltonian description. He also connected this motion with the conserved quantities, which are crucial for particle confinement in magnetic fusion \cite{Kaufman1972}. In an applied context, White, Goldston, et al. (1983) \cite{White1983} used the canonical  Hamiltonian formalism proposed by White, Boozer, et al. (1982) \cite{White1982} for describing the guiding center drift motion of particles in electromagnetic fields using magnetic field coordinates, along with Monte Carlo simulations, to theoretically and numerically explain the radial motion and losses of injected beam particles due to their resonant interaction with a single MHD mode, as well as the mode-particle energy transfer observed in PDX tokamak experiments. In a related study, White and Chance (1984) \cite{White1984} performed a canonical transformation from the set of canonical variables in the aforementioned Hamiltonian description \cite{White1982} to a new set based on the Boozer coordinates of the magnetic field, defined by Boozer (1981) \cite{Boozer1981} to exploit the toroidal and poloidal periodicity of the torus. Using these new coordinates, they developed the numerical code ORBIT to simulate particle orbits in realistic plasma equilibria, including MHD mode perturbations. This Hamiltonian formalism, which enables the study of particle motion in arbitrarily shaped plasmas with approximate toroidal or helical symmetry, has been foundational for subsequent studies of mode-particle interactions. For theoretical completeness, Littlejohn (1985), using differential forms, rigorously extracted from the phase-space Lagrangian the variables for guiding center drift motion that were used by White and Boozer (1982) \cite{White1982} and White and Chance (1984) \cite{White1984}, and proved that these variables are canonical.

Subsequently, Hsu and Sigmar (1992) \cite{Hsu1992}, using the invariants of particle guiding center drift motion, $(E,\mu,P_{\zeta})$, as derived from the canonical Hamiltonian formalism \cite{White1984}, analytically defined the particle loss boundaries and the trapped-passing boundary in the phase space of energetic particles in a general axisymmetric tokamak equilibrium. They employed this phase-space topology, along with numerical simulations, to study the response of alpha particles to TAEs, demonstrating that strong chaos appears in the phase-space region near the trapped-passing boundary layer, resulting in the transport of energetic particles to the loss boundaries. In a next work, Sigmar, Hsu, et al. (1992) \cite{Sigmar1992} numerically simulated an $n=1$ TAE mode with different poloidal mode numbers and using the Hamiltonian guiding center code ORBIT \cite{White1984}, showed that the resonance of a single alpha particle with the mode leads to resonant energy exchange and radial drift, and that alpha particles can undergo stochastic motion within a region of phase space while remaining confined.

At this point, it is important to note that the canonical Hamiltonian formalism, derived from the particle guiding center Lagrangian \cite{Littlejohn1983}, addresses one aspect of the wave-particle interaction problem: the influence of the equilibrium field and MHD waves on the motion of energetic particles—and the evolution of their distribution function when considering their collective behavior, as discussed in Ref. \cite{Kaufman1972}. The objective of this thesis is related to this first aspect, namely the quasi-linear wave-particle interactions. The other aspect concerns the effect of particle motion on the MHD waves. Berk, Breizman, et al. (1995) \cite{Berk1995} and Pinches, Appel, et al. (1998) \cite{Pinches1998} described this using the wave Lagrangian, which incorporates both the background plasma contribution to the MHD wave Lagrangian and the electromagnetic wave Lagrangian. The system Lagrangian, defined as the sum of the particle and wave Lagrangians, models the self-consistent nonlinear interaction of energetic particles with MHD waves. To account for the collective and self-consistent behavior of the entire system (background plasma, ensemble of energetic particles, equilibrium, and MHD waves), the evolution of the energetic particle distribution function must also be evaluated \cite{Kaufman1972, Pinches1998}\footnote{There are two possible scenarios for the modification of MHD waves due to the changes in the energetic particle distribution resulting from their quasi-linear interaction with waves. The first scenario is that the mode frequency is determined by the equilibrium, while the growth and mode phase are modified by the particle distribution. This is the case for TAE modes and is treated by White (Ref. \cite{White2013}, Sec. 6.9.1), using an energy conservation procedure. In the same scenario, the method involving the system Lagrangian proposed by Pinches et al. (1998) \cite{Pinches1998} resolves the problem through a set of differential equations. Kaufman (1972) \cite{Kaufman1972} addresses the quasi-linear diffusion problem in a self-consistent manner, accounting for the time-dependent modification of normal modes amplitudes due to particle diffusion resulting from their interaction with the waves. The second, more complicated scenario is that the mode frequency is determined by the particle distribution, requiring the solution of a dispersion relation involving the particle distribution (Ref. \cite{White2013} Sec. 6.9.2).}. Berk, Breizman, et al. (1995) \cite{Berk1995} expressed the system Lagrangian mentioned above in terms of the action-angle variables derived from the particle guiding center Hamiltonian in equilibrium magnetic field, connected with the particle Lagrangian  \cite{Kaufman1972, White1984}. In this formulation, the wave-particle interaction Lagrangian is written as a Fourier expansion in terms of angles. Based on the system Lagrangian, they develop a numerical algorithm and use it to demonstrate that in TAE-alpha particle interactions, both global quasilinear diffusion and an enhancement in wave energy occur due to resonance overlap. Finally, they highlight the importance of understanding the structure of resonances—specifically, the location of resonance lines in action-angle space, their width—and the transport between resonances to infer the global picture of the nonlinear evolution of wave-particle interactions.   

In more recent theoretical works closely related to this thesis, Gobbin, White et al. (2008) \cite{Gobbin2008} introduced the concept of the "kinetic" safety factor, defined as the mean helicity of a passing particle in the torus, analogous to the magnetic safety factor $q$. They calculated this factor in terms of radial position by integrating the guiding center equations of motion as defined in White and Chance (1984) \cite{White1984}. For the case of zero magnetic moment, using the ORBIT code \cite{White1984} they numerically demonstrated that in reversed field pinch (RFP) configurations, unlike in tokamak configurations where the "kinetic" safety factor closely follows the magnetic safety factor, the two factors deviate, resulting in different radial locations for magnetic resonances and wave-particle resonances. They proposed that this deviation accounts for the improved confinement of fast ions in RFP devices, supporting their conclusion with numerical simulations. White (2012) \cite{White2012} developed a numerical method for determining the location and range of mode-particle resonances by detecting broken Kolmogorov-Arnold-Moser (KAM) \cite{Arnold1963} surfaces in the phase space of the particles' constants of motion. The method was subsequently illustrated in determining resonant domains for equilibria and toroidal Alfvén modes observed in DIII-D tokamak experiments.

Kramer, Chen, et al. (2012) \cite{Kramer2012} presented a theoretical study, supported by numerical simulations using the SPIRAL following-orbit code—which calculates single particle orbits in plasma in the presence of MHD activity—demonstrating that MHD mode-particle resonances can occur even at fractional values of the particle drift-orbit transit frequency. This results in a higher density of resonances in phase space, thereby enhancing stochastic fast ion transport and losses.

Caldas et al. (2012) \cite{Caldas2012} summarize both theoretical and experimental work related to the formation of shearless transport barriers in toroidal plasmas, which arise from non-monotonic $q$ profiles, providing a wide range of references on this topic. They highlight the significance of the Hamiltonian description for both the magnetic field lines and the guiding center motion in the study of transport barriers. They discuss the presence of transport barriers in magnetic field lines for the TCABR tokamak equilibrium, which exhibits a non-monotonic 
$q$-profile. Moreover, they reported that the shearless transport barriers, induced by a non-monotonic electric field profile in the TCABR tokamak experiment, effectively reduced radial particle flux driven by electrostatic turbulent fluctuations. Similar reductions in particle transport were also observed in the Texas Helimak tokamak experiment.

White, Gorelenkov, et al. (2018) \cite{White2018} studied the resonance of a high-energy particle with a single ideal MHD mode using kinetic Poincaré plots and a local construction of canonical variables within the resonance, which enabled them to calculate the resonance width, the particle's rotation frequency within a resonance island, and the locations of the island's elliptic and hyperbolic points. They applied this method to investigate how the resonance dynamics depend on the mode structure. As noted, their method requires the prior determination of the resonance location, either through Poincaré plots or the numerical method proposed by White (2012) \cite{White2012}.

Shinohara, Bierwage, et al. (2020) \cite{Shinohara2020}, using a Hamiltonian formalism, developed a more computationally efficient method to estimate the location and width of resonant island structures driven by a single RMP, for both magnetic field lines and passing guiding center drift orbits, compared to their previous model (Shinohara et al. 2018 \cite{Shinohara2018}). The earlier model required constructing a Poincaré map for the magnetic field lines, which, as a perturbed trajectory-following procedure, is computationally expensive, followed by a mapping from magnetic field lines to guiding center space. In contrast, the new method eliminates the need for Poincaré plots, either for magnetic field lines or particle orbits. This method was applied to investigate the location and width of magnetic and orbit islands produced by RMPs in a realistically shaped KSTAR tokamak plasma, yielding very accurate predictions in both the vacuum field case—where the plasma's response to RMPs is assumed to leave the RMPs unchanged—and the case where the self-consistent plasma response \cite{Shinohara2016}, which modifies the RMPs, is taken into account. They also showed that the width of orbit islands can predict nearby resonance overlap, leading to chaotic behavior according to the Chirikov criterion. Thus, this width can be seen as the characteristic spatial scale for RMP-induced particle transport.

Spizzo, Gobbin, et al. (2021) \cite{Spizzo2021} used the Hamiltonian particle guiding center following code ORBIT \cite{White1984} to simulate the motion of fast ions in DTT equilibria, which present significant magnetic field perturbations due to the 18 toroidal field (TF) coils the DTT device is equipped with. These kinds of perturbations are known as magnetic field ripples. They simulated different scenarios of fast ions generated by negative-ion-based neutral beam heating (NNBI) and showed that resonant interactions of the fast ions with the magnetic field ripples are the physical mechanism that leads to significant particle losses. Under particular NNBI geometry and energy, collisionless (resonant) losses due to ripples are within an acceptable range. It is highlighted that even if the NNBI beam is engineered in such a way that NNBI ions avoid the resonance regimes in DTT, resonances are still present and can potentially interact with other fast particles, leaving for future work a thorough study of resonances in DTT.

\section{Research Scope} \label{Research Objectives}

Both experimental and theoretical works, as reviewed in Secs. \ref{Experimental Studies} and \ref{Theoretical Studies}, along with many other relevant studies that were not discussed in detail, indicate that intrinsically excited MHD modes and externally applied RMPs significantly affect not only the topology of the equilibrium magnetic field lines but also single particle orbits and the collective behavior of the plasma. Through resonant wave-particle interactions, these perturbations drive particle, energy, and momentum transport, which in turn impacts the confinement properties and overall performance of toroidal fusion devices. Several theoretical studies of resonant interactions initially focused on developing a canonical Hamiltonian formalism \cite{Kaufman1972, Littlejohn1983, White1983, White1984, Littlejohn1985} to describe both magnetic field lines and the particle guiding center motion in electromagnetic fields. This formalism has since been extensively employed in the development of theoretical models suitable for numerical simulations to study the behavior of magnetic field lines and guiding center orbits \cite{White1984, Hsu1992, Sigmar1992, Pinches1998, Evans2008, White2012, White2018, Shinohara2018}.

The first crucial feature in the study of resonant interactions is identifying the location in phase space where these interactions occur. This location corresponds to the point where the resonance condition for wave-particle interactions is satisfied. In wave-particle interactions, the resonance condition is dictated by the particle's Orbital Frequency Spectrum (OFS), specifically the bounce/transit poloidal frequency $\omega_{\theta}$ and the averaged toroidal precession frequency $\omega_{\zeta}$, and is expressed as $\omega + m^{\prime}\omega_{\theta} - n\omega_{\zeta} = 0$, where $\omega$ is the mode frequency, $n$ is the toroidal mode number, and $m^{\prime}$ is an integer \cite{White1984, Heidbrink2008, White2023}. Two other key features, closely related to the first, are the width of the resonances and the relationship between the wave mode numbers $(m, n)$ and the integer $m^{\prime}$ in the resonance condition. These three factors provide essential insights into resonances overlap, which, according to the Chirikov criterion, can lead to stochasticity and transport \cite{Hsu1992, Berk1995, Pinches1998, Gobbin2008, White2012, Kramer2012, Caldas2012, White2018, Shinohara2018}, or particle diffusion under multiple wave modes \cite{Kaufman1972, White2012}.

A detailed study of these aspects demands significant computational resources for orbit-following simulation codes, in order to track a large number of particles for extended time intervals, under the presence of each specific set of perturbing modes. For instance, orbits are  followed for the construction of detailed Poincare maps which provide a picture of the resonances islands of the magnetic field lines or in particle's phase space \cite{Hsu1992, Berk1995, Gobbin2008, White2012, Caldas2012, White2018, Shinohara2018, White2023}. These requirements hinder the use of such codes in scenario simulations that necessitate extensive parameter scans, including various sets of modes. Additionally, without a prior knowledge of the phase space regions that are actually affected by each perturbation set, the initial conditions for the simulated particles have to be chosen randomly, leading to a large number of initial conditions corresponding to particle orbits that are essentially unaffected by the specific perturbations, especially in the quite common case where the effects of the perturbations are strongly localized in the phase space \cite{White2012, White2018}.

Gobbin, White, et al. (2008) \cite{Gobbin2008}, White (2012) \cite{White2012} and Shinohara et al. (2018) \cite{Shinohara2018} define the mean helicity of passing particle trajectories as $\omega_{\zeta}/\omega_{\theta}$, as a function of invariants of particle motion, by tracing the unperturbed guiding center orbits and averaging over the poloidal angle. The resonance locations are then determined by the phase space positions where the particle helicity becomes a rational number. This method is more computationally efficient than creating a Poincaré plot, as it requires following unperturbed orbits for only a single poloidal transit. However, it still involves numerical integration of the guiding center equations of motion. Additionally, to compute the resonance width, perturbed orbits must still be followed, even without generating Poincaré maps, as demonstrated by White (2012) \cite{White2012} with the phase vector rotation method.

Shinohara et al. (2020) \cite{Shinohara2020} proposed a more computationally efficient method to calculate the helicity of passing particles under specific approximations, using a line integral on the poloidal plane without the need for unperturbed orbit tracing. Furthermore, to compute the poloidal bounce frequency—necessary for studying particle resonances with time-dependent perturbations along with helicity—one still needs to numerically integrate the equations of motion for a single transit. The treatment of particle helicity in these works does not fully explain the resonant behavior of trapped particles, and in Ref. \cite{Shinohara2020}, this is left for future work.

In this thesis, we present a low-computational-cost method that provides a priori knowledge of the exact locations in phase space where resonant island chains form, the width of these resonant islands, and the number of islands in each chain, which is related to the mode numbers $(m,n)$ of the perturbation. Additionally, we identify the existence and locations of transport barriers, which affect the confinement properties of a toroidal fusion device \cite{Caldas2012}. Rather than following guiding center orbits by integrating their Hamiltonian equations of motion, we exploit the canonical Hamiltonian description of guiding center motion to transfer the study of the system into the Action-Angle (AA) space, where the orbital spectrum is fully unfolded. This transformation provides a clearer picture of the wave-particle resonance structure in particle phase space \cite{Kaufman1972, Lieberman1992, Berk1995, Goldstein2002}, making it suitable for studying both trapped and passing particles

More specifically, our calculations focus on the orbital frequency spectrum of the unperturbed guiding center (GC) motion in an axisymmetric equilibrium. This includes the bounce/transit poloidal frequency and the averaged toroidal precession frequency, the ratio of which defines the mean helicity of GC trajectories. By analogy with the magnetic $q$ (safety) factor, which defines the helicity of the magnetic field lines in a torus, we define the kinetic $q$ factor for the particles as $q_{\text{kin}} = \omega_{\zeta}/\omega_{\theta}$, and we examine its relation with particle helicity in the torus. The kinetic $q$ factor essentially differs from the magnetic $q$ (safety) factor for particles with non-negligible drifts, such as energetic ions resulting in significant differences between the magnetic and the kinetic phase space. Orbital Frequencies (OF) and $q_{kin}$ are analytically determined for a given axisymmetric equilibrium as functions of the three Constants Of Motion (COM) of the GC motion—namely, the energy, magnetic moment, and canonical toroidal momentum, which are related to the actions of the GC Hamiltonian system \cite{Kaufman1972, White2013}. Rational values of $q_{kin}$ correspond to resonance conditions \cite{Gobbin2008, White2012, Zestanakis2016, White2023}, while its local extrema indicate conditions for the existence of transport barriers \cite{Caldas2012}. Both types of conditions can be mapped in the COM particle space, along with curves characterizing particles as trapped or passing, and confined or lost \cite{Hsu1992}, providing a unique overview of resonant mode-particle interaction. The decomposition of a single perturbation as a multiply periodic function of angles allows the determination of the full range of resonances excited in particle phase space \cite{Kramer2012}, including the number of islands in each island chain and the width of each resonance. 

In the case when multiple time-varying perturbations are present, the elimination of time through the linear combination $(n\zeta-\omega t)$ by transforming to the wave frame is not applicable \cite{Lieberman1992, White2013}. As a result, an additional degree of freedom is introduced to the GC Hamiltonian system due to its time dependence, leading to significant excursions along the resonance curves. In this case, KAM surfaces do not isolate different resonances, and extended transport can take place even without resonance overlap in the phase space \cite{Lieberman1992}. Moreover, in contrast to the case of stationary perturbations resonance curves intersect in the COM space forming the so-called Arnold web \cite{Wobig2001}. In this work we analytically calculate the Arnold web in the COM space of the GC motion, along which diffusive particle energy and momentum transport takes place. This calculation allows for the a-priori knowledge of exact kinetic characteristics of particles that strongly interact with the time-dependent non-axisymmetric modes. 

Compact analytical formulas in terms of Jacobi elliptic functions and complete elliptic integrals for the passing and  bounce/transit poloidal action, angle, and frequency in the case of a Large Aspect Ratio (LAR) equilibrium have been provided by Brizard (2011) \cite{Brizard2011}, under the zero-orbit-width limit where the GC Hamiltonian approximates the pendulum Hamiltonian \cite{Brizard2013}. In this thesis, the calculation of the actions, orbital frequencies, and kinetic $q$ factor is illustrated for the case of a Large Aspect Ratio (LAR) equilibrium magnetic field \cite{White2013}, where these calculations are performed analytically, not by following a pendulum-like approximation, but through an approach even closer to the guiding center (GC) Hamiltonian. The analytical results are systematically compared and shown in a remarkable agreement with numerical findings, and their domain of validity is investigated and explained in connection to the size of the drift motion and the magnetic safety factor. Both the locations and the number of islands, as well as the locations of transport barriers in the phase space, are accurately predicted by the analytical findings, as further confirmed by numerically obtained Poincaré surfaces of section. Moreover, the presence of multiple island chains, as well as the non-trivial dependence of the number of islands on the toroidal and poloidal numbers of the perturbing modes, are elucidated and predicted based on the unperturbed motion and the action-angle formulation. It is worth emphasizing that our calculations for more general axisymmetric equilibria can be similarly performed in a semi-analytical computationally efficient fashion, enabling the utilization of this valuable tool in cases of realistic experimental equilibria.


\chapter{Resonances in Action-Angle Space}\label{Sec: Chapter 3}

In this chapter, we present a concise yet comprehensive introduction to the fundamental principles of Hamiltonian dynamics, emphasizing the Action-Angle formulation and its role in analyzing resonances in Hamiltonian systems. Our approach follows the well-established methodologies outlined in the textbooks by Lichtenberg and Lieberman (1992) \cite{Lieberman1992} and Goldstein (2002) \cite{Goldstein2002}, ensuring a rigorous yet accessible treatment of the subject.

We begin by exploring canonical transformations, which provide a powerful mathematical framework for reformulating Hamiltonian systems while preserving their essential structure. This leads to the Hamilton-Jacobi theory, which focuses on transforming a Hamiltonian system into a form that reveals conserved quantities and facilitates the derivation of Action-Angle variables. Next, we introduce integrable and separable Hamiltonian systems, emphasizing their significance in constructing action-angle variables and understanding regular motion in phase space. The derivation and properties of Action-Angle variables are then discussed, highlighting their utility in simplifying the analysis of dynamical systems, particularly in the presence of weak perturbations.

The second part of the chapter is dedicated to canonical perturbation theories, which extend the action-angle formalism to more general, non-integrable systems. We first present classical perturbation theory and canonical adiabatic theory, laying the groundwork for understanding how small perturbations modify the motion of Hamiltonian systems. We then examine secular perturbation theory, which describes the formation of resonance island structures in action-angle space. The concept of isolated resonances is examined in detail, providing insights into their location, width, and the number of islands in each resonance island chain. Finally, when resonances are non-isolated, we introduce Arnold diffusion, a mechanism by which weak perturbations can induce slow, chaotic transport in nearly integrable Hamiltonian systems.

This chapter serves as a theoretical foundation for the subsequent analysis of mode-particle resonances, offering the essential mathematical tools to study the nature of resonant interactions in Hamiltonian dynamics.

\ifpdf
    \graphicspath{{Chapter3/Figs/Raster/}{Chapter3/Figs/PDF/}{Chapter3/Figs/}}
\else
    \graphicspath{{Chapter3/Figs/Vector/}{Chapter3/Figs/}}
\fi

\section{Canonical Transformations}
The Hamiltonian of a system, the dynamics of which are described by a Lagrangian $L = T-V$, where $T$ is its kinetic energy and $V$ its potential energy, in the $N$-dimensional space of the generalized coordinates $\mathbf{q}$, is given by the Legendre transformation as 
\begin{equation} \label{Lagrangian to Hamiltonian gen}
    H(\mathbf{q},\mathbf{p},t) = \mathbf{p}\mathbf{\dot{q}} - L(\mathbf{q},\mathbf{\dot{q}},t)
\end{equation}
where the product $\mathbf{p}\mathbf{\dot{q}}$ denotes summation, i.e, $\mathbf{p}\mathbf{\dot{q}} = p_1 \dot{q}_1 + p_2 \dot{q}_2 + ... + p_N \dot{q}_N$ where 
\begin{equation} \label{conjucate momenta gen}
    p_j = \frac{\partial L(\mathbf{q}, \mathbf{\dot{q}},t)}{\partial\dot{q}_j},\quad j = 1,2,...,N
\end{equation}
Then, the canonical equations of Hamilton are:
\begin{align}
    \dot{q}_j  &= \frac{\partial H(\mathbf{q},\mathbf{p},t)}{\partial p_j}\label{dot q gen} 
    \\ 
    \dot{p}_j &= -\frac{\partial H(\mathbf{q},\mathbf{p},t)}{\partial q_j} \label{dot p gen}
\end{align}
with $j = 1,2,...,N$. Details for the derivation of the expression \eqref{Lagrangian to Hamiltonian gen} through Legendre transformations can be found in Sec. 8.1 of Goldstein (2002) \cite{Goldstein2002}, as well as the rigorous derivation of Hamilton's equations of motion from a variational principle in Sec. 8.5 of the same textbook. Eqs. \eqref{dot q gen} and \eqref{dot p gen} represent a system of $2N$ first-order ordinary differential equations. With $2N$ initial conditions (i.e., $\mathbf{q}^0 = \mathbf{q}(t=0)$ and $\mathbf{p}^0 = \mathbf{p}(t=0)$), these equations can be solved to obtain $\mathbf{q}(t)$ and $\mathbf{p}(t)$ for each specific set of initial conditions. The solution $\left(\mathbf{q
}(\mathbf{q}^0, \mathbf{p}^0,t), \mathbf{q
}(\mathbf{q}^0, \mathbf{p}^0,t)\right)$ is call orbit of the Hamiltonian system

Unlike in the Lagrangian formulation, where the evolution of a dynamical system is described in the 
$N$-dimensional configuration space of the generalized coordinates $\mathbf{q}$ through them and their time derivatives $\dot{\mathbf{q}} = d\mathbf{q}/dt$, in the Hamiltonian formulation, $N$ denotes the degrees of freedom of the Hamiltonian and the system's evolution is given in the $2N$-dimensional phase space of $(\mathbf{q}, \mathbf{p})$, where the canonical coordinates $\mathbf{q}$ and canonical momenta $\mathbf{p}$ are independent and treated with equal status. This encourages selecting new quantities as canonical coordinates and momenta that describe the system in a more abstract way, which can be suitable to reveal properties of the system, such as symmetries and constants of motion, even without the need to solve its equations of motion. The transition from one set of canonical variables to a new set, which retains the canonical form of the equations of motion, Eqs. \eqref{dot q gen}, \eqref{dot p gen}, is called a canonical transformation, and the new variables are canonical too. In this work, we use only canonical descriptions, so in some cases, the characterization "canonical" will be omitted since it is implied. \footnote{Arbitrary transformations from a set of canonical variables $(\mathbf{q}, \mathbf{p})$ that do not necessarily retain the form \eqref{dot q gen}-\eqref{dot p gen} lead to non-canonical Hamiltonian descriptions. Such a transformation was considered by Littlejohn (1983) \cite{Littlejohn1983} to rigorously derive the phase space GC Lagrangian. There he also proved that even in the non-canonical Hamiltonian descriptions, Liouville's theorem, which characterizes the canonical formalism, still holds. However, the phase space GC Lagrangian expressed in particular coordinates leads to canonical Hamiltonian descriptions, as discussed in Sec. \ref{Theoretical Studies}. Details on non-canonical Hamiltonian formalism with applications in plasma physics can also be found in the work by Cary and Brizard (2009) \cite{Cary2009}.}

Following the notation of Goldstein (2002) \cite{Goldstein2002}, a point transformation of phase space from the coordinates and momenta $q_j$, $p_j$ to a new set of variables $Q_j$, $P_j$ is expressed as:
\begin{align} \label{t1}
 Q_j  &= Q_j(\mathbf{q},\mathbf{p},t)\\
 P_j  &= P_j(\mathbf{q},\mathbf{p},t)
\end{align}
The new variables $\mathbf{Q}$ and $\mathbf{P}$ are canonical if there exists a function $K(\mathbf{Q}, \mathbf{P})$ that allows the time variation of $\mathbf{Q}$ and $\mathbf{P}$ to be expressed in Hamiltonian form, i.e.,
\begin{align} 
\dot{Q}_j  &= \frac{\partial K(\mathbf{Q},\mathbf{P},t)}{\partial P_j}\\
\dot{P}_j &= -\frac{\partial K(\mathbf{Q},\mathbf{P},t)}{\partial Q_j}
\end{align}
The function $K$ serves as the Hamiltonian in the new coordinates.

If $\mathbf{Q}$ and $\mathbf{P}$ are canonical variables, they satisfy the modified Hamilton’s variational principle, just like the original canonical variables, i.e.,
\begin{align}
    \delta\int_{t_1}^{t_2}{\mathbf{P}\mathbf{\dot{Q}} - K(\mathbf{P},\mathbf{Q},t)} & = 0 \label{vQP}
    \\
    \delta\int_{t_1}^{t_2}{\mathbf{p}\mathbf{\dot{q}} - H(\mathbf{p},\mathbf{q},t)} &= 0 \label{vqp}
\end{align}
To derive Hamilton's equations (Eqs. \eqref{dot q gen}, \eqref{dot p gen}) from the modified Hamilton's principle (or respectively the Euler-Lagrange equations from Hamilton's principle), it is assumed that the variation of $\mathbf{q}$ at the endpoints, i.e., at times $t_1$ and $t_2$, is zero: $\delta\mathbf{q}(t_1) = \delta\mathbf{q}(t_2) = 0$. For canonical transformations to be applicable, it is also assumed that the variation of the canonical momenta is zero at the endpoints, i.e., $\delta\mathbf{p}(t_1) = \delta\mathbf{p}(t_2) = 0$, although this is not required for the derivation of Hamilton's equations via the calculus of variations. The same conditions at the endpoints apply to the new canonical coordinates and momenta, assuming they are canonical. 

This allows the addition of a total time derivative, $dF(\mathbf{q}, \mathbf{Q}, \mathbf{p}, \mathbf{P}, t)/dt$, to the integrands of either Eq. \eqref{vQP} or Eq. \eqref{vqp}, and even after this, the integrals will remain unchanged because
\begin{align*}
    \delta\int_{t_1}^{t_2}{\frac{dF(\mathbf{q},\mathbf{p},t)}{dt}} &= \delta F\left(\mathbf{q}(t_2),\mathbf{Q}(t_2), \mathbf{p}(t_2), \mathbf{P}(t_2), t_2\right) -  \delta F(\mathbf{q}(t_1), \mathbf{Q}(t_1), \mathbf{p}(t_2),\mathbf{P}(t_1),t_1) \\
    &= 0
\end{align*}
Thus, both expressions (Eqs. \eqref{vQP}, \eqref{vqp}) will be satisfied if their integrands are related by the following equation: 
\begin{equation}
    \lambda(\mathbf{p}\mathbf{\dot{q}} - H(\mathbf{q},\mathbf{p},t)) = \mathbf{P}\mathbf{\dot{Q}} - K(\mathbf{P},\mathbf{Q},t) 
\end{equation}
where $\lambda$ denotes a scale transformation, which can be easily shown to be canonical through a simple application to Hamilton's equations of motion. Therefore, we can set $\lambda=1$, and as a result, the canonical transformation is determined by the following relation:
\begin{equation} \label{canonical raltaion}
 \mathbf{p}\mathbf{\dot{q}} - H(\mathbf{p},\mathbf{q},t) = \mathbf{P}\mathbf{\dot{Q}} - K(\mathbf{P},\mathbf{Q},t) + \frac{dF}{dt}
\end{equation}
Here, $F$ can be a function of the new, old, or a combination of both canonical variables and time. To specify the exact form of the canonical transformation, i.e., Eqs. \eqref{t1} or their inverse, it is useful to express $F$ partially in terms of the old and new variables. In this way, $F$ serves as a bridge between the old and new variables and is called mixed variable generating function of the transformation.

For example, assuming $F = F_2(\mathbf{q}, \mathbf{P}, t) - \mathbf{Q}\mathbf{P}$ and substituting its total time derivative into Eq. \eqref{canonical raltaion}, we obtain
\begin{equation} \label{F2 can. transf.}
    p_j = \frac{\partial F_2}{\partial q_j}, \quad  Q_j = \frac{\partial F_2}{\partial P_j}
\end{equation}
with $j=1,2,...,N$ and 
\begin{equation}\label{K can. tranf. F2}
    K = H + \frac{\partial F_2}{\partial t}
\end{equation}
In the following, we utilize this specific form of $F_2$, which depends on the old coordinates and new momenta, to generate the action-angle transformation.

\section{Hamilton-Jacobi Theory}\label{Hamilton-Jacobi Theory}
Following Goldstein (2002) \cite{Goldstein2002}, let us assume that we want to find a new Hamiltonian $K(\mathbf{Q}, \mathbf{P}, t)$ that is identically zero, i.e., $K \equiv 0$, so that the solution of Hamilton's equations in the new canonical variables is trivial, yielding canonical coordinates and momenta which are constants of the motion in the new phase space. Under this assumption, Eq. \eqref{K can. tranf. F2} becomes
\begin{equation}
    H(\mathbf{q}, \mathbf{p}, t) + \frac{\partial F_2(\mathbf{q}, \mathbf{P}, t)}{\partial t} = 0
\end{equation}
and substituting $\mathbf{p}$ from the first of Eqs. \eqref{F2 can. transf.}, we obtain
\begin{equation} \label{Hamilton-Jacobi Eq}
    H(\mathbf{q}, \frac{\partial F_2}{\partial \mathbf{q}}, t) + \frac{\partial F_2}{\partial t} = 0
\end{equation}
Equation \eqref{Hamilton-Jacobi Eq} is known as the Hamilton-Jacobi equation. This equation does not involve differentiation with respect to $\mathbf{P}$, and also from the assumption that $K=0$, we know that the new momenta involved in $F_2(\mathbf{q}, \mathbf{P}, t)$ are $N$ independent constants. Thus, the Hamilton-Jacobi equation is a first-order partial differential equation in $N+1$ dimensions, i.e., $N$ coordinates $\mathbf{q}$ plus time. We search for a solution to this equation, denoted by $F_2 = S$, to be in the form, $F_2 = S = S(\mathbf{q}, \boldsymbol{\alpha}, t)$ where in this solution we also want to appear the $N+1$ independent constants of integration, $\boldsymbol{\alpha} = \alpha_1, \alpha_2, \dots, \alpha_{N+1}$. 

A solution containing as many integration constants as there are variables is called a complete solution. $F_2$ is not explicitly present in Eq. \eqref{Hamilton-Jacobi Eq}; only its partial derivatives with respect to the $N$ coordinates and time appear. Therefore, for the Hamilton-Jacobi equation, only the $N$ constants of integration are required. The solution of the Hamilton-Jacobi equation, denoted by $S(\mathbf{q}, \boldsymbol{\alpha}, t)$, referred to as Hamilton's principal function, provides this canonical transformation.

In the case of a time-independent Hamiltonian, we only demand that the new canonical momenta are constants of the motion. For this reason, we use the generating function $F = W(\mathbf{q}, \mathbf{P}) - \mathbf{Q}\mathbf{P}$, where the role of $F_2$ is played by $W$. When $N$ functions $\bm{f} = (f_1(\bm{q}, \bm{p}), f_2(\bm{q}, \bm{p}), \dots, f_N(\bm{q}, \bm{p}))$, with functional determinant $\det(\partial f_i/\partial p_j)\ne 0$\footnote{The condition
\begin{equation}\label{Eq: Jacobian with p}
    \det\left( \frac{\partial f_i}{\partial p_j} \right) \neq 0, \quad i,j=1,2,\dots,N
\end{equation}
implies that the functions $\bm{f}(\bm{q}, \bm{p}, t)$ are functionally independent. It is sufficient but not necessary condition. 

In the general case, the functions $\bm{f}(\bm{q}, \bm{p}, t)$ are functionally independent if and only if the rank of the Jacobian matrix  
\begin{equation} 
    J = \begin{pmatrix} 
    \frac{\partial f_i}{\partial q_k} & \frac{\partial f_i}{\partial p_l} 
    \end{pmatrix},
\end{equation}
is $N$, which implies the existence of at least one non-zero $N \times N$ sub-determinant of $J$. However, in this context, when we refer to independent functions, we specifically mean functions for which condition \eqref{Eq: Jacobian with p} is satisfied

A less restrictive condition than \eqref{Eq: Jacobian with p} is that the $N \times N$ determinant  
\begin{equation}
    \det\begin{pmatrix} 
    \frac{\partial f_i}{\partial q_k} & \frac{\partial f_i}{\partial p_l} 
    \end{pmatrix} \neq 0,
\end{equation}
where $k, l = 1, 2, \dots, N$, and for each degree of freedom $j$ only the partial derivative either with resect to $q$ or to $p_j$ there exist in the determinant (e.g., derivatives with respect to $q_1$ and $p_1$ do not appear in the same determinant).  

In terms of new canonical variables $(\bm{Q}, \bm{P})$, which are given by the equations $q_k = -P_k$, $p_k = Q_k$, and $q_l = Q_l$, $p_l = P_l$, with generating function $F = q_{k}p_{k} + \sum\limits_{l}q_{l} P_{l} - Q_{l} P_{l}$, this condition can be expressed as  
$
\det\left( \frac{\partial f_i}{\partial P_j} \right) \neq 0.
$ which is identical to condition \eqref{Eq: Jacobian with p}.}, are pairwise in involution, meaning that their Poisson brackets are zero, $[f_i, f_j] = 0$, with $i,j = 1, 2, \dots, N$, then the existence of a canonical transformation to a new set of canonical variables is proven, where the canonical momenta are the functions $\bm{f}$ (see Mayer's lemma and Liouville's theorem in Ref. \cite{MEllo2007} Sec. 1.10). If these functions $\bm{f}$ are the $\bm{\alpha}$, i.e., if the independent ($\det(\partial \alpha_i/\partial p_j)\ne 0$, $i,j = 1,2,\dots N$) constants $\bm{\alpha}$ are in involution, it is ensured that the canonical transformation after which the new canonical momenta are constants of the motion exists. In this case the Hamiltonian system is integrable as we will show in the following. The solution of the Hamilton-Jacobi equation provides this canonical transformation. Applying this canonical transformation, we obtain:
\begin{equation} \label{CT with W}
    K = H, \quad p_j = \frac{\partial W(\mathbf{q}, \mathbf{P})}{\partial q_j}, \quad Q_j = \frac{\partial W(\mathbf{q}, \mathbf{P})}{\partial P_j}, \quad j=1,2,\dots,N
\end{equation}
Assuming that as a constant of the motion $H(\mathbf{q}, \mathbf{p}) = \alpha_1$ and substituting $\mathbf{p}$ from the second of the equations above, we take:
\begin{equation}\label{HW2}
    H(\mathbf{q}, \frac{\partial W}{\partial \mathbf{q}}) = \alpha_1
\end{equation}
Eq. \eqref{HW2} is a first-order partial differential equation, which, analogous to Eq. \eqref{Hamilton-Jacobi Eq}, we aim to express in terms of $N-1$ independent constants of integration, $\alpha_2, \alpha_3, \dots, \alpha_N$, which are in involution and, in this case, also independent of $\alpha_1$. Since $\alpha_2, \alpha_3, \dots, \alpha_N$ are constants and $\alpha_1$ is the Hamiltonian, it follows that $[\alpha_i, \alpha_1] = d\alpha_i/dt = 0$ for each $i$ from 2 to $N$, which indicates that $\alpha_1$ is in involution with each of $\alpha_2, \alpha_3, \dots, \alpha_N$. Thus, the solution can again be expressed in terms of $\mathbf{q}$ and $N$ independent constants, which are in involution and denoted by $\boldsymbol{\alpha}$, as $W(\mathbf{q}, \boldsymbol{\alpha})$. $W$ is called Hamilton's characteristic function.

In the entire derivation discussed above, both for time-dependent and time-independent Hamiltonians, we have taken the new momenta $\mathbf{P}$ as the $N$ constants $\bm{\alpha}$. However, in general, $\mathbf{P}$ will remain constants—also independent and in involution—if they are defined as any function of the constants of integration, $\boldsymbol{\alpha}$. This is a very useful advantage that the Hamilton-Jacobi theory offers, as we will see in the following. Particularly, $\mathbf{P}$ can be written as:
\begin{align} \label{P(a)}
    P_1 &= P_1(\alpha_1,\alpha_2,\dots,\alpha_N) \notag \\
    P_2 &= P_2(\alpha_1,\alpha_2,\dots,\alpha_N) \notag \\
    & \vdots \notag \\
    P_N &= P_N(\alpha_1,\alpha_2,\dots,\alpha_N)
\end{align}
which forms an algebraic system of $N$ independent equations with $N$ unknowns. This means that the inversion of this system can provide $\boldsymbol{\alpha}$ as a function of $\mathbf{P}$, and vice versa. Thus, it can be found that
\begin{equation} \label{K(P)_W}
    H = K = \alpha_1(\mathbf{P})
\end{equation}
As a result, the new Hamiltonian is a function only of $\mathbf{P}$, giving:
\begin{align} \label{P_W}
    \dot{\mathbf{P}} = -\frac{\partial K}{\partial Q} &= 0 \Longrightarrow \mathbf{P} = constant = \mathbf{P}(\boldsymbol{\alpha}) 
\end{align}
which agrees with our initial assumption that $\mathbf{P}$ are constants of the motion. In addition:
\begin{align} \label{Q_W}
    \dot{\mathbf{Q}} &= \frac{\partial K}{\partial \mathbf{P}} \Longrightarrow \mathbf{Q} = \mathbf{u}(\mathbf{P}) t + \bm{\beta}
\end{align}
where $\mathbf{\beta}$ are other $N$ constants that can be determined by the $2N$ initial conditions that are necessary for the integration of Hamilton equations, \eqref{dot q gen} and \eqref{dot p gen}. Specifically, for a particular initial condition $(\mathbf{q}^0,\mathbf{p}^0)$ at $t=0$, which indicates a particular orbit, from the equations of the canonical transformation we can find $\mathbf{P}^0 = \mathbf{P}(\mathbf{q}^0,\mathbf{p}^0)$. $\mathbf{P}$ will have this value at each point of the orbit since it is a constant of the motion. To determine $\bm{\beta}$, one can use the third of Eqs. \eqref{CT with W} and Eq. \eqref{Q_W} at $t=0$ to find $\bm{\beta} = \mathbf{Q}(\mathbf{q}^0,\mathbf{P})$. In a time-independent system, the time is not explicitly present in the differential equations, meaning that the initial condition can be chosen at any time instant $t=t_1$ without altering the final solution. Therefore, by selecting $(\mathbf{q}^0,\mathbf{p}^0)$ at $t=t_1$ with the same value of $\mathbf{P}$, we find from Eq. \eqref{Q_W} and the third of Eq. \eqref{CT with W} that $\mathbf{u}(\mathbf{P})t_1 + \bm{\beta} = \mathbf{Q}(\mathbf{q}^0, \mathbf{P})$, where $t_1$ can be chosen such that $\bm{\beta} = 0$. This implies that in the new canonical equations of Hamilton, Eqs. \eqref{P_W} and \eqref{Q_W}, only the selection of the independent constants $\mathbf{P}$ (or $\bm{\alpha}$) determines a particular orbit, which can be transformed back to the old variables $(\mathbf{q},\mathbf{p})$. As a result, a specific set of $N$ independent constants of integration $\bm{\alpha}$ (or $\mathbf{P}$), necessary for the solution of the Hamilton-Jacobi equation, is sufficient to uniquely specify a particular orbit. In other words, a particular set of $N$ constants of the motion serves to label an orbit. This fact will be used in the following sections of this work. From this point onward, we will refer only to time-independent systems, also known as autonomous or conservative Hamiltonian systems. The extension to non-autonomous systems can be readily achieved by introducing an extended phase space (see Ref. \cite{Lieberman1992}, Sec. 1.2).

\section{Separable and Integrable Hamiltonian} \label{Separability and Integralbility}
A full solution (integral) of Hamilton's equations of motion, i.e., obtaining $\mathbf{q}(t)$ and $\mathbf{p}(t)$ for a specific set of $2N$ initial conditions, would be achieved if we could determine $\mathbf{Q}$ and $\mathbf{P}$, Eqs. \eqref{Q_W} and \eqref{P_W}, and then invert these using the canonical transformation equations, Eqs. \eqref{CT with W}. In the previous section, we discussed that such a canonical transformation exists when the Hamiltonian system admits $N$ independent constants and when these constants are in involution. A system that satisfies both these conditions is called integrable. However, such constants do not always exist, and even if they do, the solution of the Hamilton-Jacobi equation, $W(\mathbf{q}, \boldsymbol{\alpha}(\mathbf{P}))$, is not easily obtained. Thus, up to this point, what the Hamilton-Jacobi theory offers is a transformation of the original problem—solving a system of $2N$ first-order ordinary differential equations with $2N$ initial conditions, Eqs. \eqref{dot q gen}, \eqref{dot p gen}—into the problem of solving a single first-order partial differential equation in $N$ dimensions, namely the Hamilton-Jacobi equation, Eq. \eqref{HW2}. For this solution, we require $N$ independent constants of integration in involution, which determine the new canonical momenta $\mathbf{P}(\boldsymbol{\alpha})$. In fact, by employing techniques for solving partial differential equations, such as the method of Cauchy characteristics, one can derive Hamilton's equations directly from the Hamilton-Jacobi equation \cite{MEllo2007}.

Advantages from the Hamilton-Jacobi theory can be gained when the Hamilton-Jacobi equation is completely separable in some coordinates system \cite{Lieberman1992}, which also implies the integrability of the system. For a separable system, we try a solution $W(\mathbf{q},\boldsymbol{\alpha})$ expressed in a separable form as:
\begin{equation}
    W(\mathbf{q},\boldsymbol{\alpha}) = \sum_{j=1}^{N}{W_j(q_j, \boldsymbol{\alpha})}
\end{equation}
then the split Hamilton-Jacobi equation is written as: 
\begin{equation} \label{split H-J equation}
    \sum_{j=1}^N {H_j\left(q_j, \frac{\partial W_j(q_j,\boldsymbol{\alpha})}{\partial q_j}\right)} = \alpha_1
\end{equation}
with
\begin{equation}\label{H_j}
    H_j\left(q_j, \frac{\partial W_j(q_j,\boldsymbol{\alpha})}{q_j}\right) = \alpha_j, \quad j = 1,...,N
\end{equation}
where, the constants of separation $\alpha_1, \alpha_2, \dots, \alpha_N$ are referred to as "isolating integrals" or "global invariants of the motion". Here $H_j$ are not necessarily Hamiltonian functions, but their summation is. For each $j$, each of the Eqs. \eqref{H_j} can be separately solved in terms of the respective $q_j$ giving $W_j$ as:
\begin{equation}
    W_j = W_j(q_j, \boldsymbol{\alpha})
\end{equation}
where, by expressing $\boldsymbol{\alpha}$ as a function of $\mathbf{P}$ according to Eq. \eqref{P(a)} and summing over $j$, the generating function $W(\mathbf{q}, \mathbf{P})$ is obtained.

Let's say that the Hamiltonian system can be written in a separable form, meaning the $W_j$'s have been determined for each $j$. Then, from the second of Eqs. \eqref{CT with W}, we obtain $p_j = \partial W_j(q_j, \mathbf{P}) / \partial q_j = p_j(q_j, \mathbf{P})$ for $j = 1, 2, \dots, N$. Thus, for each degree of freedom, a distinct solution is found, and each solution is independently projected in the subspace $q_j, p_j$ of the phase space as $p_j = p_j(q_j, \mathbf{P})$. This implies integrability. From the discussion above, it becomes clear that a separable system is integrable, but the reverse does not necessarily hold.

In general, there are no simple criteria to indicate that a system is separable and there is no known procedure for determining all the isolating integrals of a general Hamiltonian system, or even for finding their total
number (Ref. \cite{Lieberman1992} Sec. 1.3c). However, we can show that canonical variables that are cyclic can be separated from the rest. For instance, if $q_1$ is cyclic, then $p_1$ is a constant of the motion (let’s denote $p_1 = c_1$). Then the Hamilton-Jacobi equation is written as:
\begin{equation}\label{H-J Eq cyclic}
    H(q_2,\dots,q_N, c_1, \frac{\partial W}{\partial q_2},\dots,\frac{\partial W}{\partial q_N}) = \alpha_1
\end{equation}
where we try a solution of the form $W = W_1(q_1,\bm{\alpha}) + W^{\prime}(q_2,\dots,q_N,\bm{\alpha})$. Substituting this into Eq. \eqref{H-J Eq cyclic}, we observe that it only involves $W^{\prime}$. Thus, a partial differential equation with $N-1$ variables is formed. By analogy to the discussion above, we search for a solution $W^{\prime}$ that contains $N-2$ independent constants of integration, which are also independent of $c_1$ and $\alpha_1$. Furthermore, it can be shown that $c_1$ and $\alpha_1$ are independent fulfilling the condition \eqref{Eq: Jacobian with p}. Combining these, we form the set of $N$ independent constants, $\bm{\alpha}$, which appear in the solution $W^{\prime}$. Moreover, from the canonical transformation equation $c_1 = p_1 = \partial W/\partial q_1 = \partial W_1(q_1,\bm{\alpha})$, we find that $W_1(q_1,\bm{\alpha}) = c_1 q_1$. Finally, we can write:
\begin{equation}
    W(\mathbf{q},\bm{\alpha}) = W^{\prime}(q_2,\dots,q_N, \bm{\alpha}) + c_1 q_1
\end{equation}
where $c_1$ is an element of $\bm{\alpha}$. The same process can be applied to any cyclic canonical variables, if they exist. Thus, for an $N$ degrees of freedom time-independent Hamiltonian containing $N-1$ cyclic canonical variables, the Hamilton-Jacobi equation degenerates (consider, for example, that all $\mathbf{q}$ are cyclic except for $q_N$) as:
\begin{equation}
    H(q_N,\frac{\partial W_{N}(q_N,\bm{\alpha})}{\partial q_N}, c_1,\dots,c_{N-1}) = \alpha_1
\end{equation}
which is in a separable form, and its solution will give the $W_N(q_N, \bm{\alpha})$, where $\bm{\alpha} = (c_1, \dots, c_{N-1}, \alpha_1)$. The generating function in a separable form will be given by:
\begin{equation}
    W(\mathbf{q}, \mathbf{P}) = W_{N}(q_N,\bm{\alpha}) + \sum_{j=1}^{N-1}c_j q_j.
\end{equation}

\section{Action-Angle variables} \label{Action Angle Variables}
As shown in Sec. \ref{Separability and Integralbility}, for a conservative (time-independent), fully separable, and thus integrable Hamiltonian, the canonical momenta are obtained from the equations of the canonical transformation as:
\begin{equation} \label{p_j(q_j)}
    p_j = \frac{\partial W(q_j,\bm{\alpha})}{\partial q_j}, \quad j=1,2\dots,N
\end{equation}
which provides each $p_j$ as a function only of $q_j$ and the independent constants $\bm{\alpha}$.

When the projections of the orbit on the $(q_j, p_j)$ planes—each corresponding to a distinct degree of freedom, $j = 1, 2, \dots, N$—are periodic or closed, we can define the action integral as
\begin{equation} \label{definition of J} J_j = \frac{1}{2\pi}\oint_{q_j} p_j(q_j, \bm{\alpha}) dq_j, \quad j = 1, 2, \dots, N
\end{equation}
which, as a complete integral over $q_j$, gives $J_j$ as a function only of $\bm{\alpha}$. Then, when $p_j$ is a periodic function of $q_j$ for each $j$, we can take the actions in each degree of freedom as
\begin{equation} \label{J(a)}
    \mathbf{J} = \mathbf{J}(\bm{\alpha}).
\end{equation}
In Sec. \ref{Hamilton-Jacobi Theory}, we showed that the new canonical momenta $\mathbf{P}$ can be chosen as an arbitrary function of the independent constants $\bm{\alpha}$. With Eq. \eqref{J(a)}, we make a particular choice, namely $\mathbf{P}$ as a function of $\bm{\alpha}$ to be the actions $\mathbf{J}$. This particular choice provides specific advantages, as we show in the following. For Eq. \eqref{J(a)} to be inverted to obtain $\bm{\alpha}(\bm{J})$ as a function of the actions, the condition $\det(\partial J_i/\partial \alpha_j) \neq 0$, where $i, j = 1, 2, \dots, N$, must also be satisfied.

The generating function of the canonical transformation in action-angle variables can be obtained from Eq. \eqref{p_j(q_j)} as:
\begin{equation} \label{A-A Generating Function} W(\mathbf{q}, \mathbf{J}) = \sum_{j=1}^{N} W_j(q_j, \bm{\alpha}(\mathbf{J})) = \sum_{j=1}^{N} \int_{q_j} p_j(q^{\prime}_j, \bm{\alpha}(\mathbf{J}))  dq^{\prime}_j.
\end{equation}
In analogy to Eq. \eqref{K(P)_W}, the Hamiltonian can be written as:
\begin{equation} \label{H(J)}
    H = H(\mathbf{J})
\end{equation}
and, similarly to Eq. \eqref{Q_W}, the new canonical coordinates—denoted here by $\mathbf{w}$ instead of $\mathbf{Q}$—are given by
\begin{equation} \label{w(J)}
\mathbf{w} = \bm{\omega}(\mathbf{J}) t + \bm{\beta}
\end{equation}
where, to denote $\partial H(\mathbf{J})/\partial \mathbf{J}$, we use $\bm{\omega}(\mathbf{J})$ instead of $\mathbf{u}(\mathbf{J})$.

Since we have assumed a full separable system, the generating function will be written as: $W(\mathbf{q},\mathbf{J}) = W_1(q_1,\mathbf{J}) + W_2(q_2,\mathbf{J}),\dots,+W_N(q_N,\mathbf{J})$. Then, from the third of Eqs. \eqref{CT with W}, we take:
\begin{equation}\label{w_W}
    w_j = \sum_{i=1}^{N}\frac{\partial W_i(q_i,\mathbf{J})}{\partial J_j}
\end{equation}
This  gives each $w_j$ as $w_j = w_j(\mathbf{q}, \mathbf{J})$. As mentioned in Sec. \ref{Hamilton-Jacobi Theory}, along a particular orbit, the new canonical momenta $\mathbf{P}$ are constant, i.e., $d\mathbf{P} = 0$; similarly, $d\mathbf{J} = 0$ on a particular orbit. Thus, we can write:
\begin{equation}
dw_j = \sum_{i=1}^{N}\frac{\partial w_j}{\partial q_i}dq_i 
\end{equation}
using Eq. \eqref{w_W}, this becomes:
\begin{equation} dw_j = \sum_{i=1}^{N} \frac{\partial^2 W_{i}(q_i,\mathbf{J})}{\partial q_i \partial J_j}dq_i. \end{equation}
Applying Eq. \eqref{p_j(q_j)}, we can express this as:
\begin{equation} dw_j = \frac{\partial}{\partial J_j}\sum_{i=1}^{N}{p_i(q_i,\mathbf{J})}dq_i. \end{equation}
From this equation, we can compute the variation of each $w_j$ when each $p_j(q_i,\mathbf{J})$ completes one full circle (or period). To do this, we integrate the right-hand side with respect to each $q_i$ and take the derivative with respect to $J_j$ out of the integral, since $J_j$ is constant throughout the orbit and consequently in the limits of the integrals. Thus,
\begin{equation}
    \Delta w_j = \frac{\partial }{\partial J_j} \sum_{i=1}^{N}\oint_{q_i} p_i(q_i,\mathbf{J})d\mathbf{q}_i  \overset{\eqref{definition of J}}{=} \frac{\partial}{\partial J_j} \sum_{i=1}^{N} {2\pi J_i}
\end{equation}
which gives $\Delta w_j = 2\pi$ for each $j$. Each $q_i$ is a function of all the $w_j$'s, and since all the $w_j$'s span an interval of $2\pi$ for a circle of $q_i$, we can conclude that $q_i$ can be expressed as a Fourier series, forming a multi-periodic function of the $w_j$'s with a period of $2\pi$. Furthermore, each $p_i$, as a function of $q_i$ and the constants of integration or actions $\mathbf{J}$ (as shown in Eq. \eqref{p_j(q_j)}), can also be represented as a multi-periodic function of the $w$'s, similar to any other function of $(\mathbf{q},\mathbf{p})$. Thus, we can write:
\begin{equation} \label{q,p with w}
    q_j = \sum_{\mathbf{n}}{a^{(j)}_{\mathbf{n}}(\mathbf{J})e^{i\mathbf{n}\cdot\mathbf{w}}}, \quad p_j = \sum_{\mathbf{n}}{b^{(j)}_{\mathbf{n}}(\mathbf{J})e^{i\mathbf{n}\cdot\mathbf{w}}}.
\end{equation}
From Eq. \eqref{w(J)}, we see that $\Delta\mathbf{w} = 2\pi$ when $t = T = 2\pi/\bm{\omega}(\mathbf{J})$. This indicates that for a particular orbit the frequencies of each degree of freedom can be distinctly defined, and are given by
\begin{equation}
    \omega_j(\mathbf{J}) = \frac{\partial H(\mathbf{J})}{\partial J_j}, \quad j=1,2,\dots,N.
\end{equation}
By substituting Eq. \eqref{w(J)} into Eq. \eqref{q,p with w} and assuming that $\bm{\beta}$ can be set to zero, as explained in Sec. \ref{Hamilton-Jacobi Theory}, we obtain:
\begin{equation} \label{q,p with t}
    q_j(t) = \sum_{\mathbf{n}}{a^{(j)}_{\mathbf{n}}(\mathbf{J})e^{i\mathbf{n}\cdot\bm{\omega}(\mathbf{J}
    )t}}, \quad p_j(t) = \sum_{\mathbf{n}}{b^{(j)}_{\mathbf{n}}(\mathbf{J})e^{i\mathbf{n}\cdot\bm{\omega}(\mathbf{J})t}}.
\end{equation}
This is the main advantage of the action-angle transformation, i.e., the fact that it leads to an evaluation of all the orbital frequencies (or orbital spectrum) involved in multiply periodic motion without requiring a complete solution of the equations of motion \cite{Goldstein2002}.

Note at this point that the periodicity of $p_j$ in terms of $q_j$ that we assumed at the beginning of this section does not directly imply periodic motion of the canonical variables in terms of time. Periodic motion occurs for orbits labeled by those $\bm{J}$ values where a specific relationship exists between the orbital frequencies $\bm{\omega}(\bm{J})$ associated with motion along the independent directions, such that these frequencies satisfy a resonance condition. These orbits are called resonant orbits.

To illustrate this, consider a resonant orbit characterized by $\bm{J} = \bm{J}^{rc}$, for which the following resonance condition holds:
\begin{equation} \label{resonance condition k}
    \bm{k}\cdot\bm{\omega}(\bm{J}^{rc}) = 0.
\end{equation}
Let us separate the vector $\bm{k}$ into two parts as $\bm{k} = (k_1, \bm{k}^\prime)$, where $k_1$ is the first component of $\bm{k}$ and $\bm{k}^\prime = (k_2, \dots, k_N)$. Then, Eq. \eqref{resonance condition k} can be rewritten as:
\begin{equation}
    \bm{k}^{\prime}\bm{\omega}^{\prime} = -k_1\omega_1
\end{equation}
where $\bm{\omega}^\prime = (\omega_2, \dots, \omega_N)$. Multiplying both sides of this equation by $\bm{k}^\prime$ and then dividing by $|\bm{k}^\prime|^2$, we obtain:
\begin{equation}
    \bm{\omega}^{\prime} = -\frac{k_1\bm{k}^{\prime}}{|\bm{k}^{\prime}|^2}\omega_1.
\end{equation}
Substituting the above equation into Eq. \eqref{q,p with t}, we take:
 \begin{equation} \label{q,p with t 2}
    q_j(t) = \sum_{\bm{n}= n_1, \bm{n}^{\prime}}{a^{(j)}_{\mathbf{n}}(\mathbf{J})e^{i (n_1 - \frac{k_1}{|\bm{k}^{\prime}|^2}\bm{n}^{\prime}\cdot\bm{k}^{\prime})\omega_1(\bm{J}) t}}, \quad  p_j(t) = \sum_{\bm{n}= n_1, \bm{n}^{\prime}}{b^{(j)}_{\mathbf{n}}(\mathbf{J})e^{i (n_1 - \frac{k_1}{|\bm{k}^{\prime}|^2}\bm{n}^{\prime}\cdot\bm{k}^{\prime})\omega_1(\bm{J}) t}}.
\end{equation}
where, by rearranging the coefficients $a_{\bm{n}}^{(j)}$ and $b_{\bm{n}}^{(j)}$, the rational number $m/l = (n_1 - \frac{k_1}{|\bm{k}^{\prime}|^2}\bm{n}^{\prime}\cdot\bm{k}^{\prime})$ can replace the summation over the vector $\bm{n}$, yielding: 
\begin{equation} \label{q,p with t 3}
    q_j(t) = \sum_{m}{c^{(j)}_{m}(\mathbf{J})e^{i \frac{m}{l}\cdot\omega_1(\mathbf{J}
    )t}}, \quad p_j(t) = \sum_{m}{d^{(j)}_{\mathbf{n}}(\mathbf{J})e^{i \frac{m}{l}\cdot\omega_1(\mathbf{J})t}}.
\end{equation}
which confirms that for the resonant orbits $q_j(t)$ and $p_j(t)$, $j=1,\dots, N$ are periodic functions of time. When the orbital frequencies are not connected with a resonance condition the orbits $(q_j(t),p_j(t))$ are quasi-periodic in time.

\begin{figure} [h!]
    \centering
    \includegraphics[width=0.69\textwidth,keepaspectratio]{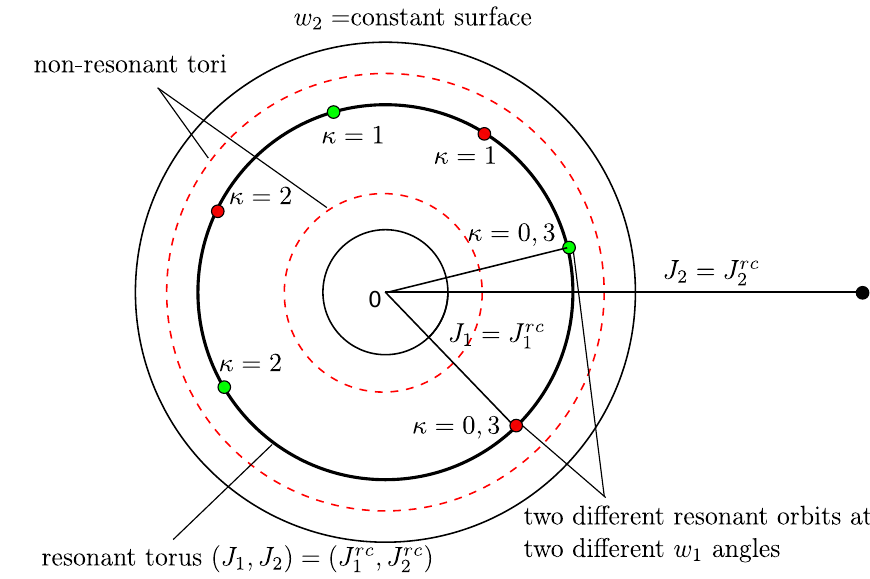}  
    \caption{A qualitative depiction of a resonant torus in a two-degree of freedom Hamiltonian system. Here $\kappa$ denotes the consecutive intersections of the orbit with the $w_2 = constant$ surface in positive or negative direction.}

    \label{fig:Fig30}
\end{figure}

In action-angle variables, each of the $N$ degrees of freedom is described by a constant action $J_j$ and an angle with a constant frequency $\omega_j(\bm{J})$, with $j = 1,\dots,N$. This defines the motion as taking place on an $N$ -dimensional torus, where the radii correspond to the constant actions, and the angular coordinates represent the angle variables. To visualize this, consider a system with two degrees of freedom, Fig. \ref{fig:Fig30}. An orbit is labeled by the constants $(J_1, J_2)$, which correspond to the radii of the torus, and by angular variables $w_1$ and $w_2$ which define its position on the torus. Starting at a point $(J_1, w_1, J_2, w_2)$ the orbit evolves in time as $(J_1, \omega_1(J_1,J_2) t, J_2, \omega_2(J_1,J_2) t)$, with $J_1$ and $J_2$ remaining constant. If the ratio $(\omega_1/\omega_2)$ is irrational, the orbit will densely fill the toroidal surface associated with $J_1$ and $J_2$. Conversely, if $(\omega_1/\omega_2)$ is rational, meaning that $\omega_1$ and $\omega_2$ satisfy a resonance condition, i.e., the orbit is a resonant orbit, the orbit will pass through only discrete points on the torus. The torus $J_1, J_2$ from which pass a resonant orbit is called a resonant torus. Orbits that start on or pass through this torus are periodic in time as we previously proved. The fulfillment of the resonance condition and consequently being an orbit periodic in time or not depends only on the actions. This means that from a resonant torus characterized by the actions can pass infinitely many periodic orbits starting from different angles, Fig. \ref{fig:Fig30}.

\section{Perturbation Theories}

\subsection{Canonical Perturbation Theory}\label{Classical Perturbation Theory}
Most multidimensional Hamiltonian systems are not integrable; that is, they lack the necessary number of independent constants of motion that are also in involution, which are required to solve the Hamilton-Jacobi equation, as discussed in Secs. \ref{Hamilton-Jacobi Theory}-\ref{Action Angle Variables}. However, for systems that differ only slightly from integrable ones—often referred to as near-integrable systems—it is possible to approximate solutions or qualitative aspects, such as the orbital frequencies, to a desired degree of accuracy by treating these systems as perturbations of an integrable system. Classical perturbation theory involves a power series expansion of the perturbation in terms of a small parameter $\epsilon$, as well as a multiple Fourier series expansion facilitated by the action-angle variables, as analyzed in the previous section. Thus, a near-integrable system, whose perturbation terms exhibit a periodic dependence on time with period $2\pi/\Omega$, can be expressed as: 
\begin{equation} \label{near integrable Hamiltonian}
    \begin{aligned}
        H &= H_0(\mathbf{J}) + \epsilon H_1(\mathbf{J},\bm{\theta},t) + \epsilon^2 H_2(\mathbf{J},\bm{\theta},t) + \dots \\
        &= H_0(\mathbf{J}) + \epsilon \sum_{\bm{m},l}H_{1(\bm{m},l)}(\bm{J})e^{i(\bm{m}\bm{\theta} + l\Omega t)} 
        + \epsilon^2 \sum_{\bm{m},l}H_{2(\bm{m},l)}(\bm{J})e^{i(\bm{m}\bm{\theta} + l\Omega t)}
    \end{aligned}
\end{equation}
Here, $H_0(\mathbf{J})$ represents the integrable Hamiltonian expressed in action-angle space $(\mathbf{J},\bm{\theta})$. By solving the Hamilton-Jacobi equation at each successive power, one can retain terms up to the desired order in $\epsilon$. This perturbation method allows for the determination of a generating function  $S(\bar{\bm{J}},\bm{\theta},t) = \bar{\bm{J}}\bm{\theta} + \epsilon S_1(\bar{\bm{J}},\bm{\theta},t) + \epsilon^2 S_2(\bar{\bm{J}},\bm{\theta},t) + \dots$ that approximately\footnote{We say 'approximately' because, to transform the system into the integrable form \eqref{bar H(bar J)}, terms in the generating function beyond the desired order in $\epsilon$ must be neglected.} transforms the system from the canonical variables $(\mathbf{J}, \bm{\theta})$ to $(\bar{\mathbf{J}}, \bar{\bm{\theta}})$. In these new variables, the transformed Hamiltonian $\bar{H}$, up to the desired order in $\epsilon$, depends only on $\bar{\bm{J}}$ and is therefore directly integrable, as described in Sec. \ref{Action Angle Variables}. For this purpose, the Hamiltonian \eqref{near integrable Hamiltonian} is expanded in a Taylor series around the barred canonical variables. The transformed Hamiltonian, expressed in these variables, is approximated up to first order in $\epsilon$\footnote{$\bar{\bm{J}}$ and $\bar{\bm{\theta}}$ appear in the expressions as a result of a Taylor expansion of functions of $\mathbf{J}$ and $\bm{\theta}$ around the point $(\bar{\bm{J}}, \bar{\bm{\theta}})$, retaining terms up to the desired order in $\epsilon$. Consequently, the functional dependence of these expressions on $(\bar{\bm{J}}, \bar{\bm{\theta}})$ mirrors the original dependence on $(\bm{J}, \bm{\theta})$.  For this reason, in these expressions, $(\bar{\bm{J}}, \bar{\bm{\theta}})$ can be directly substituted with $(\bm{J}, \bm{\theta})$, and vice versa. \label{fn:barred repalced with unbarred}}:
\begin{equation}
    \begin{aligned} \label{bar H gen}
        \bar{H}(\bar{\bm{J}},\bar{\bm{\theta}}) &= H(\mathbf{J}(\bar{\mathbf{J}},\bar{\bm{\theta}}), \bm{\theta}(\bar{\mathbf{J}},\bar{\bm{\theta}})) = H_0(\bar{\bm{J}}) + \epsilon H_{1(\bm{0},0)}(\bar{\bm{J}}) \\
        &+ \epsilon\sum_{\bm{m},l\ne 0}H_{1(\bm{m},l)}(\bar{\bm{J}})e^{i(\bm{m}\bar{\bm{\theta}} + l\Omega t)} + \epsilon \bm{\omega}(\bm{\bar{J}})\frac{\partial S_1(\bm{\bar{J}},\bar{\bm{\theta}},t)}{\partial \bar{\bm{\theta}}} + \epsilon \frac{\partial S_1(\bm{\bar{J}},\bar{\bm{\theta}},t)}{\partial t}
    \end{aligned}
\end{equation}
where:
\begin{equation}
    \frac{\partial S_1(\bm{\bar{J}},\bar{\bm{\theta}})}{\partial \bar{\bm{\theta}}} = \left.\frac{\partial S_1(\bm{\bar{J}},{\bm{\theta}})}{\partial {\bm{\theta}}}\right|_{\bm{\theta} = \bar{\bm{\theta}}}, \quad \bm{\omega}(\bar{\bm{J}}) = \left.\frac{\partial H_0(\bm{J})}{\partial \bm{J}}\right|_{\bm{J} = \bm{\bar{J}}}
\end{equation}
and $H_{1(\bm{0},0)}({\bm{J}})$ represents the Fourier series coefficients of $H_1(\bm{J},\bm{\theta},t)$ at $\bm{m}, l = 0$, or equivalently, the average of $H_1(\bm{J},\bm{\theta},t)$ over one period in $\bm{\theta}$ and $t$, denoted as $\langle H_1(\bm{J},\bm{\theta},t) \rangle_{\bm{\theta},t}$.
We chose $S_1$ to be expressed as a Fourier series expansion in terms of the angles and time and so that when substituted in Eq. \eqref{bar H gen} the last three terms which depend on $\bar{\bm{\theta}}$ and $t$ are eliminated. With such an $S_1$ the generating function is given as (Ref. \cite{Lieberman1992}, Sec. 2.2):
\begin{equation} \label{S epsilon}
    S(\bar{\mathbf{J}}, {\bm{\theta}},t) = \bar{\mathbf{J}}
    \cdot {\bm{\theta}} + \epsilon i \sum_{\mathbf{m},l\ne 0}{\frac{H_{1(\mathbf{m},l)}(\bm{\bar{J}})}{\mathbf{m}\cdot\bm{\omega}(\mathbf{\bar{J}}) + l\Omega}e^{i(\mathbf{m}\cdot {\bm{\theta}}+l\Omega t)}}.
\end{equation}
Substituting in Eq. \eqref{bar H gen} we obtain
\begin{equation} \label{bar H(bar J)}
    \bar{H}(\bar{\bm{J}}) = H_0(\bar{\mathbf{J}}) + \epsilon H_{1(
    \bm{0},0)}(\bar{\bm{J}})
\end{equation}
which is a function of $\bar{\bm{J}}$ only. Then, a solution (orbit) of the system \eqref{bar H(bar J)} is obtained directly from the relations \eqref{J(a)} and \eqref{w(J)}. Transforming back to $(\mathbf{J}, \bm{\theta})$, the resulting orbit approximates the corresponding actual orbit of the system \eqref{near integrable Hamiltonian}. Specifically, up to first order in $\epsilon$, we can take:
\begin{equation} \label{Eq: original with bar canonical variables}
    {\bm{J}} = \bar{\bm{J}} - \epsilon \frac{\partial S_1(\bar{\bm{J}}, \bar{\bm{\theta}})}{\partial \bar{\bm{\theta}}}, \quad     
    {\bm{\theta}} = \bar{\bm{\theta}} - \epsilon \frac{\partial S_1(\bar{\bm{J}}, \bar{\bm{\theta}})}{\partial \bar{\bm{J}}}
\end{equation}
where $\bar{\bm{J}}$ is constant, and $\bar{\bm{\theta}} = \bm{\omega}(\bar{\bm{J}})t + \bar{\bm{\theta}}^{init}$.

When a resonance condition is fulfilled, i.e.,
\begin{equation} \label{gen resonance condition}
    \mathbf{m}\cdot\bm{\omega}(\bm{J}) + l\Omega = 0
\end{equation}
or in the proximity of the resonance, where the denominator in Eq. \eqref{S epsilon} approaches zero, the summation does not converge. As a result, the classical perturbation method fails to describe the system in regions of resonances. In these regions, the topology of phase space trajectories changes, creating islands, as we will show in section \ref{Secular Perturbation Theory} (Ref. \cite{Lieberman1992}, Sec. 2.2). One might say that non-convergence occurs only for values of $\mathbf{J}$ where the resonance condition is exactly or nearly met. Indeed, a vector $(\mathbf{m},l)$ can always be found for any $\mathbf{J}$ such that $\mathbf{m}\cdot\bm{\omega} + l\Omega = 0$. Even if the relation between a component of $\bm{\omega}$ and the rest of the components of $\bm{\omega}$ plus $\Omega$ is irrational, since any irrational number can be closely approximated by a rational number, there will always exist a suitable $(\mathbf{m}, l)$ such that $\mathbf{m} \cdot \bm{\omega} + l\Omega$ nearly equals zero. Here, the Kolmogorov-Arnold-Moser (KAM) theorem becomes relevant (Ref. \cite{Chiricov1979}, Sec. 4.6 and Ref. \cite{Lieberman1992}, Sec. 3.2a). According to the theorem, orbits of the integrable system whose frequencies $\omega_j$ are incommensurate—i.e., quasi-periodic orbits—and not very close to rational values (meaning orbits sufficiently distant from resonances) are preserved in the near-integrable system, though they may be slightly deformed. This implies that in regions of phase space where the KAM theorem holds, the orbits obtained by transforming from $(\bar{\mathbf{J}}, \bar{\bm{\theta}})$ to $(\mathbf{J}, \bm{\theta})$ via the generating function \eqref{S epsilon} closely approximate the actual orbits of the near-integrable system (Ref. \cite{Lieberman1992}, Sec. 2.2).

\subsection{Canonical Adiabatic Theory}\label{Canonical Adiabatic Theory}
The problem of small denominators is addressed using canonical adiabatic theory (Ref. \cite{Lieberman1992}, Sec. 2.3b). This theory is applicable when all the degrees of freedom of the system, except for one that is in action-angle form, vary slowly with time, and the explicit time dependence of the system itself is also slow. Such a system can be expressed as $H(J,\theta, \bm{p}, \bm{q}, t) = H_{0}(J, \bm{p}, \bm{q}, t) + \epsilon H_1(J, \theta, \bm{p}, \bm{q}, t)$, where $(\bm{q},\bm{p})$ are the slowly (or 'adiabatically') time-varying variables, which are not necessarily in action-angle form. In this process, the fast angle is eliminated by transforming to a new set of canonical variables, and, as a result, its conjugate action becomes a constant of motion, referred to as the 'adiabatic invariant' of the motion. The accuracy of this method in describing the actual orbits of a near-integrable system depends on the strength of the perturbation $\epsilon$ and the order of $\epsilon$ to which the method is applied.

In particular, consider in Hamiltonian \eqref{near integrable Hamiltonian} that all the angles except one vary slowly with time, as also does the perturbation's explicit variation with respect to time. Denote the slow angles as $\bm{\theta_y}$ and their conjugate actions as $\bm{J_y}$, and the remaining fast angle as $\theta$ with its conjugate action as $J$. Then the Hamiltonian \eqref{bar H gen} can be expressed as
\begin{equation}
    \begin{aligned} \label{bar H gen slow}
        \bar{H}(\bar{J}, \bar{\bm{J_y}},\bar{\theta}, \bar{\bm{\theta_y}}) &= H_0(\bar{J}, \bar{\bm{J_y}}) + \epsilon H_{1(0,\bm{0},0)}(\bar{J}, \bar{\bm{J_y}}) \\ 
        & \hspace{-0.8cm} + \epsilon\sum_{k,\bm{m},l\ne 0}H_{1(k,\bm{m},l)}(\bar{J}, \bar{\bm{J_y}})e^{i(k\bar{\theta} + \bm{m}\bar{\bm{\theta_y}} + l\Omega t)}
        + \epsilon\omega (\bar{J}, \bm{\bar{J_y}})\frac{\partial S_1(\bar{J}, \bm{\bar{J_y}},\bar{\theta},\bar{\bm{\theta_y}},t)}{\partial \bar{{\theta}}} \\
        & \hspace{-0.8cm} + \epsilon^2\bm{\omega_y}(\bar{J}, \bm{\bar{J}})\frac{\partial S_1(\bar{J}, \bm{\bar{J_y}},\bar{\theta},\bar{\bm{\theta_y}},t)}{\partial \bar{\bm{\theta_y}}}
        + \epsilon^2\Omega\frac{\partial S_1(\bar{J}, \bm{\bar{J_y}},\bar{\theta},\bar{\bm{\theta_y}},t)}{\partial (\Omega t)}
    \end{aligned}
\end{equation}
where we consider such an $S_1$ so that it is a periodic function of time with period $2\pi/\Omega$, like the perturbation $H_1$. The slow time variation of $\bm{\theta_y}$ and the slow explicit time variation of the perturbation imply that $\omega_y(\bar{J}, \bar{\bm{J_y}})$ and $\Omega$ are of order $\epsilon$, and for this reason, the two last terms in Eq. \eqref{bar H gen slow} are of order $\epsilon^2$.

In accordance with Eq. \eqref{S epsilon}, the generating function will be given by:
\begin{equation}\label{gen fun S2}
    \begin{aligned} 
        S(\bar{J},\bar{\bm{J_y}}, \theta, \bm{\theta_y}, t) = \bar{J}\theta + \bar{\bm{J_y}}\bm{\theta_y} \\
        &\hspace{-1.8cm} + \epsilon i\sum_{k,\bm{m},l\ne 0}{\frac{H_{1(k,\bm{m},l)}(\bar{J},\bar{\bm{J_y}})}{k\omega(\bar{J},\bar{\bm{J_y}}) + \bm{m}\cdot \epsilon \bm{\omega_{y}}(\bar{J},\bar{\bm{J_y}}) + l \epsilon \Omega}e^{i(k\theta + \bm{m}\bm{\theta_{y}} + l\Omega t)}}.
    \end{aligned} 
\end{equation}
The resonance condition is written as:
\begin{equation}
    k\omega(J,\bm{J_y}) + \mathbf{m}\cdot\epsilon\bm{\omega_y}(J,\bm{J_y}) + l\epsilon\Omega = 0
\end{equation}
with
\begin{equation} \label{gen res condition adiabatic}
    \omega(J,\bm{J_y}) = \frac{\partial H_{0}(J,\bm{J_y})}{\partial J}, \quad \bm{\omega_y}(J,\bm{J_y}) = \frac{\partial H_{0}(J,\bm{J_y})}{\partial \bm{J_y}}.
\end{equation}
Ignoring the last two terms in Eq. \eqref{bar H gen slow}, which are second order in $\epsilon$, or equivalently ignoring the terms $\bm{m}\cdot\epsilon \bm{\omega_{y}}$ and $l\epsilon \Omega$ in the denominator of Eq. \eqref{gen fun S2}, removes the problem of small denominators since $k\omega \ne 0$. In this case, we rewrite Eqs. \eqref{bar H gen} and \eqref{gen fun S2} as:
\begin{equation}
    \begin{aligned} \label{bar H gen slow simplified}
        \bar{H}(\bar{J}, \bar{\bm{J_y}},\bar{\theta}, \bar{\bm{\theta_y}}) &= H_0(\bar{J}, \bar{\bm{J_y}}) + \epsilon H_{1(0)}(\bar{J}, \bar{\bm{J}}_{\bm{y}}, \bar{\bm{\theta}}_{\bm{y}},t) \\ 
        & \hspace{-0.8cm} + \epsilon\sum_{k \ne 0} H_{1(k)}(\bar{J}, \bar{\bm{J}}_{\bm{y}}, \bar{\bm{\theta}}_{\bm{y}},t) e^{i k\bar{\theta}}
        + \epsilon\omega (\bar{J}, \bm{\bar{J_y}})\frac{\partial S_1(\bar{J}, \bm{\bar{J_y}},\bar{\theta},\bar{\bm{\theta_y}},t)}{\partial \bar{{\theta}}}
    \end{aligned}
\end{equation}
and
\begin{equation}\label{gen fun S2 simplified}
    \begin{aligned} 
        S(\bar{J},\bar{\bm{J_y}}, \theta, \bm{\theta_y}, t) = \bar{J}\theta + \bar{\bm{J_y}}\bm{\theta_y} \\
        &\hspace{-1.8cm} + \epsilon i\sum_{k\ne0}{\frac{H_{1(k)}(\bar{J},\bar{\bm{J_y}},\bar{\bm{\theta}}_{\bm{y}},t)}{k\omega(\bar{J},\bar{\bm{J_y}})}e^{i k\theta }}.
    \end{aligned} 
\end{equation}
where
\begin{equation} \label{H_1,k}
    H_{1(k)}(\bar{J}, \bar{\bm{J}}_{\bm{y}}, {\bm{\theta}}_{\bm{y}},t) = \sum_{\bm{m},l}H_{1(k,\bm{m},l)}
        (\bar{J}, \bar{\bm{J_y}})e^{i(\bm{m}\cdot{\bm{\theta_y}} + l\Omega t)}, \quad k = \pm1, \pm2, \pm, 3, \dots
\end{equation}
Then, substituting Eqs. \eqref{gen fun S2 simplified} and \eqref{H_1,k} into Eq. \eqref{bar H gen slow simplified}, we observe that the two last terms in Eq. \eqref{bar H gen slow simplified} cancel each other out, eliminating the dependence of the new, barred system on the fast angle $\bar{\theta}$. Consequently, $\bar{J}$ becomes the adiabatic invariant of the motion. In this way, the problem of small denominators is overcome up to first order in $\epsilon$, providing an invariant of the motion, although the new system retains dependence on the slow time-varying angles and time. However, in cases where the Fourier series amplitudes $H_{1(k,\bm{m},l)}(J,\bm{J_y})$ for $(J, \bm{J_y})$ near the resonance condition \eqref{gen res condition adiabatic} do not approach zero for large values of $\bm{m}$ or for very large times (which can correspond to very large $l$), the terms $\bm{m}\cdot\epsilon \bm{\omega_{y}}$ and $l\epsilon \Omega$ are not negligible. Consequently, the problem of small denominators persists. As a result, we cannot completely remove the dependence of the new barred Hamiltonian \eqref{bar H(bar J)} on $\bar{\theta}$, and $\bar{J}$ cannot be considered a constant of motion, as it will vary over time. In such cases, particularly for systems with three or more degrees of freedom, the Arnold diffusion phenomenon arises and will be discussed in section \ref{Sec: Arnold Diffusion}.

\subsection{Secular Perturbation Theory (Resonant Islands formation)} \label{Secular Perturbation Theory}
The problem of small denominators is addressed using secular perturbation theory (Ref. \cite{Lieberman1992}, Sec. 2.4). According to this theory, a canonical transformation to a rotating coordinate system is employed. This transformation distinguishes the canonical coordinates that vary slowly (or 'adiabatically') with time from the 'fast' ones, particularly in the region of a resonance. This distinction enables the application of the canonical adiabatic theory discussed in the previous section.

Consider the near-integrable Hamiltonian \eqref{near integrable Hamiltonian}, up to first order in $\epsilon$, in the case of three degrees of freedom and time independence. We aim to examine the behavior of the system in the vicinity of a particular resonance, i.e., in the region of $\mathbf{J}$ for which the resonance condition for a specific $\bm{m_{\bm{rc}}} = (s, r, k)$ ($l = 0$ in the time-independent case) given by Eq. \eqref{gen resonance condition} is satisfied, i.e., 
\begin{equation} \label{rsk resonance}
    r\omega_1(\bm{J}) + s\omega_2(\bm{J}) + k\omega_3(\bm{J}) = 0,
\end{equation}
where the frequencies $\omega_1$, $\omega_2$, $\omega_3$ and as result the resonance condition are referred to the integrable Hamiltonian $H_0(\bm{J
})$. For this purpose, the canonical transformation to the rotating coordinate system associated with this particular resonance, using the generating function
\begin{equation} \label{J with hat J}
    F_2(\bm{\theta}, \hat{\bm{J}}) = (r\theta_1 + s\theta_2 + k\theta_3)\hat{J}_1 + \theta_2\hat{J}_2 + \theta_3\hat{J}_3
\end{equation}
is given by:
\begin{equation} \label{J to hat J}
    J_1 = r\hat{J}_1, \quad J_2 = \hat{J}_2 + s\hat{J}_1, \quad J_3 = \hat{J}_3 + k\hat{J}_1
\end{equation}
and
\begin{equation} \label{theta to hat theta}
    \hat{\theta}_1 = r\theta_1 + s\theta_2 + k\theta_3, \quad \hat{\theta}_2 = \theta_2, \quad \hat{\theta_3} = \theta_3.
\end{equation}
Substituting Eqs. \eqref{J to hat J} and \eqref{theta to hat theta}  in Eq. \eqref{near integrable Hamiltonian} we take:
\begin{equation} \label{hat J with m}
    \hat{H}(\hat{\bm{\theta}}, \hat{\bm{J}}) = \hat{H}_0(\hat{\bm{J}}) + \epsilon \sum_{\bm{m}}{\hat{H}_{1(\bm{m})}(\bm{\hat{J}})e^{\frac{i}{r}[m_1\hat{\theta}_1 + (m_2r-m_1s)\hat{\theta}_2 + (m_3 r-m_1 k)\hat{\theta}_3]}}.
\end{equation}
Writing:
\begin{equation} \label{p's with m's}
    m_1 = p_1r, \quad m_2 = p_2 + p_1s, \quad m_3 = p_3 + p_1k
\end{equation}
the Eq. \eqref{hat J with m} is expressed as:
\begin{equation} \label{hat J with p}
    \hat{H}(\hat{\bm{\theta}}, \hat{\bm{J}}) = \hat{H}_0(\hat{\bm{J}}) + \epsilon \sum_{\bm{p}}{\hat{H}_{1(p_1 r, p_2 + p_1s,p_3+p_1k)}(\bm{\hat{J}})e^{i(p_1\hat{\theta}_1 + p_2\hat{\theta}_2 + p_3\hat{\theta}_3)}}.
\end{equation}

Subsequently, we apply a canonical transformation from the $(\hat{\bm{\theta}}, \hat{\bm{J}})$ variables to the $(\bar{\bm{\theta}}, \bar{\bm{J}})$ variables, such that, in analogy to Eq. \eqref{bar H gen slow}, Eq. \eqref{hat J with p} is transformed into
\begin{equation} \label{hat H to bar H}
    \begin{aligned}
        \bar{H}(\bar{\bm{J}},\bar{\bm{\theta}}) = \hat{H}_{0}(\bar{\bm{J}}) + \epsilon \hat{H}_{1(0,0,0)}(\bar{\bm{J}}) + \epsilon \sum_{\bm{p}\ne 0}{\hat{H}_{1(p_1 r, p_2 + p_1s,p_3+p_1k)}(\bm{\bar{J}})e^{i(p_1\bar{\theta}_1 + p_2\bar{\theta}_2 + p_3\bar{\theta}_3)}} \\
        &\hspace{-9.0cm} + \epsilon^2\hat{\omega}_1\frac{\partial S_1(\bar{\bm{J}}, \bar{\bm{\theta}})}{\partial \bar{\theta}_1} + \epsilon \hat{\omega}_2\frac{\partial S_1(\bar{\bm{J}}, \bar{\bm{\theta}})}{\partial \bar{\theta}_2} + \epsilon \hat{\omega}_3\frac{\partial S_1(\bar{\bm{J}}, \bar{\bm{\theta}})}{\partial \bar{\theta}_3}.   
    \end{aligned}
\end{equation}
In analogy with Eq. \eqref{gen fun S2}, the generating function for this transformation is given by the expression:
\begin{equation} \label{S from hat to bar with p}
    \begin{aligned}
        S(\bar{\mathbf{J}}, \hat{\bm{\theta}},t) = \bar{\mathbf{J}}
        \cdot \hat{\bm{\theta}}\\
        &\hspace{-1.5cm}+ \epsilon i \sum_{\mathbf{p}\ne 0}{\frac{H_{1(p_r, p_2 + p_1s,p_3+p_1k)}(\bm{\bar{J}})\cdot e^{i(p_1\cdot\epsilon\hat{\theta}_1 + p_2\hat{\theta}_2 + p_3\hat{\theta}_3)}}{p_1\cdot \epsilon\hat{\omega}_1 + p_2\hat{\omega}_2 + p_3\hat{\omega}_3}}
    \end{aligned}
\end{equation}
where, using Eqs. \eqref{J with hat J}, along with $\hat{H}_0(\hat{\bm{J}}) = H_{0}(\bm{J}(\hat{\bm{J}}))$ and $\hat{\omega}_j = \partial \hat{H}_0 / \partial \hat{J}_j$ for $j = 1, 2, 3$, we can show that:
\begin{equation} \label{hat omega with omega}
    \hat{\omega}_1 = r\omega_1 + s\omega_2 + k\omega_3, \quad \hat{\omega}_2 = \omega_3, \quad \hat{\omega}_3 = \omega_3.
\end{equation}
From the first of Eqs. \eqref{hat omega with omega}, it is evident that in the vicinity of the resonance \eqref{rsk resonance}, $\hat{\omega}_1$ becomes very small. For this reason, it is multiplied by $\epsilon$ in Eqs. \eqref{hat H to bar H} and \eqref{S from hat to bar with p}. In other words, $\hat{\theta}_1$ behaves as a slowly varying angle over time, analogous to the $\bm{\theta}_{\bm{y}}$ angles in the canonical adiabatic theory discussed in Sec. \ref{Canonical Adiabatic Theory}. As a result, according to the canonical adiabatic theory, $\hat{\omega}_1$ can be ignored in the denominator of Eq. \eqref{S from hat to bar with p}. By doing so, small denominators in the resonance region arise only when the two remaining fast frequencies, $\hat{\omega}_2$ and $\hat{\omega}_3$, satisfy $p_2\hat{\omega}_2 + p_3\hat{\omega}_3 = 0$ or, from Eqs \eqref{hat omega with omega}, $p_2 \omega_2 + p_3 \omega_3 = 0$. 

\subsubsection{Isolated resonance}\label{Sec: isolated resonance}
In this section, we examine the case where the resonance \eqref{rsk resonance} is isolated. This means that no integers $(p_2, p_3)$ exist such that, for $\bm{J}$ in the resonance region described by Eq. \eqref{rsk resonance}, the condition $p_2 \omega_2 + p_3 \omega_3 = 0$ is satisfied. For this to hold, the Fourier amplitude $H_{1(p_r, p_2 + p_1s, p_3 + p_1k)}$ corresponding to $(p_2, p_3)$, for which $p_2 \hat{\omega}_2 + p_3 \hat{\omega}_3 = 0$, is assumed to vanish. Alternatively, we can describe this scenario as focusing on regions of $\bm{J}$ where the conditions of the KAM theorem (Ref. \cite{Lieberman1992}, Sec. 3.2a) are satisfied for $\hat{\omega}_2$ and $\hat{\omega}_3$. Specifically, these are domains where the ratio $\hat{\omega}_2 / \hat{\omega}_3$ is sufficiently irrational, ensuring that there are no integers $p_2, p_3$ for which $p_2 \omega_2 + p_3 \omega_3$ approaches zero, or that this only occurs for very large values of $p_2, p_3$, for which $H_{1(p_r, p_2 + p_1s, p_3 + p_1k)}$ approaches zero. As a result, the problem of small denominators does not arise in this context.

In such cases, the generating function given in Eq. \eqref{S from hat to bar with p} is well-defined. In consistency with the analysis in Sec. \ref{Canonical Adiabatic Theory}, substituting this function into Eq. \eqref{hat H to bar H} removes the dependence on $\hat{\theta}_2$ and $\hat{\theta}_3$, leaving $\hat{\theta}_1$ as the slow coordinate. Specifically, in analogy to Eqs. \eqref{bar H gen slow simplified} and \eqref{gen fun S2 simplified}, Eqs. \eqref{hat H to bar H} and \eqref{S from hat to bar with p} can be written as:
\begin{equation} \label{hat H to bar H simplified}
    \begin{aligned}
        \hspace{-2.3cm}\bar{H}(\bar{\bm{J}},\bar{\bm{\theta}}) = \hat{H}_{0}(\bar{\bm{J}}) + \epsilon \hat{H}^{p}_{1(0,0)}(\bar{\bm{J}}, \bar{\theta}_1) + \epsilon \sum_{p_2,p_3\ne 0}{\hat{H}^{p}_{1(p_2,p_3)}(\bar{\bm{J}}, \bar{\theta}_1)e^{i(p_2\bar{\theta}_2 + p_3\bar{\theta}_3)}} \\
        &\hspace{-9.0cm} + \epsilon \hat{\omega}_2\frac{\partial S_1(\bar{\bm{J}}, \bar{\bm{\theta}})}{\partial \bar{\theta}_2} + \epsilon \hat{\omega}_3\frac{\partial S_1(\bar{\bm{J}}, \bar{\bm{\theta}})}{\partial \bar{\theta}_3}   
    \end{aligned}
\end{equation}
and
\begin{equation}\label{gen fun S2 simplified p}
    \begin{aligned} 
        S(\bar{\bm{J}}, \hat{\bm{\theta}}) = \bar{\bm{J}}\hat{\bm{\theta}}  + \epsilon i\sum_{p_2,p_3\ne0}{\frac{\hat{H}^{p}_{1(p_2,p_3)}(\bar{\bm{J}},\hat{\theta}_1)}{p_2\hat{\omega}_2(\bar{\bm{J}}) + p_3\hat{\omega}_3(\bar{\bm{J}})}e^{i (p_2\hat{\theta}_2  + p_3\hat{\theta}_3) }}.
    \end{aligned} 
\end{equation}
where
\begin{equation} \label{H_1,p2,p3}
    \hat{H}^{p}_{1(p_2,p_3)}(\bar{\bm{J}}, \bar{\theta}_1) = \sum_{p_1}\hat{H}_{1(p_1 r, p_2 + p_1s,p_3+p_1k)}
        (\bar{\bm{J}})e^{i p_1\bar{\theta}_1}, \quad p_2,p_3 = \pm1, \pm2, \pm, 3, \dots
\end{equation}
Then, substituting Eq. \eqref{gen fun S2 simplified p} in Eq. \eqref{hat H to bar H simplified} the three last terms of Eq. \eqref{hat H to bar H simplified}
cancel each other out turning out, up to first order in $\epsilon$, the system dependent only on the $\bar{\theta}_1$ coordinate as 
\begin{equation} \label{Eq: bar H(hat theta_1)}
\begin{aligned}
             \bar{H}(\bar{\bm{J}},\bar{\bm{\theta}}) = \hat{H}_{0}(\bar{\bm{J}}) + \epsilon \hat{H}^{p}_{1(0,0)}(\bar{\bm{J}}, \bar{\theta}_1) \overset{\eqref{H_1,p2,p3}}{=}\\
         &  \hspace{-5cm}\hat{H}_{0}(\bar{\bm{J}}) + \epsilon \sum_{p_1 = -\infty}^{\infty}\hat{H}_{1(p_1 r, p_1s, p_1k)}
        (\bar{\bm{J}})e^{i p_1\bar{\theta}_1}
\end{aligned}
\end{equation}
In general, Fourier amplitudes decay rapidly as $p_1$ increases. Therefore, a good approximation of Eq. \eqref{Eq: bar H(hat theta_1)} can be obtained by considering only $p_1 = 0, \pm 1$ (see \cite{Lieberman1992}, Sec. 2.4a). Consequently, Eq. \eqref{Eq: bar H(hat theta_1)} simplifies to:
\begin{equation} \label{Eq: pendulum like H bar}
    \begin{aligned}
            \bar{H}(\bar{\bm{J}},\hat{\theta}_1) = \hat{H}_0(\bar{\bm{J}}) +          \epsilon \hat{H}_{1(0,0,0)}(\bar{\bm{J}}) \\
          & \hspace{-2.0cm}+ \epsilon  \hat{H}_{1(-r, -s, -k)}
        (\bar{\bm{J}}) e^{-i \bar{\theta}_1} +  \epsilon  \hat{H}_{1(r, s, k)}
        (\bar{\bm{J}}) e^{i\bar{\theta}_1}.
    \end{aligned}
\end{equation}

Using the canonical transformation from the $(\hat{\bm{J}}, \hat{\bm{\theta}})$ variables to the $(\bar{\bm{J}}, \bar{\bm{\theta}})$ variables with the generating function \eqref{gen fun S2 simplified p}, the coordinates $\bar{\theta}_2$ and $\bar{\theta}_3$ become cyclic. Consequently, their conjugate actions, $\bar{J}_2$ and $\bar{J}_3$, are adiabatic constants of motion. As discussed in Sec. \ref{Hamilton-Jacobi Theory}, a system with three degrees of freedom, such as the one studied here, is integrable if it admits three independent constants of motion that are in involution. Each specific set of these constants of motion defines a unique orbit of the system. In the case of system \eqref{Eq: bar H(hat theta_1)} or \eqref{Eq: pendulum like H bar}, the three constants of motion are the Hamiltonian $\bar{H}$ (since the system is autonomous) and the adiabatic constants $\bar{J}_2$ and $\bar{J}_3$. Hence, the system is integrable.

To analyze the qualitative behavior of the system \eqref{Eq: pendulum like H bar}, we calculate its fixed points for a specific set of constants of motion $(\bar{J}_1, \bar{J}_2)$ and varying $\bar{H}$ values, which label different orbits in phase space. Specifically, from the fix points analysis in Hamiltonian \eqref{Eq: pendulum like H bar} we obtain.
\begin{equation} \label{Eq: J fix point for bar H}
\begin{aligned}
            \dot{\bar{\theta}}_1=\frac{\partial \hat{H}_0(\bar{\bm{J}})}{\partial \bar{J}_1} + \epsilon \frac{\partial \hat{H}_{1(0,0,0)}(\bar{\bm{J}})}{\partial \bar{J}_1} 
             + \epsilon  \frac{\hat{H}_{1( -r, -s, -k)}
            (\bar{\bm{J}})}{\partial \bar{J}_1} e^{-i \bar{\theta}_1} +  \epsilon  \frac{\partial\hat{H}_{1(r, s, k)}
            (\bar{\bm{J}})}{\partial \bar{J}_1} e^{i\bar{\theta}_1}  = 0
    \end{aligned}
\end{equation}
and
\begin{equation} \label{Eq: theta fix point for bar H}
     \begin{aligned}
         \dot{\bar{J}}_1=-i\epsilon  \hat{H}_{1(-r, -s, -k)}
        (\bar{\bm{J}}) e^{-i \bar{\theta}_1} +  i\epsilon  \hat{H}_{1(r, s, k)}
        (\bar{\bm{J}}) e^{i\bar{\theta}_1} = 0.
     \end{aligned}
\end{equation}
From Eq. \eqref{Eq: theta fix point for bar H}, we observe that the fixed points along the $\bar{\theta}_1$ axis are located at $\bar{\theta}_1^0 = 0$ and $\bar{\theta}_1^0 = \pi$. Substituting these values into Eq. \eqref{Eq: J fix point for bar H} and solving for $\bar{J}_1$, the fixed points $\bar{J}_1^0$ along the $\bar{J}_1$ axis can be determined.

\textit{Pendulum approximation}.
Solving Eq. \eqref{Eq: J fix point for bar H} for $\bar{J}_1$ is rather complex. For this reason, an additional simplification of the Hamiltonian \eqref{Eq: pendulum like H bar} is applied. Specifically, from the equations of motion of Hamiltonian \eqref{Eq: pendulum like H bar} we easily observe that:
\begin{equation} \label{order of bar J_1 and bar theta_1}
    \dot{\bar{J}}_1 = \mathcal{O}(\epsilon|\hat{H}_{1(r,s,k)}(\bar{\bm{J}})|), \quad  \dot{\bar{\theta}}_1 = \mathcal{O}(1)
\end{equation}
Also, we can write $d \bar{J}_1 =  \dot{\bar{J}}_1/\dot{\bar{\theta}}_1d\bar{\theta}_1 = \mathcal{O}(\epsilon|\hat{H}_{1(r,s,k)}|).$ Integrating, we take 
\begin{equation} \label{Eq: Delta bar J_1}
\left.\Delta \bar{J}_1(\bar{\theta}_1)\right|_{\Delta\bar{\theta}_1} =  \mathcal{O}(\epsilon|\hat{H}_{1(r,s,k)}(\bar{\bm{J}})|)
\end{equation}

Which means that, in the plane $(\bar{\theta}_1, \bar{J}_1)$, the variation of $\bar{J}_1(\bar{\theta}_1)$ along an orbit will be of order $\epsilon$. This implies that $\bar{J}_1$ will always remain very close to its initial value throughout the entire motion. Taking into account that the Hamiltonian system \eqref{Eq: pendulum like H bar} can be regarded as a one degree of freedom system in $(\bar{\theta}_1, \bar{J}_1)$, we can show, using Eqs.\eqref{q,p with t}, that $(\bar{\theta}_1(t)$ and $\bar{J}_1(t))$ are periodic functions of time with same period and as a result the $\bar{J}_{1}(\bar{\theta}_1)$ of an orbit will be periodic function of $\bar{\theta}_1$.
Consequently, $\bar{J}_1$ both with respect to time and $\bar{\theta}_1$ oscillates periodically, remaining very close to its initial value. Additionally, since the actions $\bar{J}_2$ and $\bar{J}_3$ are constants of motion, they will retain their initial values throughout the evolution of the motion. 

\begin{figure}
    \centering
    \includegraphics[width=0.5\textwidth,keepaspectratio]{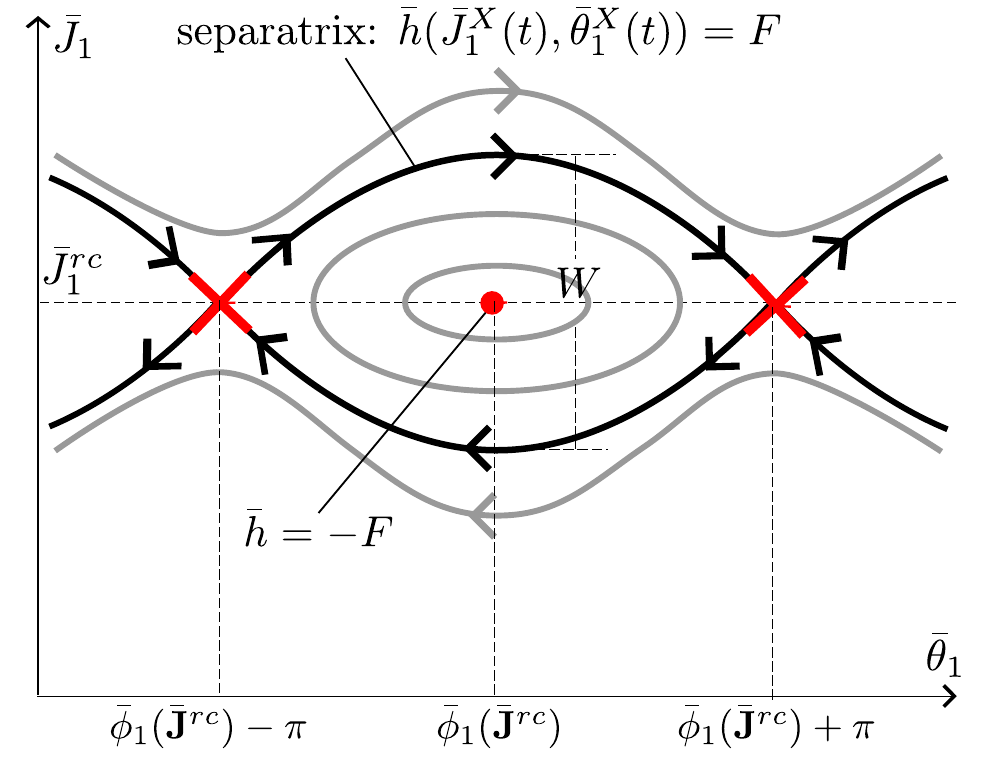}  
    \caption[Standard Hamiltonian phase space.]{The phase space $(\bar{\theta}_1,\bar{J}_1)$ of the pendulum-like Hamiltonian $h$, given in Eq. \eqref{Eq: pendulum Hamiltonian h}. The $x$-point and $o$-point are depicted in red.  The phase space is periodically repeated in $\bar{\theta}_1$ with a period of $2\pi$. $W$ is the island width in $(\bar{\theta_1},\bar{J}_1)$ space, Eq. \eqref{Eq: W in hat}}
    \label{fig:Fig31}
\end{figure}

We examine the local behavior of the system \eqref{Eq: pendulum like H bar} in the vicinity of a specific point denoted by $\bar{\bm{J}}^{rc}$. This is achieved by considering orbits with initial values at (or equivalently, orbits passing through) $\bar{\bm{J}}^{rc}$. As will be shown in the following (see Eq. \eqref{Eq: hat with bar canonical variables simplified}), $\bar{\bm{J}}^{rc}$, when transformed back to the original system, corresponds to $\bm{J}^{rc}$, which is a value of $\bm{J}$ that satisfies the resonance condition \eqref{rsk resonance}. Additionally, we allow for a free initial value of $\bar{\theta}_1$ (see Fig. \ref{fig:Fig31}). Then according to Eq. \eqref{Eq: Delta bar J_1} $\Delta \bar{J}_1 = \bar{J}_1 - \bar{J}_1^{rc} \sim \mathcal{O}(\epsilon|\hat{H}_{1(r,s,k)}(\bar{\bm{J}})|)$. The small magnitude of $\Delta \bar{J}_1$ throughout the entire $\bar{J}_1(\bar{\theta}_1)$ orbit, as taken by Hamiltonian \eqref{Eq: pendulum like H bar}, permits a Taylor expansion of the Hamiltonian around the initial value $\bar{\bm{J}}^{rc}$. By retaining terms up to a specified order, the expanded Hamiltonian closely approximates the original Hamiltonian. Consequently, the orbit derived from the expanded Hamiltonian serves as an  satisfactory approximation of the original orbit 

Subsequently, applying the Taylor expansion to each term on the right-hand side of Eq. \eqref{Eq: pendulum like H bar}, the first term becomes (note that $\Delta \bar{J}_2 = \bar{J}_2 - \bar{J}_2^{rc} = 0$ and $\Delta \bar{J}_3 = \bar{J}_3 - \bar{J}_3^{rc} = 0$ in Hamiltonian \eqref{Eq: pendulum like H bar}, since $\bar{\theta}_2$ and $\bar{\theta}_3$ are cyclic.):
  \begin{equation} \label{Eq: Taylor expanted hat H_0}
      \hat{H}(\bar{\bm{J}}) = \hat{H}_0(\bar{\bm{J}}^{rc}) + \left.\frac{\partial \hat{H}_0}{\partial \bar{J}_1}\right|_{\bar{\bm{J}} = \bar{\bm{J}}^{rc}}\cdot\Delta \bar{J}_1 +  \frac{1}{2}\left.\frac{\partial^2 \hat{H}_0}{\partial \bar{J}_1^2}\right|_{\bar{\bm{J}} = \bar{\bm{J}}^{rc}}\cdot(\Delta \bar{J}_1)^2 
  \end{equation}
where terms of order $(\Delta \bar{J}_1)^3$ and higher are neglected due to their small magnitude. The coefficient of the second term, based on the relation $\hat{H}_0(\hat{\bm{J}}) = H_{0}(\bm{J}(\hat{\bm{J}}))$ and substituting $\hat{\bm{J}}$ with $\bar{\bm{J}}$, is expressed as:
\begin{equation}
    \begin{aligned}
                \left.\frac{\partial \hat{H}_0}{\partial \bar{J}_1}\right|_{\bar{\bm{J}} = \bar{\bm{J}}^{rc}} = \left.\frac{\partial H_0(\bm{J})}{\partial J_1}\right|_{{\bm{J}} = {\bm{J}}^{rc}}\cdot\frac{\partial J_1}{\partial \bar{J}_1} \\
            & \hspace{-2.0cm}+ \left.\frac{\partial H_0(\bm{J})}{\partial J_2}\right|_{{\bm{J}} = {\bm{J}}^{rc}}\cdot\frac{\partial J_2}{\partial \bar{J}_1} + \left.\frac{\partial H_0(\bm{J})}{\partial J_3}\right|_{{\bm{J}} = {\bm{J}}^{rc}}\cdot\frac{\partial J_3}{\partial \bar{J}_1}
    \end{aligned}
\end{equation}
which using Eq. \eqref{J with hat J}  and replacing $\hat{J}$'s with $\bar{J}$ 's becomes \footnote{As discussed in Section \ref{Classical Perturbation Theory}, after the transformation from $(\hat{\bm{J}}, \hat{\bm{\theta}})$ to $(\bar{\bm{J}}, \bar{\bm{\theta}})$, the functional dependence of the Hamiltonian on either the barred or hatted canonical variables remains identical, allowing direct substitution.}:
\begin{equation}\label{Eq: first pd of hat H_0 with bar J}
    \begin{aligned}
        \left.\frac{\partial \hat{H}_0}{\partial \bar{J}_1}\right|_{\bar{\bm{J}} = \bar{\bm{J}}^{rc}} = r\omega_1(\bm{J}^{rc}) +s\omega_2(\bm{J}^{rc}) + k\omega_3(\bm{J}^{rc}) = 0
    \end{aligned}
\end{equation}
since we have assumed that $\bm{J}_{rc}$ represents the values of $\bm{J}$ that satisfy Eq. \eqref{rsk resonance}. The vanishing of this term, i.e., the second term in Eq. \eqref{Eq: Taylor expanted hat H_0}, is the reason we chose to consider orbits passing through the point $\bar{J}_1^{rc}$ and to expand the Hamiltonian around it. However, even for orbits passing through a point very close to $\bar{J}_1^{rc}$ and performing the expansion at that point, this term would still approach zero.

Expanding the second term on the right-hand side of Eq. \eqref{Eq: pendulum like H bar}, we take:
\begin{equation} \label{Eq: expanded hat H_1,0,0,0}
     \epsilon \hat{H}_{1(0,0,0)}(\bar{\bm{J}}) = \epsilon \hat{H}_{1(0,0,0)}(\bar{\bm{J}}^{rc}) + \epsilon \left.\frac{\partial \hat{H}_{1(0,0,0)}}{\partial \bar{J}_1}\right|_{\bar{J} = \bar{J}^{rc}}\cdot\Delta \bar{J}_1 + \dots
\end{equation}
Then, by keeping only the terms to the lowest order in $\epsilon$ and in $\Delta \bar{J}_1$ (see also Ref. \cite{Lieberman1992}, Sec. 2.4a.), the second term in Eq. \eqref{Eq: expanded hat H_1,0,0,0}, along with terms of higher order, are neglected. We can then consider the second term of the Hamiltonian in Eq. \eqref{Eq: pendulum like H bar} as contributing to the motion as a constant, $\epsilon \hat{H}_{1(0,0,0)}(\bar{\bm{J}}^{rc})$. The addition of a constant in a Hamiltonian does not affect the equations of motion. For this reason, we neglect the constant $\epsilon \hat{H}_{1(0,0,0)}(\bar{\bm{J}}^{rc})$ as well as the constant term $\hat{H}_0(\bar{\bm{J}}^{rc})$ raised in Eq. \eqref{Eq: Taylor expanted hat H_0} from the Hamiltonian \eqref{Eq: pendulum like H bar} after the expansion around $\bar{\bm{J}}^{rc}$.

Applying the same expansion, with the same ordering, to the coefficients of the two last terms of Eq. \eqref{Eq: pendulum like H bar}, these terms become $\epsilon \hat{H}_{1(-r, -s, -k)}(\bar{\bm{J}}^{rc}) e^{-i \bar{\theta}_1} + \epsilon \hat{H}_{1( r, s, k)}(\bar{\bm{J}}^{rc}) e^{i \bar{\theta}_1}$. Which, as Fourier series terms, can be written as
\begin{equation}
    \epsilon  \hat{H}_{1(-r, -s, -k)}
    (\bar{\bm{J}}^{rc}) e^{-i \bar{\theta}_1} +  \epsilon  \hat{H}_{1(r, s, k)}
    (\bar{\bm{J}}^{rc}) e^{i\bar{\theta}_1} = 2\epsilon|\hat{H}_{1(r, s, k)}
    (\bar{\bm{J}}^{rc})|\cos\left(\bar{\theta}_1 - \bar{\phi}_1(\bar{\bm{J}}^{rc})\right), 
\end{equation}
where the phase $\phi_1$ is $\phi_{1} = \arctan \left(-Im(\hat{H}_{1(r, s, k)}(\bar{\bm{J}}^{rc})/ Re(\hat{H}_{1(r, s, k)}(\bar{\bm{J}}^{rc})\right)$.
        
Finally, applying the above-mentioned simplifications to Hamiltonian \eqref{Eq: pendulum like H bar}, its approximation near the resonance, denoted here by $h$, is given by:
\begin{equation} \label{Eq: pendulum Hamiltonian h}
    h(\bar{\theta}_1, \bar{J}_1) = \frac{1}{2}G\cdot(\bar{J}_1 - \bar{J}_1^{rc})^2 - F\cdot \cos\left(\bar{\theta}_1 - \bar{\phi}_1(\bar{\bm{J}}^{rc})\right),
\end{equation}
where, having kept the notation of Lichtenberg and Lieberman \cite{Lieberman1992},
\begin{equation} \label{Eq: G bar J}
    G(\bar{\bm{J}}^{rc}) =   \left.\frac{\partial^2 \hat{H}_0}{\partial \bar{J}_1^2}\right|_{\bar{\bm{J}} = \bar{\bm{J}}^{rc}}
\end{equation}
and 
\begin{equation} \label{Eq: F bar J}
    F(\bar{\bm{J}}^{rc}) = -2\epsilon|\hat{H}_{1(r, s, k)}
        (\bar{\bm{J}}^{rc})|.  
\end{equation}

The Hamiltonian $h$ shows that, in the vicinity of an isolated resonance, the motion of a near-integrable system, as given by Eq. \eqref{near integrable Hamiltonian}, up to first order in $\epsilon$ and neglecting terms proportional to in $\epsilon\Delta \bar{J}_1$ (which is of order $\mathcal{O}(\epsilon^{3/2})$ as will be shown in the following), behaves similarly to that of a pendulum, a qualitative picture of which is seen in Fig. \ref{fig:Fig31}. Like the pendulum, this motion involves separatrix, libration and rotational motion in phase space. This framework has been widely used to describe the generic motion of Hamiltonian systems near resonance \cite{Chiricov1979, Lieberman1992}. For Lichtenberg and Lieberman \cite{Lieberman1992}, $h$ serves as the foundation for analyzing chaotic motion near the separatrices associated with resonances. Eq. \eqref{Eq: pendulum Hamiltonian h} provides a universal description of motion near all resonances, which is why it is referred to as "the standard Hamiltonian" (Ref. \cite{Lieberman1992}, Sec. 2.4a).

From a fixed-point analysis of the Hamiltonian $h$, similarly to the pendulum, we determine the locations of the fixed points by solving the following system of equations
\begin{equation}
    \dot{\bar{J}}_1 = -\frac{\partial h}{\partial \bar{\theta}_1} = 0, \quad     \dot{\bar{\theta}}_1 = \frac{\partial h}{\partial \bar{J}_1} = 0.
\end{equation}
The solutions for the stable or elliptic ($o$-point) and unstable or hyperbolic ($x$-point) are shown in Table \ref{tab: fix points}. Note at this point that the locations of the $o$ and $x$ - points calculated using the Hamiltonian $h$ closely approximate the location of the $o$ and $x$ - points as it would be given by the original Hamiltonian \eqref{Eq: pendulum like H bar}. Specifically if one wants to find the exact location of these points has to solve the system of equations \eqref{Eq: theta fix point for bar H} and \eqref{Eq: J fix point for bar H}.

\begin{table}[h!]
    \centering
    \begin{tabular}{l l l}
        \toprule
        GF & $o$ - point & $x$ - point \\
        \midrule
        $\geq 0$ & $\bar{J}_1 = \bar{J}_1^{rc}$, \quad $\bar{\theta}_1 = \bar{\phi}_1(\bar{\bm{J}}^{rc})$ & $\bar{J}_1 = \bar{J}_1^{rc}$, \quad $\bar{\theta}_1 = \pi + \bar{\phi}_1(\bar{\bm{J}}^{rc})$ \\
        $\leq 0$ & $\bar{J}_1 = \bar{J}_1^{rc}$, \quad $\bar{\theta}_1 = \pi + \bar{\phi}_1(\bar{\bm{J}}^{rc})$ & $\bar{J}_1 = \bar{J}_1^{rc}$, \quad $\bar{\theta}_1 = \bar{\phi}_1(\bar{\bm{J}}^{rc})$ \\
        \bottomrule
    \end{tabular}
    \caption[Fixed-points of the standard Hamiltonian.]{The locations of the elliptic ($o$-point) and the hyperbolic ($x$-point) of the Hamiltonian Eq. \eqref{Eq: pendulum Hamiltonian h}.}
    \label{tab: fix points}
\end{table}

Similar to the pendulum, the orbits around the $o$-point in the $(\bar{\theta}_1, \bar{J}_1)$ plane have an elliptic shape (see Fig. \ref{fig:Fig31}). The outermost elliptic orbit, i.e., the one with the maximum excursion of $\Delta \bar{J}_1$ from the center of the ellipse located at the $o$-point (see Table \ref{tab: fix points}), is the separatrix. The separatrix is a special orbit passing through the $x$-point at infinitely large times and can be determined by substituting the coordinates of the $x$-point into $h$. The maximum width of the separatrix is twice the maximum excursion in $\bar{J}_1$, which occurs at $\bar{\theta}_1 = 0$, and is given by (Fig. \ref{fig:Fig31}):
\begin{equation}\label{Eq: W in hat}
    W = 2\Delta \bar{J}_1 =  4\left(\frac{F}{G}\right)^{1/2} = \mathcal{O}(\epsilon^{1/2})
\end{equation}

\textit{Backward transformation from $(\bar{\bm{\theta}}, \bar{\bm{J}})$ to $(\bm{\theta}, \bm{J})$}. After determining the phase-space structure near the resonance in the $(\bar{\theta}_1, \bar{J}_1)$ phase space, a qualitative depiction of which is shown in Fig. \ref{fig:Fig31}, where $\bar{J}_2$ and $\bar{J}_3$ remain constants of motion equal to $\bar{J}_2^{rc}$ and $\bar{J}_3^{rc}$, respectively, the next question addresses the behavior in the original phase space $(\bm{\theta}, \bm{J})$. To address this, we will utilize the two canonical transformations that lead from the near-integrable Hamiltonian \eqref{near integrable Hamiltonian} to the Hamiltonian \eqref{Eq: pendulum like H bar}, but in reverse order. Starting with the second canonical transformation, we move from  $(\bar{\bm{\theta}}, \bar{\bm{J}})$ to $(\hat{\bm{\theta}}, \hat{\bm{J}})$ using the generating function \eqref{S from hat to bar with p}. To first order in $\epsilon$, this generating function provides the transformation equations as:
\begin{equation} \label{Eq: hat with bar canonical variables}
    \hat{\bm{J}} = \bar{\bm{J}} - \epsilon \frac{\partial S_1(\bar{\bm{J}}, \bar{\bm{\theta}})}{\partial \bar{\bm{\theta}}}, \quad     \hat{\bm{\theta}} = \bar{\bm{\theta}} - \epsilon \frac{\partial S_1(\bar{\bm{J}}, \bar{\bm{\theta}})}{\partial \bar{\bm{J}}}
\end{equation}
where:
\begin{equation}
    \frac{\partial S_1(\bm{\bar{J}},\bar{\bm{\theta}})}{\partial \bar{\bm{\theta}}} = \left.\frac{\partial S_1(\bm{\bar{J}},{\bm{\theta}})}{\partial {\bm{\theta}}}\right|_{\bm{\theta} = \bar{\bm{\theta}}}.
\end{equation}
For orbits passing through the resonant actions $\bar{\bm{J}}^{rc}$, $S_1$ in Eqs. \eqref{Eq: hat with bar canonical variables} is expanded in the same way that the Hamiltonian \eqref{Eq: pendulum like H bar}, $\bar{H}$, was expanded around $\bar{\bm{J}}^{rc}$, yielding:
\begin{equation} \label{Eq: hat with bar canonical variables expanded}
    \hat{\bm{J}} = \bar{\bm{J}} - \epsilon\Delta\bar{J}_1 \left.\frac{\partial^{2} S_1(\bar{\bm{J}}, \bar{\bm{\theta}})}{\partial \bar{J}_1 \partial \bar{\bm{\theta}}}\right|_{\bar{\bm{J}}=\bar{\bm{J}}^{rc}}, \quad     \hat{\bm{\theta}} = \bar{\bm{\theta}} - \epsilon\Delta\bar{J}_1 \left.\frac{\partial^{2} S_1(\bar{\bm{J}}, \bar{\bm{\theta}})}{\partial \bar{J}_1 \partial \bar{\bm{J}}}\right|_{\bar{\bm{J}}=\bar{\bm{J}}^{rc}}
\end{equation}
Here, the terms with $\Delta\bar{J}_{2,3}$ become zero because $\Delta\bar{J}_{2,3} = \bar{J}_{2,3} - \bar{J}_{2,3}^{rc} = 0$, as $\bar{J}_2$ and $\bar{J}_3$ are constants of motion and the orbit passes through $\bar{J}_2^{rc}$ and $\bar{J}_3^{rc}$. Additionally, as in the derivation of $h$ from $\bar{H}$, we neglect the small terms $\epsilon\Delta\bar{J}_1 = \mathcal{O}(\epsilon^{3/2})$. Therefore, Eqs. \eqref{Eq: hat with bar canonical variables expanded} simplify to:
\begin{equation} \label{Eq: hat with bar canonical variables simplified}
    \hat{\bm{J}} = \bar{\bm{J}} , \quad     \hat{\bm{\theta}} = \bar{\bm{\theta}}. 
\end{equation}
This proves that very close to the resonance, the transformation from $(\bar{\bm{\theta}}, \bar{\bm{J}})$ to $(\hat{\bm{\theta}}, \hat{\bm{J}})$, and vice versa, can be considered an identity transformation. Consequently, using the canonical transformation equations from $(\bm{\theta}, \bm{J})$ to $(\hat{\bm{\theta}}, \hat{\bm{J}})$, Eqs. \eqref{theta to hat theta} and \eqref{J to hat J}, we can directly express the original action-angle variables in terms of the barred canonical variables.

The $o$-point orbit, provided in Table \ref{tab: fix points}, is a fixed point, meaning that an orbit starting from it remains at the same point $(\bar{\theta}_1, \bar{J}_1)$ at all times. We aim to transform this orbit into the $(\bm{\theta}, \bm{J})$ variables. For $GF > 0$ (a similar procedure applies for $GF < 0$) and using Eqs. \eqref{Eq: hat with bar canonical variables simplified} and \eqref{theta to hat theta}, we obtain the following expressions for $\bm{\theta}$ of the $o$-point:
\begin{equation} \label{Eq: theta of o and  x points}
     \bar{\phi}_1(\bar{\bm{J}}^{rc}) = r\theta_1(t) + s\theta_2(t) + k\theta_3(t), \quad \bar{\theta}_2(t) = \theta_2(t), \quad \bar{\theta}_3(t) = \theta_3(t)
\end{equation}
From Eq. \eqref{Eq: pendulum like H bar}, we can compute $\dot{\bar{\theta}}_{2,3}$ for the $o$-point orbit as:
\begin{equation}
    \dot{\bar{\theta}}_{2,3} = \left.\frac{\partial\bar{H}}{\partial\bar{J}_{2,3}}\right|_{\bm{\bar{J}} = \bar{\bm{J}}^{rc}, \bar{\theta}_1 = 2\kappa\pi + \bar{\phi}_1(\bar{\bm{J}}^{rc}) }.
\end{equation}
The right-hand side of this equation, depends only on the constant actions $\bar{J}_{2,3}^{rc}$ and $\bar{J}_1^{rc}$, which is also constant since the orbit is a fixed point. Thus, we can write $\dot{\bar{\theta}}_{2,3} = \bar{\omega}_{2,3}(\bar{\bm{J}}^{rc})$, which is constant. Additionally, by definition, for the fixed point, we have $\dot{\bar{\theta}}_1 = 0$. Then, substituting in Eqs. \eqref{theta to hat theta}, we obtain::
\begin{equation}\label{Eq: dot theta at o  point}
    \dot{\theta}_1 = -\frac{s}{r}\bar{\omega}_2(\bar{\bm{J}}^{rc}) - \frac{k}{r}\bar{\omega}_3(\bar{\bm{J}}^{rc}), \quad \dot{\theta}_2 = \bar{\omega}_2(\bar{\bm{J}}^{rc}), \quad \dot{\theta}_2 = \bar{\omega}_3(\bar{\bm{J}}^{rc}).
\end{equation}
From this and transforming the $\bar{\bm{J}}^{rc}$ to $\bm{J}^{rc}$ with Eqs. \eqref{J to hat J}, it can be seen that the following resonance condition is satisfied:
\begin{equation}\label{Eq: rsk resonance o-point}
     r\dot{\theta}_1(\bm{J}^{rc}) + s\dot{\theta}_2(\bm{J}^{rc}) + k\dot{\theta}_3(\bm{J}^{rc}) = 0
\end{equation}

Note that, in contrast to the resonance condition in Eq. \eqref{rsk resonance}, which applies to the integrable system and holds for an infinite number of orbits starting from different angles and on the torus $\bm{J}^{rc}$, the coordinates $\theta_1$, $\theta_2$ and $\theta_3$ in the resonance equation \eqref{Eq: rsk resonance o-point} correspond to the near-integrable system. This resonance condition pertains to a single orbit, i.e., the $o$-point orbit. Additionally, note that $\dot{\theta}_1$, $\dot{\theta}_2$, and $\dot{\theta}_3$ are constant in time.

\begin{figure}
    \centering
    \includegraphics[width=1.0\textwidth,keepaspectratio]{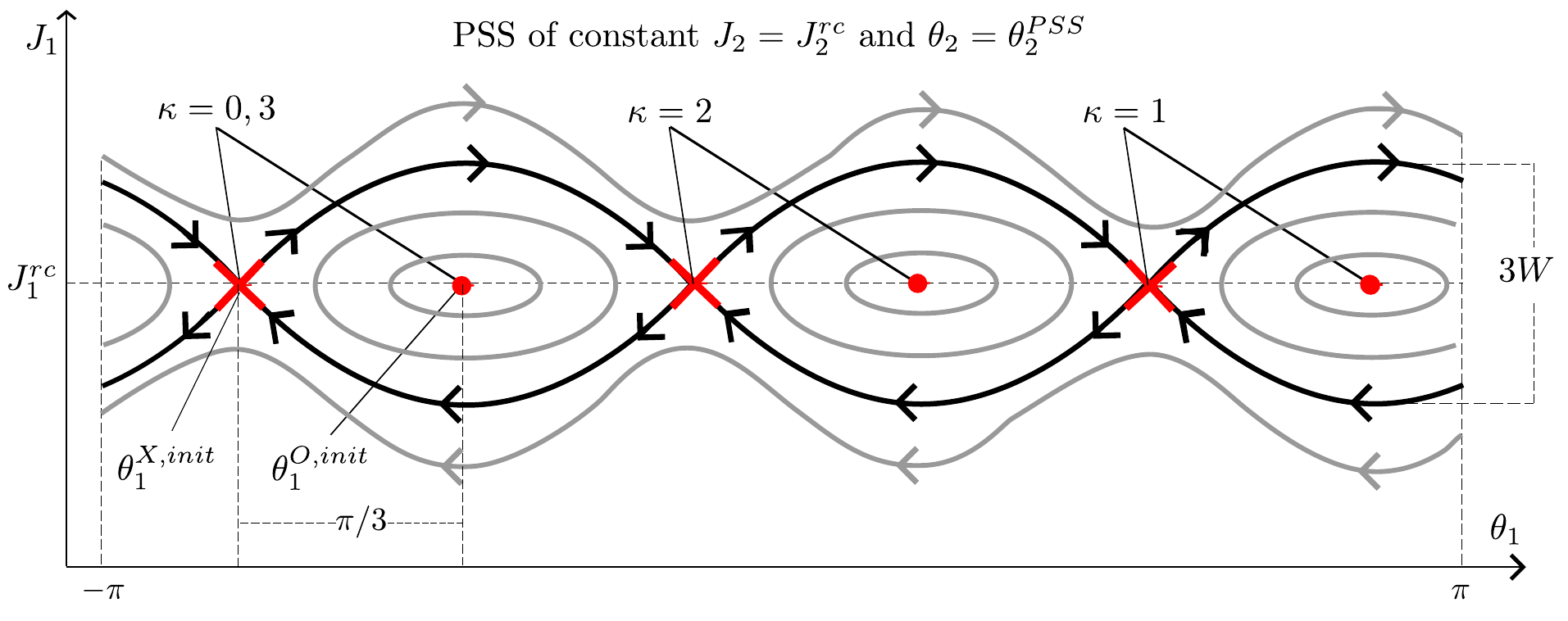}  
    \caption[A qualitative depiction of a resonance island chain in a Poincaré surface of section.]{A qualitative representation of a Poincaré surface of section (PSS) for a constant $J_2$ equal to the resonant action $J_2^{rc}$ and a fixed angle $\theta_2 = \theta_2^{PSS}$ in the near-integrable system. On this surface, $\dot{\theta}_2$ is chosen to be either positive or negative. Here, we consider a resonance with $r = 3$ and $s = 1$. On the resonant torus $\bm{J}^{rc} = (J_1^{rc}, J_2^{rc})$, the two resonant orbits that persist after the perturbation are depicted as red $x$- and $o$-points, respectively. This plot can be viewed as Fig. \ref{fig:Fig31} transformed back to the $(\bm{\theta}, \bm{J})$ phase space via the canonical transformation equations Eqs. \eqref{Eq: hat with bar canonical variables simplified}, \eqref{J to hat J}, and \eqref{theta to hat theta}. The width of the resonance islands is $r$ times the width of the island in the $(\bm{\bar{\theta}}, \bm{\bar{J}})$ phase space, as dictated by Eq. \eqref{Eq: width of island in action angle}.}
    \label{fig:Fig32}
\end{figure}

For the $\bm{J}$ of the $o$ - point:
\begin{equation}\label{Eq: J for o and x points}
    J_1(t) = r\bar{J}_1^{rc}, \quad J_2(t) = \bar{J}_2^{rc} + s\bar{J}_1^{rc}, \quad J_3(t) = \bar{J}_3^{rc} + k\bar{J}_1^{rc} 
\end{equation}
which is exactly the definition of $\bm{J}^{rc}$ in terms of $\bar{\bm{J}}^{rc}$ given above, yielding:
\begin{equation} \label{J of o point}
    J_1(t) = J_1^{rc}, \quad J_2(t) = J_2^{rc}, \quad J_3(t) = J_3^{rc} 
\end{equation}
with 
\begin{equation} \label{dot J of o point}
    \dot{J}_1 = 0, \quad \dot{J}_2 = 0, \quad \dot{J}_3 = 0 
\end{equation}
From Eqs. \eqref{Eq: rsk resonance o-point}, \eqref{J of o point}, \eqref{dot J of o point}, and considering that $\dot{\theta}_1$, $\dot{\theta}_2$, and $\dot{\theta}_3$ are constant orbital frequencies in time, we can conclude that the orbit of the $o$-point corresponds to a resonant orbit of the near-integrable system on the resonant torus ${\bm{J}}^{rc}$, satisfying the resonance condition \eqref{rsk resonance} of the integrable system $H_0(\bm{J})$. Such an orbit corresponds to periodic motion in time, as discussed in Sec. \ref{Action Angle Variables}. The same procedure can be applied to the $x$-point, leading to the same conclusion. Thus, we have shown that the two orbits of the near-integrable system corresponding to the singular points of the Hamiltonian $h$—the stable ($o$-point) and the unstable ($x$-point)—which start from a resonant torus, remain resonant orbits. As a result, the $(\bm{q}(t), \bm{p}(t))$ canonical variables corresponding to these orbits are periodic in time. This is aligned with the Poincaré-Birkhoff theorem 
\cite{Meletlidou2015}, which states that \textit{under the influence of the perturbation $\epsilon H_1$, from the infinite set of periodic orbits carried by a resonant torus, Fig. \ref{fig:Fig30}, only a finite even number of these persist as periodic orbits. Typically, only two periodic orbits survive (Fig. \ref{fig:Fig32}): one stable and the other unstable. Additional periodic orbits persist only under conditions involving symmetries in the system} (See also Sec 3.2b in Ref. \cite{Lieberman1992}).

The specific orbit in the phase space of $h$ that passes through the $x$-point is the separatrix (Fig. \ref{fig:Fig31}). In the phase space of $h$, the separatrix orbit, denoted as $(\bar{J}^{X}_1(t), \bar{\theta}^{X}_1(t))$, is the outermost orbit passing through $\bar{J}_1^{rc}$ and does so at $\bar{\theta}_1 = \pm\pi$ at infinitely large times. This orbit asymptotically approaches the $x$-point as time progresses. Substituting the $x$-point into Eq. \eqref{Eq: pendulum Hamiltonian h}, it is found that $h = F$, meaning that the separatrix satisfies the following equation:
\begin{equation} \label{Eq: separatrix of h} 
    F = \frac{1}{2}G(\bar{J}^{X}_1(t) - \bar{J}_1^{rc})^2 - F\cos{(\bar{\theta}^{X}_1(t) - \bar{\phi}_1)}.
\end{equation}

The canonical transformations allow us to take the desired initial conditions in the space $(\bm{\theta}, \bm{J})$ and transform them into the space $(\bar{\bm{\theta}}, \bar{\bm{J}})$. Instead of solving the equations of motion of the near-integrable system \eqref{near integrable Hamiltonian}, we solve the equations of motion derived from $h$ to obtain the solution $(\bar{\bm{\theta}}(t), \bar{\bm{J}}(t))$. Then, through the equations of the canonical transformation, we determine the solution in the original $(\bm{\theta}, \bm{J})$ space.

Denote the orbit that arises from the backward canonical transformation of the separatrix orbit $(\bar{J}^{X}_1(t), \bar{\theta}^{X}_1(t))$ of $h$ as $(\bm{\theta}^{X}(t), \bm{J}^{X}(t))$.  
The initial conditions for the orbit of the near-integrable system that $(\bm{\theta}^{X}(t), \bm{J}^{X}(t))$ approximates\footnote{We use the term "approximates" because, during the derivation from the near-integrable Hamiltonian \eqref{near integrable Hamiltonian} to $h$ via canonical transformations, terms of order $\epsilon^2$ and higher, as well as terms proportional to $\epsilon\Delta\bar{J}_1$, were neglected. Thus, while in a time-independent system the relationship $H = h$ should hold at every point of the orbit, in this case, what holds is $H \simeq h$, as the canonical transformation is only approximate. Additionally, the inversion of the canonical transformation equation giving the new canonical variables in terms of the old, or reversely, involves lower-order in $\epsilon$ approximations.
} are as follows:  
\begin{itemize}
    \item ${\bm{J}}^{X,init} = {\bm{J}}^{rc}$, such that $\bar{\bm{J}}^{X,init} = \bar{\bm{J}}^{rc}$  
    \item $\theta^{X,init}_1$, $\theta^{X,init}_2$, and $\theta^{X,init}_3$ are such that    
     $   r \theta^{X,init}_1 + s\theta^{X,init}_2 + k\theta^{X,init}_3 = \bar{\theta}^{X,init}_1 = \pi + \bar{\phi}_1(\bar{\bm{J}}^{rc})$,
    which is the value of $\bar{\theta}_1$ for the separatrix as it passes through $\bar{J}_1^{rc}$, i.e., the $x$-point (see Fig. \ref{fig:Fig31}).
\end{itemize}

By calculating $\bar{\theta}_2^{X}(t)$ and $\bar{\theta}_3^{X}(t)$ using the equations of motion of the Hamiltonian \eqref{Eq: pendulum like H bar} and transforming back to $(\bm{\theta}, \bm{J})$ variables via Eqs. \eqref{Eq: hat with bar canonical variables simplified}, \eqref{theta to hat theta}, and \eqref{J to hat J}, Eq. \eqref{Eq: separatrix of h} can be rewritten as:
\begin{equation} \label{Eq: separatrix of h in J, theta}
    F = \frac{1}{2}\frac{G}{r^2}(J_1^{X}(t) - J_1^{rc})^2 - F\cos{(r\theta^{X}_1(t) + s\theta^{X}_2(t) + k\theta^{X}_3(t) - \bar{\phi}_1)}.
\end{equation}

To visualize in two dimensions the behavior of the six dimensional orbit $(\bm{\theta}^{X}, \bm{J}^{X})$, and consequently the orbit of the near integrable system with the above initial conditions that is approximated by $(\bm{\theta}^{X}, \bm{J}^{X})$, a Poincaré map is needed. For simplicity, consider a system with two degrees of freedom (four dimensions), i.e., variables with the index 3 are excluded from the derivation. To examine orbits starting from the resonant torus ${\bm{J}}^{rc}$, we choose a Poincaré surface of section (PSS) of constant $J_2 = J_2^{rc}$ and consider a constant section at $\theta_2 = \theta_2^{PSS}$. The Poincaré map is constructed using points on this surface whenever the orbits cross it in a specific direction (positive or negative), i.e., whenever $\theta_2$ completes a full circle (Ref. \cite{Meletlidou2015}, Sec. 8.4.1). The structure of the $(J_1, \theta_1)$ PSS diagram repeats periodically with period $2\pi$ in $\theta_1$, similar to the phase space of $h$ (Fig.~\ref{fig:Fig31}). This periodicity follows from the equations of the canonical transformations, Eqs.~\eqref{Eq: hat with bar canonical variables simplified}, \eqref{J to hat J}, and \eqref{theta to hat theta}, for fixed values of $J_2$ and $\theta_2$, as required for the PSS diagram.  

On this surface of section, orbits corresponding to the fixed points of $h$—with initial conditions for the $x$- and $o$-points given by  
\begin{equation}\label{Eq: theta x and o init}
    \theta_1^{X,init} = \frac{\pi}{r} + \frac{1}{r} \bar{\phi}_1(\bar{\bm{J}}^{rc}) - \frac{s}{r} \theta_2^{X,init}, \quad 
    \theta_1^{O,init} = \frac{1}{r} \bar{\phi}_1(\bar{\bm{J}}^{rc}) - \frac{s}{r} \theta_2^{O,init}
\end{equation}
—correspond to resonant orbits on the resonant torus. We choose $\theta_2^{X,init} = \theta_2^{O,init} = \theta_2^{PSS}$. The relation \eqref{Eq: rsk resonance o-point} connecting the constant frequencies of the angles of these orbits shows that, on the Poincaré surface, these orbits will periodically pass through particular points each time $\theta_2$ completes a full circle. Specifically, starting from $(\theta_1^{X,O,init}, \theta_2^{PSS}, J_1^{rc}, J_2^{rc})$, the $x$- and $o$-points will pass, for each time, at $J_1(t) = J_1^{rc}$ and $J_2 =(t) J_2^{rc}$ (Eqs. \eqref{J of o point}) through:  
\begin{equation} \label{Eq: o and x points at J theta}
    \theta_{1(\kappa)}^{X,O} = \theta_{1}^{X,O,init} - 2s\frac{\kappa}{r}\pi, \quad \kappa = 0,1,2,\dots
\end{equation}  
after each successive circle of $\theta_2$ denoted with $\theta_{2,\kappa}^{X,O} = 2\kappa\pi + \theta_2^{PSS}$ (Eqs. \eqref{Eq: theta of o and  x points}, \eqref{Eq: theta x and o init}). Notably, when $\kappa = r$, the points return to their initial values, $\theta_{1(r)} = \theta_1^{X,O,init}$. For this reason, the Poincaré surface of section will show $r$ points corresponding to the $x$-point of $h$ and another $r$ points corresponding to the $o$-point of $h$ located at the resonant torus $(J_1, J_2) = (J_1^{rc}, J_2^{rc})$ (see Fig. \ref{fig:Fig32}).

In the phase space of $h$, the separatrix fulfilling Eq. \eqref{Eq: separatrix of h} starts from the $x$-point and ends at the $x$-point, and then is periodically repeated in the $(\bar{\theta}_1, \bar{J}_1)$ plane. This orbit in the original action-angle phase space $(\bm{\theta}, \bm{J})$ must satisfy Eq. \eqref{Eq: separatrix of h in J, theta}. Thus, the form of the curve corresponding to the points marked on the Poincare surface by the orbit taken from the equations of motion of the near-integrable system, with initial conditions $(\theta_2^{X,init}, \theta_2^{X,init}, J_1^{X,init}, J_2^{X,init})$, is approximately dictated by Eq.\eqref{Eq: separatrix of h in J, theta}, where $\theta_2^{X}(t) = 2\kappa\pi + \theta_2^{X,init}$, i.e., the $\theta_2$ coordinate of the Poincare surface. 

Eq. \eqref{Eq: separatrix of h in J, theta} also indicates that $(\bm{\theta}^{X}(t), \bm{J}^{X}(t))$ passes through $J_1^{rc}$ when $r\theta_1 + s(2\kappa\pi + \theta_2^{X,init}) - \bar{\phi}_1 = \pi$, i.e., using Eq. \eqref{Eq: o and x points at J theta}, through the resonant orbits of the $x$-point $\theta_1^{X}$. Since $r$ $(\theta_1^{X}, J_1^{rc})$ points exist on the Poincare surface and are periodically repeated, the form of Eq. \eqref{Eq: separatrix of h in J, theta} will also be periodically repeated $r$ times. The form of Eq. \eqref{Eq: separatrix of h in J, theta} has the same elliptic shape as the separatrix of $h$ (Eq. \eqref{Eq: separatrix of h}), with the difference that it is scaled by $r$ both with respect to $J_1$ and $\theta_1$. This elliptic form, with the tips of the ellipse at the unstable ($x$-points) resonant orbits repeated $r$ times, forms the so-called "island chain" on the Poincare surface of section. Each stable ($o$-point) resonant orbit is called an "island" and is surrounded by the separatrix. Since intersections of orbits on a Poincare surface are not allowed, the orbits of the original system, as seen on the Poincare surface of section with initial conditions on the Poincare surface and inside the separatrix, will remain inside the islands. With similar arguments as for the separatrix, these orbits will retain a similar elliptic form for the first order in $\epsilon$ approximation, as indicated by the orbits inside the separatrix in the phase space of $h$, and will also surround the stable ($o$-point) resonant orbit. 

The behavior of orbits with initial conditions outside the separatrix, i.e., far from the resonance, where the problem of small denominators does not exist, is characterized, as we saw in Sec. \ref{Classical Perturbation Theory}, by Eqs. \eqref{Eq: original with bar canonical variables}, which, on the Poincare surface of section, means orbits that are densely filled by points. The same, but with different topology, holds for orbits with initial conditions inside the separatrix. These orbits are called invariant curves. Close to the unstable resonant orbits ($x$-points), the separatrix has the form of a saddle. There, two manifolds exist: the stable and the unstable. On the stable manifold, the points on the Poincare surface of section asymptotically approach the $x$-points representing the resonant orbit, while on the unstable manifold, they move away from them \cite{Meletlidou2015}. 

In the case of an integrable Hamiltonian like $h$, stable and unstable manifolds are smoothly connected with the separatrix. In general, for instance, in the near-integrable system, the stable and the unstable manifolds intersect, creating homoclinic lobes. Homoclinic lobes are formed by the stable manifold as it asymptotically approaches the periodic orbit for $t \to \infty$. The lobes of the unstable manifold intersect with the lobes of the stable manifold, forming, around the separatrix, the so-called homoclinic tangle. This complex structure leads to the sensitive dependence of trajectories on initial conditions and chaotic evolution. This structure can be analyzed in detail using Smale's horseshoe, which is the discrete mapping that describes it (Ref. \cite{Meletlidou2015}, Sec. 8.4.2). 

The chaotic nature of the region near the $x$-points and the separatrix is the reason why we are interested in the location in phase space where they appear, which, as we proved, is the location of the resonance. When two separatrixes come very close and the lobes of two different resonances intertwine with each other, then, according to the Chirikov criterion \cite{Chiricov1979}, extended chaotic behavior arises, which, in the field of plasma physics, means particle transport in phase space far from its initial position and possibly particle loss. For this reason, except for the location of the resonance, the width of islands in a resonance island chain in phase space is also crucial for the study of particle transport. 

On the Poincare surface, the difference $J_1^{X}(t) - J_1^{rc}$ reaches its maximum at the time $t$ when $\theta_1 = \theta_1^{O}$, where $\theta_1^{O}$ is the $\theta_1$ coordinate of the $o$-point. This provides the width of the island in the original action-angle variables $(\bm{\theta}, \bm{J})$, given by:
\begin{equation}\label{Eq: width of island in action angle}
    W = 2\Delta J_1 = 4r\left(\frac{F}{G}\right)^{1/2}.
\end{equation}
In this work, having first determined the location of resonances in terms of the actions, $\bm{J}^{rc}$, we will also use $h$ and particularly the quantities $G$, $F$, and $\bar{\phi}_1$ in order to compute the width of the resonance islands in the system, as well as the number of islands in each island chain, using the integrable, unperturbed particle's guiding center Hamiltonian.

\subsection{Arnold Diffusion}\label{Sec: Arnold Diffusion}

In this section, we aim to provide a qualitative description of the phenomenon of Arnold diffusion, which occurs in systems with three or more degrees of freedom. To this end, we follow Lichtenberg and Lieberman (Ref. \cite{Lieberman1992}, Sec. 6.1) and utilize the secular perturbation theory outlined in Section \ref{Secular Perturbation Theory}. Arnold diffusion was first investigated in 1964 by Arnold (Ref. \cite{Arnold1964}), who described it in the context of a specific example involving two independent perturbations. A rigorous mathematical proof of a form of Arnold diffusion is presented in Ref. \cite{Bernard2016}, which also references numerous studies on this phenomenon since Arnold’s seminal work in 1964. The phenomenon is typically characterized by the Arnold diffusion rate (see Ref. \cite{Lieberman1992}, Sec. 6.2, and Ref. \cite{Chiricov1979}, Sec. 7), though this aspect is not covered in the present work. Instead, our focus is on identifying the regions in phase space where Arnold diffusion can occur. These regions are closely related to the locations of resonances and are distinct from the strong chaotic diffusion caused by resonance overlap. In Sec. \ref{Sec: GC Arnold diffusion}, we will demonstrate how Arnold diffusion manifests in the time-dependent particle guiding center Hamiltonian system.

Consider a three degrees of freedom time-independent, near-integrable Hamiltonian, expressed to first order in $\epsilon$:  
\begin{equation}
    H = H_0(\bm{J}) + \epsilon H_1(\bm{J}, \bm{\theta}),
\end{equation}  
where $H_0$ is the integrable part, and $H_1$ represents the perturbation. Since the Hamiltonian is time-independent, it remains constant along an orbit. Considering the variation of $H$ along an orbit, we can write  
\begin{equation}
    \frac{\Delta H}{\Delta t} = \frac{\Delta H_0}{\Delta t} + \epsilon \frac{\Delta H_1}{\Delta t} = 0,
\end{equation}  
which implies  
\begin{equation}\label{Eq: order of dH0}
    \Delta H_0 = \mathcal{O}(\epsilon).
\end{equation}  

Additionally, we can expand $\Delta H_0$ as  
\begin{equation}\label{Eq: order of dH0 2}
    \Delta H_0 (\bm{J}) = \left.\frac{\partial H_0}{\partial J_1}\right|_{\bm{J}^{init}} \Delta J_1 + \left.\frac{\partial H_0}{\partial J_2}\right|_{\bm{J}^{init}} \Delta J_2 + \left.\frac{\partial H_0}{\partial J_3}\right|_{\bm{J}^{init}} \Delta J_3 = \mathcal{O}(\epsilon),
\end{equation}  
which, using Eq. \eqref{Eq: order of dH0} yields: 
\begin{equation} \label{Eq: order of dH0 vector}
    \Delta \bm{J} \cdot \left.\bm{\nabla} H_0\right|_{\bm{J}^{init}} = \mathcal{O}(\epsilon),
\end{equation}  
where $\left.\bm{\nabla} H_0\right|_{\bm{J}^{init}}$ is the gradient of $H_0$ with respect to $\bm{J}$ at $\bm{J} = \bm{J}^{init}$, where $\bm{J}^{init}$ is the initial condition of the orbit. This gradient is normal to the two-dimensional surface defined by  $H_0(\bm{J})=$constant. Since the orbits start from $\bm{J}^{init}$ then the constant $H_0(\bm{J})$ will be $H_0(\bm{J}) = H_0(\bm{J}^{init})$. In general, $\bm{\nabla} H_0 = \mathcal{O}(1)$, so the above equation implies that the perpendicular deviation of $\Delta \bm{J} = \bm{J}(t) - \bm{J}^{init}$ from the surface $H_0(\bm{J}) = H_0(\bm{J}^{init})$ is of order $\epsilon$. Thus, for a near-integrable system, the actions $\bm{J}(t)$ remain close to this surface.  

\begin{figure} [h!]
    \centering
    \includegraphics[width=0.69\textwidth,keepaspectratio]{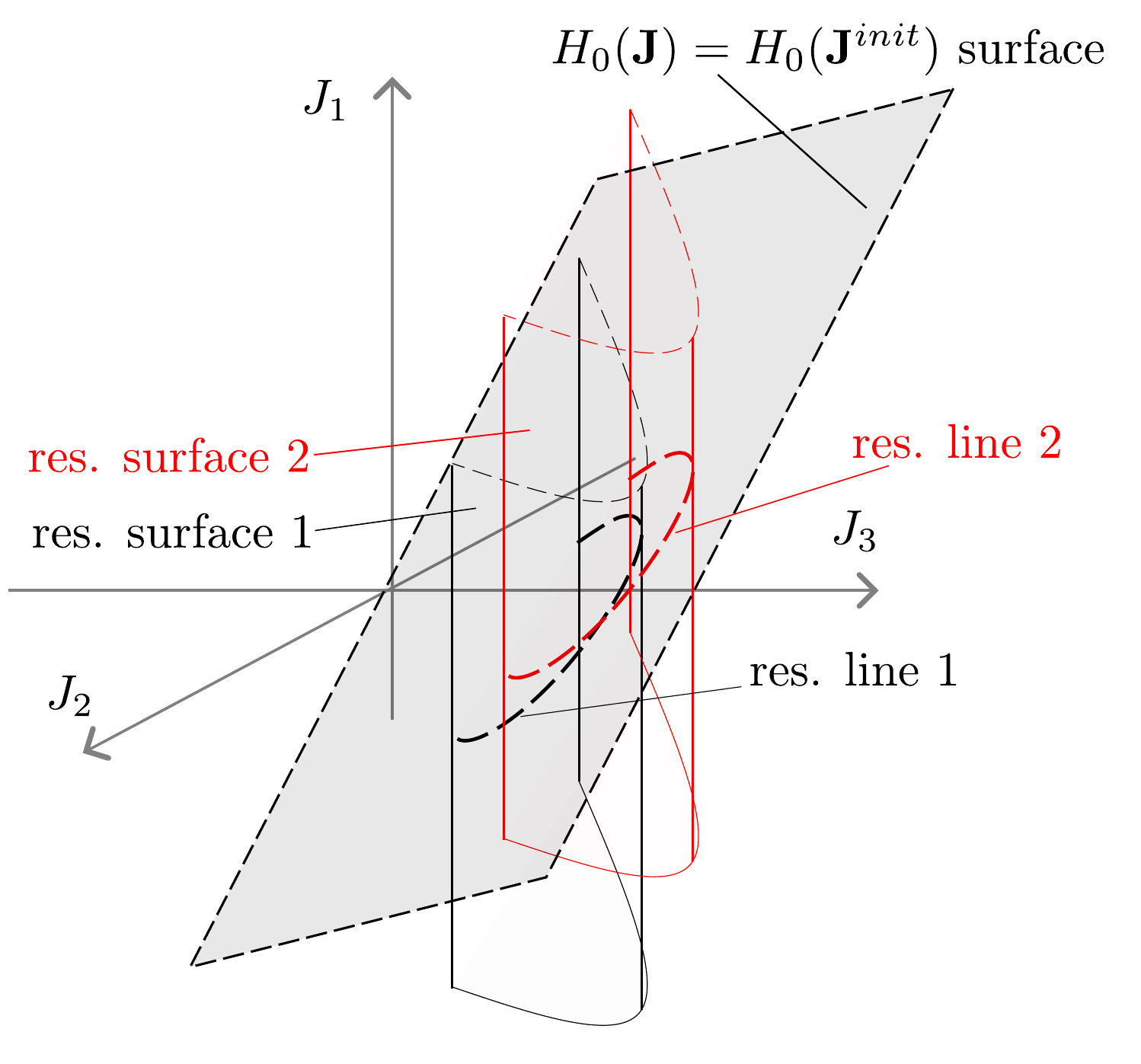}  
    \caption[A qualitative depiction of resonance surfaces.]{  
    Intersection of two resonance surfaces with a constant $H_0(\bm{J}) = H_0(\bm{J}^{\text{init}})$ surface.  
    The resonance lines (black and red dashed lines) on the constant $H_0(\bm{J})$ surface are formed by this intersection. For systems with three or more degrees of freedom, resonance lines on a constant $H_0(\bm{J})$ surface may intersect. }  
    \label{fig:Fig33}
\end{figure}

Moreover, each resonance relation $\bm{m} \cdot \bm{\omega}(\bm{J}) = 0$ (e.g., Eq. \eqref{rsk resonance}) defines a two-dimensional surface in the three-dimensional action space. The intersection of such resonance surfaces for different $\bm{m}$ with the surface $H_0(\bm{J}) = H_0(\bm{J}^{init})$ forms resonance lines on the constant-energy ($H_0$) surface, as shown in Fig. \ref{fig:Fig33}.

If the system has two degrees of freedom, the resonance surfaces as well as the constant energy surface reduce to curves in the two-dimensional $\bm{J}$-space, and their intersections become isolated points. It is easily seen that in three degrees of freedom or more, different resonance lines, i.e., resonances $\bm{m}_{\bm{rc}} \cdot \bm{\omega}(\bm{J}) = 0$ corresponding to different linearly independent $\bm{m}_{\bm{rc}}$'s, on a constant energy surface can intersect, forming the so-called Arnold web. In contrast, intersections of different resonance lines in two dimensions cannot occur.

Suppose $\bm{J}^{init}$ lies near the resonance defined by Eq. \eqref{rsk resonance}. At some time $t$, if the orbit reaches a $\bm{J}$-position far from the resonance—i.e., the orbit travels in a direction perpendicular to the resonance surface by a distance in $\bm{J}$ of order greater than $\epsilon$, where the KAM theorem applies (see Ref. \cite{Lieberman1992}, Sec. 3.2a), and no other resonances exist—we can transform to the barred system described by the Hamiltonian in Eq. \eqref{hat H to bar H}. In this transformation, small denominators are absent, and the dependence of $\bar{H}$ on $\bar{\bm{\theta}}$ is removed, rendering $\bar{\bm{J}}$ constant. The variable $\bm{J}$ is then determined by the first equations of Eq. \eqref{Eq: original with bar canonical variables}. Substituting $S_1$ from Eq. \eqref{S epsilon} and using $\bar{\bm{\theta}} = \bm{\omega}(\bar{\bm{J}})t + \bar{\bm{\theta}}^{init}$, and noting that $\bar{\bm{J}}$ remains constant, we find that $\bm{J}(t)$ depends on time through the term $e^{i\bm{m}(\bm{\omega}(\bar{\bm{J}})t + \bar{\bm{\theta}}^{init})}$. Consequently, $\bm{J}(t)$ oscillates around the constant $\bar{\bm{J}}$ with an amplitude of order $\Delta \bm{J} = \mathcal{O}(\epsilon)$. As a result, if the orbit enters such a region, it cannot return to $\bm{J}^{init}$ when time is reversed, thereby violating time-reversal symmetry. Thus, starting near the resonance, the orbit remains confined to this region as well as to the constant-$H_0$ surface, as shown in Eq. \eqref{Eq: order of dH0 vector}.

For systems with two degrees of freedom, the orbit remains confined near the intersection point of the resonance line with the constant-$H_0$ line. Specifically, for $\bm{J}^{init}$ near a resonance, the maximum deviation in $J_1$ is $\Delta J_1 = \mathcal{O}(\epsilon^{1/2})$ (Eq. \eqref{Eq: width of island in action angle}). Additionally, since $\bar{J}_2$ is constant, the first of Eqs. \eqref{Eq: hat with bar canonical variables simplified} and the second of Eqs. \eqref{J with hat J} yield $\Delta J_2 = \mathcal{O}(\epsilon^{1/2})$. This result can be also obtained by Eq. \eqref{Eq: order of dH0 2} considering that $\Delta J_1 = \mathcal{O}(\epsilon^{1/2})$ in the absence of a third degree of freedom $J_3$. Consequently, large excursions in actions are not possible in systems with two degrees of freedom.

\begin{figure} [h!]
    \centering
    \includegraphics[width=0.59\textwidth,keepaspectratio]{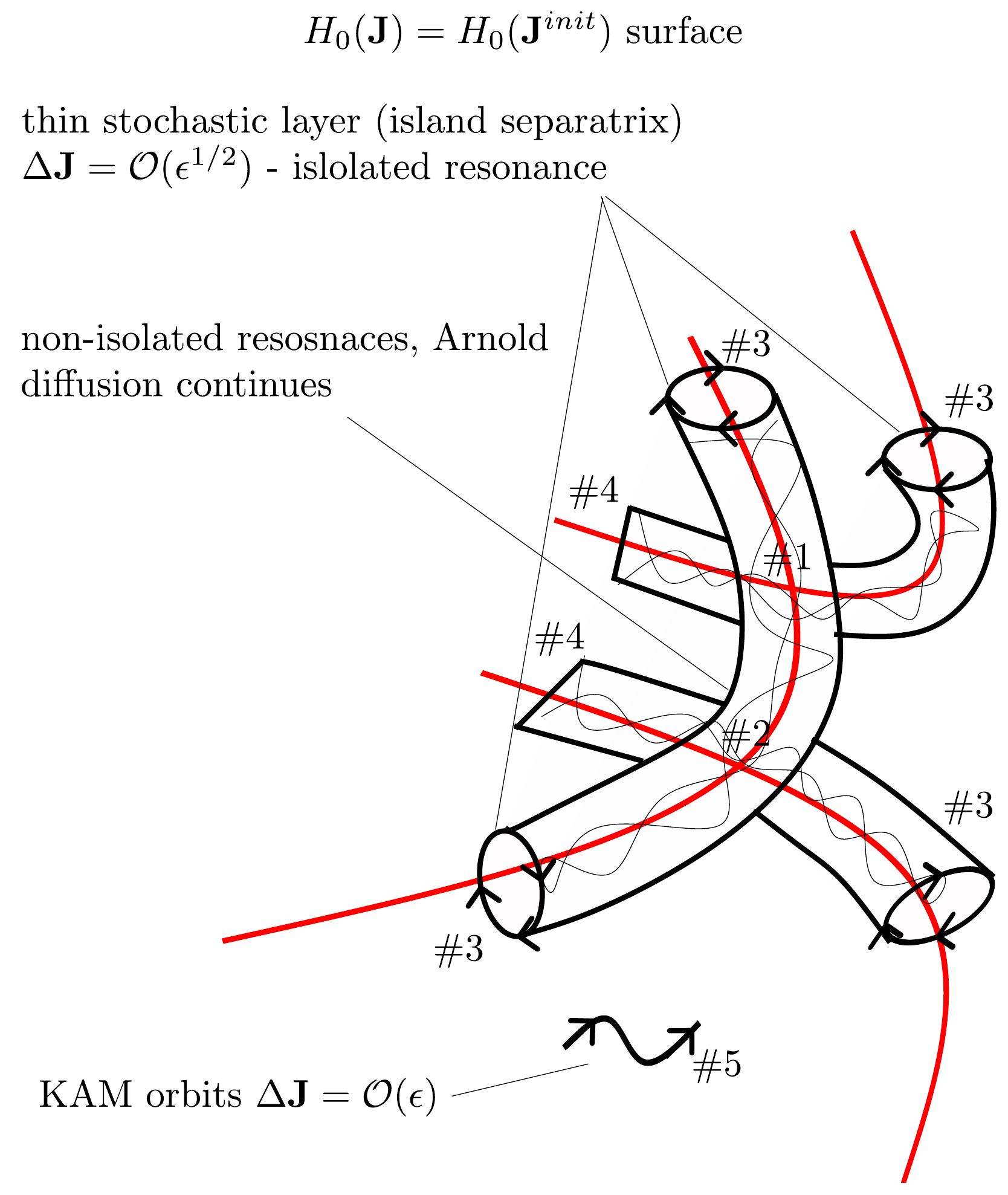}  
    \caption[A qualitative depiction of Arnold wed and diffusion.]{The constant unperturbed energy surface $H_0(\bm{J}) = H_0(\bm{J}^{\text{init}})$, depicted in Fig. \ref{fig:Fig33}, with the resonance lines (marked in red) forming the Arnold web. Far from resonances, KAM curves are formed, and the variation in $\bm{J}$ is of order $\mathcal{O}(\epsilon)$ (point \#5). Away from the intersection points of resonances, i.e., in regions where a resonance is isolated, resonance islands form, with the maximum excursion in $\bm{J}$ occurring at the separatrix (thin chaotic layer), where $\Delta\bm{J} = \mathcal{O}(\epsilon^{1/2})$ (points \#3). Near the intersection points of resonance lines, Arnold diffusion can occur, allowing the orbit to diffuse along resonance lines. If the orbit encounters another intersection point during diffusion, the diffusion continues (from point \#1 to point \#2). If not, it moves away from the intersection point to an isolated resonance region, where diffusion terminates (points \#3). Orbit diffusion may also stop before the orbit moves away from the intersection point due its specific dynamical behavior (points \#4).}

    \label{fig:Fig34}
\end{figure}

Conversely, for systems with three or more degrees of freedom, significant excursions in $\bm{J}$ can occur while the orbit remains near both the resonance surface and the constant-$H_0$ surface. In three dimensions, the intersection of these surfaces forms lines along which the orbit may evolve on the constant-$H_0$ surface (Fig. \ref{fig:Fig34}). However, as shown in Section \ref{Sec: isolated resonance}, if the resonance is isolated, then $\bar{J}_2$ and $\bar{J}_3$ remain constant, and $\Delta J_1 = \mathcal{O}(\epsilon^{1/2})$. From the first of Eqs. \eqref{Eq: hat with bar canonical variables simplified} and Eqs. \eqref{J with hat J}, it follows that $\Delta J_2 = \mathcal{O}(\epsilon^{1/2})$ and $\Delta J_3 = \mathcal{O}(\epsilon^{1/2})$. Thus, in this case, excursions in actions beyond $\mathcal{O}(\epsilon^{1/2})$ cannot occur.

The situation changes if the resonance is not isolated. According to the definition in Section \ref{Sec: isolated resonance}, the resonance \eqref{rsk resonance} in the vicinity of a particular point, $\bm{J} = \bm{J}^{rc}$, is non-isolated in two cases. 

The first case occurs when there exist integers $\bm{m}^{\prime}_{\bm{rc}} = (s^{\prime}, r^{\prime}, k^{\prime})$ such that the resonance condition
\begin{equation} \label{Eq:rc'}
    r^{\prime}\omega_1(\bm{J}) + s^{\prime}\omega_2(\bm{J}) + k^{\prime}\omega_3(\bm{J}) = 0
\end{equation}
is also satisfied at $\bm{J} = \bm{J}^{rc}$. Simultaneously,
\begin{equation} \label{Eq:rc}
    r\omega_1(\bm{J}) + s\omega_2(\bm{J}) + k\omega_3(\bm{J}) = 0
\end{equation}
holds at $\bm{J} = \bm{J}^{rc}$. This implies that the line defined by the intersection of the surface $H_0$ with the surface \eqref{Eq:rc}, and the line defined by the intersection of $H_0$ with the surface \eqref{Eq:rc'}, intersect at $\bm{J} = \bm{J}^{rc}$.

Consider the denominator in the generating function from Eq. \eqref{gen fun S2 simplified p}, which was used to address the small denominator problem arising near the resonance \eqref{Eq:rc}. Using Eqs. \eqref{Eq:rc}, we observe that for $p_1 = s^{\prime}r - r^{\prime}s$ and $p_2 = k^{\prime}r - r^{\prime}k$, the denominator $p_2\hat{\omega}_2 + p_3\hat{\omega}_3$ in Eq. \eqref{gen fun S2 simplified p} vanishes at $\bm{J}^{rc}$. Consequently, in the vicinity of $\bm{J}^{rc}$, we cannot eliminate the dependence of the $\bar{H}$ Hamiltonian \eqref{hat H to bar H simplified} on $\bar{\theta}_2$ and $\bar{\theta}_3$. As a result, $\bar{J}_2$ and $\bar{J}_3$ are not constants of motion. Moreover, the pendulum-like Hamiltonian \eqref{Eq: pendulum Hamiltonian h}, which governs $\bar{J}_1$, is not valid. Consequently, $J_1$, $J_2$, and $J_3$ are not restricted to evolve within $\mathcal{O}(\epsilon^{1/2})$ around $\bm{J}^{rc}$ but are instead free to arbitrarily move along the resonance line.

In the second case, the resonance is non-isolated if, at $\bm{J} = \bm{J}^{rc}$, the ratio $\hat{\omega}_2 / \hat{\omega}_3$ is insufficiently irrational. This implies that integers $p_2$ and $p_3$ exist such that $p_2\hat{\omega}_2 + p_3\hat{\omega}_3$ approaches zero, giving rise to the problem of small denominators. Like in the first case, in this case also, $J_1$, $J_2$, and $J_3$ are free to vary arbitrarily along the resonance line. Using the last two of Eqs. \eqref{hat omega with omega}, we see that $0\omega_1(\bm{J}) + p_2\omega_2(\bm{J}) + p_3\omega_3(\bm{J})$ approaches zero, indicating that the resonance
\begin{equation}
    0\omega_1(\bm{J}) + p_2\omega_2(\bm{J}) + p_3\omega_3(\bm{J}) = 0
\end{equation}
also occurs at $\bm{J} = \bm{J}^{rc}$, intersecting the resonance \eqref{Eq:rc} at this point.

Thus, the intersection of resonances in non-isolated cases for systems with three or more degrees of freedom constitutes a mechanism that facilitates Arnold diffusion. This process is analogous to the persistence of small denominator problem, as discussed in Section \ref{Canonical Adiabatic Theory}.

Arnold diffusion can drive an orbit away from the intersection of two resonance lines, while still remaining close to one resonance. Away from an intersection point, the resonance becomes isolated, and the analysis in Section \ref{Sec: isolated resonance} applies. In such regions, variations in $J_1$, $J_2$, and $J_3$ are limited to $\mathcal{O}(\epsilon^{1/2})$, preventing Arnold diffusion from occurring. If the resonance line along which diffusion occurs (e.g., resonance \eqref{Eq:rc}) intersects only one other resonance line (e.g., \eqref{Eq:rc'}), the diffusion remains localized around the intersection point, i.e., within the range of $\bm{J}$ where the small denominators problem exists. However, if additional resonances intersect resonance \eqref{Eq:rc} within the local diffusion range, the small denominator problem extends over a broader region, allowing for more extensive diffusion. As a result, a sequence of consecutive intersections between resonance lines can lead to global Arnold diffusion along the Arnold web (Fig. \ref{fig:Fig34}).

When an orbit undergoing Arnold diffusion along a resonance line reaches a region where the resonance is isolated (point \#3 in Fig. \ref{fig:Fig34}), i.e., a domain far from intersection points (points \#1 and \#2 in Fig.~\ref{fig:Fig34}) of resonances lines, it encounters the outermost elliptic orbit of the isolated resonance island—the separatrix discussed in Sec. \ref{Sec: isolated resonance} (circles in Fig.~\ref{fig:Fig34}). Conversely, within an isolated resonance island, the orbit that can approach a non-isolated resonance region (where Arnold diffusion occurs) is the outermost orbit, the separatrix. Thus, the interface between a diffusive orbit and the orbits within an isolated resonance island is the separatrix, which exhibits chaotic dynamics, as discussed in Section~\ref{Sec: isolated resonance}.  

In the absence of overlapping resonances, the separatrix manifests as a thin chaotic layer. In the way described in the paragraph above, the weak chaotic motion within the separatrix of a resonance, even in the absence of resonances overlapping (i.e., for very small $\epsilon$), can lead to chaotic diffusive motion (Arnold diffusion) in systems with three or more degrees of freedom (Fig. \ref{fig:Fig34}), Ref. \cite{Lieberman1992}, Sec. 6.1b, Ref. \cite{Chiricov1979}, Sec. 7).

It is important to note that, compared to the faster chaotic motion due to resonance overlap at larger perturbation strengths, $\epsilon$, Arnold diffusion occurring at very small $\epsilon$ is much slower. As shown in the near-integrable Hamiltonian, $\dot{\bm{J}} = -\partial H/\partial \bm{\theta} = \mathcal{O}(\epsilon)$. Furthermore, while the conditions discussed here enable diffusion along resonance lines, they do not guarantee it. Finally, during its diffusion along a resonance line, an orbit may be transferred to another resonance line at an intersection point, continuing its diffusion along the new line.
\chapter{Canonical description of the Guiding Center motion} \label{Sec: Chapter 4}

The guiding center approximation provides a reduced description of charged particle motion in a magnetized plasma, separating the fast gyromotion from the slower drift dynamics. A Hamiltonian formulation of this motion offers a systematic approach to studying guiding center behavior, particularly in the presence of perturbations.

This chapter first introduces the equilibrium magnetic field, which defines the landscape within which guiding center motion occurs. We begin with the Hamiltonian description of magnetic field lines and the role of magnetic coordinates, followed by a discussion of equilibrium field configurations relevant to toroidal magnetic confinement systems. Next, we examine the guiding center (GC) Hamiltonian system, which describes the GC drift motion in a magnetic field. Based on this Hamiltonian description, orbits are classified in the Constants Of the Motion (COM) space. We also discuss the structure of the non-axisymmetric perturbations (modes) with which the guiding center interacts via resonances.  

At the end of this chapter, we present the results of our work \textit{"Kinetic versus Magnetic Chaos in Toroidal Plasmas: A Systematic Quantitative Comparison"} (Ref.~\cite{Moges2024}). There, in the presence of symmetry-breaking perturbations, we investigate the chaoticity of GC drift motion and compare it to the chaoticity of the magnetic field lines Hamiltonian system (Sec.~\ref{Sec: Magnetic field representation in general curvilinear coordinates}). The former arises due to mode-particle resonances, while the latter results from mode-mode resonances. Mode-particle resonances will be discussed in detail in Chapter \ref{Ch:Mode-Particle GC resonances analysis based on Hamiltonian Formalism}.

\ifpdf
    \graphicspath{{Chapter4/Figs/Raster/}{Chapter3/Figs/PDF/}{Chapter4/Figs/}}
\else
    \graphicspath{{Chapter4/Figs/Vector/}{Chapter4/Figs/}}
\fi
\section{Magnetic filed lines Hamiltonian and Equilibrium Magnetic field}

In this section, we present two general representations of the magnetic field in magnetic field coordinates. The contravariant form follows from $\nabla \cdot \bm{B} = 0$, Eq. \eqref{MHD M4} of the single fluid MHD model, while the covariant form follows from $\nabla \times \bm{B} = \bm{j}$, Eq. \eqref{MHD M2} of the same model. The first representation leads to a Hamiltonian description of the magnetic field lines, which will be presented in Sec. \ref{Sec: Magnetic field representation in general curvilinear coordinates}, following Ref. \cite{White2013}, Secs. 1.5 and 1.6. Subsequently, and also based on Ref. \cite{White2013}, Secs. 2.4, 2.6, 2.8, and 2.9, both the contravariant and covariant expressions in straight field line coordinates are used to derive the form of the axisymmetric equilibrium magnetic field. This field satisfies the force balance equation of the single fluid MHD model, $\bm{j} \times \bm{B} = \nabla p$, Eq. \eqref{force balance eq}. A subcategory of this equilibrium is the large aspect ratio (LAR) configuration, provided in analytical form and utilized in this work.

\subsection{Magnetic Coordinates and Hamiltonian description of magnetic field lines} 
\label{Sec: Magnetic field representation in general curvilinear coordinates}

A general expression for the magnetic field, derived from Gauss's law for magnetism, $\nabla \cdot \bm{B} = 0$ (Eq. \eqref{MHD M4}), which facilitates the Hamiltonian formulation of guiding-center (GC) motion, can be written in general magnetic coordinates $(\psi, \theta, \zeta)$ \cite{Boozer1981, White1982, White1984, White2013}. Specifically, in contravariant form, the magnetic field is expressed as (see Ref. \cite{White2013}, Sec. 1.5):
\begin{equation}\label{Eq: gen magnetic field}
    \mathbf{B} = \mathbf{\nabla}\psi \times \mathbf{\nabla}\theta - \mathbf{\nabla}\psi_p(\psi, \theta, \zeta) \times \mathbf{\nabla}\zeta, 
\end{equation}
where $\zeta$ and $\theta$ are topologically toroidal and poloidal angular variables, respectively. These variables are not conventional geometric angles but periodic coordinates with a period of $2\pi$, representing spatial positions. The volume defined by $(\psi, \theta, \zeta)$ is topologically toroidal, akin to the structure of magnetic field lines.

The coordinates $(\psi(\bm{r}), \theta(\bm{r}), \zeta(\bm{r}))$ and the function $\psi_p(\psi, \theta, \zeta)$ are arbitrary functions defined with respect to a reference coordinate system, such as Cartesian or cylindrical coordinates, where $\bm{r}$ represents the position vector in this coordinate system. By selecting specific forms for $(\psi, \theta, \zeta)$ and $\psi_p(\psi, \theta, \zeta)$, a particular magnetic field can be constructed using Eq. \eqref{Eq: gen magnetic field}. For example, if $\psi_p$ is chosen as a function of $\psi$ only, the resulting magnetic field exhibits particular symmetries. Conversely, for a given magnetic field expressed in Cartesian or cylindrical coordinates, it is possible to determine $(\psi, \theta, \zeta)$ and $\psi_p$ as functions of the original coordinates so that the magnetic field can be represented in the form of Eq. \eqref{Eq: gen magnetic field}. In cases where the magnetic field exhibits symmetries, $\psi_p$ will depend solely on $\psi$.

From the gradient of $\psi_p(\psi, \theta, \zeta)$ in magnetic coordinates, expressed as
\begin{equation} \nabla \psi_p = \frac{\partial \psi_p}{\partial \psi} 
    \nabla \psi + \frac{\partial \psi_p}{\partial \theta} \nabla \theta + \frac{\partial \psi_p}{\partial \zeta} \nabla \zeta,
\end{equation}
we derive the following relationships: 
\begin{equation} \label{Eq: dB1} 
    \frac{\bm{B} \cdot \nabla \psi}{\bm{B} \cdot \nabla \zeta} = -\frac{\partial \psi_p}{\partial \theta}, \quad \frac{\bm{B} \cdot \nabla \theta}{\bm{B} \cdot \nabla \zeta} = \frac{\partial \psi_p}{\partial \psi}. 
\end{equation}
We write $\bm{B} = (|\bm{B}|/|d\bm{B}|)d\bm{B}$, where $d\bm{B}$ is a differential vector in the direction of $\bm{B}$, and therefore tangent to the magnetic field line. In this sense, $d\bm{B}$ represents the infinitesimal displacement along the magnetic field line in the direction of $\bm{B}$. Using the contravariant representation of $d\bm{B}$, along with Eqs. \eqref{Eq: e with nabla} and \eqref{Eq: e dot nabla}, we then obtain $d\bm{B}\cdot\nabla\psi = d\psi$, where $d\psi$ denotes the differential variation of $\psi$ along the magnetic field line. Similarly, we have $d\bm{B}\cdot\nabla\zeta = d\zeta$ and $d\bm{B}\cdot\nabla\theta = d\theta$, with $d\zeta$ and $d\theta$ representing the differential changes of the field line in $\nabla\theta$ and $\nabla\zeta$ directions respectively .
Thus, from Eqs. \eqref{Eq: dB1}, we obtain the Hamiltonian system:
\begin{equation} \label{Eq: eqs of motion of magnetic field lines} 
    \frac{d\psi}{d\zeta} = -\frac{\partial \psi_p}{\partial \theta}, \quad \frac{d\theta}{d\zeta} = \frac{\partial \psi_p}{\partial \psi}, 
\end{equation}
where the Hamiltonian is $\psi_p(\psi,\theta,\zeta)$, the time is $\zeta$ and $\theta$ is the canonical coordinate and $\psi$ its conjugate momentum (Ref. \cite{White2013}, Sec. 1.5).

Under symmetry conditions, there exist two-dimensional surfaces on which the magnetic field lies, satisfying $\bm{B} \cdot \nabla f = 0$, where $f(\bm{r}) = constant$ defines such a surface. For fixed $\theta$ and $\zeta$, $\psi$ and $\psi_p$ represent two-dimensional nested surfaces in the $(\psi, \theta, \zeta)$ coordinate system. If $\psi_p$ is independent of $\theta$ and $\zeta$ (i.e., the Hamiltonian system exhibits two symmetries), Eq. \eqref{Eq: gen magnetic field} implies $\bm{B} \cdot \nabla \psi = 0$ and $\bm{B} \cdot \nabla \psi_p = 0$. Hence, both $\psi$ and $\psi_p$ are magnetic surfaces, whose existence is directly linked to the presence of symmetries. In this case, $\psi_p = \psi_p(\psi)$, and from the right-hand equations of Eqs. \eqref{Eq: dB1} and \eqref{Eq: eqs of motion of magnetic field lines}, the safety factor $q$, defined as $1/q = (\bm{B} \cdot \nabla \theta)/(\bm{B} \cdot \nabla \zeta)$, is given by:
\begin{equation} \label{Eq: q factor}
    \frac{1}{q} = \frac{d\psi_p(\psi)}{d\psi},
\end{equation}
and the magnetic field form Eq. \eqref{Eq: gen magnetic field} becomes:
\begin{equation}\label{Eq: gen magnetic field straight}
    \mathbf{B} = \mathbf{\nabla}\psi \times \mathbf{\nabla}\theta - \frac{1}{q(\psi)}\mathbf{\nabla}\psi \times \mathbf{\nabla}\zeta.
\end{equation}
In addition in this case the magnetic potential $\bm{A}$ for which $\bm{B} = \nabla\times\bm{A}$ 
\begin{equation}\label{Eq: A in straight}
    \bm{A} = \psi\nabla\theta - \frac{1}{q(\psi)}\nabla\zeta.
\end{equation}
Under these symmetries, it can also be shown that the toroidal flux through a $\zeta = \text{constant}$ surface and the poloidal flux through a $\theta = \text{constant}$ surface are given by $2\pi\psi$ and $2\pi\psi_p$, respectively. For this reason, $\psi$ and $\psi_p(\psi)$ are also referred to as magnetic flux surfaces. Additionally, when $d\zeta/d\theta = (\bm{B} \cdot \nabla \zeta)/(\bm{B} \cdot \nabla \theta) = 1/q(\psi)$, the magnetic coordinates are called 'straight field line coordinates' (see Ref. \cite{White2013}, Sec. 2.4).

Finally, it can be observed that the Hamiltonian of the system becomes $\psi_p = \psi_p(\psi)$, which is in action-angle form, where $\psi$ represents the action and $\theta$ the angle. The frequencies are given by $\omega_\theta = d\theta/d\zeta = q(\psi)$ and $\omega_\zeta = d\zeta/d\zeta = 1$. Thus, the resonance condition for the magnetic field lines, $m\omega_\theta - n\omega_\zeta = 0$, becomes:
\begin{equation} \label{Eq: resoance conition for mfl}
    q(\psi) = \frac{m}{n},
\end{equation}
where $m$ and $n$ are integers. Which means that it is dictated by safety factor or magnetic $q$ profile, $q(\psi)$. The structure of magnetic field lines is analogous to the orbits of a Hamiltonian system expressed in action-angle variables (see Fig. \ref{fig:Fig30}). In this representation, $J_1 = \psi$, $w_1 = \theta$, and $(w_2 = \zeta, J_2 = \psi_p)$. \footnote{This analogy is valid because $\psi_p$ serves as the Hamiltonian, and $\zeta$ represents the "time." In general, the conjugate momentum of a time coordinate is the Hamiltonian.}

\subsection{Equilibrium magnetic field}\label{Sec: Equilibrium magnetic field}

\subsubsection{Force balance equation}

Following Ref. \cite{White2013}, Sec. 2.4, when the pressure is $p = p(\psi)$, the current $\bm{j}$ can be written, using Eq. \eqref{force balance eq}, in a contravariant form analogous to Eq. \eqref{Eq: gen magnetic field} as:
\begin{align} \label{Eq: j contravariant}
    \bm{j} &= \left(\frac{\partial\bar{I}(\psi)}{\partial\psi} + \frac{\partial \alpha}{\partial \theta}\right)\nabla\psi \times \nabla \theta + \left(\frac{\partial \bar{g}(\psi)}{\partial \psi} + \frac{\partial \alpha}{\partial \zeta}\right)\nabla\psi \times \nabla\zeta,
\end{align}
where $\bar{I}(\psi)$, $\bar{g}(\psi)$, and $\alpha(\psi,\theta,\zeta)$ are arbitrary functions of the magnetic coordinates. $\alpha(\psi,\theta,\zeta)$ is periodic in $\theta$ and $\zeta$ and due to this it can be shown that $2\pi\bar{I}(\psi)$ and $2\pi\bar{g}(\psi)$ correspond to the toroidal and poloidal currents, respectively.

Given Eq. \eqref{Eq: j contravariant} for the current, the magnetic field that satisfies Eq. \eqref{MHD M2}, i.e., $\bm{j} = \nabla\times\bm{B}$, can be expressed as:
\begin{equation} \label{Eq: covariant general equilibrioum magnetic field}
    \bm{B} = g(\psi,\theta,\zeta)\nabla\zeta + I(\psi,\theta,\zeta)\nabla\theta + \delta(\psi,\theta,\zeta)\nabla\psi,
\end{equation}
where
\begin{equation} \label{Eq: I,g,delta gen}
    I(\psi,\theta,\zeta) = \bar{I}(\psi) + \frac{\partial\sigma}{\partial \theta}, \quad g(\psi,\theta,\zeta) = \bar{g}(\psi) + \frac{\partial\sigma}{\partial \zeta}, \quad \delta(\psi,\theta,\zeta) = \frac{\partial\sigma}{\partial\psi} - \alpha(\psi, \theta, \zeta),
\end{equation}
and $\sigma(\psi,\theta,\zeta)$ is an arbitrary function periodic in $\theta$ and $\zeta$. It is easily seen that \eqref{Eq: covariant general equilibrioum magnetic field} can be also written in the form:
\begin{equation}\label{Eq: B covariant with nabla sigma}
    \bm{B} = \bar{g}(\psi)\nabla\zeta + \bar{I}(\psi)\nabla\theta + \nabla\sigma - \alpha\nabla\psi
\end{equation}

From the contravariant expression, Eq. \eqref{Eq: gen magnetic field straight}, it is immediately evident that the component of $\bm{B}$ in the direction of $\nabla \psi$ is zero, i.e., $\nabla\cdot\bm{B} = 0$. Since the expressions for $\bm{B}$ in Eqs. \eqref{Eq: gen magnetic field straight} and \eqref{Eq: covariant general equilibrioum magnetic field} must be equated, the covariant expression also satisfies this property. This is consistent with the condition $\bm{j} \times \bm{B} = \nabla p(\psi) = \frac{\partial p}{\partial \psi} \nabla \psi$, that the covariant form of $\bm{B}$ satisfies, which implies that $\nabla \psi$ is perpendicular to both $\bm{j}$ and $\bm{B}$. Consequently, requiring $\bm{B} \cdot \nabla \psi = 0$, with $\bm{B}$ given by Eq. \eqref{Eq: covariant general equilibrioum magnetic field}, allows $\delta$ to be expressed as a function of $g$ and $I$ as:
\begin{equation} \label{Eq: delta with g, I}
    \delta(\psi, \theta, \zeta) = -\frac{I\nabla\theta\cdot\nabla\psi + g\nabla\zeta\cdot\nabla\psi}{|\nabla\psi|^2}.
\end{equation}

Thus far, we have two distinct general representations for the magnetic field in straight field line coordinates, given by Eqs. \eqref{Eq: gen magnetic field straight} and \eqref{Eq: covariant general equilibrioum magnetic field}. Both representations must be equated, as they satisfy the following equations of the single-fluid MHD model, $\nabla\cdot\bm{B} = 0$ and $\bm{j} = \nabla\times\bm{B}$, where $\bm{j}$ is given by Eq. \eqref{Eq: j contravariant}. To determine an equilibrium magnetic field, the arbitrary functions $q(\psi)$, $\bar{I}(\psi)$, $\bar{g}(\psi)$, $\sigma(\psi,\theta,\zeta)$, and $\alpha(\psi,\theta,\zeta)$ appearing in the magnetic field expressions, Eqs. \eqref{Eq: gen magnetic field} and \eqref{Eq: covariant general equilibrioum magnetic field}, as well as the pressure profile $p(\psi)$ in the equilibrium condition, Eq. \eqref{force balance eq}, must be appropriately chosen to satisfy the force balance condition $\bm{j}\times\bm{B} = \nabla p$, Eq. \eqref{force balance eq}. 

\subsubsection{Boozer Coordinates}
Boozer coordinates are a specific type of magnetic coordinates that are particularly relevant to this work, as they enable a canonical Hamiltonian formulation of the guiding center motion, as will be discussed in Sec. \ref{Particle Guiding Center Drift Motion in Magnetic field}. In 1981, Boozer \cite{Boozer1981} expressed the magnetic field in a covariant form as:
\begin{equation}\label{Eq: B covariant Boozer}
    \bm{B} = \bar{g}(\psi)\nabla\zeta^{\prime} + \bar{I}(\psi)\nabla\theta^{\prime} + \delta^{\prime}\nabla\psi.
\end{equation}
Following Ref. \cite{White2013}, Sec. 2.4.2, we can derive Eq. \eqref{Eq: B covariant Boozer} by applying the transformations $\zeta^{\prime} = \zeta + q(\psi)w(\psi,\theta,\zeta)$ and $\theta^{\prime} = \theta + w(\psi,\theta,\zeta)$, setting
\begin{equation}\label{Eq: sigma in Boozer}
    \sigma = \left(\bar{g}(\psi)q(\psi) + \bar{I}(\psi)\right)w(\psi,\theta,\zeta),
\end{equation}
and substituting into Eq. \eqref{Eq: B covariant with nabla sigma}. Doing so, we find:
\begin{equation}\label{Eq: delta in Boozer}
    \delta^{\prime} = w\left((\bar{g} - 1)\frac{dq}{d\psi} + q\frac{d\bar{g}}{d\psi} + \frac{d\bar{I}}{d\psi}\right) - \alpha.
\end{equation}

Moreover, applying the aforementioned transformation, we observe that the contravariant form of the magnetic field in straight field line coordinates remains unchanged:
\begin{equation}\label{Eq: B contravariant Boozer}
    \bm{B} = \mathbf{\nabla}\psi \times \mathbf{\nabla}\theta - \frac{1}{q(\psi)}\mathbf{\nabla}\psi \times \mathbf{\nabla}\zeta = \mathbf{\nabla}\psi \times \mathbf{\nabla}\theta^{\prime} - \frac{1}{q(\psi)}\mathbf{\nabla}\psi \times \mathbf{\nabla}\zeta^{\prime}.
\end{equation}
Thus, the magnetic potential, $\bm{A}$, in relation to Eq. \eqref{Eq: A in straight}, can be expressed as
\begin{equation}\label{Eq: A in straight Boozer}
    \bm{A} = \psi\nabla\theta^{\prime} - \frac{1}{q(\psi)}\nabla\zeta^{\prime}.
\end{equation}

Taking the dot product of Eqs. \eqref{Eq: B covariant Boozer} and \eqref{Eq: B contravariant Boozer}, we obtain  
\begin{eqnarray}\label{Eq: magntitude of B in Boozer}
    B^2 = \frac{\bar{g} q + \bar{I}}{q J^{\prime}},
\end{eqnarray}  
where $J^{\prime} = \nabla\psi \cdot (\nabla\theta^{\prime} \times \nabla\zeta^{\prime})$.

\subsubsection{Axisymmetric-cylindrical configuration}

To proceed further in determining the functions that define the equilibrium, $q(\psi)$, $\bar{I}(\psi)$, $\bar{g}(\psi)$, $\sigma(\psi,\theta,\zeta)$, and $\alpha(\psi,\theta,\zeta)$, specific expressions for the magnetic coordinates $\psi(\bm{r})$, $\theta(\bm{r})$, $\zeta(\bm{r})$ are required. In this work, we focus on a particular class of straight field line equilibrium magnetic fields, the axisymmetric configurations, where all scalar quantities are independent of $\zeta$ ($\partial/\partial \zeta = 0$). Due to axisymmetry, it follows directly from Eqs. \eqref{Eq: I,g,delta gen} that $g = g(\psi)$.

Furthermore, a right-handed cylindrical coordinate system $(X,Z,\phi)$ is considered, where $X$ is the radial distance, $Z$ is the vertical axis, and $\phi$ is the toroidal angle. The corresponding unit vectors are $(\hat{X}, \hat{Z}, \hat{\phi})$, with $\hat{X} \times \hat{\phi} = -\hat{Z}$. Using the transformation equations from Cartesian to cylindrical coordinates, we obtain
\begin{equation}\label{Eq: nabla phi}
    \nabla \phi = \frac{\hat{\phi}}{X}, 
\end{equation}
where $\hat{\phi} = \partial \bm{r}/\partial \phi/|\partial \bm{r}/\partial \phi|$, and $\bm{r} = x(X,Z,\phi)\hat{x} + y(X,Z,\phi)\hat{y} + z(X,Z,\phi)\hat{z}$ is a vector in Cartesian coordinates. From this, we also derive that $\bm{e}_{\phi} = \partial \bm{r}/\partial \phi$ as
\begin{equation}\label{Eq: e_phi}
    \bm{e}_{\phi} = X \hat{\phi}. 
\end{equation}

The equilibrium is assumed to be toroidally symmetric, meaning that the unit vector $\hat{\phi}$ is perpendicular to both $\nabla\psi$ and $\nabla\theta$, i.e.,
\begin{equation}\label{Eq: orthogonality of phi}
    \nabla\phi\cdot\nabla \psi = 0, \quad \nabla\phi\cdot\nabla \theta = 0.
\end{equation}
Additionally, it is assumed that 
\begin{equation}\label{Eq: phi(psi, theta ,zeta)}
    \phi = \zeta + \nu(\psi, \theta),
\end{equation}
where $\nu$ is an arbitrary function of $(\psi, \theta)$.

Taking $\bm{B}$ from Eq. \eqref{Eq: gen magnetic field straight}, we find $\bm{B} \cdot \nabla \theta = 1/Jq$, where $J = \nabla \psi \cdot (\nabla \theta \times \nabla \zeta)$ (Ref. \cite{White2013}, Sec. 1.3). On the other hand, using $\bm{B}$ from Eq. \eqref{Eq: covariant general equilibrioum magnetic field} and employing Eqs. \eqref{Eq: orthogonality of phi} and \eqref{Eq: nabla phi}, we obtain $\bm{B} \cdot \nabla \phi = g/X^2$. Furthermore, as previously discussed, regardless of whether $\bm{B}$ is taken from Eq. \eqref{Eq: gen magnetic field straight} or Eq. \eqref{Eq: covariant general equilibrioum magnetic field}, it holds that $\bm{B} \cdot \nabla \psi = 0$. Additionally, from Eq. \eqref{Eq: phi(psi, theta ,zeta)}, we have $\nabla \phi =  \nabla \zeta + (\partial \nu/\partial \psi) \nabla \psi + (\partial \nu/\partial \theta) \nabla \theta$. Combining all these results with $\bm{B}\cdot\nabla\zeta/\bm{B}\cdot\nabla\theta = q$ we obtain
\begin{equation} \label{Eq: dnu/dtheta}
    \frac{\partial v}{\partial \theta} = \frac{gqJ}{X^2} - q
\end{equation}

\subsubsection{Contravariant representations of $\bm{B}$ and $\bm{j}$ in axisymmetric-cylindrical configuration}

We employ the following expressions that relate the contravariant basis vectors $\bm{e}_{\psi}, \bm{e}_{\theta}, \bm{e}_{\zeta}$ with the covariant basis vectors $\nabla \psi, \nabla \theta, \nabla \zeta$ in a curvilinear coordinate system $(\psi, \theta, \zeta)$ (Ref. \cite{White2013}, Sec. 1.3):
\begin{equation}\label{Eq: e with nabla}
    \bm{e}_{a} \times \bm{e}_{\beta} = J \epsilon_{a \beta \gamma} \nabla \gamma, \quad \quad \nabla a \times \nabla \beta = \frac{1}{J} \epsilon^{a \beta \gamma} \bm{e}_{\gamma}, 
\end{equation}
and

\begin{equation}\label{Eq: e dot nabla}
    \bm{e}_{a} \cdot \nabla \beta = \delta_{a}^{\beta},
\end{equation}
where $\epsilon_{a \beta \gamma}$ is antisymmetric under odd permutations of indices, with $\epsilon_{\psi \theta \zeta} = 1$. The Kronecker delta $\delta_{a}^{\beta}$ is defined as $\delta_{a}^{\beta} = 0$ for $a \ne \beta$ and $\delta_{a}^{\beta} = 1$ for $a = \beta$, with indices $a, \beta, \gamma$ referring to $\psi, \theta, \zeta$.

It follows that the Jacobian determinant of the transformation from $(x, y, z)$ to $(\psi, \theta, \zeta)$, $J$, is given by:

\begin{equation}\label{Eq: Jacobian psi, theta, zeta}
    J = \bm{e}_{\psi} \cdot (\bm{e}_{\theta} \times \bm{e}_{\zeta}) = \left(\nabla \psi \cdot (\nabla \theta \times \nabla \zeta) \right)^{-1}. 
\end{equation}
Using Eq. \eqref{Eq: phi(psi, theta ,zeta)}, this Jacobian equals the Jacobian of the transformation from $(x, y, z)$ to $(\psi, \theta, \phi)$:

\begin{equation}\label{Eq: Jacobian psi, theta, phi}
    J = \bm{e}_{\psi} \cdot (\bm{e}_{\theta} \times \bm{e}_{\zeta}) = \bm{e}_{\psi} \cdot (\bm{e}_{\theta} \times \bm{e}_{\phi}).
\end{equation}

Using Eqs. \eqref{Eq: phi(psi, theta ,zeta)} and \eqref{Eq: e with nabla}, we can rewrite Eq. \eqref{Eq: gen magnetic field straight} as:

\begin{align}
    \bm{B} &= \left(1 + \frac{1}{q}\frac{\partial v}{\partial \theta}\right) \nabla\psi \times \nabla\theta + \frac{1}{q} \nabla\phi \times \nabla\psi \label{Eq: B contravariant nu a} \\
    &= \frac{1}{J}\left(1 + \frac{1}{q}\frac{\partial v}{\partial \theta}\right) \bm{e}_{\phi} + \frac{1}{Jq} \bm{e}_{\theta}, \label{Eq: B contravariant nu b} 
\end{align}
which, after substituting $\partial v/\partial \theta$ from Eq. \eqref{Eq: dnu/dtheta}, becomes:

\begin{equation}\label{Eq: magnetic field contravariant}
    \bm{B} = \frac{g}{X^2}\bm{e}_{\phi} + \frac{1}{Jq}\bm{e}_{\theta}. 
\end{equation}

Similarly, the current density $\bm{j}$ given by Eq. \eqref{Eq: j contravariant}, after assuming axisymmetry by setting $\partial \alpha/\partial \zeta = 0$ and $g(\psi) = \bar{g}(\psi)$, can be written as
\begin{align}
    \bm{j} &= \left(\frac{\partial\bar{I}(\psi)}{\partial\psi} + \frac{\partial \alpha}{\partial \theta} - \frac{\partial \bar{g}(\psi)}{\partial \psi} \frac{\partial \nu}{\partial\theta} \right)\nabla\psi \times \nabla\theta - \frac{\partial \bar{g}(\psi)}{\partial \psi} \nabla\phi \times \nabla\psi \label{Eq: j contravariant a}\\
    &= \frac{1}{J} \left(\frac{\partial\bar{I}(\psi)}{\partial\psi} + \frac{\partial \alpha}{\partial \theta} - \frac{\partial \bar{g}(\psi)}{\partial \psi} \frac{\partial \nu}{\partial\theta} \right)\bm{e}_{\phi} - \frac{1}{J} \frac{\partial \bar{g}(\psi)}{\partial \psi}\bm{e}_{\theta}. \label{Eq: j contravariant b}
\end{align}

From Eqs. \eqref{Eq: magnetic field contravariant}, \eqref{Eq: j contravariant a}, and \eqref{Eq: j contravariant b}, it is observed that $j^{\psi} = B^{\psi} = 0$. Moreover, $B^{\phi}$, $B^{\theta}$, and $j^{\theta}$ depend only on the coordinate system quantities ($\psi$, $X$, and $J$) and functions $q(\psi)$, $g(\psi)$, whereas $j^{\phi}$ contains $\bar{I}(\psi)$ and $\alpha(\psi,\theta)$. To eliminate this dependency, we can derive $j^{\phi}$ using Eq. \eqref{Eq: e dot nabla} which gives $j^{\phi} = \bm{j} \cdot \nabla\phi$. Since, $\bm{j} = \nabla\times\bm{B}$ this becomes $j^{\phi} = (\nabla \times \bm{B}) \cdot \nabla\phi$. By substituting $\bm{B}$ from Eq. \eqref{Eq: B contravariant nu a}, we obtain $j^{\phi} = \nabla \cdot (\nabla\psi/(qX^2))$. Thus,
\begin{equation} \label{Eq: j contravarinat c}
    \bm{j} = \nabla \cdot \left(\frac{\nabla\psi}{qX^2}\right) \bm{e}_{\phi} - \frac{1}{J} \frac{dg(\psi)}{d\psi} \bm{e}_{\theta}.
\end{equation}
Comparing Eqs. \eqref{Eq: j contravariant b} and \eqref{Eq: j contravarinat c} and substituting $\partial \nu/\partial \theta$ from Eq. \eqref{Eq: dnu/dtheta} , we find that
\begin{equation} \label{Eq: j toroidal coeficient}
    \frac{1}{J} \left(\frac{\partial\bar{I}(\psi)}{\partial\psi} + \frac{\partial \alpha}{\partial \theta} - \frac{\partial \bar{g}(\psi)}{\partial \psi} \frac{\partial \nu}{\partial\theta} \right) = \nabla \cdot \left(\frac{\nabla\psi}{qX^2}\right)
\end{equation}
This same expression arises if one computes the toroidal flux of $\bm{j}$ given by Eq. \eqref{Eq: j contravariant} through a surface with constant $\zeta$, which equals $2\pi\bar{I}(\psi)$, and equates it to the flux obtained using $\bm{j}$ from Eq. \eqref{Eq: j contravarinat c}.

\subsubsection{Adjustment to Tokamak Geometry}

The toroidal angle $\phi$ is related to magnetic coordinates $(\psi, \theta, \zeta)$ through Eq. \eqref{Eq: phi(psi, theta ,zeta)}. As discussed in Sec. \ref{Sec: Magnetic field representation in general curvilinear coordinates}, to determine the general expression of the magnetic field, Eq. \eqref{Eq: gen magnetic field straight}, we also need two additional relations to connect $X$ and $Z$ with the magnetic coordinates $(\psi, \theta, \zeta)$. Following Ref.~\cite{White2013}, Sec.~2.8, we write the cylindrical coordinates in relation to the toroidal geometry of the tokamak, as illustrated in Fig. 2.3 of Ref. \cite{White2013}:
\begin{equation}\label{Eq: X,Z with r,theta}
    X = 1 + r\cos{\theta} - \Delta(r), \quad Z = r\sin{\theta}
\end{equation}
with
\begin{equation}\label{Eq: psi with r}
    \psi = \frac{r^2}{2} (1 + \mathcal{O}(\epsilon^2))
\end{equation}
where $\theta$ is the poloidal angle, and $\epsilon = a/R$ is the inverse aspect ratio, i.e., the ratio of the minor radius $a$ to the major radius $R$ of the tokamak torus. As shown in Eq.~\eqref{Eq: psi with r}, $\psi$ contains an ordering with respect to $\epsilon$. $\Delta(r)$ is the Shafranov shift. All quantities representing distances, such as $X$, $Z$, and $r$, are normalized to the major radius $R$ (see Table \ref{tab:Tab41}).

\begin{table}[h!]
    \centering
    \begin{tabular}{l l l l}
        \toprule
        Quantity & Norm. Factor & Units (SI) & Quantity after norm. \\
        \midrule
        Magnetic field $\bm{B}$ & $B_0$ & Tesla & $\bm{B}$ \\
        
        Magnetic potential $\bm{A}$ & $B_0 R_0$ & Tesla$\cdot$meters & $\bm{A}$ \\
        
        Magnetic fluxes $\psi$, $\psi_p$ & $B_0 R_0^2$ & Tesla$\cdot$meters$^2$ & $\psi$, $\psi_p$ \\
        
        Angular variables $\theta$, $\zeta$, $\phi$, $w$, $\nu$ & - & dimensionless & $\theta$, $\zeta$, $\phi$, $w$, $\nu$ \\
        
        Safety factor $q$ & - & dimensionless & $q$ \\
        
        Functions $g$, $I$, $\sigma$ & $B_0 R_0$ & Tesla$\cdot$meters & $g$, $I$, $\sigma$ \\
        
        Functions $\delta$, $\alpha$, $c$ & $1/R_0$ & 1/meters & $\delta$, $\alpha$, $c$ \\
        
        Distances $X$, $Z$, $r$ & $R_0$ & meters & $X$, $Z$, $r$ \\ 
        
        Tensor $\nabla$ & - & 1/meters & $\nabla$ \\       
        \bottomrule
    \end{tabular}
    \caption[Normalizations table 1.]{Quantities involved in the derivation of the magnetic field, along with their normalization factors and SI units. Here, $R_0$ denotes the location of the magnetic axis, given in $\SI{}{\metre}$, and $B_0$ is the magnetic field strength on the magnetic axis, given in $\SI{}{\tesla}$.}
    \label{tab:Tab41}
\end{table}

The system of Eqs.~\eqref{Eq: X,Z with r,theta} and \eqref{Eq: psi with r} can be solved for $\theta$ and $\psi$, yielding $\theta(X,Z,\phi)$ and $\psi(X,Z,\phi)$. Along with $\zeta(X,Z,\phi)$, which can be obtained from Eq.~\eqref{Eq: phi(psi, theta ,zeta)}, we have constructed specific curvilinear coordinates $\psi, \theta, \zeta$ in terms of the cylindrical coordinates $(X, Z, \phi)$, providing a magnetic field given by Eq.~\eqref{Eq: gen magnetic field straight}.

If $\bm{r}$ is the position vector in cylindrical coordinates, i.e., $\bm{r} = X\hat{X} + Z\hat{Z} + \phi\hat{\phi}$, it can be shown that $|\bm{e}_{\theta}| = \left|\partial \bm{r}/\partial \theta\right| = r$. From the definition of $\hat{\theta}$, $\hat{\theta} = (\partial \bm{r}/\partial \theta)/\left|\partial \bm{r}/\partial \theta\right| = -\sin{\theta}\hat{X} + \cos{\theta}\hat{Z}$, we obtain
\begin{align}
    \bm{e}_{\theta} &= r\hat{\theta} \label{Eq: e theta with hat theta} \\
    &= -r\sin{\theta}\hat{X} + r\cos{\theta}\hat{Z} \label{Eq: e theta with hat XZphi}
\end{align}

Using Eqs.~\eqref{Eq: X,Z with r,theta} and \eqref{Eq: psi with r} and neglecting terms of order $\mathcal{O}(\epsilon)$ in $\psi$, we can also find $\bm{e}_{\psi}$ from $\bm{e}_{\psi} = \partial \bm{r} / \partial \psi = (\partial \bm{r}/\partial r)(\partial r/\partial \psi)$, yielding
\begin{equation}\label{Eq: e psi with XZphi}
  \bm{e}_{\psi} = \frac{1}{r}(\cos{\theta} - \Delta^{\prime})\hat{X} + \frac{1}{r}\sin{\theta}\hat{Z}
\end{equation}
where $\Delta^{\prime} = d\Delta(r)/dr$. Combining these with $\bm{e}_{\phi}$ from Eq.~\eqref{Eq: e_phi}, we find
\begin{equation}\label{Eq: J with X,r,theta}
    J = X(1 - \Delta^{\prime}\cos{\theta}).
\end{equation}

From Eq. \eqref{Eq: magnetic field contravariant}, the magnetic field $\bm{B}$ is written as
\begin{align}
    \bm{B} &= \frac{g(\psi)}{X^2}\bm{e}_{\phi} + \frac{1}{qX(1 - \Delta^{\prime}\cos{\theta})}\bm{e}_{\theta} \label{Eq: B contravariant b} \\
    &=  -\frac{r\sin{\theta}}{qX(1 - \Delta^{\prime}\cos{\theta})}\hat{X} + \frac{r\cos{\theta}}{qX(1 - \Delta^{\prime}\cos{\theta})}\hat{Z} +\frac{g(\psi)}{X}\hat{\phi} \label{Eq: B contravariant c}
\end{align}

Using $\bm{e}_{\phi}$, $\bm{e}_{\theta}$, and $\bm{e}_{\psi}$ as given by Eqs. \eqref{Eq: e_phi}, \eqref{Eq: e theta with hat XZphi}, and \eqref{Eq: e psi with XZphi} along with the first equation of Eqs. \eqref{Eq: e with nabla}, we can write $\nabla\phi$, $\nabla\theta$, and $\nabla\psi$ in the form $\nabla a = a_{X}\hat{X} + a_{Z}\hat{Z} + a_{\phi}\hat{\phi}$, where $a$ refers to $\psi$, $\theta$, or $\phi$. Inverting, we obtain $\beta = \beta^{\phi}\nabla\phi + \beta^{\theta}\nabla\theta + \beta^{\psi}\nabla\psi$, where $\beta$ refers to $\hat{X}$, $\hat{Z}$, or $\hat{\phi}$. Substituting $\hat{X}$, $\hat{Z}$ and $\hat{\phi}$ in Eq. \eqref{Eq: B contravariant c}, we derive the covariant representation of $\bm{B}$ as:
\begin{align}
    \bm{B} &= g(\psi)\nabla\phi + \frac{r^2}{qX(1 - \Delta^{\prime}\cos{\theta})}\nabla\theta + \frac{\Delta^{\prime}\sin{\theta}}{qX(1 - \Delta^{\prime}\cos{\theta})}\nabla\psi  \label{Eq: B covariant a}\\
    &= g(\psi)\nabla\phi + \frac{r^2}{qJ}\nabla\theta + \frac{\Delta^{\prime}\sin{\theta}}{qJ}\nabla\psi \label{Eq: B covariant b}
\end{align}

We replace $\nabla\phi$ by $\nabla\phi = \nabla\zeta + (\partial \nu/\partial \theta)\nabla\theta + (\partial \nu/\partial \psi)\nabla\psi,
$ and use $\partial \nu/\partial \theta$ from Eq. \eqref{Eq: dnu/dtheta}. We also rewrite the coefficient of $\nabla\theta$ as:
$r^2/qJ = r^2/q + r^2(1 - J)/qJ$. Substituting these expressions back into Eq. \eqref{Eq: B covariant b}, we obtain Eq. \eqref{Eq: B covariant with nabla sigma} in the form:
\begin{align} \label{Eq: B covariant with nabla sigma a}
    \bm{B} &= g(\psi)\nabla\zeta + \frac{r^2}{q}\nabla\theta + \nabla\sigma - \alpha\nabla\psi,
\end{align}
where
\begin{align}
    \frac{\partial \sigma}{\partial \theta} &= g\left(\frac{gqJ}{X^2} - q\right) + \frac{r^2(1 - J)}{qJ},\label{Eq: dsigma/dtheta} \\
    \frac{\partial \sigma}{\partial \zeta} &= 0 \quad \text{(due to axisymmetry)}, \label{Eq: dsigma/dzeta} \\
    \frac{\partial \sigma}{\partial \psi} &= g\left(\frac{\partial \nu}{\partial \psi}\right) + c(\psi,\theta) \label{Eq: dsigma/dpsi},
\end{align}
and
\begin{equation}\label{Eq: alpha(psi,theta)}
    \alpha = -\frac{\Delta^{\prime}\cos{\theta}}{qJ} + c(\psi,\theta).
\end{equation}
The function $\partial\nu/\partial\psi$ can be freely chosen. Moreover, the function $c(\psi, \theta)$ cancels out in the representation of $\bm{B}$ when the above expressions are substituted back into Eq. \eqref{Eq: B covariant with nabla sigma a}. It is introduced to ensure that $\alpha$ satisfies Eq. \eqref{Eq: j toroidal coeficient}. In other words, $c(\psi, \theta)$ is derived by solving the partial differential equation given by Eq. \eqref{Eq: j toroidal coeficient}. Since only the term $\partial c(\psi, \theta)/\partial \theta$ appears in this equation, its integration yields a family of solutions for $c(\psi, \theta)$ parameterized by an arbitrary function of $\psi$. Finally, using the first two of Eqs. \eqref{Eq: I,g,delta gen} along with Eqs. \eqref{Eq: dsigma/dtheta}-\eqref{Eq: alpha(psi,theta)}, and performing the necessary calculations, we confirm that $\delta(\psi,\theta)$ as given by the last of Eqs. \eqref{Eq: I,g,delta gen} matches $\delta(\psi,\theta)$ as given by Eq. \eqref{Eq: delta with g, I}, as expected.

Note that throughout the derivation in this section, we have assumed $\psi(r) = r^2/2$, neglecting terms of order $\epsilon^2$ and higher from Eq. \eqref{Eq: psi with r}. Including these higher-order terms would modify the expressions accordingly; however, the procedure remains the same. This simplification was made because, in this work, when considering analytically expressed equilibria, we are primarily interested in the large aspect ratio axisymmetric cylindrical equilibrium, which will be discussed in following section.

\subsubsection{Grad-Shafranov Equation}\label{Sec: Grad-Shafranov Equation}

Until now, we have expressed all the quantities involved in the magnetic field and current representations in terms of the magnetic field coordinates $(\psi, \theta, \zeta)$, where $J$ is a function of these coordinates, along with the functions $g(\psi)$ and $q(\psi)$. While we have constructed specific expressions for $(\psi, \theta, \zeta)$ that lead to particular magnetic field configurations (axisymmetric and cylindrical, straight field line coordinates), we have made no particular assumptions regarding the functions $g(\psi)$ and $q(\psi)$. In this section, we will demonstrate that for the magnetic field to satisfy the equilibrium condition, the functions $g(\psi)$, $q(\psi)$, and the pressure $p(\psi)$ must satisfy a differential equation: the Grad-Shafranov equation.

We consider an axisymmetric-cylindrical configuration as described in the previous section. To derive the Grad-Shafranov equation, we employ the contravariant representations of $\bm{B}$ and $\bm{j}$ given by Eqs. \eqref{Eq: magnetic field contravariant} and \eqref{Eq: j contravarinat c}, respectively, within the force balance equation $\bm{j} \times \bm{B} = \nabla p(\psi)$.

Expanding the force balance equation, we have  
\begin{equation}
    (j^{\phi}\bm{e}_\phi + j^{\theta}\bm{e}_\theta + j^{\psi}\bm{e}_\psi)\times(B^{\phi}\bm{e}_\phi + B^{\theta}\bm{e}_\theta + B^{\psi}\bm{e}_\psi) = (\partial p(\psi)/\partial\psi)\nabla\psi.
\end{equation}
Using the first of Eqs. \eqref{Eq: e with nabla} to express $\nabla \psi = -(1/J) \bm{e}_{\theta} \times \bm{e}_{\phi}$, the above equation becomes
\begin{align} \label{Eq: Grad-Shafranov 0}
    (j^{\phi}B^{\theta} - j^{\theta}B^{\phi} + \frac{\partial p(\psi)}{\partial \psi})\bm{e}_{\phi}\times\bm{e}_{\theta} &= 0\Rightarrow\\
    & (j^{\phi}B^{\theta} - j^{\theta}B^{\phi} + \frac{1}{J}\frac{\partial p(\psi)}{\partial \psi}) = 0
\end{align}
By substituting the expressions for $j^{\phi}$, $B^{\phi}$, $j^{\theta}$, and $B^{\theta}$, we derive the Grad-Shafranov equation in magnetic coordinates:
\begin{equation}\label{Eq: Grad-Shafranov equation}
    \nabla \cdot \left( \frac{\nabla \psi}{q X^2} \right) + q(\psi) \frac{d p(\psi)}{d \psi} + \frac{g(\psi) q(\psi)}{X^2} \frac{d g(\psi)}{d \psi} = 0. 
\end{equation}
The gradient $\nabla F(x', y', z')$ can be evaluated in any curvilinear coordinate system $(x', y', z')$ as  
\begin{equation}\label{Eq: general nabla definition}
    \nabla F = \frac{\partial F}{\partial x'} \nabla x' + \frac{\partial F}{\partial y'} \nabla y' + \frac{\partial F}{\partial z'} \nabla z', 
\end{equation}
where $F(x', y', z')$ is an arbitrary scalar function. For instance, for Cartesian coordinates $(x, y, z)$, this becomes $\nabla F = \frac{\partial F}{\partial x} \hat{x} + \frac{\partial F}{\partial y} \hat{y} + \frac{\partial F}{\partial z} \hat{z}$, where $(\hat{x}, \hat{y}, \hat{z})$ are the Cartesian unit vectors. For cylindrical coordinates $(X, Z, \phi)$, applying Eq. \eqref{Eq: general nabla definition} gives $\nabla F = \frac{\partial F}{\partial X} \hat{X} + \frac{\partial F}{\partial Z} \hat{Z} + \frac{1}{X} \frac{\partial F}{\partial \phi} \hat{\phi}$, where $(\hat{X}, \hat{Z}, \hat{\phi})$ are the cylindrical unit vectors.

Note that, although the equilibrium condition $\bm{j} \times \bm{B} = \nabla p(\psi)$ is a vector equation, which could theoretically yield three independent equations, only one meaningful equation emerges: Eq. \eqref{Eq: Grad-Shafranov equation}, aligned with the direction of $\nabla \psi$. This occurs because in the $\nabla \phi$ and $\nabla \theta$ directions, the resulting equations are automatically satisfied due to $p = p(\psi)$ and $j^{\psi} = B^{\psi} = 0$. Consequently, we are free to choose two of the three functions $q(\psi)$, $p(\psi)$, and $g(\psi)$, while the third must satisfy Eq. \eqref{Eq: Grad-Shafranov equation}. To solve the Grad-Shafranov equation in cylindrical coordinates $(X, Z, \phi)$, substitute $\psi$ up to the desired order in $\epsilon$ and $X$ as defined in the previous section. Then, separate the equation into two parts: one containing terms independent of $\theta$, and another containing terms proportional to $\cos{\theta}$. In both cases, the variable is $\psi$ (or $r$). The solution of the first equation determines $q$, $g$, and $p$ as functions of $\psi$ (or $r$), while the second equation provides $\Delta$ as a function of $\psi$ (or $r$). Details on the solution procedure for the Grad-Shafranov equation can be found in Ref. \cite{White2013}, Sec. 2.8.

\subsubsection{Large Aspect Ratio Axisymmetric Cylindrical Equilibrium} \label{Sec: LAR equilibrium}
In this section, we derive a simplified equilibrium magnetic field configuration—the Large Aspect Ratio (LAR) magnetic field—which will be used in subsequent sections. In the LAR equilibrium, the normalized radius $r$ is assumed to be of order $\epsilon$, meaning the minor radius of the torus is much smaller than its major radius. Additionally, we assume a low $\beta$ plasma, $\beta = 2p/B^2 = \mathcal{O}(\epsilon^2)$, which implies $p = \mathcal{O}(\epsilon^2)$ with $B^2 = \mathcal{O}(1)$ and $q(\psi) = \mathcal{O}(1)$ as will be confirmed later. Based on these assumptions, we neglect terms of order $\epsilon^2$ and higher in the derivation. We also ignore the Shafranov shift, which is of order $\epsilon^2$ (see Eq. (2.84) in Ref. \cite{White2013}), and assume $\alpha(\psi,\theta) = c(\psi,\theta) = 0$.  

As discussed in Section \ref{Sec: Magnetic field representation in general curvilinear coordinates}, using Eq. \eqref{Eq: gen magnetic field}, we can find that the toroidal flux of the magnetic field, i.e., the magnetic flux through a surface of constant $\zeta$ (with infinitesimal area given by $d\bm{S}_{\zeta} = \bm{e}_{\psi} \times \bm{e}_{\theta} d\psi d\theta$), equals $2\pi\psi$. Computing this flux using the expression for $\bm{B}$ from Eq. \eqref{Eq: magnetic field contravariant}, we obtain:
\begin{align} \label{Eq: psi with g}
    2\pi\psi &= \int_{S_{\zeta}} \bm{B} \cdot d\bm{S}_{\zeta} \\
    &= \int_{S_{\zeta}} \frac{g(\psi)}{X^2} \bm{e}_{\phi} \cdot (\bm{e}_{\psi} \times \bm{e}_{\theta}) d\psi d\theta 
    = \int_{S_{\zeta}} \frac{g(\psi)}{X^2} J d\psi d\theta,
\end{align}
where $J = \bm{e}_{\phi} \cdot (\bm{e}_{\psi} \times \bm{e}_{\theta}) = \bm{e}_{\zeta} \cdot (\bm{e}_{\psi} \times \bm{e}_{\theta})$ is given by Eq. \eqref{Eq: J with X,r,theta}. Applying the LAR simplifications, the previous equation becomes:
\begin{align}
    2\pi\psi &= \int_{r}g(r)(2\pi r - r^2\sin{\theta})dr \overset{\cdot \psi = r^2/2}{\Rightarrow}\\
    \int_{r}\left[g(r)(2\pi r - r^2\sin{\theta}) - 2\pi r\right]dr &= 0\overset{\forall r}{\Rightarrow}
    g(r)(2\pi r - r^2\sin{\theta}) = 2\pi r \label{Eq: g(r) = 0}
\end{align}
In the relation above, Eq. \eqref{Eq: g(r) = 0}, if $g(r)$ were of order $\epsilon$ or lower (i.e., $\epsilon^2, \epsilon^3, \dots$), then Eq. \eqref{Eq: g(r) = 0} could not be satisfied since $r = \mathcal{O}(\epsilon)$. The same holds if $g(r) = \mathcal{O}(1/\epsilon), \mathcal{O}(1/\epsilon^2), \dots$. Eq. \eqref{Eq: g(r) = 0} can be satisfied if $g(r) = \mathcal{O}(1)$. In this case, the term $g(r)r^2\sin{\theta}$ on the left-hand side of Eq. \eqref{Eq: g(r) = 0} is of order $\epsilon^2$, and according to the LAR approximation, it can be neglected. As a result, the remaining terms give 
\begin{equation}\label{Eq: g}
    g(\psi) = 1.
\end{equation}
Substituting $g(r)$ in the Grad-Shafranov equation yields a differential equation that relates $q$ with $p$, considering only the $\theta$-independent terms. Terms dependent on $\theta$, appearing as coefficients of $\cos{\theta}$, are of order higher than $\epsilon^2$ and are neglected.

Note that $g(\psi)$, like all quantities discussed here, is expressed in normalized units (see Table \ref{tab:Tab41}). Additionally, if higher-order terms in $\epsilon$ were kept for $\psi$ (Eq. \eqref{Eq: psi with r}), as well as for the other quantities in Eq. \eqref{Eq: psi with g} (non-LAR case) then, higher-order terms in $g$ would arise in the form $g = 1 + g_2(r)$, with $g_2(r) = \mathcal{O}(\epsilon)$.

From Eq. \eqref{Eq: B covariant with nabla sigma a}, we directly find:
\begin{equation}\label{Eq: I(psi)}
    I(\psi) = \frac{r^2}{q(r)}.
\end{equation}
Substituting this result into Eq. \eqref{Eq: j toroidal coeficient}, performing the calculations, and neglecting terms of order $\epsilon^2$ and higher, we find that the equation is satisfied. Additionally, from Eq. \eqref{Eq: dnu/dtheta}, we observe that:
\begin{equation}
    \frac{\partial\nu}{\partial\theta} = qr\cos{\theta} = \mathcal{O}(\epsilon).
\end{equation}
Using this, we infer from Eq. \eqref{Eq: dsigma/dtheta} that $\partial\sigma/\partial\theta = \mathcal{O}(\epsilon)$. Note also that $\partial\nu(\psi,\theta)/\partial\psi$ can be freely chosen. We select it so that $\partial\sigma/\partial\psi$, from Eq. \eqref{Eq: dsigma/dpsi}, is of order $\mathcal{O}(1)$ or lower, ensuring that $\sigma = \mathcal{O}(\epsilon)$ from Eqs. \eqref{Eq: dsigma/dtheta} and \eqref{Eq: dsigma/dpsi}.

Transforming to Boozer coordinates, Eqs. \eqref{Eq: B covariant Boozer} - \eqref{Eq: B contravariant Boozer}, we obtain from Eqs. \eqref{Eq: sigma in Boozer} and \eqref{Eq: delta in Boozer} that $\delta^{\prime} = \mathcal{O}(\epsilon)$. Given that $\nabla\psi = \mathcal{O}(\epsilon)$, the last term in Eq. \eqref{Eq: B covariant Boozer} is of order $\epsilon^2$, and thus can be neglected under the LAR approximation.

Summarizing, using Eq. \eqref{Eq: B covariant Boozer}, the covariant form of the LAR axisymmetric cylindrical magnetic field equilibrium in Boozer coordinates becomes:
\begin{equation}\label{Eq: LAR B}
    \bm{B} = \nabla\zeta + \frac{r^2}{q(r)} \nabla\theta,
\end{equation}
where the prime has been omitted for simplicity. From Eq. \eqref{Eq: B contravariant Boozer}, the contravariant form is given by Eq. \eqref{Eq: magnetic field contravariant}. Substituting $\bm{e}_{\phi}$, $\bm{e}_{\theta}$, $X$, and $J$ from Eqs. \eqref{Eq: e_phi}, \eqref{Eq: e theta with hat theta}, \eqref{Eq: X,Z with r,theta}, and \eqref{Eq: J with X,r,theta}, and keeping terms up to first order in $\epsilon$, we find the magnitude of the LAR axisymmetric cylindrical equilibrium in Boozer coordinates:
\begin{equation}\label{Eq: LAR B amplitude}
    B = 1 - \sqrt{2\psi} \cos{\theta},
\end{equation}
where
\begin{equation}\label{LAR psi}
    \psi = \frac{r^2}{2}.
\end{equation}
We also choose a class of equilibria for which magnetic $q(\psi)$ profile is given by (Ref. \cite{White2013}, Sec. 2.9)
\begin{equation} \label{gen q factor}
    q(\psi) = q_{0} \Bigg[1 + \bigg(\bigg(\frac{q_{w}}{q_{a}}\bigg)^\nu - 1\bigg)(\psi/\psi^{(w)} - \lambda)^\nu \Bigg]^{1/\nu},
\end{equation}
where $\lambda$ denotes the location of the local minimum of the $q$ profile at $\psi/\psi^{(w)} = \lambda$, corresponding to a flux surface with zero magnetic shear. Moreover, $\psi^{(w)}$ (we also denote it as $\psi_w$) represents the outermost closed toroidal magnetic flux surface. When $\lambda = 0$, the minimum occurs at the magnetic axis ($\psi/\psi^{(w)} = 0$), and $q_a$ and $q_w$ represent the values of $q$ at the magnetic axis and the wall, respectively. Finally, $\nu$ controls the radial shape of the $q$-profile. These analytic expressions will be used in subsequent sections.

According to the analysis in Sec. \ref{Sec: Magnetic field representation in general curvilinear coordinates}, in straight field line coordinates where $\psi_p = \psi_p(\psi)$, Eqs. \eqref{Eq: eqs of motion of magnetic field lines} indicate that the toroidal magnetic flux surface $\psi$ remains constant as a function of $\theta$ along magnetic field lines. These constant-$\psi$ toroidal magnetic flux surfaces can be visualized in the $X,Z$ plane using Eqs. \eqref{Eq: X,Z with r,theta} and \eqref{Eq: psi with r} for different values of $\psi$, with $\theta$ varying from $0$ to $2\pi$. \footnote{Similarly, constant-$\theta$ surfaces can be constructed by varying $\psi$ from $0$ to $\psi^{(w)}$, where the index $(w)$ denotes the wall (the outermost closed toroidal magnetic flux surface).}

Note that in LAR equilibria, keeping terms up to $\mathcal{O}(\epsilon^1)$, the equilibrium constant $\psi$ surfaces in the $X,Z$ plane are concentric circles since the Shafranov shift $\Delta$ was neglected. In non-LAR cases corresponding to realistic equilibria, where second-order terms are retained, the surfaces become shifted circles ($\Delta \ne 0$). At third order, they exhibit elliptical and triangular distortions. In this work, the tools initially developed and demonstrated for LAR equilibria will also be applied to realistic equilibria obtained numerically, which exhibit both ellipticity and triangularity.

\section{Particle Guiding Center Drift Motion in Magnetic field} \label{Particle Guiding Center Drift Motion in Magnetic field}
The Lagrangian that describes the motion of a charged particle in an electromagnetic field, in normalized units (see Table \ref{tab:Tab42}), is given in the three-dimensional configuration space $\bm{x}$ by:
\begin{align}\label{Eq: particle Lagrangian}
    L(\bm{x},\dot{\bm{x}},t) &= \frac{1}{2}|\dot{\bm{x}}|^2 + \dot{\bm{x}}\cdot\bm{A}(\bm{x},t) - \Phi(\bm{x},t)\\
    &=  \left(\bm{A}(\bm{x},t) + \dot{\bm{x}}\right)\cdot\dot{\bm{x}} - H(\bm{x},\dot{\bm{x}},t),
\end{align}
where $\bm{A}(\bm{x},t)$ and $\Phi(\bm{x},t)$ are the magnetic and electric potentials, respectively, and $\bm{x}$ denotes the particle's position. Note that all quantities are in normalized units (see Table \ref{tab:Tab42}), and this convention will be followed unless stated otherwise. The Hamiltonian is given by:
\begin{equation}
    H(\bm{x},\dot{\bm{x}},t) = \frac{|\dot{\bm{x}}|^2}{2} + \Phi(\bm{x},t).
\end{equation}

To transform to phase space, we define the conjugate momentum $\bm{p}$ and coordinate $\bm{q}$ as:
\begin{align}
   \bm{p} &= \frac{\partial L}{\partial \dot{\bm{x}}} = \dot{\bm{x}} + \bm{A}(\bm{x},t) \label{Eq: transformation to phase space p} \\
   \bm{q} &= \bm{x}, \label{Eq: transformation to phase space q}
\end{align}
which leads to $\dot{\bm{x}} = \bm{p} - \bm{A}(\bm{x},t)$ (see Ref. \cite{Goldstein2002}, Sec. 8.1). Substituting into Eq. \eqref{Eq: particle Lagrangian}, the Lagrangian in the six-dimensional phase space $(\bm{q},\bm{p})$ is written as(see Ref. \cite{Goldstein2002}, Sec. 8.5):
\begin{equation}\label{Eq: phase space particle Lagrangian}
    L(\bm{q},\bm{p},\dot{\bm{q}}, \dot{\bm{p}},t) = \bm{p}\cdot{\dot{\bm{q}}} - H(\bm{q},\bm{p},t).
\end{equation}
with 
\begin{eqnarray}
    H(\bm{q},\bm{p},t) = \frac{|\bm{p} - \bm{A}(\bm{q},t)|^2}{2} + \Phi(\bm{q},t)
\end{eqnarray}

While Eq. \eqref{Eq: phase space particle Lagrangian} provides a canonical Hamiltonian description of particle motion, the derivation of the guiding-center (GC) Lagrangian requires a different transformation. Instead of transforming to $(\bm{q},\bm{p})$, we use the variables $(\bm{x}, \bm{u})$ with $\bm{u} = \dot{\bm{x}}$, treating $(\bm{x}, \bm{u})$ as independent variables \cite{Littlejohn1983}. This leads to the alternative Lagrangian:
\begin{equation}\label{Eq: extended Lagrangian}
   L(\bm{x},\bm{u},\dot{\bm{x}}, \dot{\bm{u}},t) = \left(\bm{A}(\bm{x},t) + \bm{u}\right)\cdot\dot{\bm{x}} - H(\bm{x},\dot{\bm{x}},t).
\end{equation}
Using this formulation and variational principles—such as gauge transformations that preserve the Euler-Lagrange equations under $L \rightarrow L + d S(\bm{x},\bm{u},t)/dt$—Littlejohn rigorously derives the guiding-center Lagrangian \cite{Littlejohn1983}.  This derivation is based on neglecting second-order terms in $\epsilon$ and higher from the Lagrangian. Three small parameters of order $\mathcal{O}(\epsilon^2)$ characterize the guiding-center motion (Ref. \cite{White2013}, Sec. 3.1):
\begin{equation}\label{Eq: small parameters}
    \epsilon_1 : \frac{w}{\dot{\xi}} = \mathcal{O}(\epsilon^2) \ll 1, \quad \epsilon_2 : \frac{\partial\bm{F}/\partial t}{\dot{\xi}}=\mathcal{O}(\epsilon^2) \ll 1, \quad \epsilon_3 : \frac{\dot{X}}{\dot{\xi}}=\mathcal{O}(\epsilon^2) \ll 1,
\end{equation}
where $\bm{F}$ represents either $\bm{B}$ or $\bm{A}$, $\bm{X}$ is the guiding-center position, $\xi$ is the gyro-angle, and $\dot{\xi} = \mathcal{O}(1/\epsilon)$ (or $\Omega_g$) is the gyrofrequency. The perpendicular velocity due to gyromotion is $w = \rho\dot{\xi}$, where $\rho = \mathcal{O}(\epsilon)$ is the Larmor radius. Additionally, from the transformation of particle position $\bm{x}$ to GC position $\bm{X}$, we have that $\dot{\bm{F}} = \partial\bm{F}/\partial t + (\dot{\bm{X}}\cdot\nabla)\bm{F}$ which from the second and the third of Eqs. \eqref{Eq: small parameters} gives:
\begin{equation}\label{Eq: small parameters 2}
    \frac{\dot{\bm{F}}}{\dot{\xi}}=\mathcal{O}(\epsilon^2) \ll 1, \quad \bm{F} = \bm{A},\bm{B}.
\end{equation}
In non-normalized units, these conditions translate to (Ref. \cite{Bierwage2022}):
$\epsilon_1$: The Larmor radius must be small compared to the magnetic field scale length. $\epsilon_2$: The wave frequency (e.g., MHD waves) must be much smaller than the cyclotron frequency, $\omega \ll \Omega_g$. $\epsilon_3$: The guiding center's parallel displacement in one gyration must be smaller than the parallel wavelength.

The guiding-center Lagrangian is finally given by:
\begin{equation}\label{Eq: GC Lagrangian gen}
    L(\bm{X}, \rho_{||},\xi,\mu,\dot{\bm{X}},\dot{\rho}_{||},\dot{\xi},\dot{\mu}) = \left(\bm{A}(\bm{X},t) + \rho_{||}\bm{B}(\bm{X},t)\right)\cdot\dot{\bm{X}} + \mu\dot{\xi} - H,
\end{equation}
where
\begin{equation}\label{Eq: mu definition}
    \mu = \frac{w^2}{2 B} = \frac{\rho^2\dot{\xi}^2}{2 B}, \quad \rho_{||} = \frac{u_{||}}{B},
\end{equation}
with $u_{||}$ being the velocity parallel to the magnetic field and:
\begin{equation}\label{Eq: GC Hamiltonian 1}
    H = \frac{\rho_{||}^2}{2}B^2(\bm{X},t) + \mu B(\bm{X},t) + \Phi(\bm{X},t).
\end{equation}

As expected, the GC Lagrangian, Eq.\eqref{Eq: GC Lagrangian gen}, is not generally in canonical form. A canonical Hamiltonian is not immediately extracted from Eq. \eqref{Eq: GC Lagrangian gen}, unlike in the case of the particle Lagrangian, Eq. \eqref{Eq: particle Lagrangian}, from which we easily obtained the phase space Lagrangian Eq. \eqref{Eq: phase space particle Lagrangian}. In other words, $H$ in its general form given by Eq. \eqref{Eq: GC Hamiltonian 1} is not a canonical Hamiltonian \cite{Cary2009}. However, expressing the magnetic field in Boozer magnetic coordinates (Sec. \ref{Sec: Magnetic field representation in general curvilinear coordinates}), we can bring the system into a canonical form. Specifically, by substituting into Eq. \eqref{Eq: GC Lagrangian gen} the expression for $\bm{B}$ from Eq. \eqref{Eq: B covariant Boozer} and $\bm{A}$ from Eq. \eqref{Eq: A in straight Boozer}, where $\bm{X} = \bm{X}(\psi, \theta, \zeta)$, and using that $\nabla a \cdot \dot{\bm{X}} = \dot{a}$ (with $\alpha$ referring to $\psi$, $\theta$, and $\zeta$, which have no explicit time dependence), we obtain  
\begin{equation}\label{Eq: Gc Lagrangian in Boozer 1}
    L = (\psi + \rho_{||}I(\psi))\dot{\theta} + (\rho_{||}g(\psi) - \psi_{p}(\psi))\dot{\zeta} + \rho_{||}q(\psi) \delta\dot{\psi}_{p} - H 
\end{equation}
where
\begin{equation}\label{GC H}
    H = \frac{\rho_{||}^2}{2}B^2(\psi,\theta) + \mu B(\psi,\theta).
\end{equation}
There exist coordinate transformations in which the term $\rho_{||}q(\psi) \delta\dot{\psi}_{p}$ can be removed through a gauge transformation of the Lagrangian (Ref. \cite{Littlejohn1985}). Equivalently, neglecting $\delta$ in the Euler-Lagrange equations derived from Eq. \eqref{Eq: Gc Lagrangian in Boozer 1} results in only a nonsecular modification of the guiding center motion. Thus, we omit this term from Eq. \eqref{Eq: Gc Lagrangian in Boozer 1} and take (Ref. \cite{White2013}, Sec. 3.2):
\begin{equation}\label{Eq: GC Lagrangian in Boozer}
    L = (\psi + \rho_{||}I(\psi))\dot{\theta} + (\rho_{||}g(\psi) - \psi_{p}(\psi))\dot{\zeta} - H.
\end{equation}

From this form, it is immediately evident that the canonical momenta conjugate to the coordinates $\theta$ and $\zeta$ are given in terms of $\rho_{||}$ and $\psi$ as \cite{White1984}:
\begin{align}\label{canon moments}
    P_{\theta} = \psi + \rho_{||} I(\psi), \qquad P_{\zeta} = \rho_{||} g(\psi) - \psi_{p}(\psi),
\end{align}
while $\mu$ is the canonical momentum conjugate to the gyro-angle $\xi$. Consequently, the Hamiltonian \eqref{GC H}, in the absence of an electric field (i.e., $\Phi=0$), can be rewritten as:
\begin{equation} \label{GC H 1}
    H = \frac{\left[P_{\zeta} + \psi_{p}(P_{\zeta},P_{\theta})\right]^2}{2 g^2(P_{\zeta},P_{\theta})} B^2(P_{\zeta}, P_{\theta}, \theta) + \mu B(P_{\zeta}, P_{\theta}, \theta).
\end{equation}
Using Eqs. \eqref{canon moments}, $\psi$ can be expressed as a function of $(P_{\zeta}, P_{\theta})$ by inverting the following equation:
\begin{equation}\label{Eq: Ptheta with psi 1}
    P_{\theta} = \psi + \frac{P_{\zeta} + \psi_p(\psi)}{g(\psi)}I(\psi).
\end{equation}
Since the Hamiltonian does not explicitly depend on time, the energy $E$ of the system is a conserved quantity. Moreover, the absence of the angles $\zeta$ (due to axisymmetry) and $\xi$ (due to gyro-averaging) in the Hamiltonian \eqref{GC H 1} implies that their conjugate momenta, $P_{\zeta}$ and $\mu$, respectively, are also conserved. Therefore, the Hamiltonian system is integrable, and its orbits can be described in terms of the three constants of motion $(E, \mu, P_{\zeta})$. Note that $E$ corresponds to the Hamiltonian $H$ in Eq. \eqref{GC H 1}. Each point in this three-dimensional Constants-Of-the-Motion (COM) space uniquely labels a GC orbit, with the shape of each orbit, as projected onto the poloidal plane $(\theta, P_\theta)$, being given by the level curves of the Hamiltonian (\ref{GC H 1}).

\begin{table}[h!]
    \centering
    \begin{tabular}{l l l l}
        \toprule
        Quantity & Norm. Factor & Units (SI) & Quantity after norm. \\
        \midrule
         Lagrangian $\&$ Hamiltonian $L$, $H$ & $M R_0^2\Omega_{g0}^2$ & joule & $L$, $H$ \\
         
         Velocities: $\bm{u}$, $\dot{\bm{x}}$, $w$, $\dot{\bm{X}}$ & $R_0 \Omega_{g0}$ & meters/seconds & $\bm{u}$, $\dot{\bm{x}}$, $w$, $\dot{\bm{X}}$ \\
         
         Gyro angle $\xi$ & - & dimensionless & $\xi$\\
         
         Gyro frequency $\dot{\xi}$ & $\Omega_{g0}$ & rad/seconds & $\dot{\xi}$\\

         Time $t$ & $1/\Omega_{g0}$ & seconds & $t$ \\

         Parallel to $\bm{B}$ velocity $u_{||}$ & $R_0\Omega_g$ & meters/second & $\rho_{||}$ \\
         
         $\mu B_{0}$ & $M R_0^2 \Omega_{g0}^2$ & joule & $\mu$ \\ 

         $P_{\zeta}$ & - & dimensionless & $P_{\zeta}$\\

         $P_{\theta}$ & - & dimensionless & $P_{\theta}$\\ 
        \bottomrule
    \end{tabular}
    \caption[Normalizations table 1.]{Quantities involved in the derivation of the canonical GC Hamiltonian, along with their normalization factors and SI units. Each quantity is normalized by the corresponding factor in the second column. Here, $M$ denotes the particle's mass in kilograms, and $\Omega_{g} = ZeB/M$ ($\Omega_{g0} = ZeB_0/M$) is the gyrofrequency, where $Ze$ is the particle's charge in coulombs, and $B$ is the magnetic field strength in tesla, with units of 1/seconds. $R_0$ represents the location of the magnetic axis in meters, while $B_0$ is the magnetic field strength at the magnetic axis, given in tesla. To extract the normalized relations discussed in this chapter, this table should be considered alongside Table \ref{tab:Tab41}.}  
\label{tab:Tab42}

\end{table}

\subsection{Orbit classification in the COM space} \label{Sec: Orbits in COM space}
Orbits can be classified as either trapped or passing (circulating) and as either confined or lost (Ref. \cite{Hsu1992}, Ref. \cite{White2013}, Sec. 3.3), depending on their location relative to well-defined curves in COM space. The corresponding diagrams provide a valuable tool for studying the orbital spectrum and the impact of resonances on particle transport, as will be demonstrated in the following sections.

To classify the orbits, we need to determine the locations of the fixed points of the GC Hamiltonian, Eq. \eqref{GC H 1}, as functions of the COMs, $(E,\mu,P_{\zeta})$. To do so, we perform a fixed-point analysis on the Hamiltonian \eqref{GC H 1}. From the equations of GC motion, we obtain:
\begin{eqnarray}\label{Eq: GC equations of motion}
    \dot{P}_{\zeta} &=& 0 \label{dot Pzeta}\\
    \dot{\zeta} &=& \frac{\rho_{||}}{g^2(P_\zeta,P_\theta)}B^2 \label{dot zeta} \\
    \dot{P}_
    \theta &=& -\frac{\partial B}{\partial \theta}\left[\frac{\rho_{||}^2}{g^2(P_\zeta,P_\theta)} B + \mu\right] \label{dot Ptheta} \\
    \dot{\theta} &=& \left[\frac{1}{q(P_{\theta})}\frac{\rho_{||}}{g^2(P_\zeta,P_\theta)} - \frac{\partial g}{\partial P_{\theta}}\frac{\rho_{||}^2}{g^3(P_\zeta,P_\theta)}\right]B^2 + \frac{\partial B}{\partial P_{\theta}}\left[\frac{\rho_{||}^2}{g^2(P_\zeta,P_\theta)} B + \mu\right] \label{dot theta}
\end{eqnarray}
where $\rho_\parallel = P_\zeta + \psi_p(\psi)$.

To find the fixed points on the $(\theta, P_{\theta})$ plane, we solve the system of algebraic equations, $\dot{P}_{\theta} = 0$ and $\dot{\theta} = 0$, with $\dot{P}_{\theta}$ and $\dot{\theta}$ taken by Eqs. \eqref{dot Ptheta} and \eqref{dot theta} respectively. A solution of the first equation is $\partial B/\partial \theta = 0$. A solution of the second equation, to first order in $\rho$ (or equivalently, to first order in $w$), which—due to the definition of $\mu$ in Eq. \eqref{Eq: mu definition}—implies $\mu \approx 0$, gives $\rho_{||} = 0$. Thus, to leading order in $\rho$, a class of fixed points is given by
\begin{align}
    \frac{\partial B(P_{\theta},\theta,P_{\zeta})}{\partial \theta} &= 0 \label{Eq: dB/dtheta}\\
    \psi_p(P_{\zeta},P_{\theta}) &= -P_{\zeta}.\label{Eq: rho parallel zero} 
\end{align}
Solving Eqs. \eqref{Eq: dB/dtheta} and \eqref{Eq: rho parallel zero} for $(\theta, P_{\theta})$ gives the fixed points in the $(\theta, P_{\theta})$ plane as functions of $P_{\zeta}$ and $\mu$. Substituting these points into Eq. \eqref{GC H 1} then yields the energy of the fixed points in terms of $P_{\zeta}$ and $\mu$. For a particular value of $\mu$, we can obtain the curves $E(P_{\zeta})$ in the constant-$\mu$ plane corresponding to the fixed points.

To derive analytical expressions, we apply this method to the LAR equilibrium discussed in Sec. \ref{Sec: LAR equilibrium}, for which $g(\psi) = 1$ (Eq. \eqref{Eq: g}). Additionally, since $I(\psi) = r^2/q(r) = \mathcal{O}(\epsilon)$ (Eq. \eqref{Eq: I(psi)}), for relatively small values of $\rho_{||}$, the term $\rho_{||}I(\psi)$ in the first equation of \eqref{canon moments} is much smaller than $\psi$ and can be neglected (Ref. \cite{White2013}, Sec. 6.2). Thus, we take $P_{\theta} = \psi$. Substituting this into Eq. \eqref{Eq: LAR B amplitude}, we express the magnetic field amplitude in canonical coordinates as
$B = \left(1 - \sqrt{2 P_{\theta}}\cos{\theta}\right)$.

Equation \eqref{Eq: dB/dtheta} then gives $\theta = 0, \pm\pi$. One of the fixed points ($\theta = 0$) is an elliptic center, while the other ($\theta = \pm\pi$) is a hyperbolic saddle. These fixed points determine the topology of the GC motion and the number of distinct orbit families. The elliptic center is surrounded by trapped (banana-shaped) orbits, whereas the hyperbolic saddle defines a heteroclinic orbit that connects the fixed points at $\theta = \pm \pi$ and separates trapped and co-passing (or counter-passing) orbits.

The $E(P_{\zeta})$ curves in the constant-$\mu$ plane corresponding to these fixed points, also known as the Trapped-Passing Boundary (TPB), are given by
\begin{equation} \label{trapped passing boundary} 
E = \mu\left(1 \mp \sqrt{2\psi_{p}^{-1}(-P_{\zeta})}\right),
\end{equation}
where $\psi_p^{-1}$ is the inverse function of $\psi_p(\psi)$.

Note that in the case of more complex equilibria or the presence of an electric field, additional fixed points and orbit families may arise by solving Eqs. \eqref{dot Ptheta} and \eqref{dot theta} (Ref. \cite{Anastassiou2024}). These existence of additional heteroclinic orbits (separatrixes) corresponds to additional TPB curves in the COM space.

Orbits can also be classified based on whether they are confined or lost. To this end, the left wall (LW) and right wall (RW) curves in the COM space are defined as the sets of $(E,\mu,P_{\zeta})$ points that label GC orbits (described by Eq. \eqref{GC H 1}) that pass through $\psi^{(w)}$ at $\theta = \pi$ and $\theta = 0$, respectively. Substituting these conditions into the Hamiltonian \eqref{GC H} and using Eqs. \eqref{canon moments}, the general form of the $E(P_{\zeta})$ curves in the constant-$\mu$ plane is parabolic and is given by:
\begin{align}
    E &= \frac{(P_{\zeta} + \psi_p(\psi^{(w)}))^2}{g^2(\psi^{(w)})}B^2(\psi^{(w)}, \pi) + \mu B(\psi^{(w)}, \pi) \quad \text{(left wall)}\label{Eq: left wall gen}\\
    E &= \frac{(P_{\zeta} + \psi_p(\psi^{(w)}))^2}{g^2(\psi^{(w)})}B^2(\psi^{(w)}, 0) + \mu B(\psi^{(w)}, 0) \quad \text{(right wall)}\label{Eq: right wall gen}
\end{align}

Additionally, the magnetic axis (MA) parabola is defined by the set of $(E,\mu,P_{\zeta})$ points corresponding to GC orbits that pass through $\psi = 0$. To determine the MA parabola, we require the expression for $B(\psi=0,\theta)$. Even for a general axisymmetric equilibrium magnetic field (as discussed in Sec. \ref{Sec: Equilibrium magnetic field}), taking into account that $\Delta(r=0) = \Delta^{\prime}(r=0) = 0$ (see Ref. \cite{White2013}, Sec. 2.8) and using Eqs. \eqref{Eq: sigma in Boozer}, \eqref{Eq: dsigma/dtheta}, and \eqref{Eq: dsigma/dzeta}, we find, using Eq. \eqref{Eq: magntitude of B in Boozer}, that $B$ at $\psi=0$ is independent of $\theta$. Substituting this into Eq. \eqref{GC H 1}, we obtain the general expression for the magnetic axis parabola:
\begin{equation}\label{Eq: magnetic axis gen}
    E = \frac{(P_{\zeta} + \psi_p(\psi=0))^2}{g^2(\psi=0)}B^2(\psi = 0) + \mu B(\psi=0).
\end{equation}

Applying this framework to the LAR equilibrium, the walls parabolas are given by:
\begin{equation}\label{walls}
    E = \frac{\left(P_{\zeta} + \psi_{p}(\psi_{w})\right)^2}{2}\left(1\mp\sqrt{2\psi_{w}}\right)^2 + \mu\left(1\mp\sqrt{2\psi_{w}}\right),
\end{equation}
where the signs $(-)$ and $(+)$ correspond to GC orbits that touch the right and left walls at the midplane, respectively. Furthermore, orbits passing through the magnetic axis ($\psi=0$) satisfy the equation:
\begin{equation}\label{magnetic axis}
    E = \frac{P_{\zeta}^2}{2} + \mu.
\end{equation}
Eqs. \eqref{trapped passing boundary}, \eqref{walls}, and \eqref{magnetic axis} define parabolic curves in the $(E, P_{\zeta})$ plane, enabling the classification of each orbit in a constant-$\mu$ slice of the three-dimensional COM space (Ref. \cite{White2013}, Sec. 3.3).

\subsection{Non-Axisymmetric perturbations}\label{Sec: Non-Axisymmetric perturbations}

As discussed in Section \ref{MHD Modes}, the linear MHD modes, arising from an ideal displacement $\bm{\xi}$ of the plasma, perturb it from its equilibrium state, leading to the following perturbation of the equilibrium magnetic field:
\begin{equation} \label{delta B ideal 1}
    \delta\bm{B} = \nabla\times(\bm{\xi}\times\mathbf{B})
\end{equation}
where $\bm{B}$ is the equilibrium magnetic field. 

In this work, to demonstrate the resonance analysis—conducted in the unperturbed GC Hamiltonian system — we consider flute-type magnetic perturbations written in the form (Ref.  \cite{White2013}, Sec.~3.9.2):
\begin{equation}\label{Eq: flute modes}
    \delta\bm{B} = \nabla\times(\alpha\bm{B})
\end{equation}
where $\alpha$ is given by 
\begin{equation} \label{alpha}
\alpha(\psi,\theta,\zeta,t)=\sum_{m,n} \alpha_{m,n}(\psi)e^{i(n\zeta - m\theta - \omega t)}
\end{equation}
with $m, n$ being the poloidal and toroidal mode numbers, respectively, $\omega$ the mode frequency and $\alpha_{m,n}(\psi)$ representing the radial profile of the mode amplitude. This perturbation not only introduces time dependence in the GC system, due to its dependence on $t$, but also breaks the axisymmetry of the magnetic field.

Perturbations such as tearing and shear Alfv\'en modes are primarily perpendicular to $\bm{B}$ and are considered flute modes. In Ref. \cite{White2013}, it is justified that within ideal MHD, whether the MHD modes are described through the linearization method (Eq. 
\eqref{delta B ideal}) or the Lundquist description (Eq. \eqref{Lundquist B}), mode-particle resonances with ideal MHD modes can be effectively investigated using the magnetic field perturbation representation in Eq. \eqref{Eq: flute modes}. This form is also used to describe externally applied resonant magnetic perturbations (RMPs) (Ref. \cite{Zhang2024}).

We now examine how this magnetic field perturbation affects the GC motion. The perturbation of the magnetic potential $\bm{A}$ is given by $\delta\bm{A} = \alpha\bm{B}$. Substituting $\bm{A} \to \bm{A} + \delta\bm{A}$ in the Lagrangian \eqref{Eq: extended Lagrangian}, an additional perturbation term, $\delta\bm{A} \cdot \dot{\bm{x}} = \alpha\bm{B} \cdot \dot{\bm{x}}$, appears. Assuming that the perturbation $\delta \bm{B}$ (Eq. \eqref{Eq: flute modes}) is given in the position of the GC, i.e., $\delta \bm{B} = \delta \bm{B}(\bm{X},t)$, and following the procedure outlined in Ref. \cite{White2013} (Sec. 3.1) for deriving the GC Lagrangian, including this additional term, we find that $\alpha\bm{B} \cdot \dot{\bm{x}} = \alpha\bm{B} \cdot \dot{\bm{X}}$. As a result, the Lagrangian \eqref{Eq: GC Lagrangian in Boozer} is modified as:
\begin{align}
    L &= (\psi + \rho_{||}I(\psi))\dot{\theta} + (\rho_{||}g(\psi) - \psi_{p}(\psi))\dot{\zeta} + \alpha g(\psi)\dot{\zeta} + \alpha I(\psi)\dot{\theta} - H.\\
    &= \left(\psi + (\rho_{||}+\alpha)I(\psi)\right)\dot{\theta} + \left((\rho_{||}+\alpha)g(\psi) - \psi_{p}(\psi)\right)\dot{\zeta} - H.
\end{align}\label{Eq: GC Lagrangian in Boozer}
Thus, the canonical momenta are modified as:
\begin{align}\label{canon moments perturbed}
    P_{\theta} = \psi + (\rho_{||}+\alpha) I(\psi), \qquad P_{\zeta} = (\rho_{||}+\alpha) g(\psi) - \psi_{p}(\psi).
\end{align}
Using this, the magnetic perturbation is incorporated into the GC Hamiltonian by modifying the normalized parallel velocity $\rho_{\parallel}$ to $\rho_{\parallel} + \alpha$ in Eq. \eqref{GC H}, and introducing the electric potential $\Phi(P_{\zeta}, P_{\theta}, \zeta, \theta, t)$  due to the time dependence of the perturbed magnetic field. This leads to the modified GC Hamiltonian \cite{White1982, White1984, White2013c}:
\begin{equation} \label{Per GC H 1}
    \begin{aligned}
        H &= \frac{\left[P_{\zeta} + \psi_{p}(P_{\zeta},P_{\theta}) - \alpha(P_{\zeta},P_{\theta}, \zeta, \theta, t)\right]^2}{2 g(P_{\zeta},P_{\theta})} B^2(P_{\zeta}, P_{\theta}, \theta)\\
        &+ \mu B(P_{\zeta}, P_{\theta}, \theta) + \Phi(P_{\zeta},P_{\theta}, \zeta, \theta, t).
    \end{aligned}
\end{equation}
The function $\psi$ can be expressed in terms of $(P_{\zeta}, P_{\theta})$ by solving the second of Eqs. \eqref{canon moments perturbed} for $(\rho_{||} + a)$ and substituting the result into the first equation of \eqref{canon moments perturbed}. This leads to  
\begin{equation}\label{Eq: Ptheta with psi 2}
    P_{\theta} = \psi + \frac{P_{\zeta} + \psi_p(\psi)}{g(\psi)} I(\psi).
\end{equation}  
which has the same form as in the unperturbed system, given by Eq. \eqref{Eq: Ptheta with psi 1}.

\section{Drift motion of GC - Kinetic versus Magnetic Chaos in Toroidal Plasma}

As discussed in Sec.~\ref{Sec: Magnetic field representation in general curvilinear coordinates}, magnetic field lines form an integrable Hamiltonian system in action-angle coordinates, $(\theta,\psi)$, when both $\psi$ and $\psi_p$ are magnetic surfaces, i.e., when $\psi_p = \psi_p(\psi)$. In this case, $\psi$ remains constant with respect to both $\theta$ and $\zeta$, meaning that each constant-$\psi$ surface corresponds to a well-defined, concentric, and closed flux surface of the equilibrium magnetic field.  

For axisymmetric and time independent magnetic fields, both Hamiltonian systems, describing the magnetic field lines (Eq. \eqref{Eq: eqs of motion of magnetic field lines})  and the GC motion (Eq. \eqref{GC H 1}), are integrable. However, the presence of axisymmetry-breaking perturbations (Eq. \eqref{alpha}) results in the destruction of an invariant of each system and renders them both non-integrable (Eq. \eqref{Per GC H 1}). The effect of the perturbation on the phase space of the two systems is strongly inhomogeneous due to its resonant character, and therefore, chaos appears only in regions where resonance conditions are fulfilled. For the system of the magnetic field lines, the resonance condition is given by Eq. \eqref{Eq: resoance conition for mfl} while the resonance condition for the GC Hamiltonian system will be discussed in the next chapter. Perturbations with bigger amplitude may result in resonance overlap and extended chaotic regions, but still these regions usually occupy only a part of the phase space. The localized character of phase space chaoticity gives rise to the question of the specific location of chaos. It is worth emphasizing that for the GC system the phase space location does not refer only to a spatial location, such as the radial distance from the magnetic axis (related to the magnetic flux $\psi$ and canonical poloidal particle momentum $P_\theta$), but also to the values of the kinetic characteristics of the particles, namely their energy $E$, magnetic moment $\mu$, and canonical toroidal momentum $P_\zeta$, being Constants Of the Motion (COM) of the unperturbed GC Hamiltonian (Sec. \ref{Sec: Orbits in COM space}). 

A visual identification of chaotic field lines and particle orbits can be performed by constructing Poincar{\'e} surfaces of section \cite{Lieberman1992, Meletlidou2015}. Relatively low-energy particles tend to follow magnetic field lines so that kinetic and magnetic Poincar{\'e} surfaces of section are similar. However, for more energetic particles, there is no obvious relation between kinetic and magnetic chaos; in fact, kinetic chaos can be stronger in spatial locations where the magnetic chaos is weaker, and vice versa. The systematic investigation of the relation between kinetic and magnetic chaos necessitates the utilization of an index for efficient chaos detection and quantification.

In this section, we present our work in Ref. \cite{Moges2024}, where we systematically compare the chaoticity of magnetic field lines to that of particle orbits under non-axisymmetric and time-independent magnetic perturbations. Chaos detection is based on the calculation of the Smaller ALignment Index (SALI) \cite{Skokos2001}, which, with this work, was applied for the first time in the context of plasma physics. This quantitative measure of chaos enables the characterization of particle orbits in terms of their chaoticity in the system's center-of-mass (COM) space, providing a compact overview of the impact of a specific perturbation on particles with different kinetic characteristics. Details of the SALI method and its comparison with the Lyapunov exponents method for detecting chaotic orbits can be found in Ref. \cite{Moges2024} and references therein. Here, we present the results of applying this method.  

The results were obtained by numerically solving the GC equations of motion (Eqs. \eqref{Eq: GC equations of motion}) in canonical coordinates, as derived from the perturbed GC Hamiltonian in Eq. \eqref{Per GC H 1}. Moreover, it can be shown that the guiding center velocity of the unperturbed system in normalized units (see Tables \ref{tab:Tab41} and \ref{tab:Tab42}) is given by \cite{Boozer1980}
\begin{equation} \label{Eq: GC drift velocity}
    \boldsymbol{u}_{g} = u_{||}\boldsymbol{\hat{b}} +  \rho_{||}^2\boldsymbol{B} \times (\boldsymbol{\hat{b}} \cdot \nabla) \boldsymbol{\hat{b}} + \mu\frac{\boldsymbol{B} \times \nabla B}{B^2},
\end{equation}  
where the first term represents the GC velocity parallel to the magnetic field lines. The other two terms correspond to GC drift velocities due to the inhomogeneity of the magnetic field: the second term arises from the curvature of the field lines, while the third term is due to the transverse gradient of the magnetic field.  

By writing the GC equations of motion and neglecting drifts—specifically setting $\rho_{||}^2 \to 0$ and $\mu \to 0$—the resulting solution represents GC motion where the GC exactly follows the magnetic field lines (first term in Eq. \eqref{Eq: GC drift velocity}). In this way, we effectively obtain the "motion" of the magnetic field lines.  

In the following, we work with the LAR equilibrium magnetic field and consider a typical $q$-factor profile given by Eq. \eqref{gen q factor} with parameters $\nu=2$, $q_{ma}=1.1$, $q_{w} = 4.0$, and $\psi_{w} = 0.05$. We investigate time-independent perturbations of the form given in Eq. \eqref{Eq: flute modes}.  For $\alpha$ (Eq. \eqref{Eq: alpha(psi,theta)}), we consider two perturbation modes with mode numbers $(m,n)=(3,2)$ and $(5,2)$. For simplicity, we neglect the radial profile of the perturbations, assuming that the coefficients $\alpha_{m,n}$ are independent of $\psi$ and have equal values, $\alpha_{3,2}=\alpha_{5,2}=\epsilon=7.5\times 10^{-5}$. Moreover, in the following cases, we consider hydrogen particles in a LAR magnetic field with $B_0=1$~T.  
 
\begin{figure}[h!]
	\centering
	\includegraphics[width=1.0\textwidth]
    {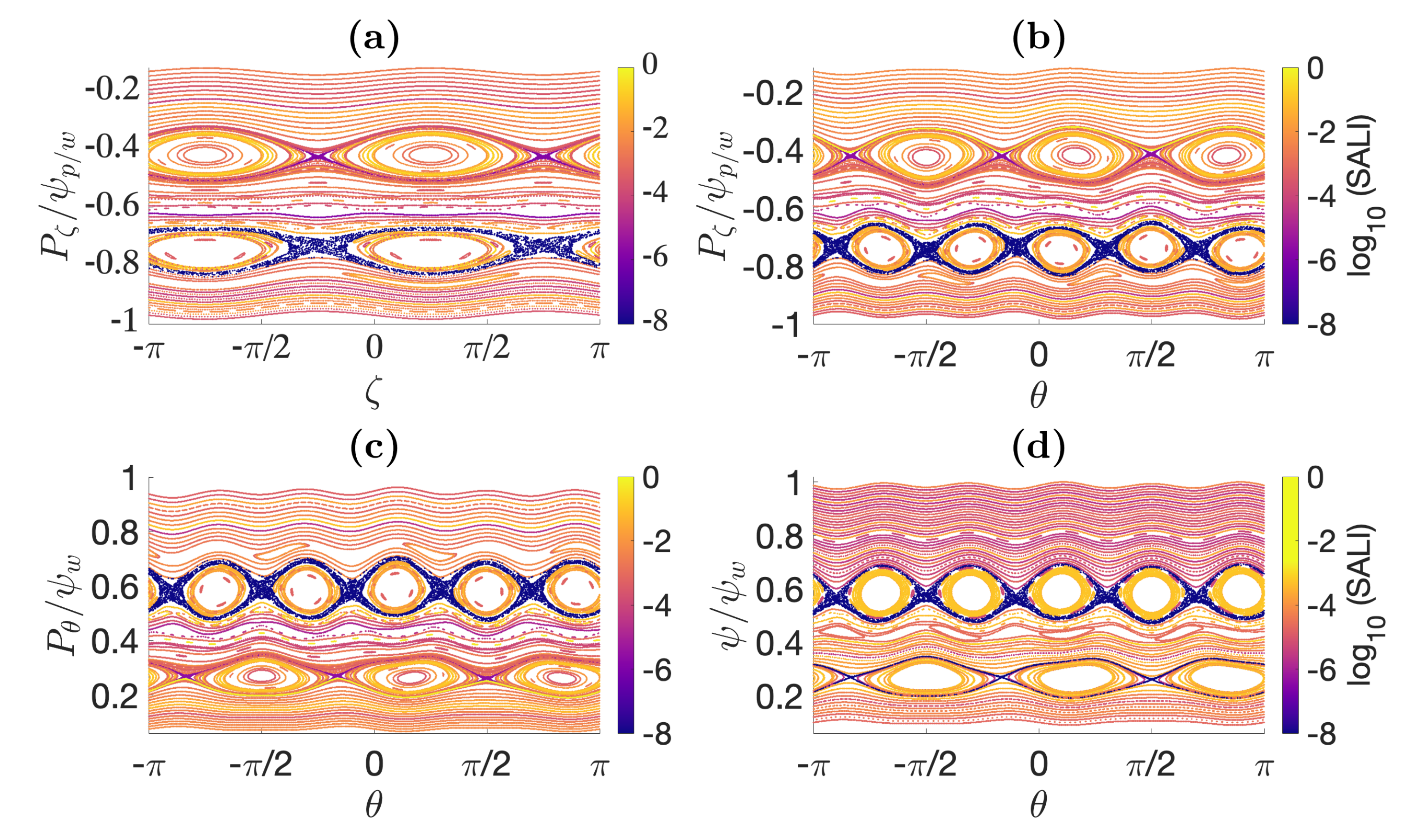}
    \centering
	\caption[Poincar{\'e} surfaces of section for low-energy particles showing the magnetic versus kinetic chaos.]{(a), (b) and (c) Kinetic Poincar{\'e} surfaces of section for low-energy particles with $E=3.4$eV and $\mu B_{0}=2.6$eV. (d) Magnetic Poincar{\'e} surface of section. All orbits are coloured according to their SALI value calculated for $10^{8}$ time units. Two perturbation modes with mode numbers $(m,n)=(3,2), (5,2)$ and equal amplitudes are considered $\alpha_{3,2}=\alpha_{5,2}=\epsilon=7.5\times 10^{-5}$. }
\label{fig:Fig44}
\end{figure}

The case of low-energy particles with $E=3.4$eV and $\mu B_{0}=2.6$eV is depicted in Fig.~\ref{fig:Fig44}. Kinetic Poincar{\'e} surfaces of section $(P_\zeta, \zeta)$ and $(P_\zeta, \theta)$ are shown in Figs. \ref{fig:Fig44}(a) and (b), with the orbits being coloured according to their final SALI value. The number of islands in each island chain is determined by the respective mode numbers. The kinetic Poincar{\'e} surface of section $(P_\theta, \theta)$ is shown in  Fig. \ref{fig:Fig44}(c). Since $P_\theta=\psi$ (LAR approximation) this kinetic Poincar{\'e} surface of section can be directly compared to the magnetic one shown in Fig. \ref{fig:Fig44}(d). The comparison clearly shows that kinetic and magnetic chaos appear close to the separatrices of the island chains and confirms that chaos is located in the same radial ($\psi$) position, with the degree of chaoticity, as quantified by SALI, being similar. 

\begin{figure}[h!]
	\centering
	\includegraphics[width=1.0\textwidth]{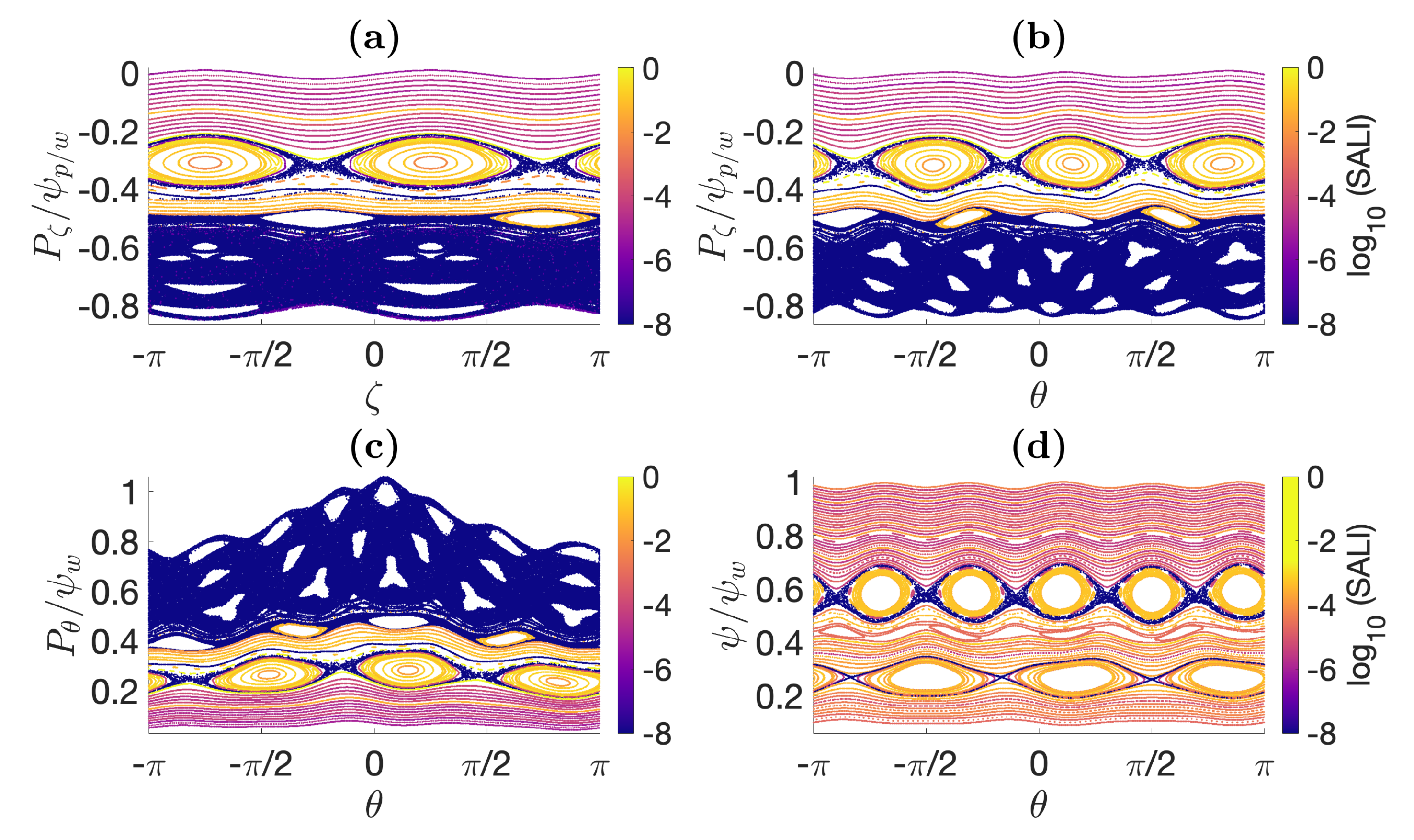}
	\caption[Poincar{\'e} surfaces of section for thermal particles showing the magnetic versus kinetic chaos.]{(a), (b) and (c) Kinetic Poincar{\'e} surfaces of section for thermal particles with $E=2.9$keV and $\mu B_{0}=2.0$keV. (d) Magnetic Poincar{\'e} surface of section. All orbits are coloured according to their SALI value calculated for $10^{8}$ time units. Two perturbation modes with mode numbers $(m,n)=(3,2), (5,2)$ and equal amplitudes are considered $\alpha_{3,2}=\alpha_{5,2}=\epsilon=7.5\times 10^{-5}$. }
 \label{fig:Fig45} 
\end{figure}

The case of thermal particles with $E=2.9$keV and $\mu B_{0}=2.0$keV is depicted in Fig.~\ref{fig:Fig45}. As shown in the kinetic Poincar{\'e} surfaces of section in Figs. \ref{fig:Fig45}(a) and (b), the resonant islands appear in different values of $P_\zeta$ and the lower chaotic region is significantly more extended in comparison to Figs. \ref{fig:Fig44}(a) and (b). Moreover, as shown in Figs. \ref{fig:Fig45}(c) and (d), the kinetic chaos related to the upper primary resonant island chain extends to a significantly larger range of radial positions in comparison to magnetic chaos. In fact, this chaotic region connects to the wall ($P_\theta / \psi_w =1$), suggesting perturbation-induced stochastic particle loss. Also, secondary island chains appear within the chaotic regions of the kinetic Poincar{\'e} surface of section.        

\begin{figure}[h!]
    \centering
    \includegraphics[width=1.0\textwidth,keepaspectratio]{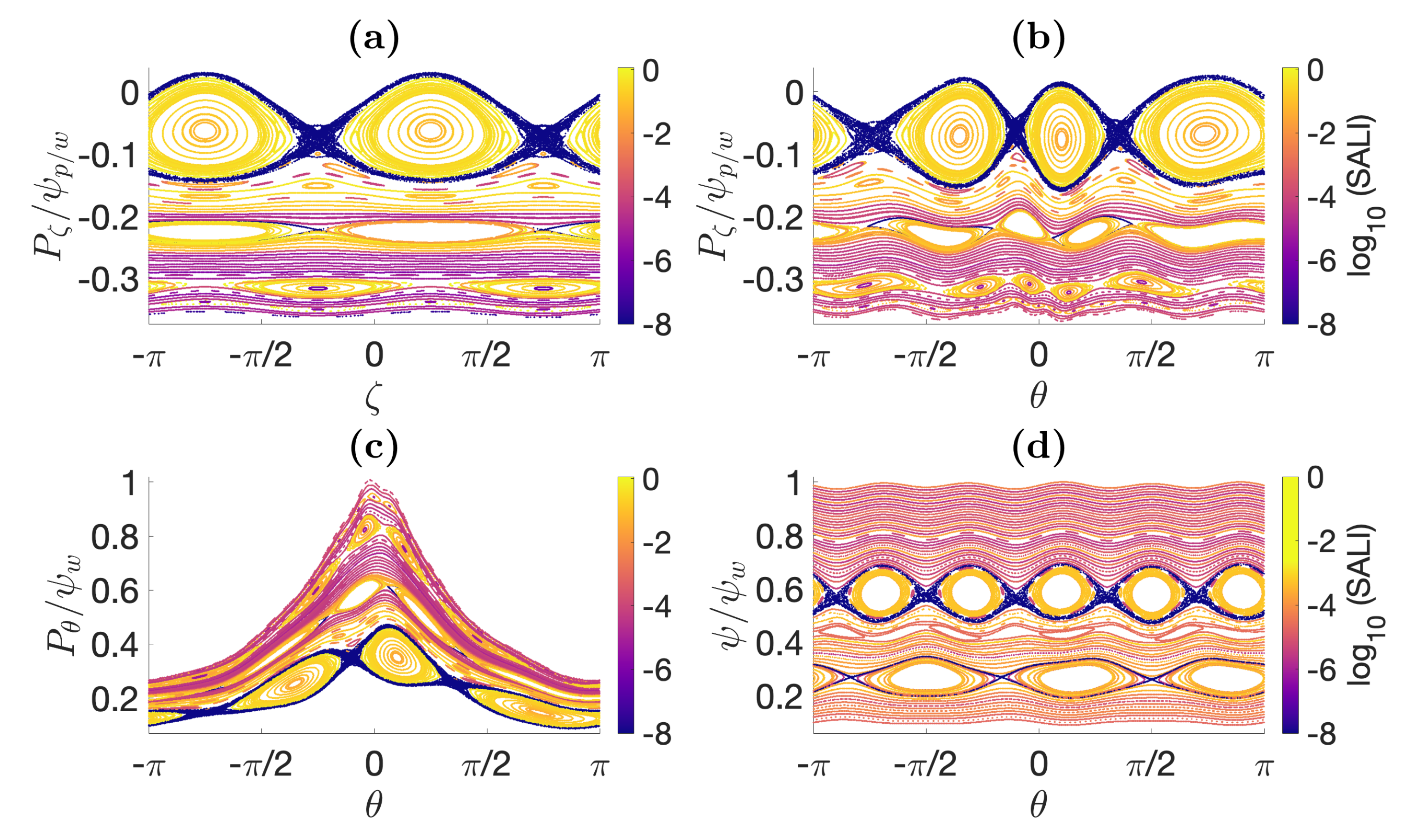}
    \caption[Poincar{\'e} surfaces of section for energetic particles showing the magnetic versus kinetic chaos.]{(a), (b) and (c) Kinetic Poincar{\'e} surfaces of section for energetic particles with $E=39.1$keV and $\mu B_{0}=33.1$keV. (d) Magnetic Poincar{\'e} surface of section. All orbits are coloured according to their SALI value calculated for $10^{8}$ time units. Two perturbation modes with mode numbers $(m,n)=(3,2), (5,2)$ and equal amplitudes are considered $\alpha_{3,2}=\alpha_{5,2}=\epsilon=7.5\times 10^{-5}$. }
    \label{fig:Fig46}
\end{figure}

The case of higher-energy particles with $E=39.1$keV and $\mu B_{0}=33.1$keV is depicted in Fig.~\ref{fig:Fig46}, where the kinetic Poincar{\'e} surfaces of section clearly show that the resonance chains are located at quite different values of $P_\zeta$ in comparison to the previous lower-energy particles \cite{Zestanakis2016, Antonenas2021}. In this case, the extent of kinetic chaos is markedly reduced in comparison to the case of thermal particles with $E=2.9$keV and $\mu B_{0}=2.0$keV (Fig.~\ref{fig:Fig45}) and it is located close to the separatrices of the primary island chains. 

The kinetic and magnetic Poincar{\'e} surfaces of section in Figs.~\ref{fig:Fig44}-\ref{fig:Fig46} clearly show that the same magnetic perturbations, and therefore the same magnetic field chaos, has completely different effects on particles with different kinetic characteristics. Although, Poincar{\'e} surfaces of section can serve for illustrating characteristic cases, a systematic global overview of the effect of the magnetic field chaos, due to a specific set of modes, on the entire range of particles can be provided in the space of the constants of the motion uniquely labeling each GC orbit and characterizing it in terms of being trapped or passing, and confined or lost. Moreover, such a description necessitates chaos quantification enabling the assignment of a chaos measure at each point (orbit) in the space of the constants of the motion. 

\begin{figure}[h!]
	\centering
 	\includegraphics[width=0.475\textwidth]{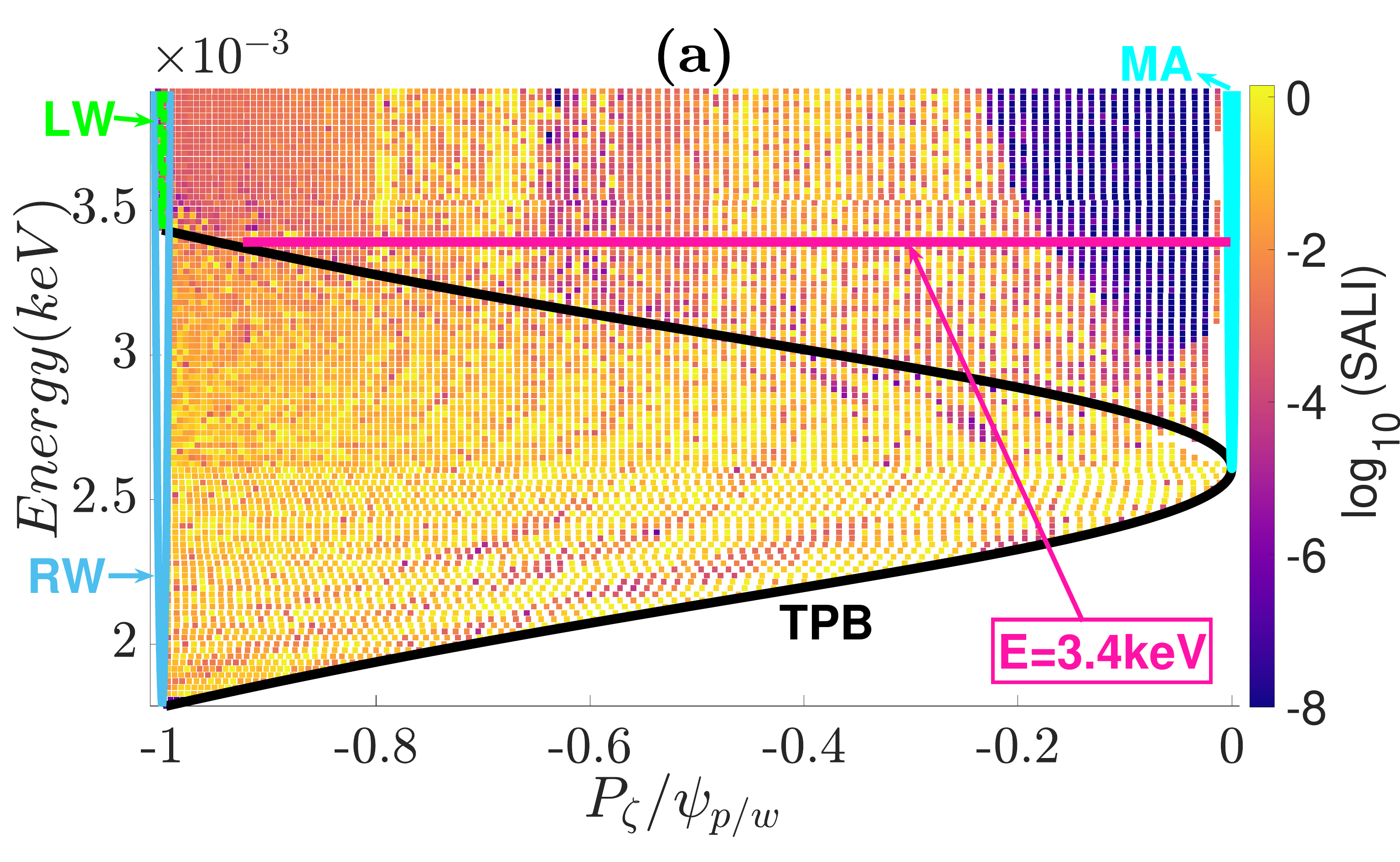}
      \includegraphics[width=0.475\textwidth]{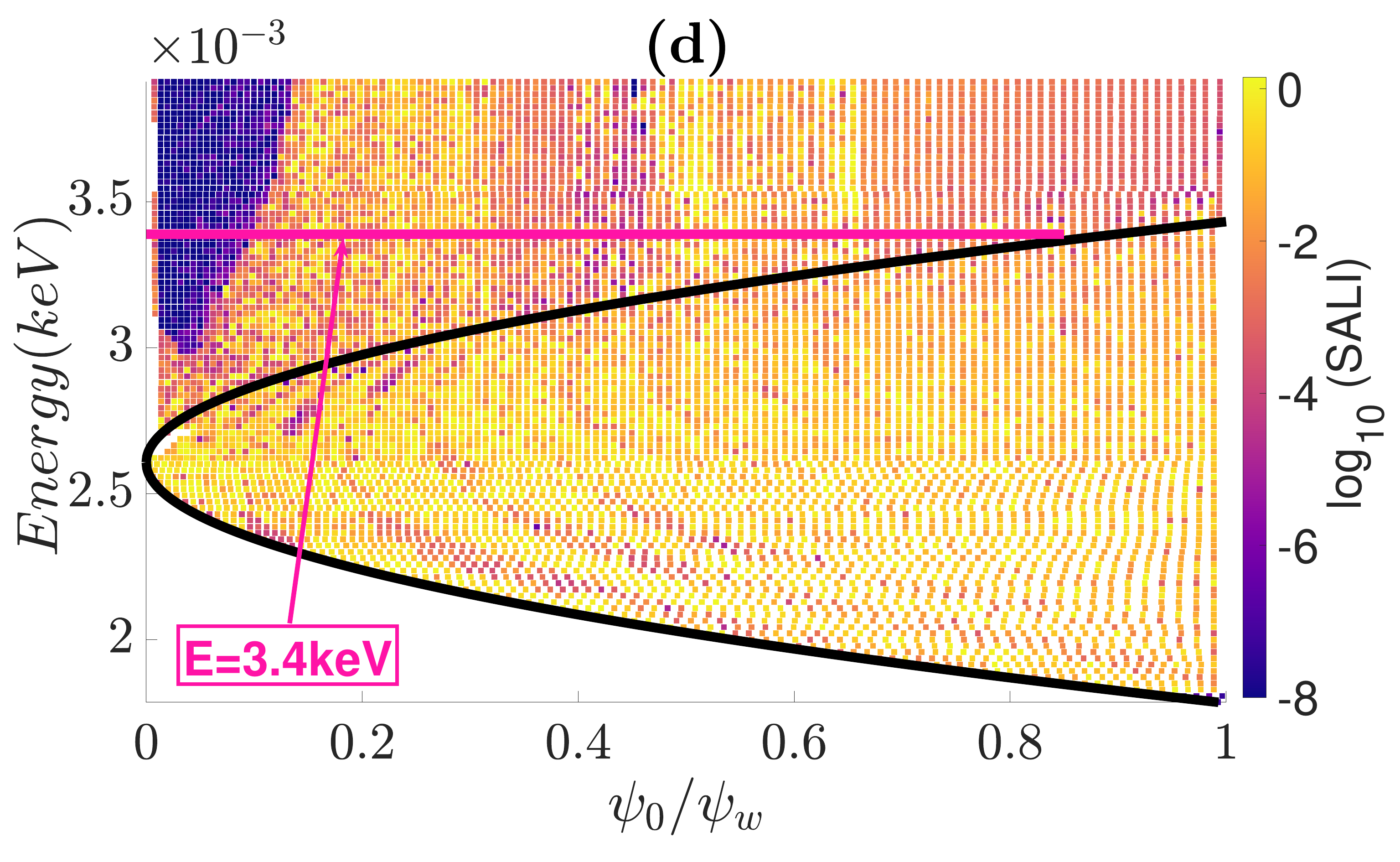}
   \includegraphics[width=0.475\textwidth]{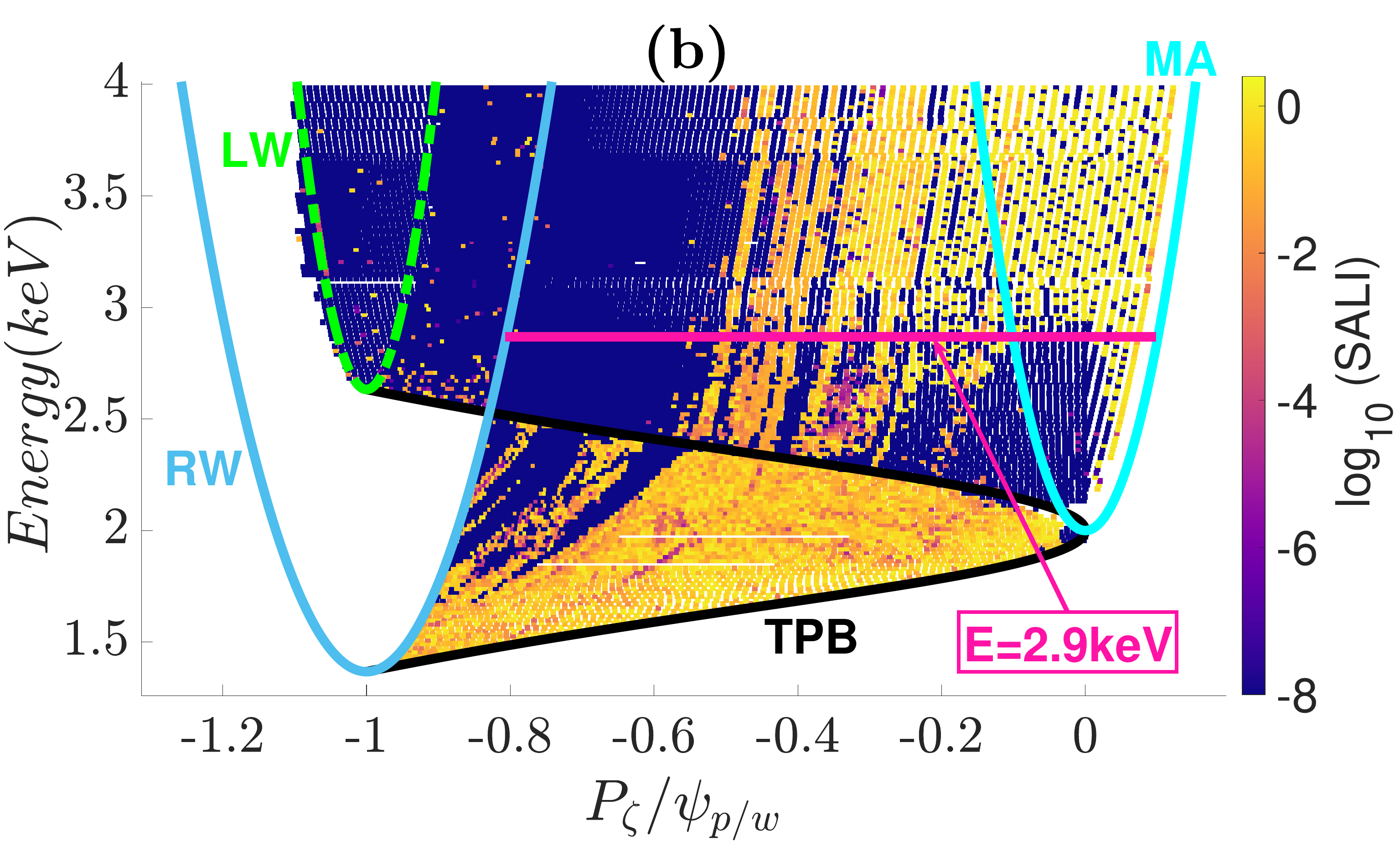}   
       \includegraphics[width=0.475\textwidth]{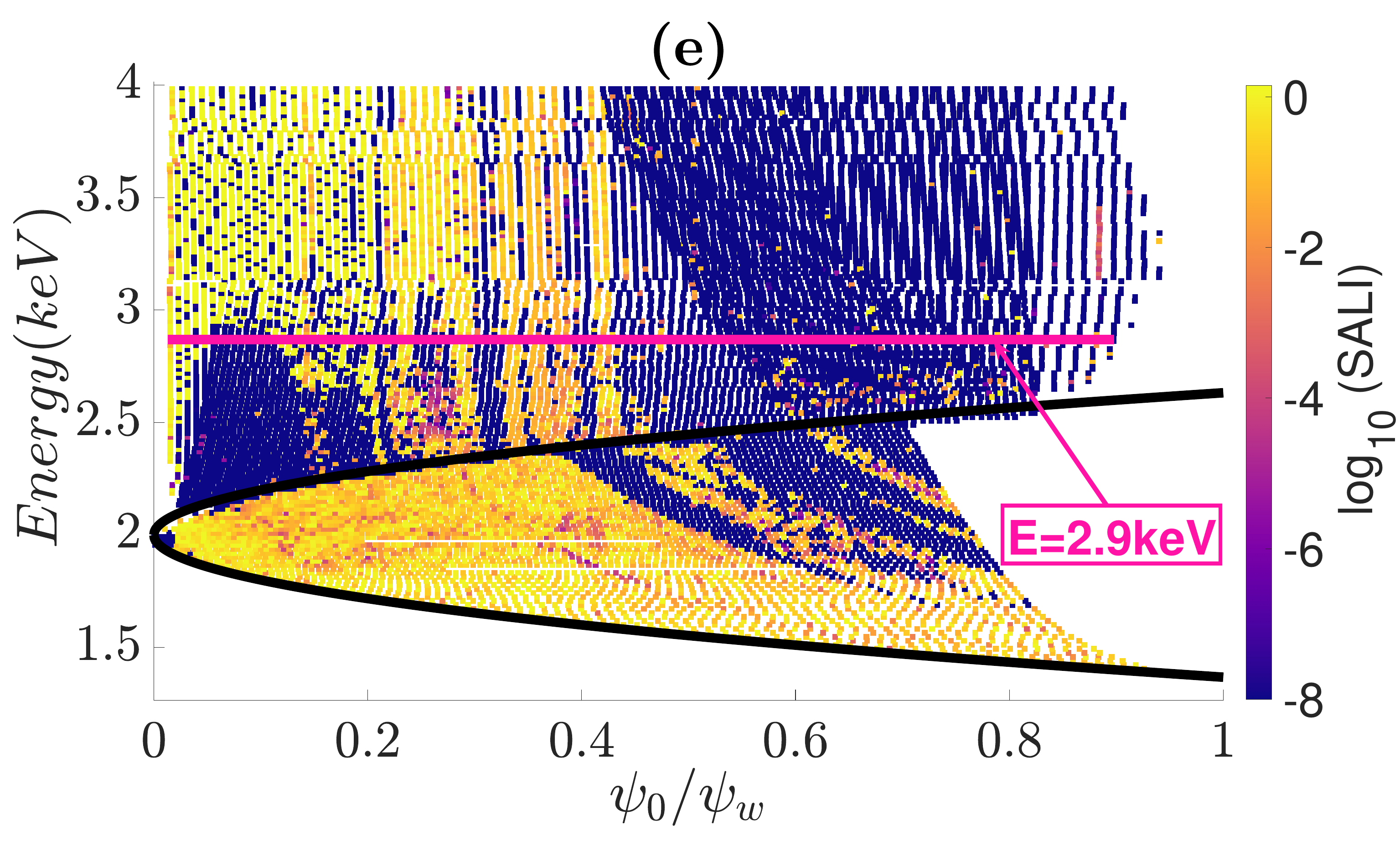}
 	  \includegraphics[width=0.475\textwidth]{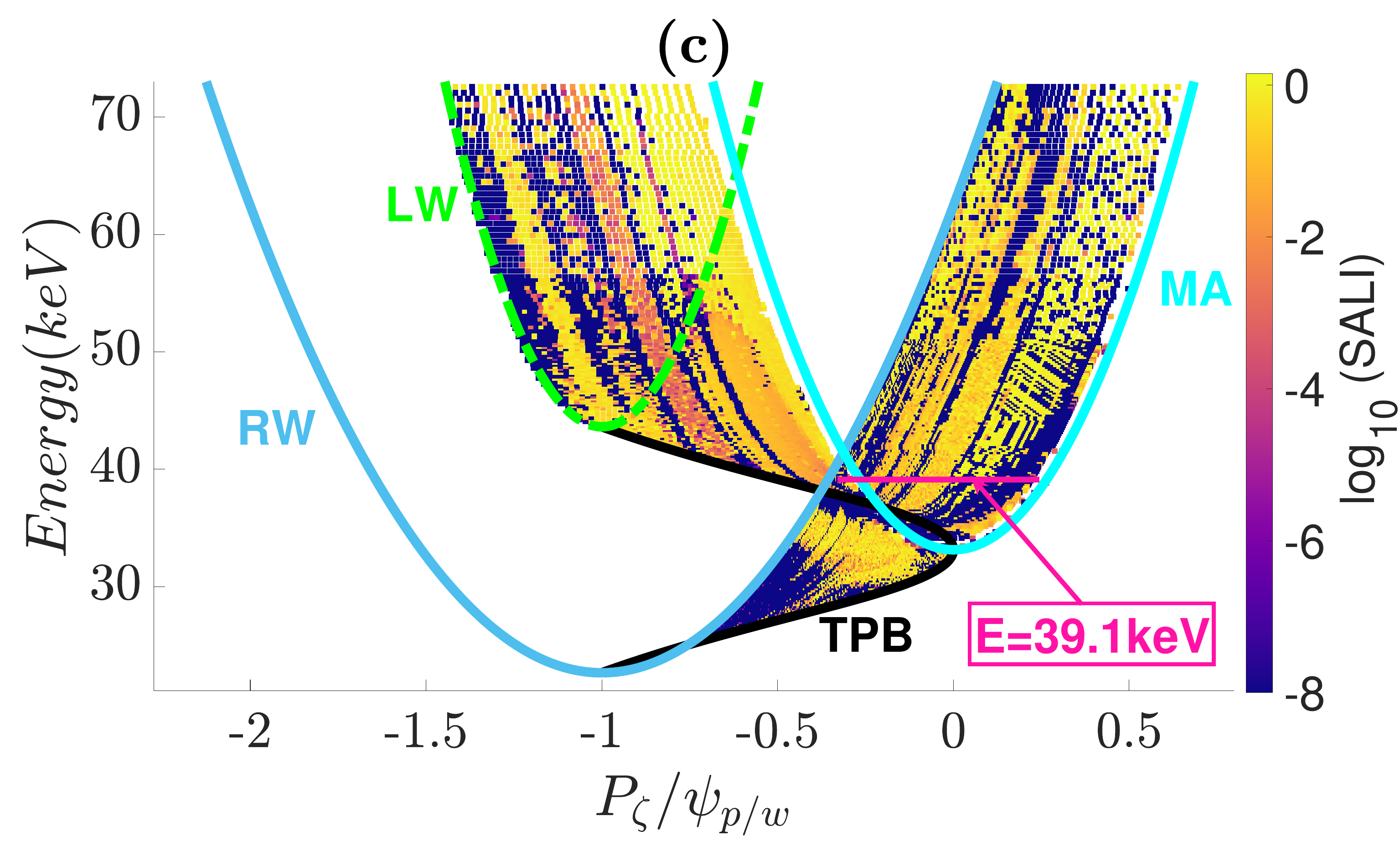}
    \includegraphics[width=0.475\textwidth]{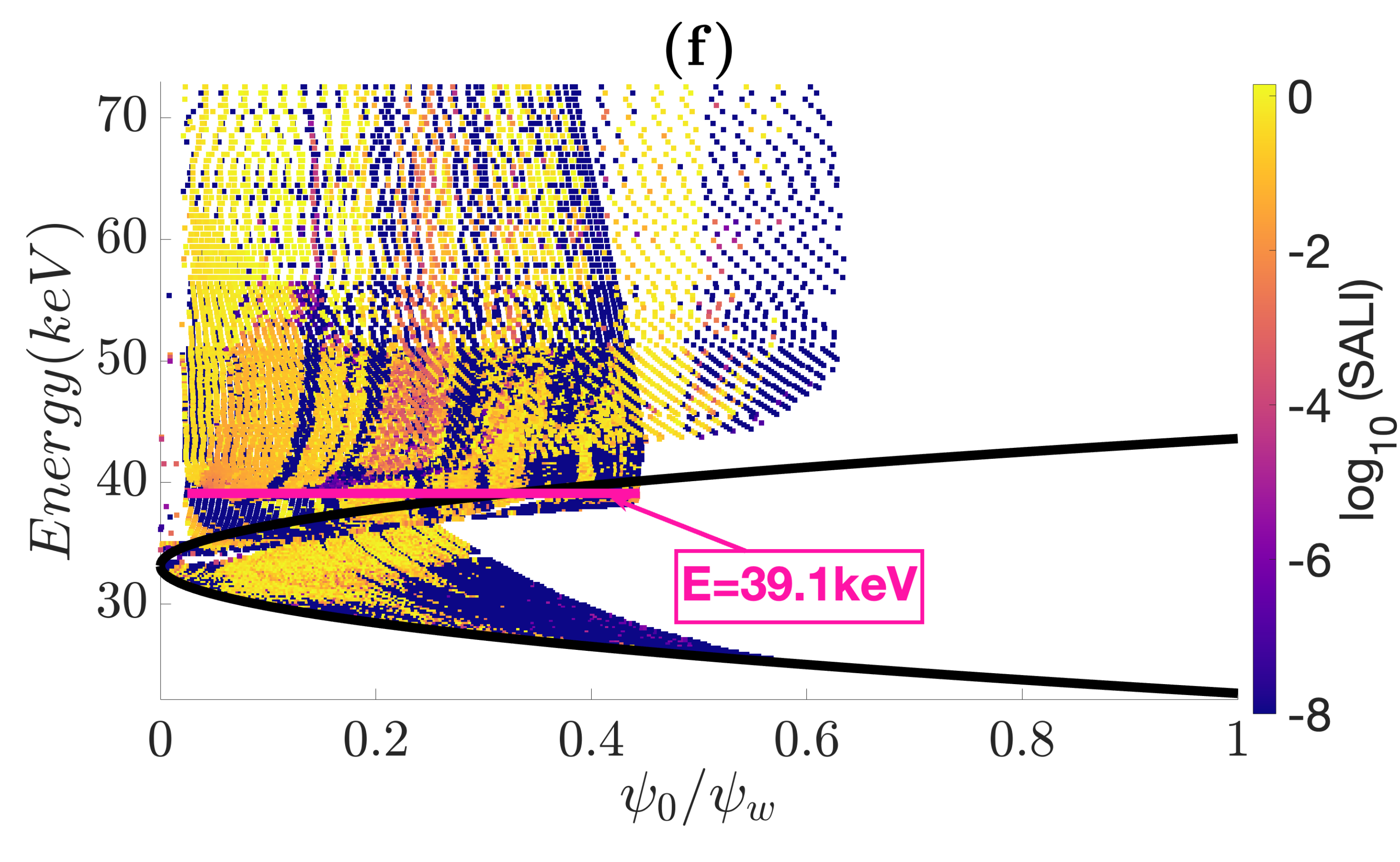}
    
	\caption[Smaller ALignement Index (SALI) application in particle constants of the motion space.]{Constant $\mu$ plane cuts of the 3D constants-of-the-motion space  $(E,P_\zeta,\mu)$ (a-c) or $(E,\psi_0,\mu)$ (d-f) with $\mu B_0=2.6$eV (a, d), $\mu B_0=2.0$keV (b, e), and $\mu B_0=33.1$keV (c, f) corresponding to perturbation modes as in Figs. \ref{fig:Fig44}-\ref{fig:Fig46}. Each point defines an orbit and it is colored according to the calculated SALI value, with dark blue and yellow areas representing chaotic and regular motion, respectively. The specific energy level considered in the Poincar{\'e} surfaces of section in Figs. \ref{fig:Fig44}-\ref{fig:Fig46} is denoted by a horizontal pink line. The right wall (RW), left wall (LW), magnetic axis (MA), and trapped-passing boundary (TPB) [Eqs. \eqref{walls}, \eqref{magnetic axis},  \eqref{trapped passing boundary}] are depicted by solid light blue, dashed light green and solid cyan parabolas respectively, and provide an overview of the specific kinetic characteristics of particle orbits with different chaoticity. The radial position where kinetic chaos occurs can be obtained in (d-f) where the reference flux surface $\psi_0$ of its orbit is used instead of $P_\zeta$ according to Eq.~(\ref{psi0}).}
	\label{fig:Fig47}
\end{figure}

The three-dimensional COM space $(E, P_\zeta, \mu)$ is systematically dissected in Fig.~\ref{fig:Fig47} (a-c), where the SALI value, calculated at each point, quantifies the chaoticity of the respective orbit. Moreover, these diagrams reveal whether the chaotic orbits correspond to trapped or passing particles as well as their positions with respect to the magnetic axis and the wall, as defined by Eqs. \eqref{walls}, \eqref{magnetic axis},  \eqref{trapped passing boundary}. It is evident that the same magnetic chaoticity (same perturbative modes) corresponds to completely different kinetic chaoticity depending strongly on the kinetic characteristics of the particles. The diagrams provide a clear and detailed overview of the effect of a specific set of perturbative modes on the chaotic particle, energy and momentum transport and confinement in toroidal fusion devices. Narrow chaotic strips correspond to localized chaos at separatrices of isolated resonant island chains (this will be discussed in detail in the next chapter), as illustrated in Figs. \ref{fig:Fig44}-\ref{fig:Fig46}, whereas wider chaotic regions correspond to extended chaos due to resonance overlap. The relative position of the chaotic areas with respect to the position of the wall and the trapped/passing boundary, provides information on  particle losses as well as complex particle trapping-detrapping dynamics due to the formation of a chaotic sea connecting both sides of the separatrix (trapped/passing boundary) of the unperturbed particle motion. It is worth noting the remarkable high degree of aggregation of the information presented here, in comparison to Poincar{\'e} surfaces of section [Figs. \ref{fig:Fig44}-\ref{fig:Fig46}], each one corresponding to a constant energy line, as well as the importance of efficiently quantifying chaos in terms of SALI that enables the assignment of a chaos measure to each orbit corresponding to a point of the dense grid used to depict the fine details of kinetic chaos. 

\clearpage

The radial position and extent of kinetic chaos cannot be rigorously defined due to the fact that even the unperturbed particle orbits, especially the more energetic ones, are not restricted on a single flux surface due to GC drifts. However, a flux surface of reference $\psi_0$, around which the particle drifts, can be defined for each unperturbed orbit in order to have an estimation for the approximate radial distance where kinetic chaos actually takes place. Such a flux surface $\psi_0$ can be defined in general according to  
\begin{equation}
    \psi_p(\psi_0)=\left< \psi_p (\psi) \right >=\left<\frac{\pm g(\psi)}{B(\psi,\theta)}\sqrt{2\left(E-\mu B(\psi,\theta)\right)} \right> - P_\zeta,
\end{equation}
where the brackets denote orbit averaging. For trapped particles, it suffices to consider $\psi_0$ as the value of $\psi$ at the points of velocity reversal, where the trapped particles spend the most of their time. To lowest order in a LAR configuration, $\psi_0$ can be defined for all types of orbits, in terms of the constant variables $(E, \mu, P_\zeta)$, as follows 
\begin{equation}
    \psi_{p}(\psi_{0}) = \left\{
    \begin{array}{ll}
     -P_\zeta    & \text{, trapped}\\
     \pm \sqrt{2(E-\mu)}-P_\zeta   & \text{, co/counter-passing}
    \end{array}
    \right.
    \label{psi0}
\end{equation}
 and can be used as an alternative to $P_\zeta$ constant of the motion uniquely labeling each orbit along with $E$ and $\mu$ and characterizing the radial position and extent of kinetic resonances, islands and chaos \cite{Pinches1998, Antonenas2021} (for other alternative sets of COM see Refs. \cite{Bierwage2022b, Benjamin2023}). The respective information of the kinetic chaoticity in the $(E,\psi_0,\mu)$ space is depicted in Figs.~\ref{fig:Fig47}(d-f).

In summary, low-energy particles closely follow magnetic field lines and have approximately the same chaoticity with them. The effect of the non-axisymmetric perturbation on the topology of the magnetic field lines is localized to regions where magnetic resonant islands or chaotic regions, under conditions for resonance overlap, are formed. Higher-energy particles undergo large drifts across the magnetic field lines, and the effect of the non-axisymmetric perturbations on their orbits is radically different from the effect of the same perturbations on the magnetic field lines. In this general case, the locations of kinetic and magnetic resonant island chains, the conditions for kinetic and magnetic chaos, as well as the degrees of kinetic and magnetic chaoticities are drastically different. Chaotic transport of high-energy particles determines the confinement limitations and the performance of the fusion devices and necessitates the understanding of the relation between magnetic and kinetic resonances and the respective magnetic and kinetic chaos. A systematic study of this relation requires: (a) the identification as well as the quantification of chaos, and (b) the compact representation of particle orbits in a kinetic parameter space. 

In our work in Ref. \cite{Moges2024}, we introduced the SALI (Smaller ALignement Index) as an efficient measure for detecting and quantifying both magnetic and kinetic chaos and discuss its advantages in comparison with other standard measures, such as the mLE (maximum Lyapunov Exponent). Kinetic and magnetic chaos are compared, in terms of Poincar{\'e} surfaces of section for indicative cases of low- and high- energy particles to justify the need for a systematic detailed comparison which is performed in the three-dimensional kinetic space of the constants of motion. Each point of this space uniquely represents an unperturbed GC orbit and classifies it in terms of being trapped or passing, and confined or lost. Under the presence of perturbations, a specific SALI value is assigned to each point, providing detailed information regarding the specific kinetic characteristics of the orbits that actually become chaotic due to perturbations, as well as their position with respect to the wall of the torus. These diagrams provide a detailed phase space resolution of kinetic chaos, enabling the physical intuition on the role of specific sets of perturbative modes in terms of particle, energy and momentum transport, and can be employed as a valuable tool for understanding the interplay between internal instabilities and intentionally induced external perturbations, and investigate synergetic effects and mode-engineering strategies for mitigating high-energy particle losses. Moreover, they are of particular importance in cases of multiple time-dependent perturbations where Poincar{\'e} surfaces of section are difficult to be defined and used for the visualization of the phase space topology. Overall, the quantification of kinetic chaos contains distilled information of the complex phase space dynamics that can be further used in integrated tokamak transport simulation codes, as well as for comparison with fast ion and other diagnostics.

\chapter{Mode-Particle GC resonances analysis based on Hamiltonian Action-Angle Formalism} \label{Ch:Mode-Particle GC resonances analysis based on Hamiltonian Formalism}

In the first part of this chapter, we present the results of our studies on the analytical calculation of the orbital frequencies (OF) of the guiding center motion in toroidally confined plasmas. These works, detailed in Refs. \cite{Antonenas2021, Antonenas2024}, employ the theoretical tools developed in Chapter \ref{Sec: Chapter 3} to investigate mode-particle guiding center (GC) resonant interactions, whose significance was introduced in Chapter \ref{Sec: Chapter 2}. We implement these tools in the guiding center Hamiltonian system, formulated in Chapter \ref{Sec: Chapter 4}.

Our approach is based on the Hamiltonian Action-Angle (A-A) formalism, which provides a compact and efficient framework for studying the orbital dynamics of GC motion. Specifically, we derive analytical expressions for the orbital frequencies (OF) of GC trajectories and express the kinetic $q$-factor as a function of the particle’s Constants of Motion (COM). The kinetic $q$-factor plays a crucial role in identifying the conditions for mode-particle resonances, enabling the precise determination of resonance locations in COM space. Additionally, we demonstrate that local extrema of $q_{kin}$ correspond to Transport Barrier (TB) locations, which influence plasma transport properties. Since the entire derivation is based on the A-A formalism, it allows not only the prediction of the location and number of islands in each island chain but also the width of the islands, as well as all resonances induced by a perturbative mode.  

Extending beyond the LAR framework, we extend the applicability of our OF analysis methodology from LAR to numerically reconstructed equilibria. We employ a computationally efficient semi-analytical geometrical method applied on any given unperturbed equilibrium. The calculated resonance diagrams contain all the essential
information for the response of all particle species - including energetic particles – in the
specific equilibrium under the presence of any type of multi-scale symmetry-breaking
perturbations. Systematic comparisons with numerical particle tracing (Poincaré diagrams) show an excellent agreement with the predicted locations of the resonances and the Transport Barriers.

In the second part of this chapter, based on analytically calculated orbital frequencies, we construct the Arnold web of GC motion for the LAR equilibrium and demonstrate the phenomenon of Arnold diffusion in GC motion under multiple time-dependent perturbative modes.

The organization of this chapter is as follows. In Section \ref{Action-Angle Transformation}, we present the action-angle formulation of the guiding center Hamiltonian system. Section \ref{Analytical} introduces our Drift Center (DC) approximation and presents the analytical derivation of $q_{kin}$. The resonant response to non-axisymmetric perturbations in LAR equilibrium is analyzed in Section \ref{Resonse}. The number of islands that are formed in a resonance island chain in a Poincaré surface of section is discussed in Sec. \ref{Sec: Number of islands}. In Section \ref{Sec: realistic equilibrium}, we extend our methodology to realistic equilibrium magnetic fields, introducing a semi-analytical geometric approach and performing systematic comparisons with numerical particle-tracing results. Finally in Sec. \ref{Sec: GC Arnold diffusion} we discuss the particle energy and momentum transport on the Arnold
web of the GC motion.

\ifpdf
    \graphicspath{{Chapter5/Figs/Raster/}{Chapter3/Figs/PDF/}{Chapter5/Figs/}}
\else
    \graphicspath{{Chapter5/Figs/Vector/}{Chapter5/Figs/}}
\fi

\section{Action-Angle formulation of the GC Hamiltonian System} \label{Action-Angle Transformation}
The GC Hamiltonian can be formulated in Action-Angle (AA) variables through a canonical transformation. As discussed in Sec. \eqref{Action Angle Variables}, the AA formulation provides unique advantages for the study of complex particle dynamics, by exploiting the full range of concepts and methods of Hamiltonian systems. When expressed in AA variables, the different time scales of multiply periodic particle dynamics are distinctly separated into different degrees of freedom. The respective orbital frequencies can be readily calculated, without requiring a complete solution to the equations of the motion, facilitating the formulation of resonance conditions that dictate particle, momentum and energy transport, under the influence of external perturbations to the integrable axisymmetric configuration. Moreover, the AA variables offer a powerful means for a systematic dynamical contraction of the GC dynamics in a hierarchy of descriptions with reduced dimensionality \cite{Goldstein2002, White2013, Brizard2014, Antonenas2021, Anastassiou2024}.

For the GC Hamiltonian (\ref{GC H 1}) the three Action variables are defined as 
\begin{equation} \label{gen Action Integral}
    J_{i} = \frac{1}{2\pi}\oint {p_{i}(q_{i}; E,\mu,P_\zeta)dq_{i}}
\end{equation}
where $(q_i,p_i)=(\xi,\mu), (\zeta,P_\zeta), (\theta,P_\theta)$. The integration is performed over a complete cycle of the respective canonical position $q_i$, with the variables describing the other degrees of freedom kept constant \cite{Goldstein2002}. For the first two degrees of freedom, the canonical positions $\xi$ and $\zeta$ are cyclic, completing a full circle within the range of $0$ to $2\pi$; their respective canonical momenta $\mu$ and $P_\zeta$ are invariants, and the integration simply yields
\begin{equation}\label{gen Jzeta}
    J_\xi= \mu, \quad J_\zeta= \sigma P_\zeta
\end{equation}
with $\sigma =+1$ and $\sigma=-1$ for co- and counter-moving orbits, with respect to the magnetic field, respectively. For the third (poloidal) degree of freedom we have
\begin{equation} \label{gen Jtheta}
    J_{\theta} = \frac{1}{2\pi}\oint{P_{\theta}(\theta^{\prime}; E, \mu, P_{\zeta})}d\theta^{\prime},
\end{equation}
where $P_\theta$ is obtained from Eq. (\ref{GC H 1}) as a function of $\theta$ and the three COM. It is worth noting that the explicit appearance of the (non-cyclic) canonical position $\theta$ in the Hamiltonian (\ref{GC H 1}) results in its non-trivial variation; consequently, $\theta$ may be restricted to vary in subsets of the interval $(0,2\pi)$ or may even be stationary. The canonical transformation to the AA variables is provided by the mixed-variable generating function
\begin{equation} \label{genarating function}
   F(\theta, \zeta, \xi, J_{\theta}, J_{\zeta}, J_{\xi}) = J_{\zeta}\zeta + J_{\xi}\xi+ f(\theta,J_{\theta}, J_{\zeta}, J_{\xi})
\end{equation}
where 
\begin{equation}
    f(\theta,J_{\theta}, J_{\zeta}, J_{\xi})=\int^{\theta}{P_{\theta}(\theta^{\prime}, E, \mu, P_{\zeta})}d\theta^{\prime} 
\end{equation}
and the three COM, $(E,\mu,P_\zeta)$, are now considered as functions of the three Actions $\mathbf{J}=(J_\theta,J_\zeta,J_\xi)$.

The new angles are related to the old ones through the canonical transformation as \cite{Kaufman1972, Zestanakis2016}
\begin{equation}\label{eq:Angles}
    \hat{\theta}=\frac{\partial f(\mathbf{J,\theta})}{\partial J_\theta}, \qquad \hat{\zeta}=\zeta+\frac{\partial f(\mathbf{J,\theta})}{\partial J_\zeta}, \qquad \hat{\xi}=\xi+\frac{\partial f(\mathbf{J,\theta})}{\partial J_\xi}. 
\end{equation}
Expressed in AA variables, the Hamiltonian is independent of all new Angles, and the orbital frequencies are given as $\hat{\omega}_i=\partial H(\mathbf{J})/\partial J_i$, $i=\theta,\zeta,\xi$. It is worth emphasizing that, in contrast to the new Angles $(\hat{\theta},\hat{\zeta},\hat{\xi})$, the original angles $(\theta,\zeta,\xi)$ do not have a linear time dependence, and therefore, their time derivatives $(\dot{\theta},\dot{\zeta},\dot{\xi})$ do not correspond to any orbital frequency. The first of Eqs. (\ref{eq:Angles}) implies that the time dependence of the Angle $\hat{\theta}$ is solely through the old angle $\theta$, so that it is also a periodic function of time with the same frequency
\begin{equation}\label{gen hat omega theta}
\omega_\theta=\hat{\omega}_\theta=\frac{\partial H(\mathbf{J})}{\partial J_\theta}. 
\end{equation}
It is also worth noting that $\hat{\theta}$ corresponds to an unbounded phase coordinate with a linear time dependence, whereas $\theta$ can be either unbounded (for passing orbits) or bounded (for trapped orbits) and exhibits a more complex time dependence. The case of the other two Angles, $\hat{\zeta}$ and $\hat{\xi}$ is qualitatively different, since they depend, not only on the corresponding original angles, $\zeta$ and $\xi$, but also on $\theta$ (see also Sec. \ref{Sec: Number of islands}). 

By implicit differentiation it can be readily shown \cite{Antonenas2021, Anastassiou2024, Antonenas2024} that 
\begin{equation}\label{gen hat omega zeta}
    \hat{\omega}_{\zeta} = \frac{\partial H(\mathbf{J})}{\partial J_\zeta}=-\frac{\partial H}{\partial J_{\theta}}\frac{\partial J_{\theta}}{\partial J_{\zeta}} = -\hat{\omega}_{\theta}\frac{\partial J_{\theta}}{\partial J_{\zeta}}
\end{equation}
and, similarly,
\begin{equation}\label{gen hat omega xi}
    \hat{\omega}_{\xi} =\frac{\partial H(\mathbf{J})}{\partial J_\xi} = -\frac{\partial H}{\partial J_{\theta}}\frac{\partial J_{\theta}}{\partial J_{\xi}} = -\hat{\omega}_{\theta}\frac{\partial J_{\theta}}{\partial J_{\xi}}.
\end{equation}
The form of these equations suggests that $(-J_\theta)$ acts as a new Hamiltonian in the remaining Action-Angle variables, where the new time variable is measured in units of the bounce/transit period, similarly to the definition of the dynamical contraction discussed in (Ref. \cite{White2013}, p. 97).\

Focusing on the drift motion in the slower degrees of freedom $(\hat{\theta},\hat{\zeta})$, we have  
\begin{equation} \label{precession}
 \hat{\omega}_{\zeta} = -\hat{\omega}_{\theta} \frac{1}{2\pi}\oint{\frac{\partial P_{\theta}(\theta', E, P_{\zeta}, \mu)}{{\partial P_{\zeta}}}} d \theta'= \frac{(\Delta \zeta)_{T_{\theta}}}{T_\theta},
\end{equation}
where we have used a change of variables from $\theta$ to $\zeta$, with $d\zeta / d\theta= - \partial P_{\theta}/\partial P_{\zeta}$, and $(\Delta \zeta)_{T_\theta}$ denotes the variation of $\zeta$ over a complete period $T_\theta$ of the poloidal angle $\theta$, indicating that the orbital frequency $\hat{\omega}_\zeta$ corresponds to the bounce/transit averaged rate of toroidal precession. Similarly, it can be shown that $\hat{\omega}_\xi$ corresponds to the bounce/transit averaged gyrofrequency. Thus, the complete Orbital Spectrum (OS) of the GC motion is determined in terms of the three Action variables $\mathbf{J}$ or the three COM $(E,\mu,P_\zeta)$. It is worth clarifying that in the following we do not consider high-frequency perturbations, introducing $\xi$ dependence, so that the non-axisymmetrically perturbed Hamiltonian system can be treated as a one-parameter $\mu$ family of two-degree of freedom Hamiltonian systems; however, we have preferred to keep a general formulation demonstrating the generality of this approach and suggesting possible extensions to high-frequency perturbations.

\section{Drift center (DC) approximation - Analytical calculation of the kinetic $q$-factor} \label{Analytical}
To calculate the Action $J_{\theta}$, as defined by Eq. \eqref{gen Jtheta}, it is necessary to solve Eq. \eqref{GC H 1} in order to obtain $P_\theta$ as a function of $\theta$ and the three COM $(E,\mu,P_\zeta)$, defining each orbit along which the integration is performed. Even for the simplest case of a LAR equilibrium, an analytical expression is not always available, and even when it is, the subsequent integration may not be performed analytically. To address this issue, we employ an approximation by selecting an appropriate magnetic surface of reference, $\psi_0$, for each GC orbit, where the magnetic field is evaluated as $B(\psi_0,\theta)$. This approximation is analogous to the GC approximation (with the key difference that it does not reduce the dimensionality of the system), where the magnetic field at each position of a gyrating particle is approximated by its value at the center of its orbit. Just as the validity of the GC approximation depends on the spatial variation of the magnetic field within a gyration orbit, the validity of this approximation depends on the variation of the magnetic field within a GC drift orbit, with larger drifts making the approximation less reliable. 

We refer to this method as the Drift Center (DC) approximation. It can be applied to any equilibrium (not just LAR) if the equilibrium-related functions, $g(\psi)$ and $I(\psi)$, which appear in the GC Hamiltonian system (Eqs.~\eqref{canon moments}, \eqref{GC H 1}), are evaluated at $\psi_0$, and if $\psi_p(P_{\zeta},P_{\theta})$ is expanded as  
\begin{equation}\label{Eq: psip expanded at psi_0}
    \psi_{p}(P_{\zeta}, P_{\theta})=\psi_{p}(P_{\zeta},P_{\theta_{0}})+\frac{P_{\theta}-P_{\theta_{0}}}{q(P_{\zeta},P_{\theta_{0}})\left.\frac{\partial P_{\theta}}{\partial\psi}\right |_{\psi=\psi_0}}.     
\end{equation}
where $\psi_p(P_{\zeta},P_{\theta_{0}}) = \psi_p(\psi_0)$ and $q(P_{\zeta},P_{\theta_{0}}) = q(\psi_0)$. $P_{\theta_0}$ is obtained from Eq.~\eqref{Eq: Ptheta with psi 1} as 
\begin{equation}\label{Eq: Ptheta with psi0}
    P_{\theta_0} = \psi_0 + \frac{P_{\zeta} + \psi_p(\psi_0)}{g(\psi_0)}I(\psi_0).
\end{equation}
Then, for any equilibrium magnetic field $\bm{B}(\psi,\theta)$, the canonical momentum $P_{\theta}(\theta,E,\mu,P_{\zeta})$ is given by: 
\begin{align}\label{Eq: Ptheta(theta) gen B}
    P_{\theta}=\sigma q(\psi_0)\frac{\sqrt{2g(\psi_0) E - \mu B(\psi_0,\theta)}}{B(\psi_0,\theta)} - \left[q(\psi_0)(P_{\zeta}+\psi_p(\psi_0)) + P_{\theta_0}(\psi_0)\right]
\end{align}
where $\sigma =+1$ for trapped and co-passing orbits and $\sigma = -1$ for counter-passing orbits. An appropriate magnetic surface of reference can be defined as a function of the constants of motion, $\psi_{0}(E,\mu,P_\zeta)$, according to (see also Ref.~\cite{Pinches1998})  
\begin{equation} \label{def of psi0}
    \psi_{0} = \begin{cases} 
     \psi_{p}^{-1}(-P_\zeta),    & \text{trapped}\\
     \psi_{p}^{-1}(\pm \sqrt{2(E-\mu)}-P_\zeta),   & \text{co-/counter-passing}
    \end{cases}
\end{equation}

For the LAR equilibrium (Sec.~\ref{Sec: LAR equilibrium}), where $\psi=P_\theta$, Eq.~\eqref{Eq: Ptheta with psi0} using  Eq. \eqref{def of psi0} yields:  
\begin{equation} \label{def of Ptheta0}
    P_{\theta_{0}} = \begin{cases} 
     \psi_{p}^{-1}(-P_\zeta),    & \text{trapped}\\
     \psi_{p}^{-1}(\pm \sqrt{2(E-\mu)}-P_\zeta),   & \text{co-/counter-passing}
    \end{cases}
\end{equation}
where $\psi_0 (P_{\theta_0})$ corresponds to the flux surface at the banana tip $(\rho_\parallel=0)$ and the point $\theta=\pi/2$ of the trapped and passing orbits, respectively \cite{Pinches1998}, and $\psi_p^{-1}$ denotes the inverse function.
Moreover, $\psi_{p}(P_{\theta})$ can be approximated by a first order Taylor expansion around $P_{\theta_0}$, as
\begin{equation}\label{Eq: psip expanded at psi_0}
    \psi_{p}(P_{\theta})=\psi_{p}(P_{\theta_{0}})+\frac{P_{\theta}-P_{\theta_{0}}}{q(P_{\theta_{0}})}.     
\end{equation}
    
The Action can be calculated as follows:

\begin{equation} \label{jt}
J_{\theta} = \frac{q(P_{\theta_{0}})}{2\pi}\oint{\frac{\sqrt{2\left[E - \mu\left(1-\sqrt{2 P_{\theta_{0}}}\cos\theta\right)\right]}}{1-\sqrt{2 P_{\theta_{0}}}\cos\theta}}
      -\left[q(P_{\theta_{0}})\rho_{\parallel 0}(P_\zeta,P_{\theta_0})-P_{\theta_{0}}\right]d\theta
\end{equation}
with 
\begin{equation}
    \rho_{\parallel 0}(P_\zeta,P_{\theta_0})=P_{\zeta} + \psi_{p}(P_{\theta_{0}}),
\end{equation}

The analytical evaluation of the integral in Eq. \eqref{jt} is presented in detail in the appendix of Ref. \cite{Antonenas2021}. Taking this into account, the corresponding analytical expressions for trapped $(t)$ and passing $(p)$ particles are given, respectively, by  
\begin{equation}\label{Jtheta_t}
    J^{(t)}_{\theta} =   q(P_{\theta_{0}})\frac{8\sqrt{\mu r}}{\pi\eta(1-r)}\left[(\eta k -1)\Pi(\eta k, k)+K(k)\right]
\end{equation}
\begin{equation}\label{Jtheta_p}
    J^{(p)}_{\theta} = q(P_{\theta_{0}})\frac{4\sqrt{\mu r}}{\pi\eta(1-r)}\left[\frac{\eta k -1}{\sqrt{k}}\Pi(\eta, \frac{1}{k})+\frac{K(\frac{1}{k})}{\sqrt{k}}\right]
            - \sigma\left[q(P_{\theta_{0}})\rho_{\parallel 0}(P_\zeta,P_{\theta_0})-P_{\theta_{0}}\right],
\end{equation}
where $r=\sqrt{2P_{\theta_{0}}}$, $\eta=-2r/(1-r)$, $k = \left(E-\mu(1-r)\right)/2\mu r$ is the trapping parameter with $0<k<1$ for trapped and $k>1$ for the passing orbits, $K,\Pi$ are the complete elliptic integrals of the first and third kind, respectively, while $\sigma =+1$ for trapped and co-passing orbits and $\sigma = -1$ for counter-passing orbits \cite{Antonenas2021}.

Based on Eqs. \eqref{gen hat omega zeta}, \eqref{Jtheta_t}, \eqref{Jtheta_p}, the kinetic $q$-factor $q_{kin}$ can be analytically calculated (the resulting expression is quite lengthy and therefore ommited) as a function of the three COM as
\begin{equation} \label{q-kinetic}
    q_{kin}(E, \mu, P_{\zeta})\equiv \frac{\hat{\omega}_\zeta} {\hat{\omega}_\theta}
    =-\sigma\frac{\partial J_\theta^{(t,p)}(E, \mu, P_{\zeta})}{\partial  P_\zeta}.
\end{equation}

\begin{figure}[h!]
    \centering
    \includegraphics[width=0.49\textwidth]{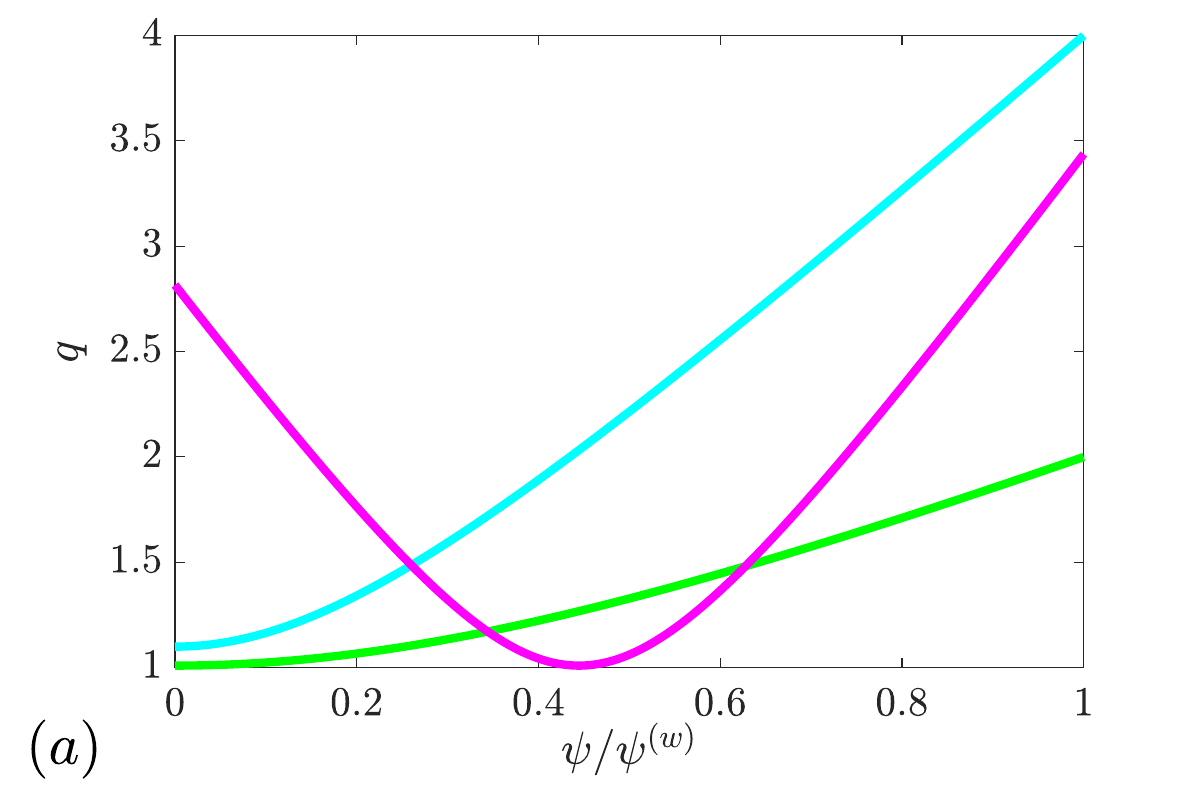}
    \centering
    \includegraphics[width=0.49\textwidth]{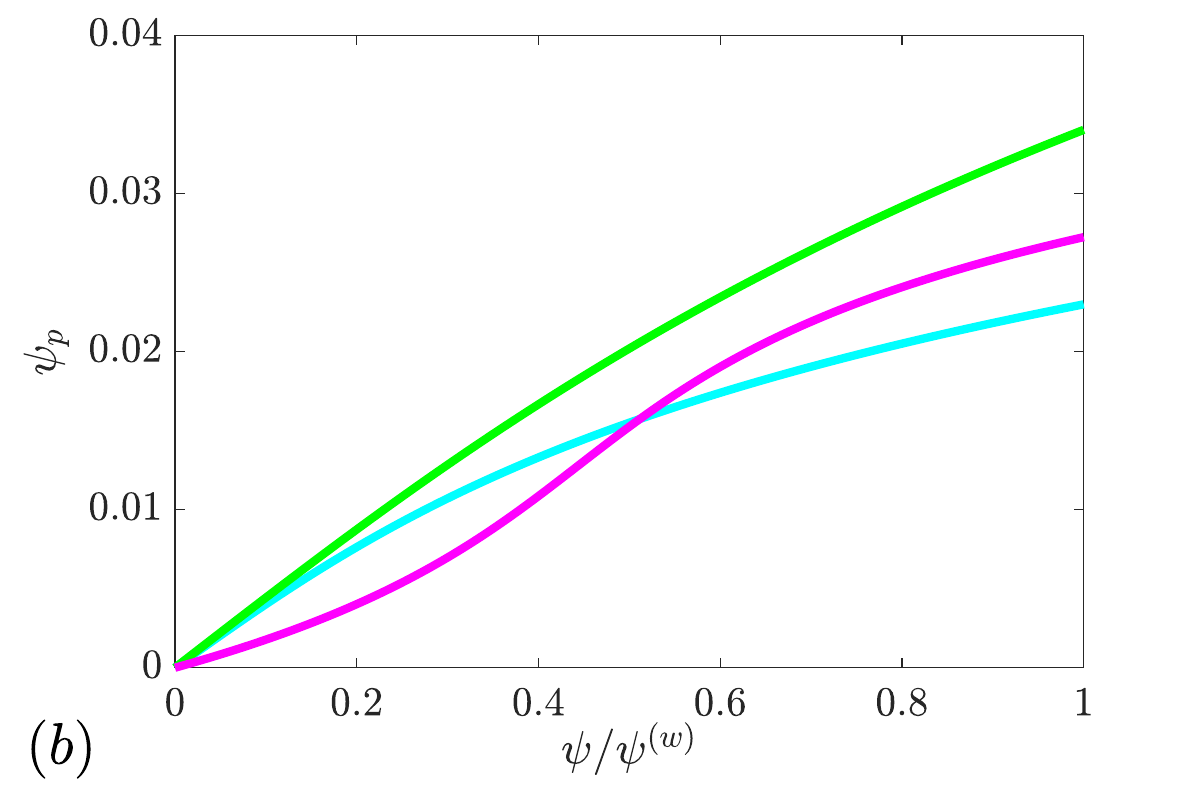}
    \caption[$q$ profiles in LAR equilibrium.]{(a): Three different $q(\psi)$ profiles.  All profiles are described by Eq. \eqref{gen q factor} with $\nu=2$. Cyan lines ($q_{1}$):  $q_{a}=1.1$, $q_{w}=4.0$, $\lambda=0$. Green lines ($q_{2}$): $q_{a}=1.01$, $q_{w}=3.0$, $\lambda=0$. Magenta lines ($q_{3}$): $q_{a}=1.01$, $q_{w}=3.0$, $\lambda=0.44$. (b): The corresponding poloidal flux $\psi_{p}(\psi)$ as obtained by integrating each $q$ profile over $\psi$.}
    \label{fig:Fig51}
\end{figure}

In the following, we present analytical calculations of the $q_{kin}$ based on Eqs. \eqref{Jtheta_t}, \eqref{Jtheta_p}, \eqref{q-kinetic} for particles with various kinetic characteristics and drift orbit widths, we investigate its dependence on the magnetic $q$ profile, and systematically compare with numerical results in order to estimate the range of validity of the analytical expressions for a LAR axisymmetric equilibrium.

The magnetic $q$ profile is given by the expression \eqref{gen q factor}. Three different $q$ profiles (with $\nu=2$) are illustrated in Fig. \ref{fig:Fig51}(a). Both $q_1$ and $q_2$ are monotonic profiles $(\lambda=0)$. The $q_{1}$ profile (cyan line) corresponds to parameters $q_{a}=1.1$, $q_{w}=4.0$, while the $q_{2}$ profile (green line) has a smaller value at the magnetic axis $(q_{a}=1.01)$ and the wall $(q_{w}=3.0$). The $q_{3}$ profile (magenta line) is non-monotonic, and is defined by $q_{a}=1.01$, $q_{w}=6.0$, and $\lambda=0.44$, with its local minimum (at $\psi/\psi^{(w)}=\lambda=0.44$) corresponding to a flux surface with zero magnetic shear (note that for this profile $q_a$ and $q_w$ do not correspond to the values of $q$ at the axis and the wall). This profile is an example of a reversed shear magnetic $q$ profile, similar to those created in  Tokamak Fusion Test Reactor (TFTR) experiments, where they have been shown to lead to reduced particle transport \cite{Levinton1995}. Reversed shear configurations have also been shown to be related to the formation of internal transport barriers and reduced transport in DIII-D discharges \cite{Strait1995} as well as in reversed field pinch (RFP) configurations \cite{Gobbin2008}, and are expected to reduce alpha particles losses  in ITER \cite{Fasoli2007}. Evidently, the poloidal flux for each $q$ profile has different dependence on the toroidal flux $\psi_p(\psi)$, which results in a different value at the wall $\psi_{p}^{(w)}$, as depicted in Fig. \ref{fig:Fig51}(b).

The role of the kinetic characteristics of the particles along with the magnetic $q$ profile with respect to the form of the $q_{kin}$ profile, and the validity of the analytical results can be investigated by systematically dissecting the three dimensional COM space. In the following we consider a LAR configuration with major radius $R_0=\SI{1.65}{\metre}$, inverse aspect ratio $r/R_0 = 0.18$ and on-axis magnetic field $B_{0}=\SI{1}{\tesla}$, and investigate hydrogen orbits (atomic number $Z=1$), focusing on four characteristic cases.  

\textbf{Case \#1:} Low-energy particles, with $\mu B_{0} = \SI{2}{\kilo\electronvolt}$ ($\mu = 7.7\times 10^{-6}$ in normalized units), in the $q_1$ profile.  

\textbf{Case \#2:} Mildly energetic particles (in comparison to Case \#1), with $\mu B_{0} = \SI{10}{\kilo\electronvolt}$ ($\mu =3.8\times 10^{-5}$ in normalized units), in the $q_1$ profile.

\textbf{Case \#3:} Mildly energetic particles with the same characteristics as in Case \#2 ($\mu B_{0}=\SI{10}{\kilo\electronvolt}$), in the $q_{2}$ profile, which results in a higher value of the poloidal flux at the wall $\psi_{p}^{(w)}$. 

\textbf{Case \#4:} Mildly energetic particles with the same characteristics ($\mu B_0=\SI{10}{\kilo\electronvolt}$) as in Cases \#2 and \#3, in the non-monotonic $q_3$ profile. This case favors the formation of transport barriers due to the existence of local minima of the kinetic $q$-factor in the COM space. 

\begin{figure}[h!]
    \centering
    \includegraphics[width=0.49\textwidth]{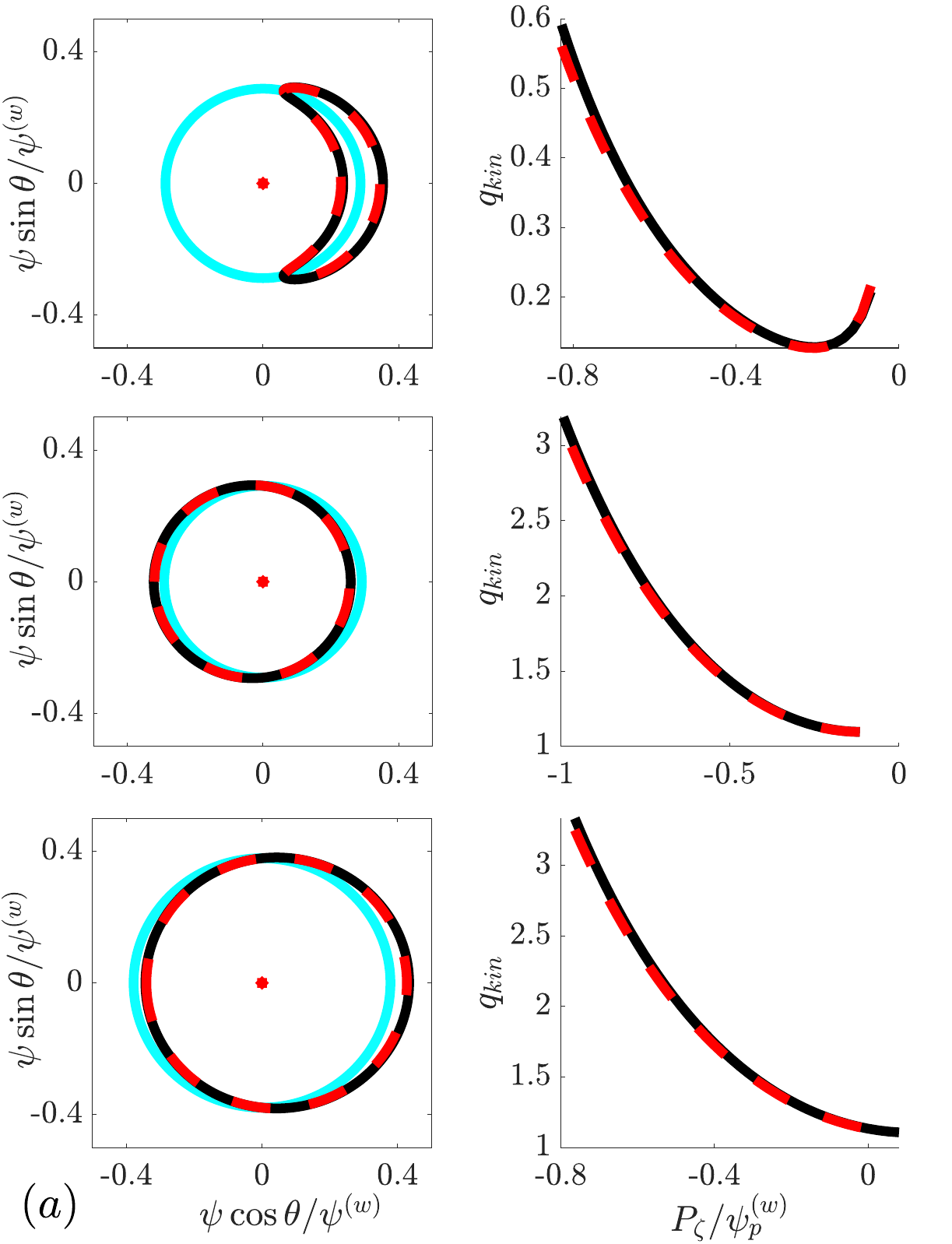}
    \centering
    \includegraphics[width=0.49\textwidth]{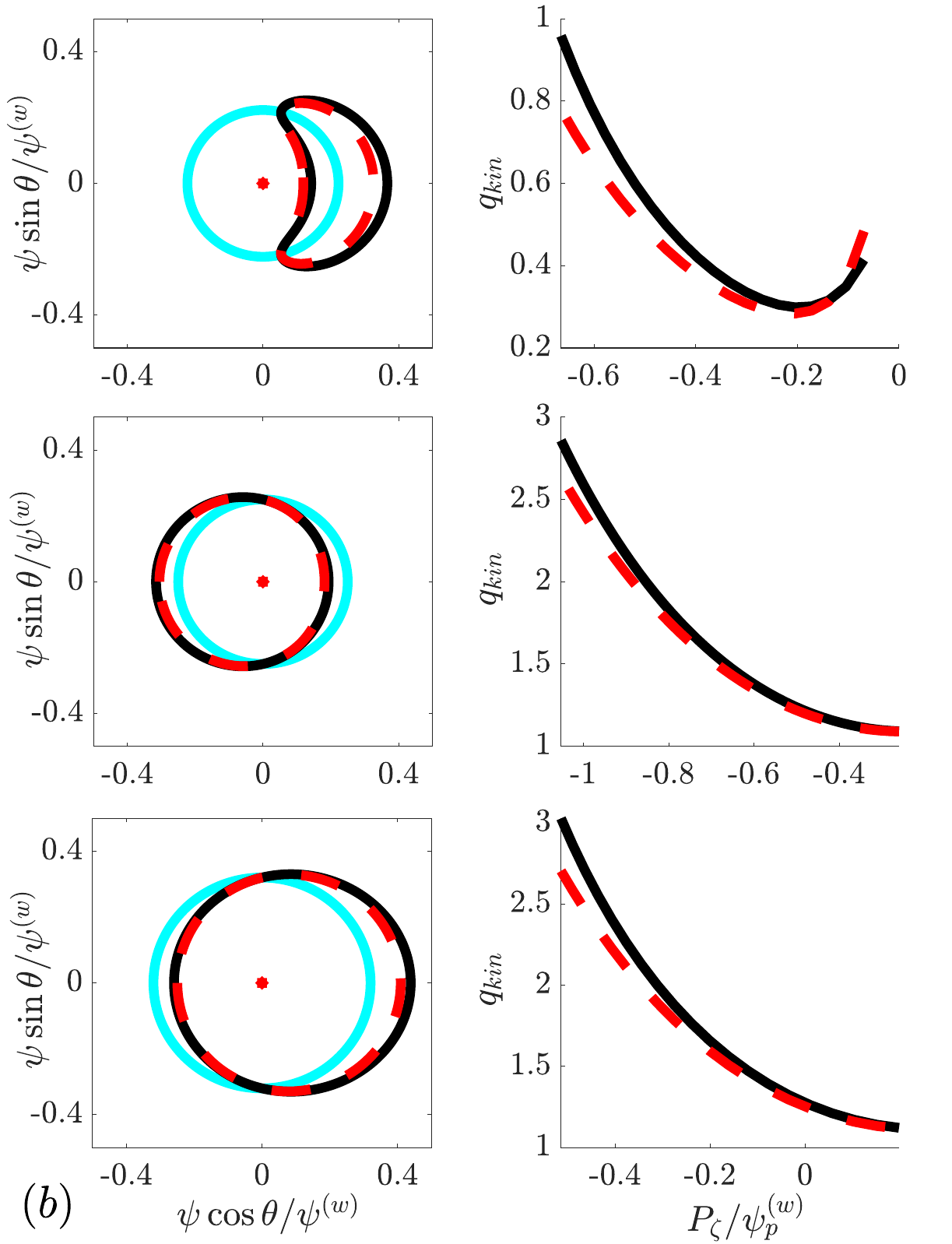}
    \caption[Unperturbed GC orbits in GC phase space and $q_{kin}$ as a function of $P_{\zeta}$ in LAR equilibrium.]{(a) Case \#1: Low-energy particles, with $\mu B_{0} = \SI{2}{\kilo\electronvolt}$, in the $q_1$ profile. (b) Case \#2: More energetic particles with $\mu B_{0} = \SI{10}{\kilo\electronvolt}$, in the $q_1$ profile. From top to bottom: trapped particles with  $E/\mu B_{0} = 0.996$, counter passing, and co-passing particles with  $E/\mu B_{0} = 1.4$. 
    Left columns: characteristic orbits in the configuration space along with their corresponding flux surface of reference $\psi_0=P_{\theta_0}$ (cyan lines).  Right columns: $q_{kin}$ as a function of the canonical momentum $P_\zeta$. Black-solid and red-dashed lines denote numerical and analytical results, respectively. The deviation between analytical and numerical calculations increases with the particle energy and the drift-orbit width.} 
    \label{fig:Fig52}
\end{figure}

\begin{figure}[h!]
    \centering
    \includegraphics[width=0.49\textwidth]{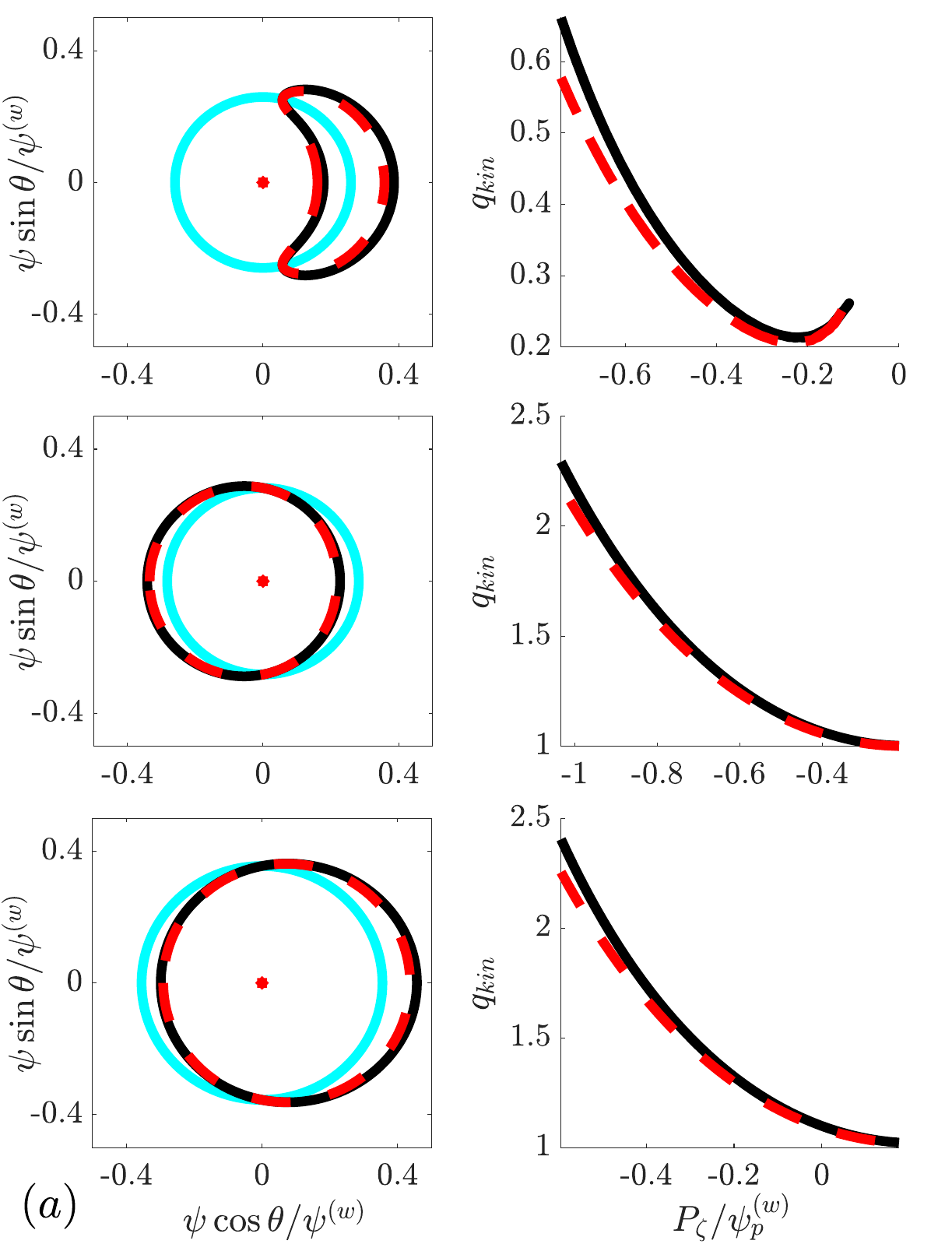}
    \centering
    \includegraphics[width=0.49\textwidth]{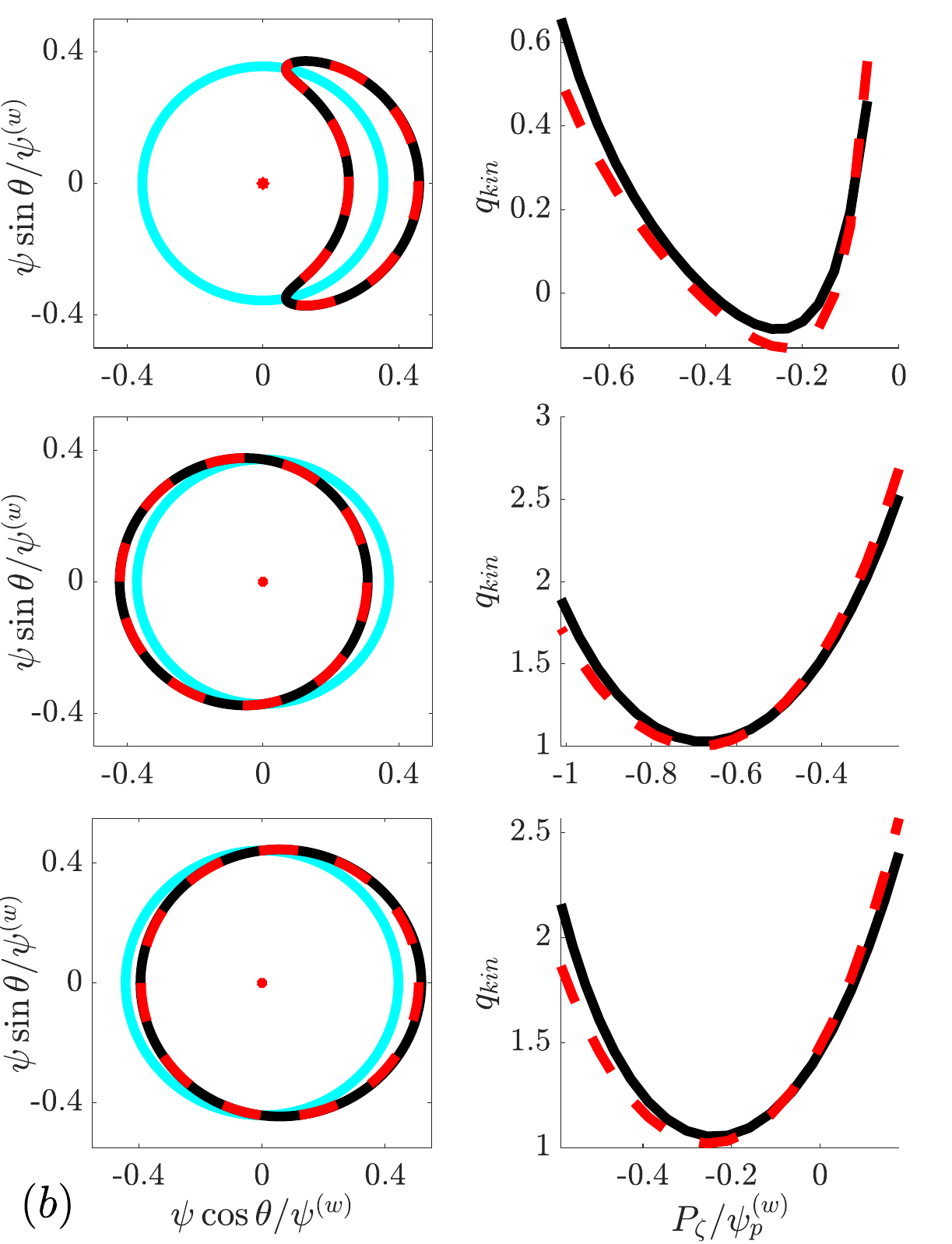}
    \caption[Unperturbed GC orbits in phase space and $q_{kin}$ as a function of $P_{\zeta}$ in LAR equilibrium.]{(a) Case \#3: Energetic particles with $\mu B_{0} = \SI{10}{\kilo\electronvolt}$ in the $q_2$ profile. (b) Case \#4: Energetic particles with $\mu B_{0} = \SI{10}{\kilo\electronvolt}$ in the $q_3$ profile.
    From top to bottom: trapped particles with $E/\mu B_{0}= 0.996$, counter passing, and co-passing particles with  $E /\mu B_{0} = 1.4$. 
    Left columns: characteristic orbits in the configuration space along with their corresponding flux surface of reference $\psi_0=P_{\theta_0}$ (cyan lines).  Right columns: $q_{kin}$ as a function of the canonical momentum $P_\zeta$. Black-solid and red-dashed lines denote numerical and analytical results, respectively.
    In Case \#3, the deviations between analytical and numerical calculations are reduced in comparison to Case \#2. In Case \#4, in contrast to the previous cases, a local minimum in $q_{kin}$ is observed for passing particles.} 
    \label{fig:Fig53}
\end{figure}

In Figs. \ref{fig:Fig52} and \ref{fig:Fig53}, the analytical calculation of $q_{kin}$ (red dashed lines), based on Eq. \eqref{q-kinetic}, is compared to calculations obtained from the numerical integration of the GC equations of motion (black solid lines), for the four cases. Each orbit is uniquely labeled by a specific set of the COM $(E, \mu, P_{\zeta})$. For a given value of $\mu$, particles with different energies $E$ can be trapped (top), counter-passing (middle), or co-passing (bottom). The first columns depict the poloidal projection of characteristic GC orbits in the configuration space. The analytically obtained orbits (red-dashed lines) are calculated as level sets of the Hamiltonian [Eq. \eqref{GC H 1}] with the magnetic field evaluated at the flux surface of reference $\psi_0=P_{\theta_0}$, given by Eq. \eqref{def of Ptheta0} and depicted by cyan lines. The second columns depict the kinetic $q$-factor as a function of the canonical momentum $P_\zeta$ for trapped and counter/co-passing particles. 
Figures Figs. \ref{fig:Fig52}, \ref{fig:Fig53} clearly show that there is a direct relation between the drift width of a GC orbit and the discrepancy between the analytically and the numerically calculated values of the kinetic $q$-factor, which is based on approximating the value of the magnetic field in the course of a GC orbit with a single value on a flux surface of reference $\psi_0$.  For low-energy particles, the agreement between analytical and numerical calculations is excellent, with maximum deviations of less than $5\%$ and $2\%$ for trapped and passing particles, respectively, as shown for the Case \#1 in Fig. \ref{fig:Fig52}(a). Under the same magnetic field configuration (same magnetic $q$ factor, $q_1$), the accuracy deteriorates for more energetic particles, as shown for the Case \#2 in Fig. \ref{fig:Fig52}(b). However, for the same energetic particles, the accuracy significantly improves when the profile of the magnetic $q$ factor changes from $q_1$ to $q_2$, as depicted in Fig.  \ref{fig:Fig53}(a). It is worth noting that, in all cases, the discrepancies are larger for trapped particles. The case of the non-monotonic $q$ profile ($q_3$) is illustrated in Fig. \ref{fig:Fig53}(b), where a local minimum of $q_{kin}$ is shown to occur for passing particles as well, in contrast to the previous cases where such a local minimum occurred only for trapped particles. As will be shown in the following section, a local extremum of $q_{kin}$ indicates the formation of a transport barrier under the presence of non-axisymmetric perturbations. In this case of passing particles, the minimum occurs for particle orbits located closer to the wall in contrast to the case of trapped particles, where the minimum occurs for particle orbits close to the magnetic axis. 

\begin{figure}[h!]
    \centering
    \includegraphics[width=0.49\textwidth,keepaspectratio]{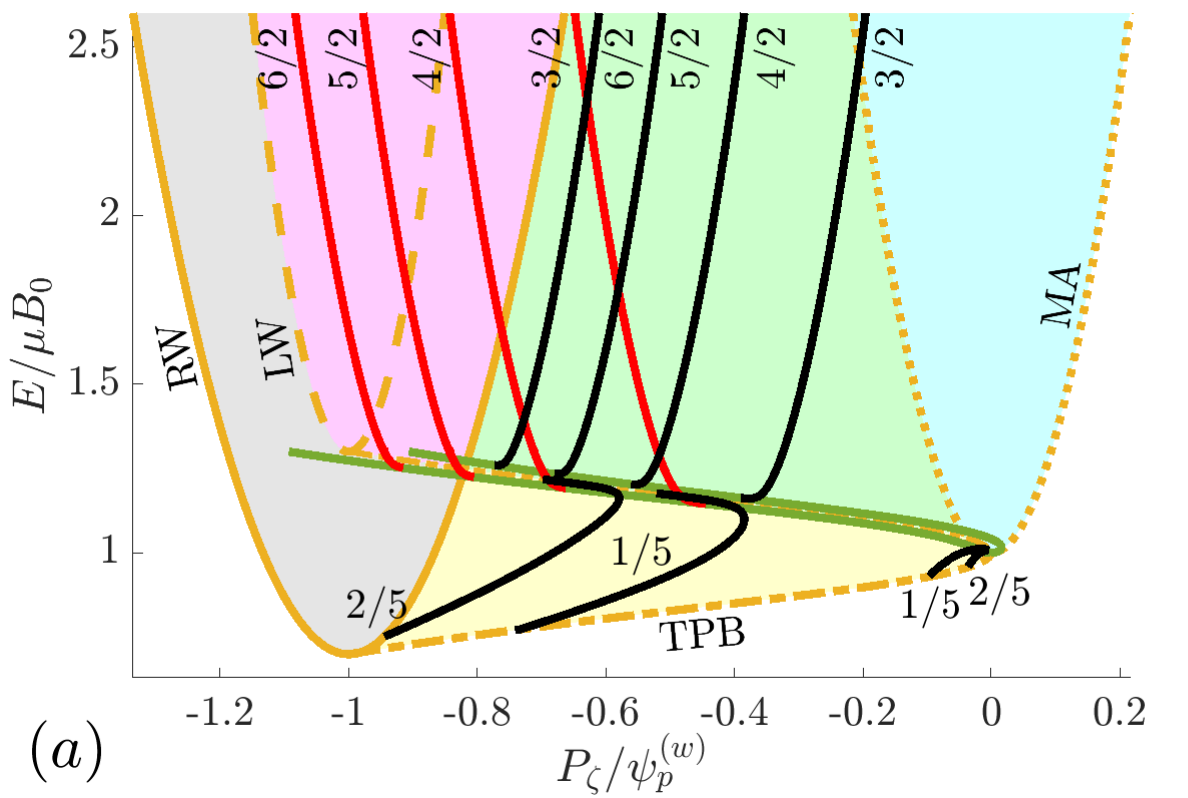}
    \includegraphics[width=0.49\textwidth,keepaspectratio]{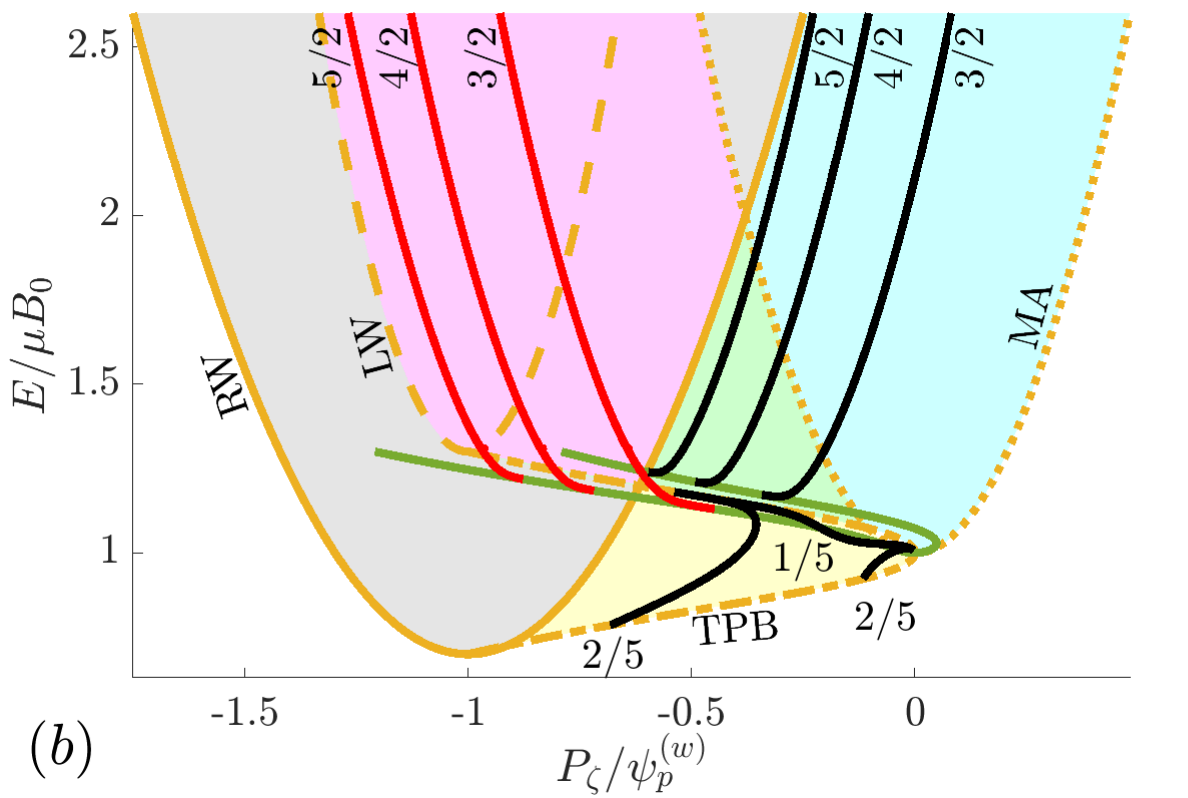}
    \includegraphics[width=0.49\textwidth,keepaspectratio]{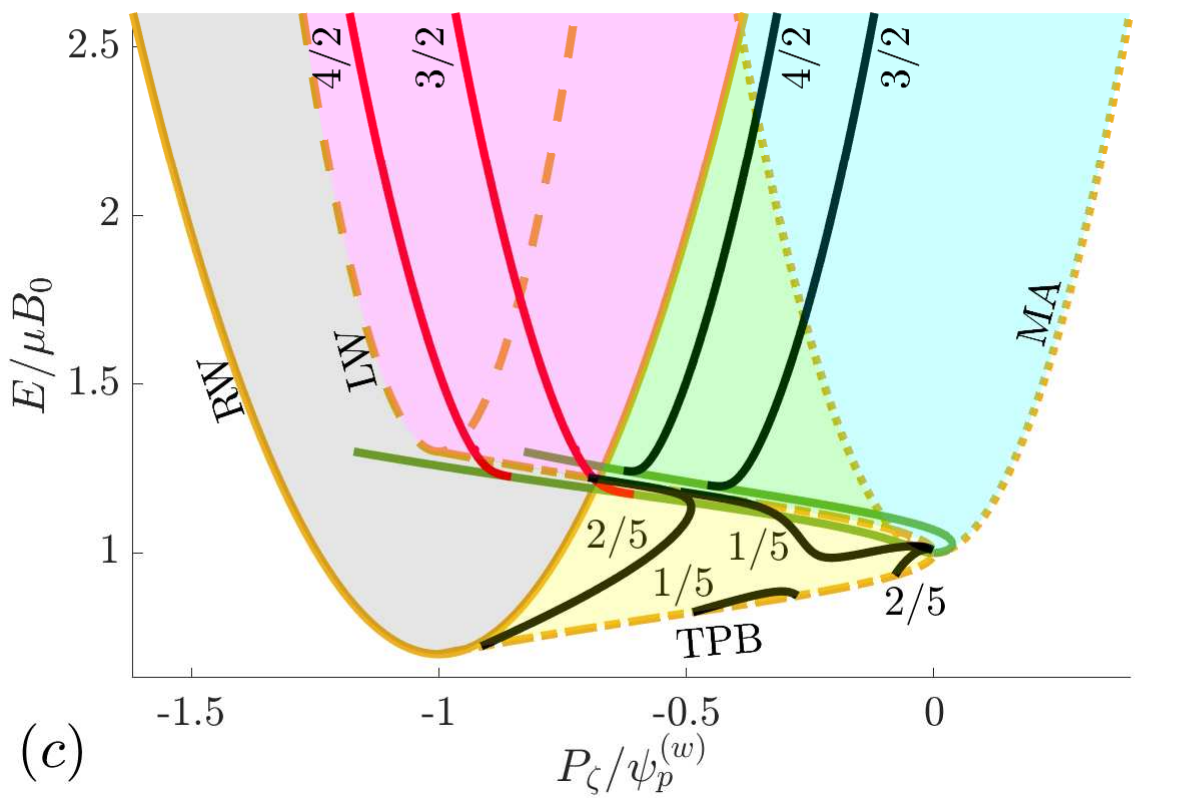}
    \includegraphics[width=0.49\textwidth,keepaspectratio]{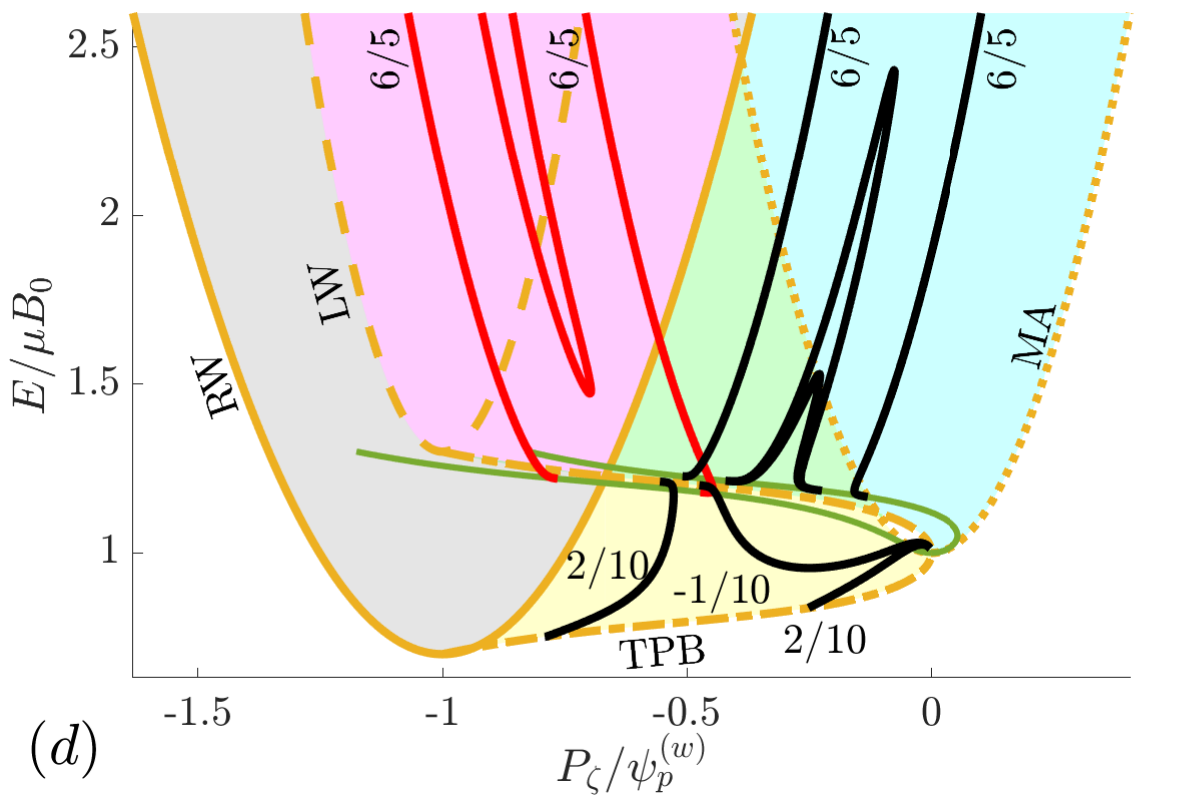}
    \caption[Resonances curves in COM space in LAR equilibrium.]{Constant-$\mu$ slices (plane cuts) of the three dimensional COM space $(E,\mu, P_{\zeta})$. Cases \#1-\#4 are depicted in (a)-(d), respectively. Yellow-colored parabolas depict the location of the magnetic axis (MA) (dotted line), the trapped-passing boundary (TPB) (dot-dashed line), the left wall (LW) (long-dashed line) and the right wall (RW) (plain line).  The color-shaded areas denote regions of trapped (yellow), co-passing (cyan), counter-passing (magenta), both co- and counter-passing (green), and lost (grey), particles. Black lines correspond to resonances with trapped and co-passing particles, and red lines correspond to resonances with counter-passing particles. Co-passing and counter-passing resonant curves are terminated on the green lines close to the TPB. A constant pitch $\Lambda=E/\mu B_{0}$ line intersects twice a passing particle resonance only for the Case \#4, whereas twice intersected trapped particle resonances are very close to the magnetic axis in all cases.}
    \label{fig:Fig54}
\end{figure}

A systematic assessment of the range of validity of the analytical results, based on the width of the drift GC orbits, can be provided in the space of the three COM where the boundaries of the magnetic axis and the walls are given in terms of the kinetic characteristics of the particles (Sec. \ref{Sec: Orbits in COM space}). Figure \ref{fig:Fig54} depicts characteristic curves representing the left (inner) and the right (outer) walls, and the magnetic axis, in a constant$-\mu$ plane of the three dimensional COM space $(E,\mu,P_\zeta)$ for the Cases \#1-\#4, based on their definitions in Eqs. \eqref{walls} and \eqref{magnetic axis}. For constant pitch $\Lambda=E/\mu B_0$, the distance $\Delta P_\zeta$ between the neighboring legs of the parabolas corresponding to the magnetic axis and the right wall can be used as a measure for the size of the drift width of the GC orbits (see also Sec. 3.3 of Ref. \cite{White2013}). Utilizing the expressions of Eqs. \eqref{walls}, \eqref{magnetic axis}, along with a first order Taylor expansion with respect to $\sqrt{2 \psi^{(w)}}$ this distance is calculated as
\begin{equation}
    \frac{\Delta P_\zeta}{\psi_{p}^{(w)}}\equiv\frac{P_\zeta^{(a)}-P_\zeta^{(rw)}}{\psi_{p}^{(w)}}=1-\frac{1+2(\Lambda-1)}{\sqrt{2(\Lambda-1)}}\left(\frac{\Delta B}{B_0}\right)\left(\frac{\sqrt{\mu}}{\psi_{p}^{(w)}}\right), \label{DPzeta} 
\end{equation}
where $P_\zeta^{(a)}, P_\zeta^{(rw)}$ are the values of $P_\zeta$ at the magnetic axis and the right wall, respectively, and $\Delta B / B_0 = \sqrt{2 \psi^{(w)}}=r$ ($r$ is the minor radius) is the variation of the magnetic field amplitude from the magnetic axis to the walls. This equation clearly shows the role of the kinetic characteristics, along with the value of the poloidal flux at the wall $\psi_{p}^{(w)}$, on the drift orbit width and the corresponding validity of the analytical calculations: for a given pitch $\Lambda$, $\Delta P_\zeta$ decreases with increasing $\mu$ and increases with increasing $\psi_{p}^{(w)}$. This dependence can be confirmed by observing the horizontal distance between the parabolas corresponding to the right wall and the magnetic axis in Fig. \ref{fig:Fig54}, and explains the validity of the analytical results and its dependence on the kinetic characteristics of the particles and on the $q$ profile. More specifically, the comparison between Figs. \ref{fig:Fig54}(b) and \ref{fig:Fig54}(c) clearly shows the role of the $\psi_{p}^{(w)}$, leading to the conclusion that a higher value of the total poloidal flux (higher plasma current) results in smaller particle drifts.  

The employment of a flux surface of reference $\psi_0$ for the magnetic field evaluation, in order to obtain the analytical expressions for the $q_{kin}$, also influences the approximation of the trapped-passing boundary. In the absence of such approximation, the trapped-passing boundary is uniquely defined by Eq. \eqref{trapped passing boundary}. However, under the above approximation, the trapped-passing boundary is defined by the condition $k=1$ or $E = \mu(1+\sqrt{2P_{\theta_{0}}})$ which, by the definition of $P_{\theta_{0}}$, as in Eq. \eqref{def of Ptheta0},  gives different values for trapped, co-passing and counter-passing particles. More specifically, for trapped particles the approximate condition aligns with the exact trapped-passing boundary, whereas for co- and counter-passing particles, the respective condition gives two different trapped-passing boundaries  (yellow dot-dashed and green lines, respectively, in Fig. \ref{fig:Fig54}). Notably, the approximate curves for co-passing and counter-passing particles are located above and bellow the exact trapped-passing boundary, respectively.   

Having calculated the kinetic $q$-factor as a function of the three COM, its level sets at rational values $q_{kin}=m/n$ can be readily obtained. Several curves corresponding to different rational values are depicted in Fig. \ref{fig:Fig54}, for trapped (black), co-passing (black) and counter-passing (red) particles. These curves show the exact locations in the COM space for the particles that can resonantly interact with specific non-axisymmetric perturbations, as will be shown in the following section, and provide an overview for the resonant response of the toroidal plasma. Moreover, it is worth noting that, in accordance to the previous discussion referring to the extrema of the $q_{kin}$, only for the non-monotonic $q$ profile of Case \#4, a constant pitch $\Lambda=E/\mu B_{0}$ line intersects twice a passing particle resonance, whereas twice intersected trapped particle resonances are very close to the magnetic axis, in all cases.     

\section{Resonant response to non-axisymmetric time-independent perturbations} \label{Resonse}

In order to analyze the resonant effects of the perturbations on the GC orbits (see Sec. \ref{Sec: Non-Axisymmetric perturbations}) with the utilization of the previously calculated kinetic $q$-factor, the expression of Eq. \eqref{alpha} has to be transformed to AA variables 
\begin{equation} \label{alpha hat}
    \alpha( \hat{\theta},\hat{\zeta},J_{\theta},J_{\zeta}, J_{\xi})= \sum_{m',n'}{\hat{a}_{m',n'}(J_{\theta},J_{\zeta}, J_{\xi})e^{i(n'\hat{\zeta}-m'\hat{\theta} )}},
\end{equation}
allowing the resonance condition to be straightforwardly expressed as 
\begin{equation}
    q_{kin}(E,\mu,P_\zeta)=\frac{m'}{n'}.
\end{equation}
It is important to emphasize that, due to the nonlinear character of the transformation from the geometrical angles $(\theta,\zeta)$ to the Angles $(\hat{\theta},\hat{\zeta})$ according to Eq.\eqref{eq:Angles}, although $n'=n$ (Sec. \ref{Sec: Number of islands}, Eq. \eqref{mode}), there exists no direct correspondence between the poloidal mode numbers $m$ and $m'$. In fact, even a single perturbative mode in Eq. \eqref{alpha} may lead to multiple modes in Eq. \eqref{alpha hat}. The impact of a single geodesic acoustic-like compressional mode on fast-ion transport and losses due to resonances overlapping at fractional harmonics of the mode frequency has been observed in DIII-D tokamak experiments \cite{Nazikian2008}. A theoretical study explaining the fractional resonances that appear under geodesic acoustic mode as well as other MHD activity, like Alfvén eigenmodes, has been presented in \cite{Kramer2012}, whereas the role of primary and sideband resonances in the confinement of energeting ions under the presence of RMPs has been considered for EAST \cite{He2020}. The relation between the two variable sets crucially determines the modification of the GC phase space due to the applied perturbations, as will be shown in the following paragraphs and further discussed in Sec. \ref{Sec: Number of islands}.

\begin{figure}[h!]
    \centering
    \includegraphics[width=0.49\textwidth,keepaspectratio]{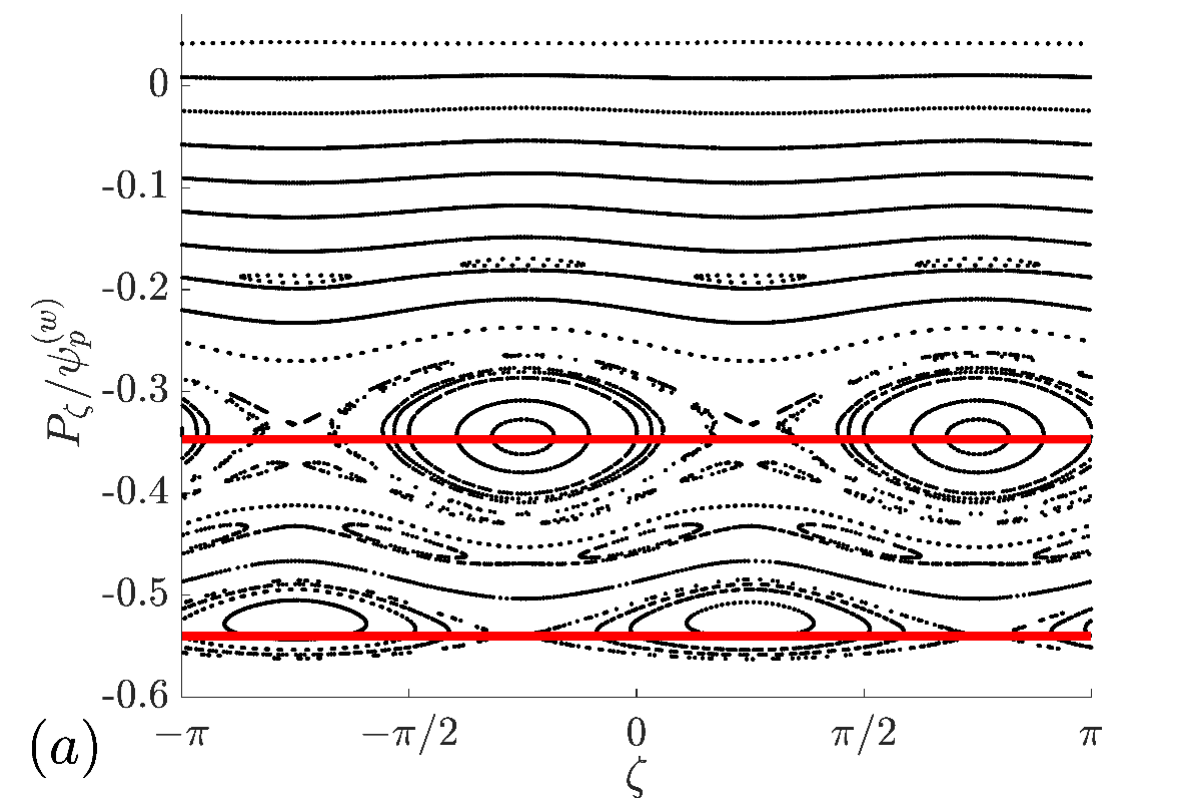}
    \includegraphics[width=0.49\textwidth,keepaspectratio]{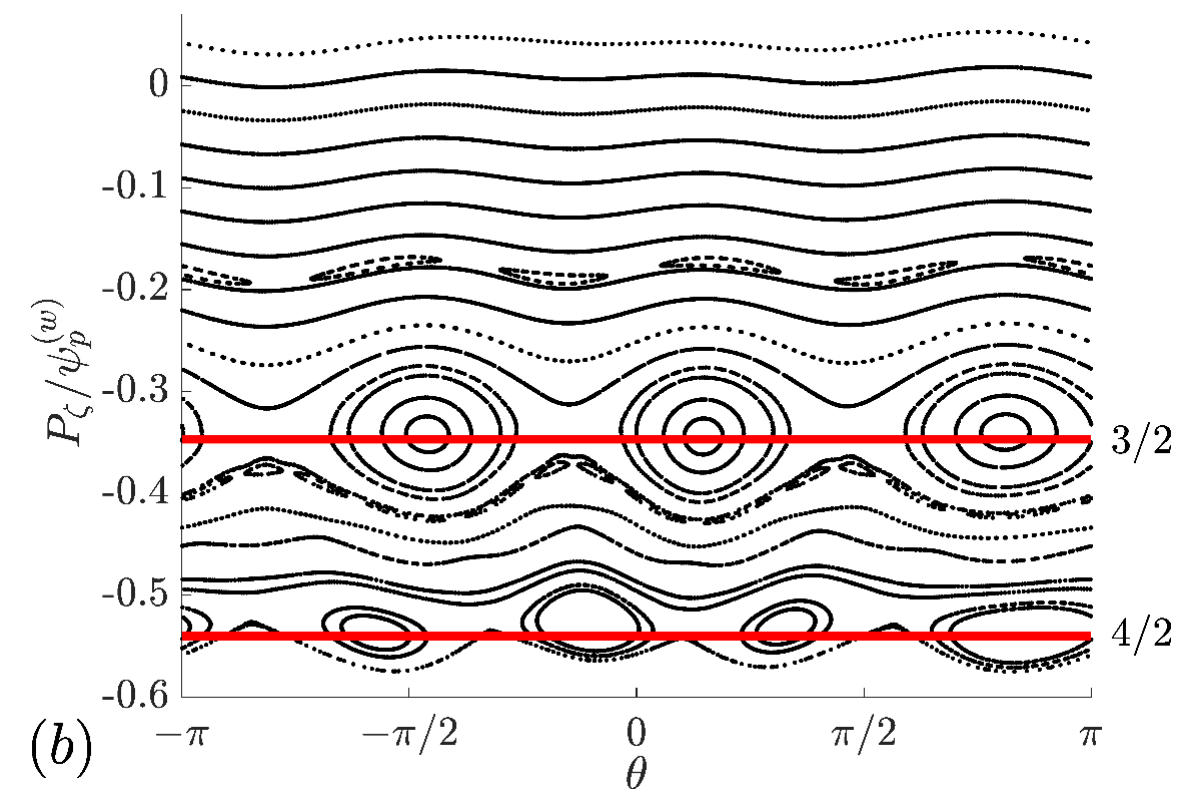}
    \includegraphics[width=0.49\textwidth,keepaspectratio]{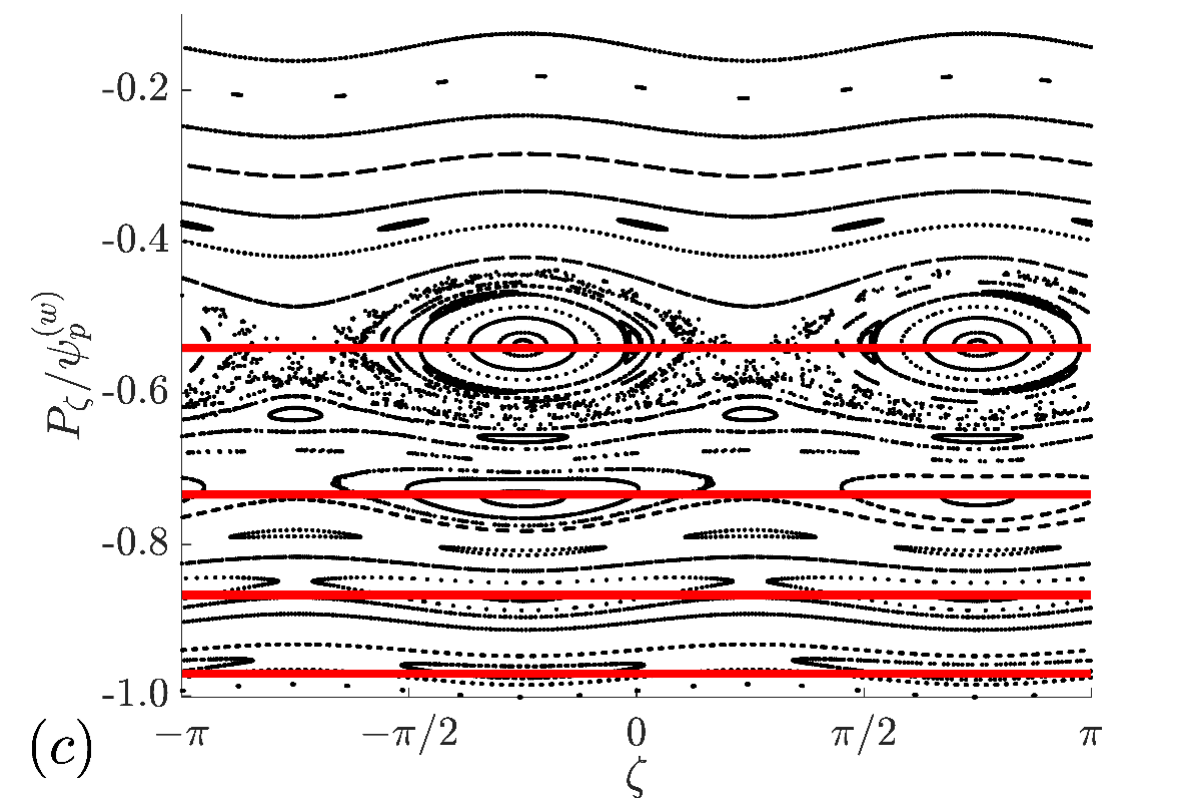}
    \includegraphics[width=0.49\textwidth,keepaspectratio]{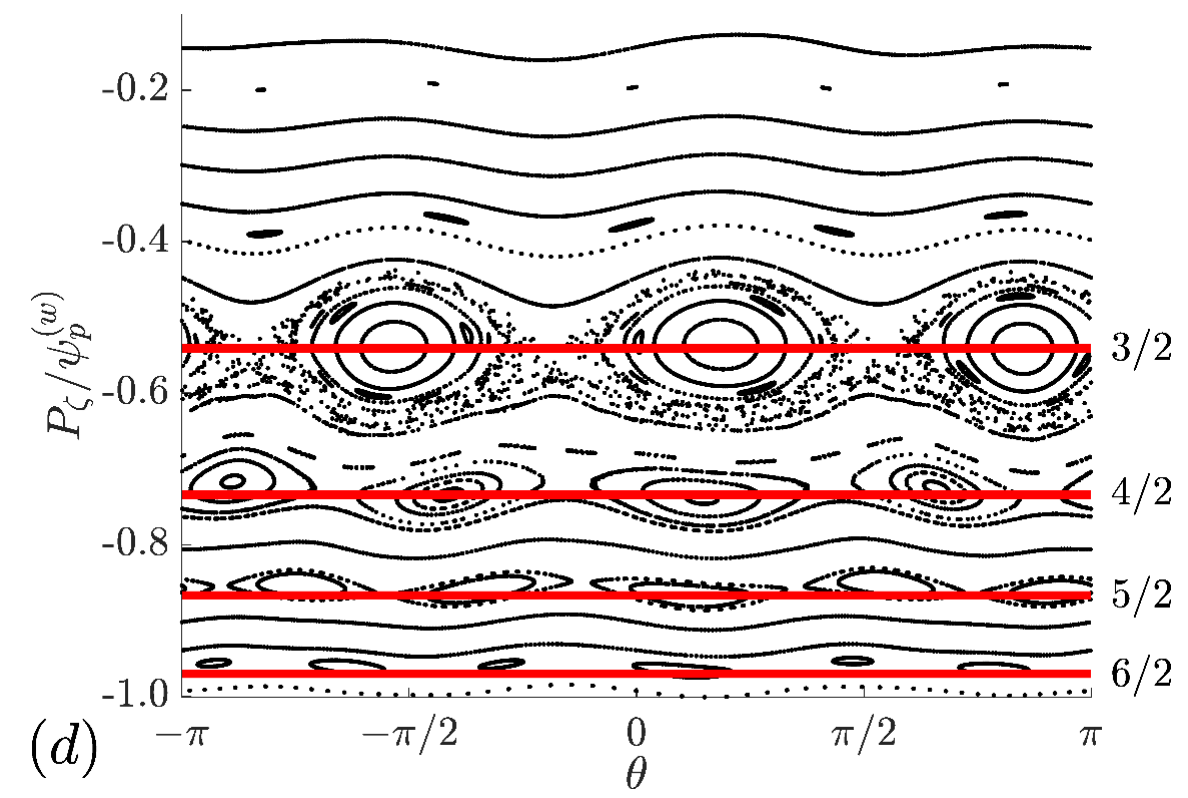}
    \caption[Poincaré surfaces of section for passing particles in LAR equilibrium]{Poincaré surfaces of section for passing particles of Case \#1 ($\mu B_{0} = \SI{2}{\kilo\electronvolt}$ in the $q_{1}$ profile), under the presence of a single perturbative mode $(m,n)=(3,2)$ with amplitude $\alpha_{3,2}=0.5\times10^{-4}$. Co-passing particles with $E/\mu B_{0} = 1.20$, crossing the surfaces of section in the positive directions are depicted in panels (a) and (b), while counter-passing particles with $E/\mu B_{0} = 1.43$, crossing the surfaces of section in the negative direction are depicted in panels (c) and (d). Panels (a) and (c) display Poincaré surfaces of section at constant poloidal angle $\theta=0$, whereas panels (b) and (d) show Poincaré surfaces at a constant toroidal angle $\zeta=0$. The red horizontal lines denote the predicted locations of the corresponding resonant island chains in accordance to Fig. \ref{fig:Fig54}(a), showing a remarkable accuracy.}
    \label{fig:Fig55}
\end{figure}

Without loss of generality, concerning the resonant character of mode-particle interactions and their effects on stochastic transport, in the following analysis, we consider perturbations with a constant mode-amplitude, that is, $\alpha_{m,n}=\epsilon$, where $\epsilon$ represents the ratio of the amplitude of the perturbative magnetic mode to the background magnetic field. The effect of the perturbative modes on the GC phase space is strongly inhomogeneous and localized in regions where the kinetic characteristics of the particles satisfy the respective resonance conditions. In order to dissect the phase space, we employ appropriate Poincaré surfaces of section at either $\theta=0$ or $\zeta=0$, for fixed values of $\mu$ and $E$. For illustrative purposes, we focus on the Case \#1. According to Fig. \ref{fig:Fig54}(a) the pitch values of $E/\mu B_{0}=1.2$ and $E/\mu B_{0}=1.43$ correspond to passing particles with different ranges of $P_{\zeta}/\psi_{p}^{(w)}$. The Poincaré surfaces of section, under the presence of a single perturbing mode with $(m,n)=(3,2)$, are depicted in  Fig. \ref{fig:Fig55}(a)-(b) for co-passing particles with pitch value $E/\mu B_{0}=1.2$,  and Fig. \ref{fig:Fig55}(c)-(d) for counter-passing particles with pitch value $E/\mu B_{0}=1.43$. It is clear that even a single perturbing mode can generate several resonant island chains centered around values of $P_\zeta$ where $q_{kin}$ factor has a rational value. The number of islands in each island chain in the $\theta=0$ surface is equal to the toroidal mode number $n$, whereas the number of islands in each chain in the $\zeta=0$ surface is equal to the different poloidal mode numbers $m'$ resulting from the transformation to Action-Angle variables according to the discussion in the Sec. \ref{Sec: Number of islands} (Fig. \ref{fig:Fig59}(b)). The analytically obtained $q_{kin}$ accurately predicts the resonance locations, indicating the paramount importance of the respective diagram shown Fig. \ref{fig:Fig54}, which provides concise information regarding the resonant response of the particles to non-axisymmetric perturbations and pinpointing the locations of the phase space where strong mode-particle interactions actually take place. 

\begin{figure}[h!]
    \centering  
    \begin{minipage}{0.49\textwidth}
      \raisebox{1cm}{  \includegraphics[width=\textwidth,keepaspectratio]{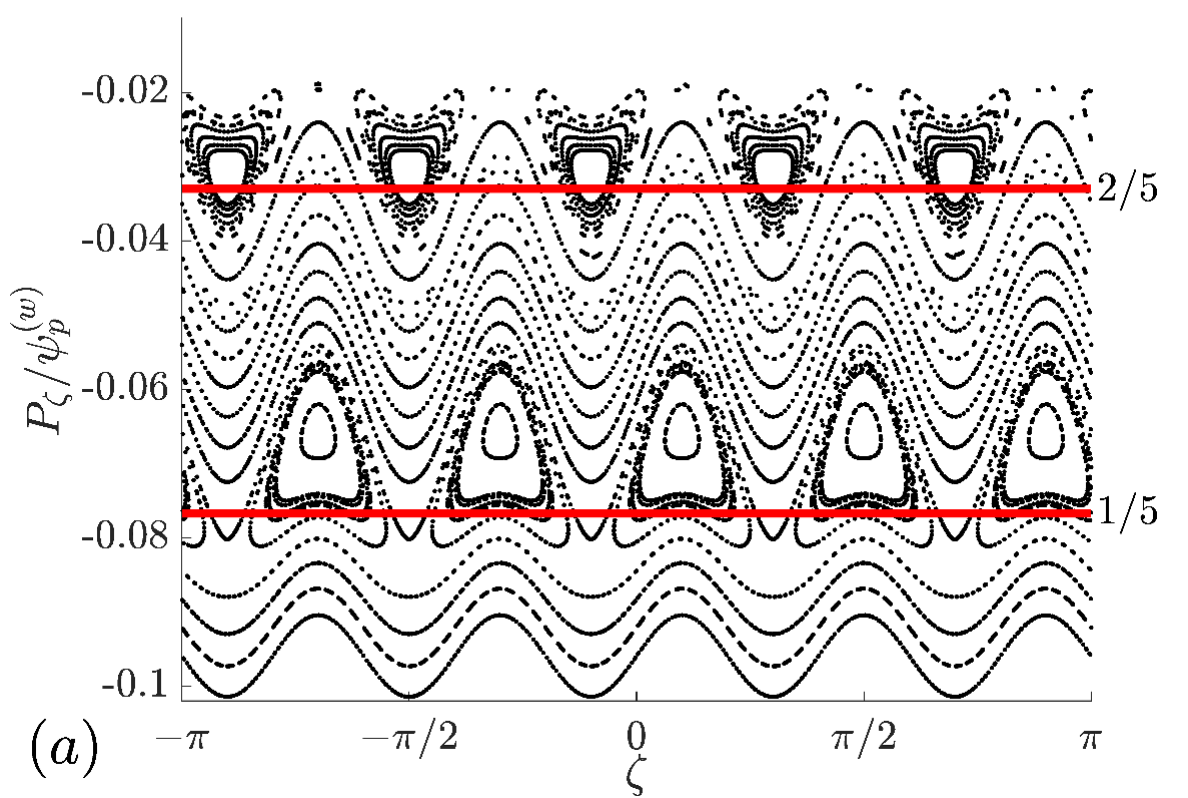}}
    \end{minipage}
    \hfill
    \begin{minipage}{0.49\textwidth}
        \includegraphics[width=\textwidth,keepaspectratio]{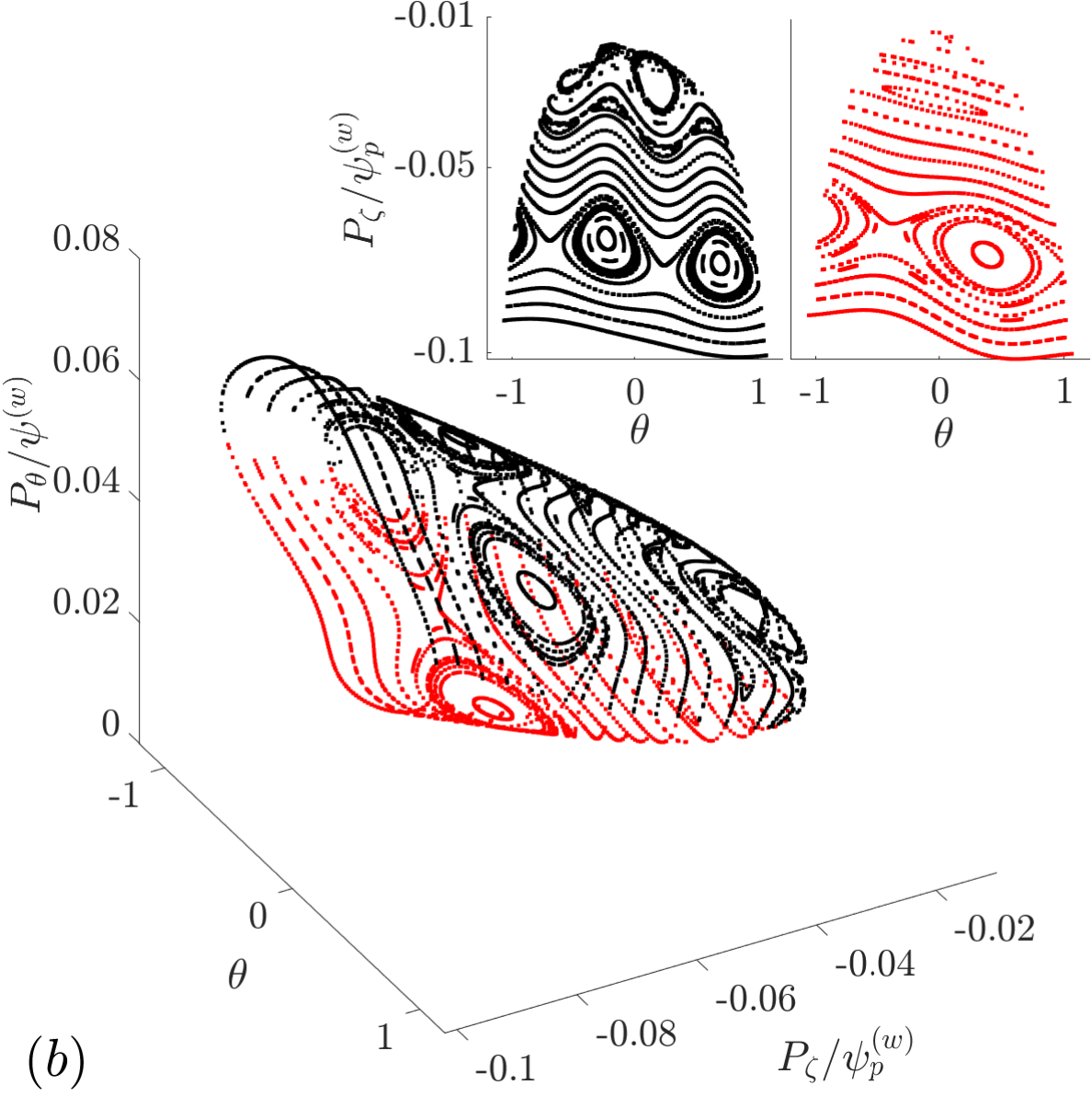}
    \end{minipage}
    \begin{minipage}{0.49\textwidth}
      \raisebox{1cm}{  \includegraphics[width=\textwidth,keepaspectratio]{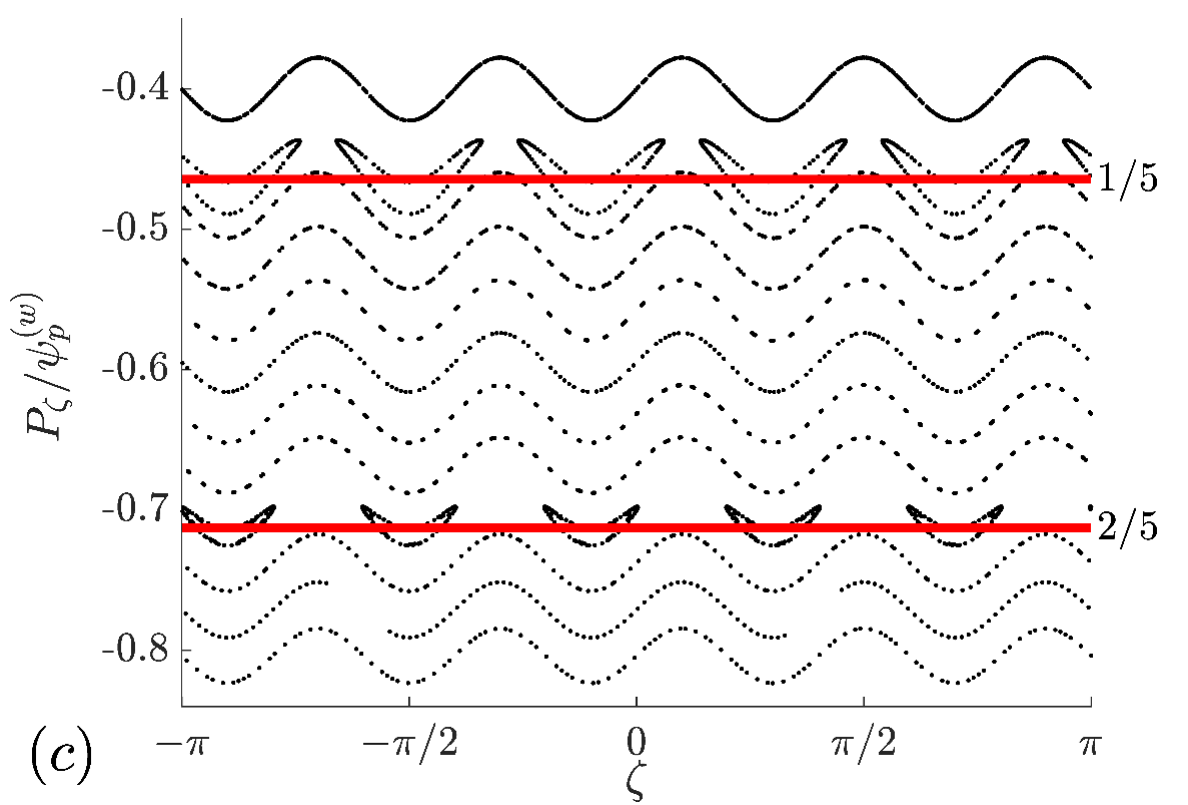}}
    \end{minipage}
    \hfill
    \begin{minipage}{0.49\textwidth}
        \includegraphics[width=\textwidth,keepaspectratio]{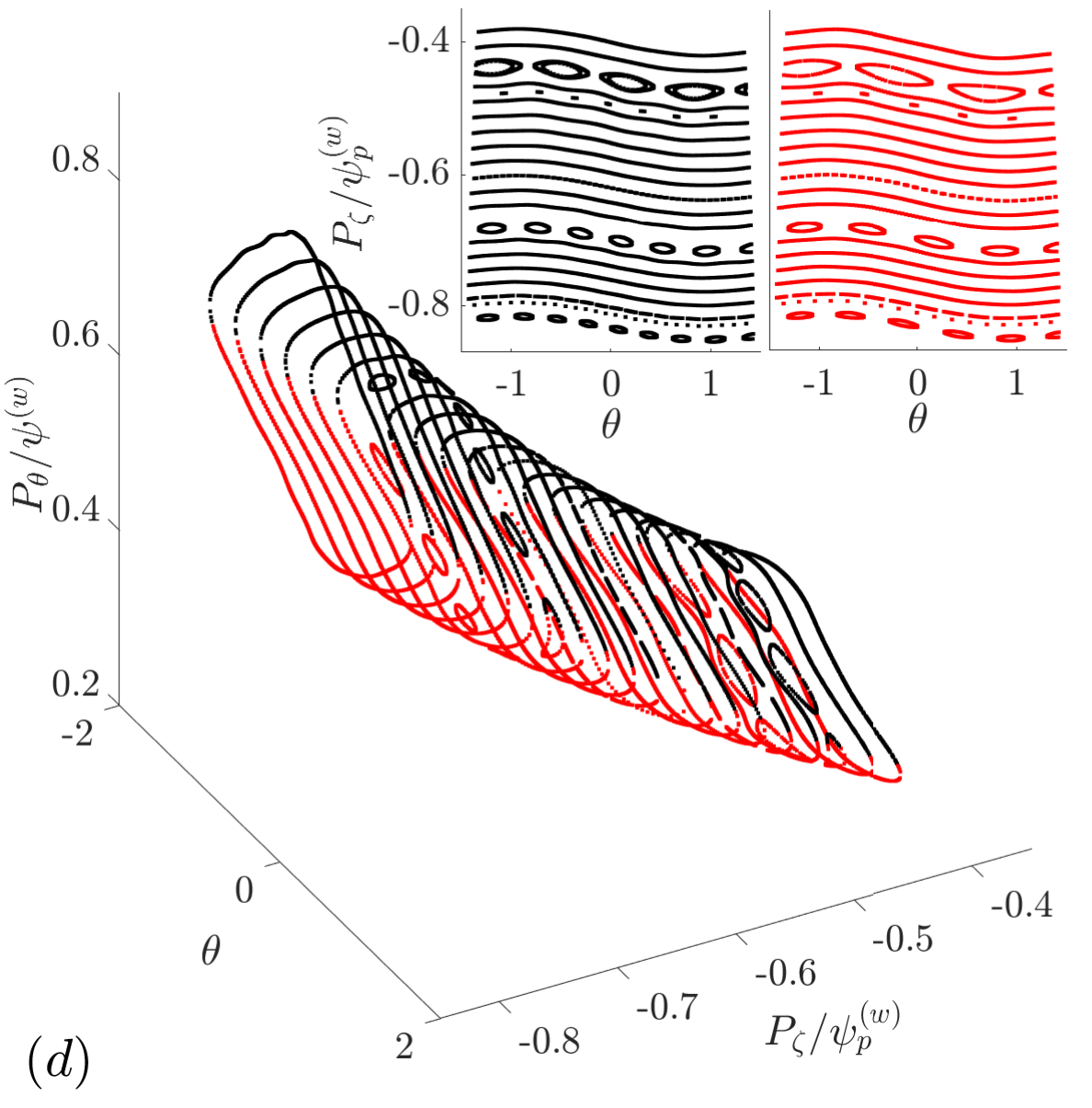}
    \end{minipage}

    \caption[Poincaré surfaces of section for trapped particles in LAR equilibrium]{Poincaré surfaces of section for trapped particles of Case \#1 ($\mu B_{0} = \SI{2}{\kilo\electronvolt}$ in the $q_{1}$ profile) with $E/\mu B_{0} = 0.996$, under the presence of two perturbative modes, $(m,n) = (1,5)$ and $(2,5)$. 
    Cases of two different perturbation amplitudes $\alpha_{1,5} = \alpha_{2,5} = 0.5 \times 10^{-4}$ and $\alpha_{1,5} = \alpha_{2,5} = 2.1 \times 10^{-4}$ are depicted in panels (a)-(b) and (c)-(d), respectively. 
    Panels (a) and (c) display Poincaré surfaces of section at the constant poloidal angle $\theta=0$, with red horizontal lines denoting the predicted locations of the corresponding resonant island chains in accordance to Fig. \ref{fig:Fig54}(a).
    Panels (b) and (d) show Poincaré surfaces at a constant toroidal angle $\zeta=0$, with black and red color indicating orbit intersections with the surface of section in positive and negative directions, respectively. The 3D diagrams depict intersections in both directions lying in a two-dimensional constant energy surface.} 
    \label{fig:Fig56}
\end{figure}

A characteristic case of trapped particles with pitch $E/\mu B_{0}=0.97$ under the presence of  two perturbative modes with mode numbers $(m,n)=(1,5),(2,5)$ is depicted in Fig. \ref{fig:Fig56}. On the $\theta=0$ Poincaré surfaces of section, shown in Fig. \ref{fig:Fig56}(a) and (c), the locations of the resonances are accurately predicted by the analytical results. In accordance to Fig. \ref{fig:Fig54}(a), each resonance appears at two locations of the phase space and the number of islands in each resonance chain corresponds to the toroidal mode number $n$. The $\zeta=0$ Poincaré surfaces of section, defined by orbit intersection in the positive and negative directions are shown in Fig. \ref{fig:Fig56}(b) and \ref{fig:Fig56}(d), respectively, with the orbits appearing to be open. A clearer view of the phase space topology can be provided by a generalized Poincaré surface of section where orbit traces of both positive and negative intersections are depicted and shown to lie in a two-dimensional constant energy surface in the three-dimensional space ($P_\zeta,P_\theta,\theta$). The number of islands in each resonant chain (in both directions) can be accurately predicted from the corresponding unperturbed orbit on the basis of the discussion presented in Sec. \ref{Sec: Number of islands}, as shown by comparing the number of crossings depicted in Fig. \ref{fig:Fig59}(a) with the number of islands in the $q_{kin}=2/5$ resonant chain shown in Fig. \ref{fig:Fig56}(d).

\begin{figure}[h!]
    \centering
    \includegraphics[width=0.49\textwidth,keepaspectratio]{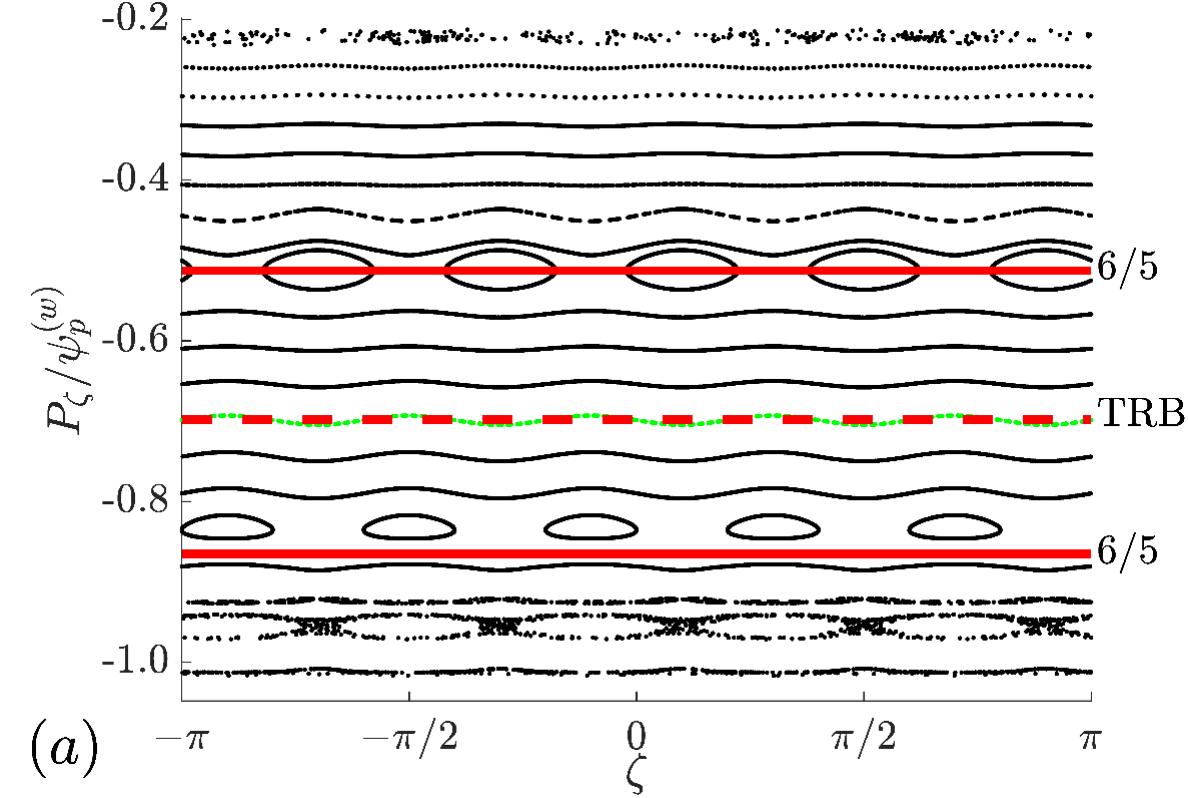}
    \includegraphics[width=0.49\textwidth,keepaspectratio]{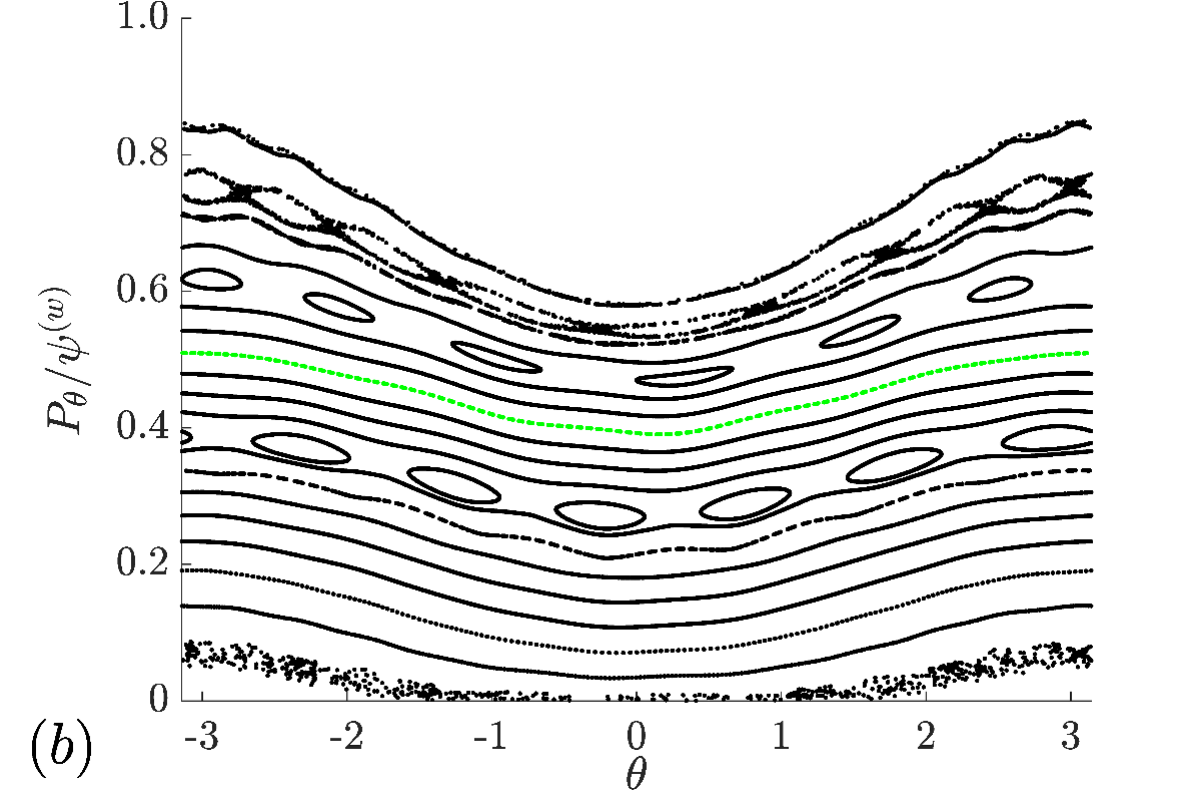}
    \includegraphics[width=0.49\textwidth,keepaspectratio]{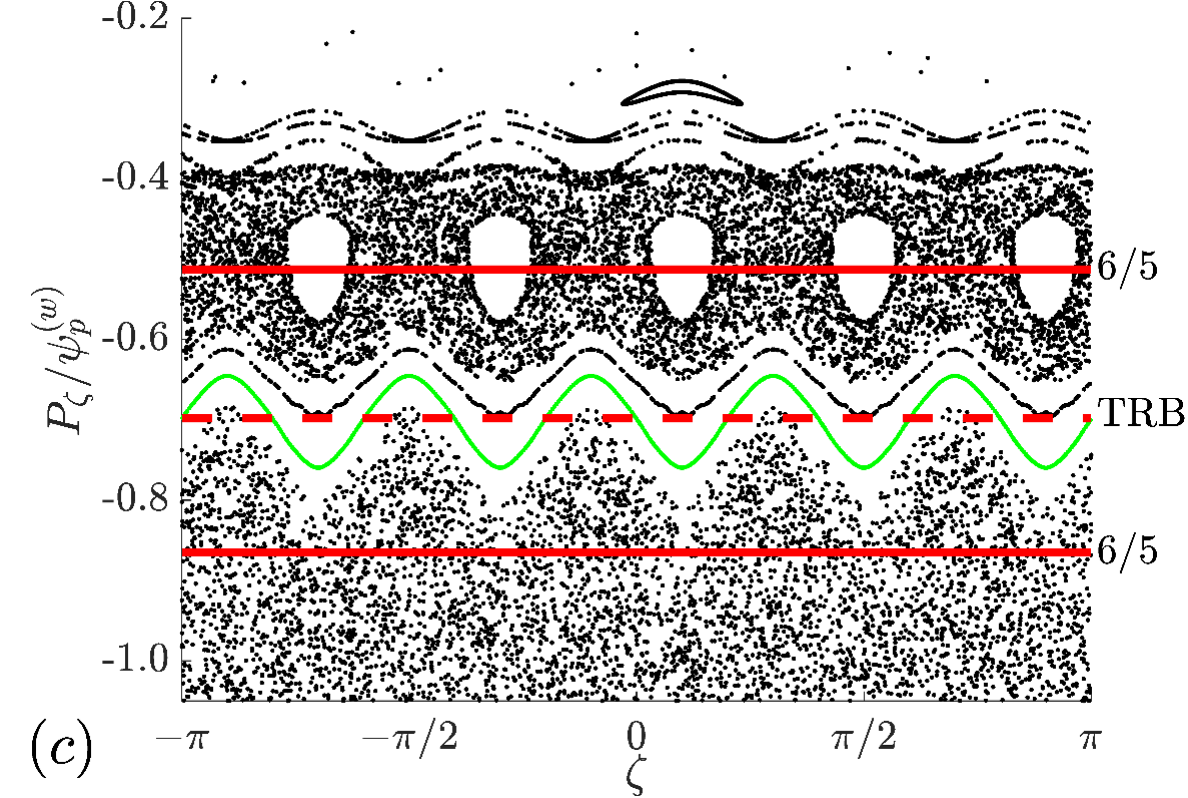}
    \includegraphics[width=0.49\textwidth,keepaspectratio]{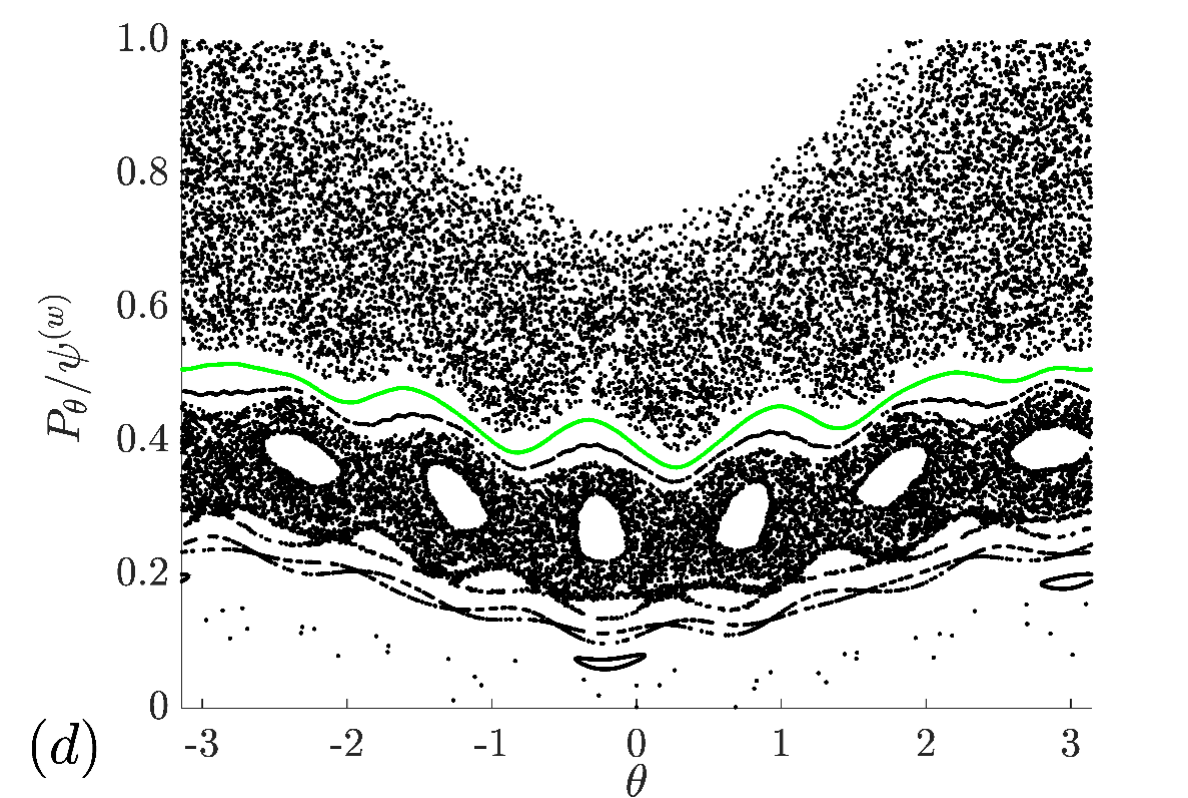}
    
    \caption[Demonstration of transport barrier in Poincaré surfaces of section in LAR equilibrium.]{Poincaré surfaces of section for counter-passing particles of Case \#4 ($\mu B_{0} = \SI{10}{\kilo\electronvolt}$ in the $q_{3}$ profile) with $E/\mu B_{0} = 1.43$, under the presence of a single perturbative mode $(m,n)=(6,5)$. Cases of two different perturbation strengths $\alpha_{6,5}=1.1\times 10^{-5}$ and $\alpha_{6,5} = 1.1\times 10^{-4}$ are depicted in panels (a)-(b) and (c)-(d), respectively.
    Panels (a) and (c) display Poincaré surfaces of section at the constant poloidal angle $\theta=0$, whereas panels (b) and (d) show Poincaré surfaces at a constant toroidal angle $\zeta=0$. The solid red horizontal lines denote the predicted locations of the corresponding resonant island chains and the dotted red horizontal line denotes the predicted location of the transport barrier, in accordance to Fig. \ref{fig:Fig54}(a) [and Fig. \ref{fig:Fig53}(b)]. The green curves indicate the transport barriers located at a value of $P_\zeta/\psi_p^{(w)}$ corresponding to the minimum of the $q_{kin}$, shown in the mid panel of Fig. \ref{fig:Fig53}(b).} 
    \label{fig:Fig57}
\end{figure}

As mentioned in a previous section, a non-monotonic $q$ profile (Case \#4) results in the presence of a local minimum in the kinetic $q$-factor for passing particles, as shown in Fig. \ref{fig:Fig53}(b) and the corresponding appearance of certain resonances in two locations in the COM space, as depicted in Fig. \ref{fig:Fig54}(d). The Poincaré surfaces of section for a characteristic case of co-passing particles with $E/\mu B_{0} = 1.43$ is depicted in Fig. \ref{fig:Fig57}, under the presence of a single perturbative mode with mode numbers $(m,n)=(6,5)$, for two different values of the perturbation strength. In addition to the location of the resonance chains, the location of a transport barrier is predicted at the value of $P_\zeta$ that corresponds to a local minimum of the $q_{kin}$ \cite{Anastassiou2024}. The significance of the transport barrier formation becomes more evident for increasing perturbation strength, as a large part of the phase space connected to the wall has become chaotic and the transport barrier prevents particles from escaping towards the wall.

\section {Relation between the angle variables $(\theta,\zeta)$ and $(\hat{\theta},\hat{\zeta})$} \label{Sec: Number of islands}

The Angle variables $(\hat{\theta},\hat{\zeta})$ essentially differ from the geometrical angle variables $(\theta,\zeta)$, since the former are linear functions of time, whereas the latter are in general quasi-periodic functions of time in the interval $[0, 2\pi)$. The Angle variables are phase-like $(\hat{\theta}=\hat{\omega}_\theta t + \hat{\theta}_0, \hat{\zeta}=\hat{\omega}_\zeta t + \hat{\zeta}_0)$ with the frequencies $\hat{\omega}_\theta, \hat{\omega}_\zeta$ providing the appropriate time scaling of the oscillation in each degree of freedom. The two variable sets are related through the canonical transformation Eq. \eqref{eq:Angles}, implying that $\hat{\theta}$ shares the same period with $\theta$ (although they differ as functions of time), whereas this is not the case for $\zeta$ and $\hat{\zeta}$. In general $\zeta$ can be a quasi-periodic function of time unless the frequencies of $\hat{\zeta}$ and $\theta$ (or $\hat{\theta}$) have a rational ratio, i.e. $\hat{\omega}_\zeta / \hat{\omega}_\theta = m' / n'$, corresponding to a resonant orbit (see Sec. \ref{Action Angle Variables}).   

From the canonical transformation Eq. \eqref{eq:Angles} it is clear that $\theta$ and $\hat{\theta}$ are related through Fourier series expansions as
\begin{eqnarray}
    \theta^{(t)}&=&\sum_k a_k(\mathbf{J})\exp(ik\hat{\theta})\\
    \theta^{(p)}&=&\hat{\theta}+\sum_l a_l(\mathbf{J})\exp(il\hat{\theta})
\end{eqnarray}
for trapped particles (libration type of motion) and passing particles (rotation type of motion), respectively \cite{Goldstein2002}. These equations facilitate the transformation of any non-axisymmetric perturbative mode in Action-Angle variables as follows:
\begin{equation} \label{mode}
\exp\left\{-i(m\theta-n\zeta)\right\}=\exp\left\{-i\left[ \left(m\theta(\hat{\theta})-n\frac{\partial f}{\partial J_\zeta}(\mathbf{J},\theta(\hat{\theta})) \right)-n\hat{\zeta}\right]  \right\}.   
\end{equation}
This clearly shows that a single $(m,n)$ mode in the original angles $(\theta,\zeta)$ results in a multitude of modes $(m',n)$ with different poloidal mode numbers $m'$ in the Angle variables $(\hat{\theta},\hat{\zeta})$ (for more details see Sec. \ref{Sec: Compute F}). 

The above analysis suggests that a single non-axisymmetric perturbative mode results, in general, in multiple resonant islands chains in different locations of the phase space where the condition $q_{kin}=m'/n$ is fulfilled, with the width of these chains depending on the corresponding mode amplitudes as expressed in Action-Angle coordinates (see Sec. \ref{Sec: Resonance islands width}). It is worth emphasizing that although the resonance condition expresses periodicity of the motion in the Angle variables, it also implies periodicity of the motion in the original variables, as indicated by the transformation Eq. \eqref{eq:Angles}. However, due to the nonlinear character of the transformation, the two periods might be different. Consequently, the number of cycles in the two degrees of freedom within a period may differ between the original and the Angle variable sets.

In particular, applying Eqs. \eqref{Eq: theta of o and  x points}, \eqref{Eq: dot theta at o  point}, \eqref{J of o point}, and \eqref{dot J of o point} to the GC Hamiltonian, we obtain the $o$-point of the resonance $(n,-m')$ island as 
\begin{equation}\label{Eq: o point in hat theta zeta}
    n\hat{\zeta}(t) - m'\hat{\theta}(t) = -\bar{\phi}(J_{\theta}^{rc},J_{\zeta}^{rc},\mu)
\end{equation}
and  
\begin{equation}\label{Eq: Jtheta and Jzeta of o point}
    J_{\theta}(t) = J_{\theta}^{rc}, \quad J_{\zeta}(t) = J_{\zeta}^{rc}
\end{equation}
From the resonance diagrams in Fig. \ref{fig:Fig54}, for a specific value of $\mu$, we can predict $E^{rc}$ for a chosen $P_{\zeta}^{rc}$ for a particular $(m',n)$ resonance line. Using Eqs. \eqref{gen Jtheta} and \eqref{gen Jzeta} that connect $J_{\theta}$ and $J_{\zeta}$ with the constants of the motion $(E,\mu,P_{\zeta})$, Eq. \eqref{Eq: Jtheta and Jzeta of o point} states that the $o$-point will always be at $(E^{rc}, P_{\zeta}^{rc})$ at any time, i.e.,
\begin{equation}\label{Eq: E and Pzeta of o point}
        E(t) = E^{rc}, \quad P_{\zeta}(t) = P_{\zeta}^{rc}
\end{equation}
Therefore, the $o$-points of first-order resonances remain on the same $P_{\zeta} = P_{\zeta}^{rc}$ for all times, as seen in the Poincar\'e diagrams \ref{fig:Fig55}-\ref{fig:Fig57}. 

To determine the values of $\theta$ and $\zeta$ where these $o$-points appear, we transform from action-angle coordinates back to the original poloidal and toroidal angles. Substituting Eqs. \eqref{eq:Angles} and \eqref{Eq: E and Pzeta of o point} into Eq. \eqref{Eq: o point in hat theta zeta}, we obtain  
\begin{equation}\label{Eq: o point at theta zeta}
   n\frac{\partial f(J_{\theta}^{rc},J_{\zeta}^{rc},\mu, \theta)}{\partial J_\zeta} + n\zeta - m'\hat{\theta}(J_{\theta}^{rc},J_{\zeta}^{rc},\mu, \theta)= -\bar{\phi}(J_{\theta}^{rc},J_{\zeta}^{rc},\mu)
\end{equation}

As discussed in Sec. \ref{Secular Perturbation Theory}, the $o$-point corresponds to a resonant orbit of the unperturbed system that persists in the perturbed system. Hence, $\theta(t)$ and $\zeta(t)$ for the $o$-point remain periodic functions of time. From this we can infer that $\theta$ and $\zeta$ will periodically return to its initial values. Thus, on a Poincar\'e surface defined by a constant $\theta = \theta^{init}$, the $\theta$ points will be given by  
\begin{equation}\label{Eq: theta_k}
    \theta_k = \theta^{init} + 2k\pi, \quad k=0,1,2,\dots
\end{equation}
where $k$ denotes complete cycles of $\theta$. Substituting this into Eq. \eqref{Eq: o point at theta zeta}, we obtain the sequence of Poincar\'e points in the $\zeta$ coordinate for the $o$-point as  \begin{equation}\label{Eq: zeta_k}
        \zeta_k = \frac{m'}{n}\hat{\theta}(J_{\theta}^{rc},J_{\zeta}^{rc},\mu, \theta_k)-\frac{\partial f(J_{\theta}^{rc},J_{\zeta}^{rc},\mu, \theta_k)}{\partial J_\zeta} - \frac{1}{n}\bar{\phi}(J_{\theta}^{rc},J_{\zeta}^{rc},\mu)
\end{equation}
where $\theta_k$ is given by Eq. \ref{Eq: theta_k}. As seen from the first of Eqs. \eqref{eq:Angles} $\theta$ and $\hat{\theta}$ share the same time period. This means that each time $\theta$ completes a full circle (i.e., at the time events of the Poincar\'e surface), $\hat{\theta}$ also completes a full circle. As a result,
\begin{equation}\label{Eq: hat theta_k}
    \hat{\theta}(J_{\theta}^{rc},J_{\zeta}^{rc},\mu, \theta_k) = \hat{\theta}_k = \hat{\theta}^{init} + 2k\pi, \quad k = 0,1,2,\dots
\end{equation}
where $\hat{\theta}^{init} = \hat{\theta}(J_{\theta}^{rc},J_{\zeta}^{rc},\mu,\theta^{init}) = \hat{\theta}_{k=0}$. Moreover, as discussed in Sec. \ref{Action Angle Variables}, any function of the original variables can be written as a multiply periodic function of the angles with period $2\pi$. This, along with the first of Eqs. \eqref{eq:Angles}, implies that the second term on the left-hand side of Eq. \eqref{Eq: o point at theta zeta} is a periodic function of $\hat{\theta}$ with period $2\pi$. Consequently,
\begin{equation}\label{Eq: df dP_zeta at theta_k}
    \frac{\partial f(J_{\theta}^{rc},J_{\zeta}^{rc},\mu, \theta_k)}{\partial J_\zeta} = \frac{\partial f(J_{\theta}^{rc},J_{\zeta}^{rc},\mu, \theta^{init})}{\partial J_\zeta}, \quad k = 0,1,2,\dots
\end{equation}
Substituting Eqs. \eqref{Eq: hat theta_k} and \eqref{Eq: df dP_zeta at theta_k} into Eq. \eqref{Eq: zeta_k}, we obtain
\begin{equation}\label{Eq: zeta_k 1}
        \zeta_k = \frac{m'}{n}2k\pi + \frac{m'}{n}\hat{\theta}(J_{\theta}^{rc},J_{\zeta}^{rc},\mu, \theta^{init})-\frac{\partial f(J_{\theta}^{rc},J_{\zeta}^{rc},\mu, \theta^{init})}{\partial J_\zeta} - \frac{1}{n}\bar{\phi}(J_{\theta}^{rc},J_{\zeta}^{rc},\mu)
\end{equation}
Since the three last terms on the left-hand side of Eq. \eqref{Eq: zeta_k 1} are a constant phase, we see that $n$ islands, for $k=1,2,\dots,n$, will appear on the Poincar\'e surface of section of constant $\theta = \theta^{init}$. A similar procedure applies to the $x$-point of the resonance, but with a phase shift of $\pi/n$ (considering table \ref{tab: fix points} substitute $\bar{\phi}(J_{\theta}^{rc},J_{\zeta}^{rc},\mu)$ with $\bar{\phi}(J_{\theta}^{rc},J_{\zeta}^{rc},\mu)+\pi$ in Eq. \eqref{Eq: o point in hat theta zeta}), leading to  
\begin{equation}\label{Eq: zeta_k 1 x-point}
        \zeta_k = \frac{m'}{n}2k\pi + \frac{m'}{n}\hat{\theta}^{init}-\frac{\partial f(J_{\theta}^{rc},J_{\zeta}^{rc},\mu, \theta^{init})}{\partial J_\zeta} - \frac{1}{n}\bar{\phi}(J_{\theta}^{rc},J_{\zeta}^{rc},\mu) - \frac{\pi}{n}
\end{equation}
indicating that the $o$-point and $x$-point are separated by $\pi/n$ in $\zeta$.  

On a Poincar\'e surface of section at constant $\zeta = \zeta^{\text{init}}$, the values of $\zeta$ are given by
\begin{equation} \label{Eq: zeta_k 2}
    \zeta_k = \zeta^{\text{init}} + 2k\pi, \quad k=0,1,2,\dots
\end{equation}
Substituting this into Eq. \eqref{Eq: o point at theta zeta}, we obtain
\begin{equation}\label{Eq: theta_k with zeta_k}
        \frac{m'}{n}\hat{\theta}(J_{\theta}^{rc},J_{\zeta}^{rc},\mu, \theta_k)-\frac{\partial f(J_{\theta}^{rc},J_{\zeta}^{rc},\mu, \theta_k)}{\partial P_\zeta}= \zeta_k + \frac{1}{n}\bar{\phi}(J_{\theta}^{rc},J_{\zeta}^{rc},\mu).
\end{equation}
Unlike Eq. \eqref{Eq: zeta_k}, where $\theta_k$ is given by Eq. \eqref{Eq: theta_k} and we determine $\zeta_k$, here $\zeta_k$ is known from Eq. \eqref{Eq: zeta_k}, and we seek to determine $\theta_k$. However, unlike Eq. \eqref{Eq: zeta_k}, where $\zeta_k$ is directly specified, here $\theta_k$ must be solved from Eq. \eqref{Eq: theta_k with zeta_k}, in which it appears in a complex manner.  As a result, the number of $o$-points appearing on a Poincar\'e surface of section at constant $\zeta$ generally differs from $m'$. By analogy, the same holds for the $x$-points.

\begin{figure} [h!]
    \centering
    \includegraphics[width=0.49\textwidth,keepaspectratio]{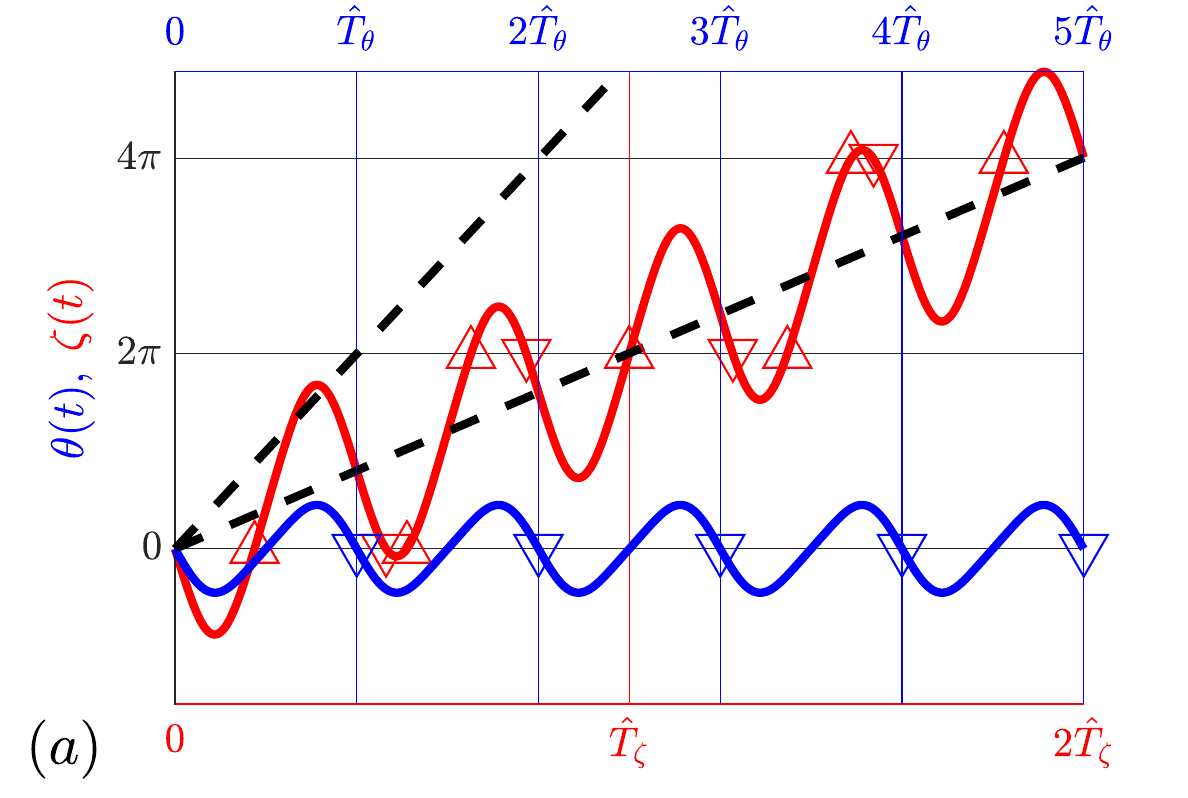}
    \includegraphics[width=0.49\textwidth,keepaspectratio]{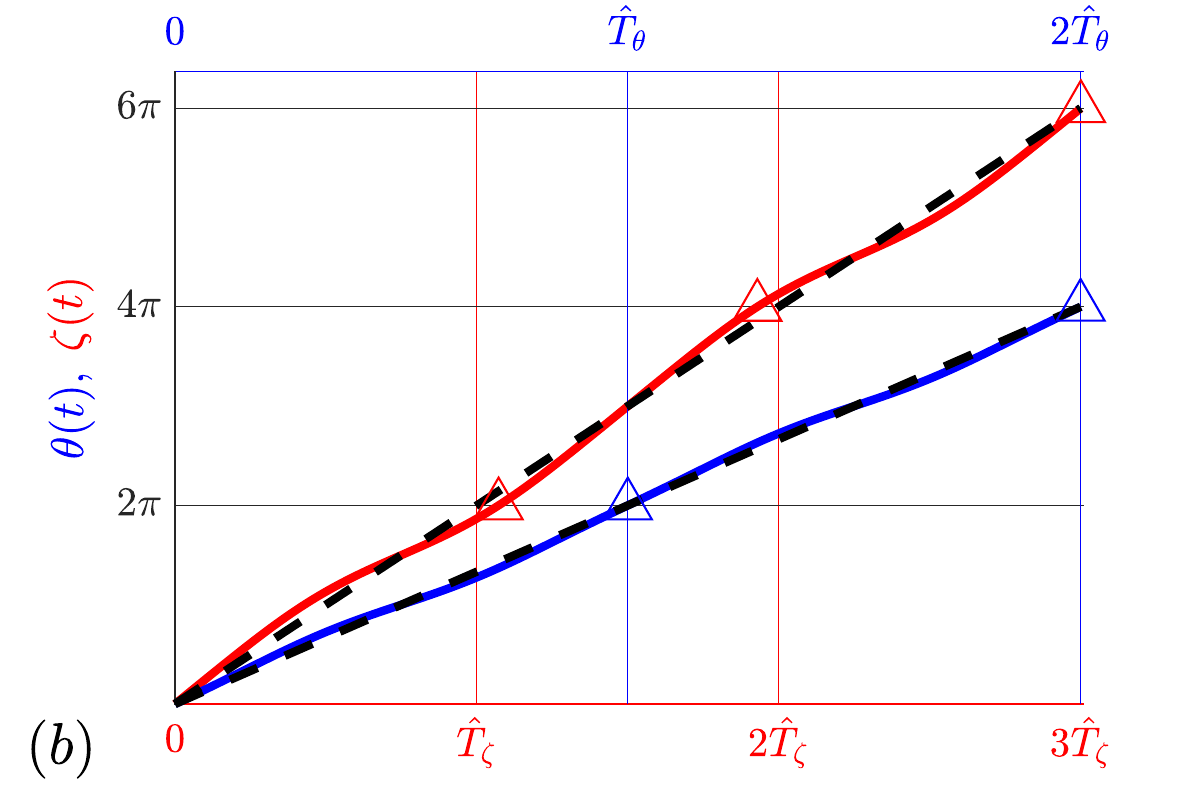}    
    \caption[Relation between the orbital frequencies and the poloidal ($\theta$) and toroidal ($\zeta$) angles of a resonant orbit.]{Time dependence of $\theta$ (blue lines) and $\zeta$ (red lines) for a periodic trapped orbit with $q_{kin}=2/5$ (left) and a passing orbit with $q_{kin}=3/2$ (right). The time dependence, measured in $\hat{T}_\zeta$ and $\hat{T}_\theta$ directly provides the dependence on $\hat{\zeta}$ and $\hat{\theta}$, respectively. Black dashed lines depict the time variations of $\hat{\zeta}$ and $\hat{\theta}$. The traces of the periodic orbit on the Poincaré surfaces of section at $\theta = 0$ and $\zeta = 0$ are indicated by upward blue and red triangles for positive directions, and downward triangles for negative directions.}
    \label{fig:Fig58}
\end{figure}

To see this consider the equations of the GC motion \eqref{dot Pzeta}-\eqref{dot theta} which imply that, for passing particles $(\rho_\parallel \neq 0)$, $\dot{\zeta}(t)\neq 0$ and $\dot{\theta}(t)\neq 0$, making $\zeta(t)$ and $\theta(t)$ monotonic functions of time. Given the linear relation of the Angles with time, they are also monotonic functions of the Angles, and therefore there is a one-to-one correspondence between the two variable sets. However, this is not the case for trapped particles for which $\rho_\parallel = 0$ at the turning points (where $\dot{\zeta}(t) = 0$), and at the banana tips where ($\dot{\theta}(t) = 0$). The relation between the two variable sets, as well as their time dependence, is depicted in Fig. \ref{fig:Fig58}, both for passing and trapped particles.

\begin{figure} [h!]
    \centering
    \includegraphics[width=0.49\textwidth,keepaspectratio]{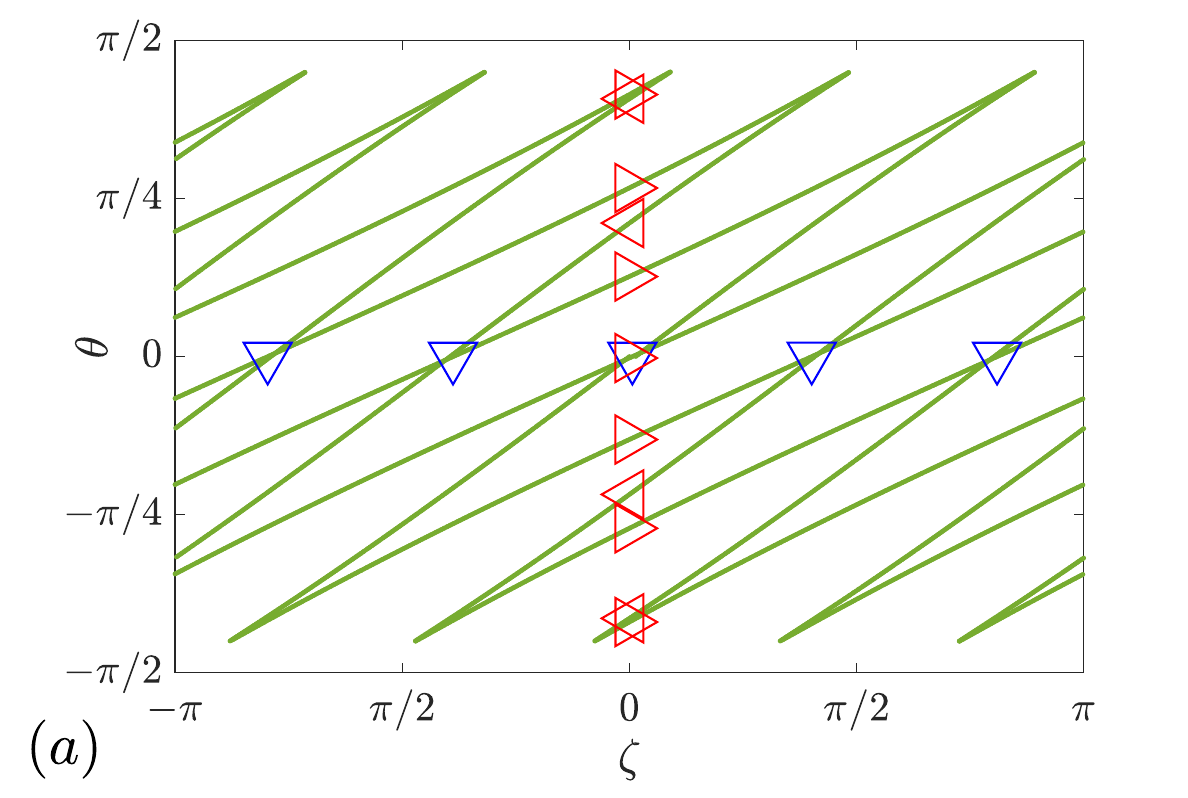}
    \includegraphics[width=0.49\textwidth,keepaspectratio]{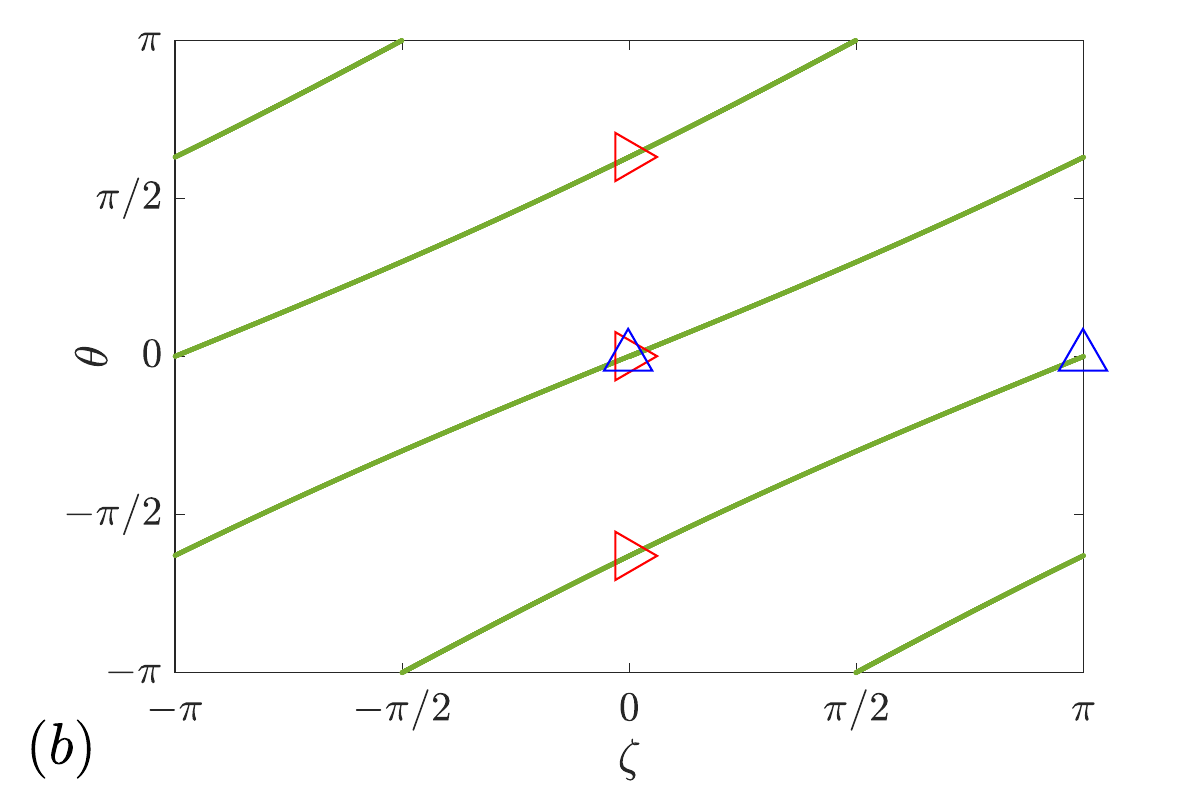}
    \caption[Relation between the poloidal ($\theta$) and toroidal ($\zeta$) angle of a resonant orbit.]{Winding of $\theta$ and $\zeta$ for a periodic trapped orbit with $q_{kin}=2/5$ (a) and a passing orbit with $q_{kin}=3/2$ (b). Upward and downward red triangles indicate the traces of the periodic orbit on the $\zeta = 0$ Poincaré surface in positive and negative directions, respectively, while left- and right-facing blue triangles denote the traces on the $\theta = 0$ Poincaré surface in positive and negative directions, respectively.}
    \label{fig:Fig59}
\end{figure}

Under the presence of perturbations, the  same resonance conditions apply for both variable sets, although a different number of resonances appear, due to Eq. \eqref{mode}, and a different number of islands appear within each resonance chain. The islands in a Poincaré surface of section of the perturbed system surround elliptical points corresponding to the traces of a periodic orbit on the Poincaré surface of section, and therefore their number can be a priori determined by counting these traces for an unperturbed periodic (resonant) orbit as shown in Fig. \ref{fig:Fig59}. For a passing periodic orbit with $q_{kin}=3/2$ the number of traces in the Poincaré surfaces of section $\zeta=0$ and $\theta=0$ are 3 and 2, respectively, suggesting that the numbers of islands in the resonant chain under the presence of a $(m,n)=(3,2)$ perturbative mode correspond to these mode numbers, and correctly predicting the number of islands in the Poincaré surfaces of section depicted in Figs. \ref{fig:Fig55}(a) and \ref{fig:Fig55}(b). However, for a trapped orbit with $q_{kin}=2/5$ the number of traces in the Poincaré surfaces of section $\zeta=0$ and $\theta=0$ are 11 and 5, respectively, suggesting that under the presence of a $(m,n)=(2,5)$ perturbative mode the number of islands in the $\zeta=0$ Poincaré surface of section will differ from the mode number, and predicting the number of islands depicted in Figs. \ref{fig:Fig56}(c) and \ref{fig:Fig56}(d).

\section{Width of resonance islands}\label{Sec: Resonance islands width}

The width of islands in a resonance island chain, expressed in terms of the system's actions, is given by Eq. \eqref{Eq: width of island in action angle}, with a qualitative depiction shown in Fig. \ref{fig:Fig32}. In this section, we apply relation \eqref{Eq: width of island in action angle} to the GC Hamiltonian system, which has already been transformed into action-angle variables in Sec. \ref{Action-Angle Transformation}.

The perturbed GC Hamiltonian \eqref{Per GC H 1} can be expressed as a near-integrable, time-independent Hamiltonian in action-angle variables of the form \eqref{near integrable Hamiltonian} (see Eqs. \eqref{Eq: GC H with time}-\eqref{perturbation_in_AA}, setting $\omega=0$):
\begin{equation}\label{Eq: near integrable GC H}
    \begin{aligned}
    H(\hat{\theta},\hat{\zeta},J_{\theta},J_{\zeta};\mu) &= H_0(J_{\theta}, J_{\zeta};\mu) + \epsilon \sum_{m',n'}\hat{H}_{1(m',n')}(J_{\theta},J_{\zeta};\mu)e^{i(n'\hat{\zeta}-m'\hat{\theta})}\\
    &+ \epsilon^2 \sum_{m',n'}\hat{H}_{2(m',n')}(J_{\theta},J_{\zeta};\mu)e^{i(n'\hat{\zeta}-m'\hat{\theta})}
\end{aligned}
\end{equation}

Here, $J_{\xi} = \mu$ is treated as a parameter since the perturbation is independent of $\hat{\xi}$, and $H_0$ is the unperturbed GC Hamiltonian \eqref{GC H 1}, given by
\begin{equation}
    H_0(J_{\theta},J_{\zeta};\mu) = E(J_{\theta},J_{\zeta};\mu)
\end{equation}
The actions in terms of the constants of motion $(E,\mu,P_{\zeta})$ are determined via the canonical transformation relations in Eqs. \eqref{gen Jzeta}-\eqref{gen Jtheta}, with the angles relations given by Eqs. \eqref{eq:Angles}.

We now aim to determine the width of the islands in $J_{\zeta}$ for a resonance characterized by $(n',-m')$ using Eq. \eqref{Eq: width of island in action angle}. This requires computing
\begin{equation}   
    \hat{H}_0
    (\bar{J}_{\theta},\bar{J}_{\zeta};\mu) = H_0(J_{\theta}(\bar{J}_{\theta},\bar{J}_{\zeta}),J_{\zeta}(\bar{J}_{\theta},\bar{J}_{\zeta});\mu)
\end{equation}
and 
\begin{equation}  \label{Eq: hat hat H1m',n'}
     \hat{\hat{H}}_{1(m',n')}(\bar{J}_{\theta},\bar{J}_{\zeta};\mu) = \hat{H}_{1(m',n')}(J_{\theta}(\bar{J}_{\theta},\bar{J}_{\zeta}),J_{\zeta}(\bar{J}_{\theta},\bar{J}_{\zeta});\mu)
\end{equation}
and from these, evaluating
\begin{equation} \label{Eq: G bar J GC}
    G(\bar{J}_{\theta}^{rc},\bar{J}_{\zeta}^{rc};\mu) =   \left.\frac{\partial^2 \hat{H}_0}{\partial \bar{J}_{\zeta}^2}\right|_{\bar{J}_{\theta} = \bar{J}_{\theta}^{rc}, \bar{J}_{\zeta} = \bar{J}_{\zeta}^{rc}}
\end{equation}
and 
\begin{equation} \label{Eq: F bar J GC}
    F(\bar{J}_{\theta}^{rc},\bar{J}_{\zeta}^{rc};\mu) = -2\epsilon|\hat{\hat{H}}_{1(m',n')}  (\bar{J}_{\theta}^{rc},\bar{J}_{\zeta}^{rc};\mu)|.  
\end{equation}

The resonance condition is
\begin{equation}
    -m'\hat{\omega}_\theta(J_{\theta},J_{\zeta};\mu) + n'\hat{\omega}_{\zeta}(J_{\theta},J_{\zeta};\mu) = 0.
\end{equation}
By analogy with Eqs. \eqref{rsk resonance}, \eqref{J to hat J}, and \eqref{theta to hat theta}, and using the equivalence of barred and hatted variables to first order in $\epsilon$ (as discussed in Sec. \ref{Secular Perturbation Theory}, Eq. \eqref{Eq: hat with bar canonical variables simplified}), we obtain
\begin{equation}\label{Eq: Jzeta Jtheta with barred ones}
    J_{\zeta} = n'\bar{J}_{\zeta}, \quad J_{\theta} = \bar{J}_{\theta} - m'\bar{J}_{\zeta}
\end{equation}

\subsection{Computation of $G(\bar{J}_{\theta}^{rc},\bar{J}_{\zeta}^{rc};\mu)$}

Eq. \eqref{Eq: first pd of hat H_0 with bar J} indicates that 
\begin{equation}
    \frac{\partial \hat{H}_0}{\partial \bar{J}_{\zeta}} =     -m'\hat{\omega}_\theta(J_{\theta}(\bar{J}_{\theta},\bar{J}_{\zeta}),J_{\zeta}(\bar{J}_{\theta},\bar{J}_{\zeta});\mu) + n'\hat{\omega}_{\zeta}(J_{\theta}(\bar{J}_{\theta},\bar{J}_{\zeta}),J_{\zeta}(\bar{J}_{\theta},\bar{J}_{\zeta});\mu)
\end{equation}
which using Eq. \eqref{Eq: Jzeta Jtheta with barred ones} gives
\begin{equation}\label{Eq: G at GC}
    \begin{aligned}
        \left.\frac{\partial^2 \hat{H}_0}{\partial \bar{J}_{\zeta}^2}\right|_{\bar{J}_{\theta} = \bar{J}_{\theta}^{rc}, \bar{J}_{\zeta} = \bar{J}_{\zeta}^{rc}} &= m'\left(m'\left.\frac{\partial \hat{\omega}_{\theta}}{\partial J_{\theta}}\right|_{J_{\theta} = J_{\theta}^{rc}, J_{\zeta} = J_{\zeta}^{rc}} - n'\left.\frac{\partial \hat{\omega}_{\theta}}{\partial J_{\zeta}}\right|_{J_{\theta} = J_{\theta}^{rc}, J_{\zeta} = J_{\zeta}^{rc}}\right)\\
        &- n'\left(m'\left.\frac{\partial \hat{\omega}_{\zeta}}{\partial J_{\theta}}\right|_{J_{\theta} = J_{\theta}^{rc}, J_{\zeta} = J_{\zeta}^{rc}} - n'\left.\frac{\partial \hat{\omega}_{\zeta}}{\partial J_{\zeta}}\right|_{J_{\theta} = J_{\theta}^{rc}, J_{\zeta} = J_{\zeta}^{rc}}\right)
    \end{aligned}
\end{equation}

In the case of the LAR equilibrium, the right-hand side of Eq. \eqref{Eq: G at GC} can be analytically computed. To achieve this, we first note that the resonance diagrams (Fig. \ref{fig:Fig54}) allow us to determine the location $E^{rc}$ of the specific resonance $(n', -m')$ in terms of the constants of motion, given a chosen $P_{\zeta}^{rc}$ for a fixed $\mu$. Substituting these values into Eqs. \eqref{gen Jtheta} and \eqref{Jtheta_t}-\eqref{Jtheta_p}, we obtain $J_{\theta}^{rc}$ and $J_{\zeta}^{rc}$.

Using Eq. \eqref{gen hat omega theta} wherein $H=E$, $\hat{\omega}_{\theta}$ is analytically determined by
\begin{equation}\label{Eq: hat omega theta analyt}
    \hat{\omega}_{\theta}(E,P_{\zeta};\mu) = \frac{1}{\frac{\partial J_{\theta}(E,P_{\zeta};\mu)}{\partial E}}
\end{equation}
where $J_{\theta}$ in terms of $(E,P_{\zeta},\mu)$ is analytically given by Eqs. \eqref{Jtheta_t} and \eqref{Jtheta_p}. Meanwhile, from Eq. \eqref{q-kinetic}, we have
\begin{equation}\label{Eq: hat omega zeta analyt}
    \hat{\omega}_{\zeta} = q_{kin}(E,P_{\zeta};\mu)\hat{\omega}_{\theta}(E,P_{\zeta};\mu).
\end{equation}
From Eqs. \eqref{gen Jzeta} and \eqref{Jtheta_t}-\eqref{Jtheta_p}, we observe that between the variable sets $(E,P_{\zeta},\mu)$ and $(J_{\theta},J_{\zeta},\mu)$, the energy is expressed as $E = E(J_{\theta},J_{\zeta};\mu)$, while $P_{\zeta} = \sigma J_{\zeta}$, i.e., a function of $J_{\zeta}$ only, with $\mu$ treated as a parameter. Thus,
\begin{align}
    \frac{\partial\hat{\omega}_{\theta}(E(J_{\theta},J_{\zeta};\mu),P_{\zeta}(J_{\zeta});\mu)}{\partial J_{\theta}} &= \frac{\partial \hat{\omega}_{\theta}}{\partial E}\frac{\partial E}{\partial J_{\theta}}, \\
    \frac{\partial\hat{\omega}_{\zeta}(E(J_{\theta},J_{\zeta};\mu),P_{\zeta}(J_{\zeta});\mu)}{\partial J_{\theta}} &= \frac{\partial \hat{\omega}_{\zeta}}{\partial E}\frac{\partial E}{\partial J_{\theta}}.
\end{align}
From Eq. \eqref{gen hat omega theta}, we know that $\partial E/\partial J_{\theta} = \omega_{\theta}$. Thus,
\begin{align}\label{Eq: hat omegas with Jtheta}
    \left.\frac{\partial \hat{\omega}_{\theta}}{\partial J_{\theta}}\right|_{J_{\theta} = J_{\theta}^{rc}, J_{\zeta} = J_{\zeta}^{rc}} &= \frac{\partial \hat{\omega}_{\theta}(E^{rc},P_{\zeta}^{rc};\mu)}{\partial E}\omega_{\theta}(E^{rc},P_{\zeta}^{rc};\mu), \\
    \left.\frac{\partial \hat{\omega}_{\zeta}}{\partial J_{\theta}}\right|_{J_{\theta} = J_{\theta}^{rc}, J_{\zeta} = J_{\zeta}^{rc}} &= \frac{\partial \hat{\omega}_{\zeta}(E^{rc},P_{\zeta}^{rc};\mu)}{\partial E}\omega_{\theta}(E^{rc},P_{\zeta}^{rc};\mu).
\end{align}
Similarly, considering that $\partial E/\partial J_{\zeta} = \hat{\omega}_{\zeta}$ (Eq. \eqref{gen hat omega zeta}) and $\partial P_{\zeta}/\partial J_{\zeta} = \sigma$ (Eq. \eqref{gen Jzeta}), we obtain
\begin{align}\label{Eq: hat omegas with Jzeta}
    \left.\frac{\partial \hat{\omega}_{\theta}}{\partial J_{\zeta}}\right|_{J_{\theta} = J_{\theta}^{rc}, J_{\zeta} = J_{\zeta}^{rc}} &= \frac{\partial \hat{\omega}_{\theta}(E^{rc},P_{\zeta}^{rc};\mu)}{\partial E}\omega_{\zeta}(E^{rc},P_{\zeta}^{rc};\mu) + \sigma\frac{\partial \hat{\omega}_{\theta}(E^{rc},P_{\zeta}^{rc};\mu)}{\partial P_{\zeta}}, \\
    \left.\frac{\partial \hat{\omega}_{\zeta}}{\partial J_{\zeta}}\right|_{J_{\theta} = J_{\theta}^{rc}, J_{\zeta} = J_{\zeta}^{rc}} &= \frac{\partial \hat{\omega}_{\zeta}(E^{rc},P_{\zeta}^{rc};\mu)}{\partial E}\omega_{\zeta}(E^{rc},P_{\zeta}^{rc};\mu) + \sigma\frac{\partial \hat{\omega}_{\zeta}(E^{rc},P_{\zeta}^{rc};\mu)}{\partial P_{\zeta}}.
\end{align}
Substituting Eqs. \eqref{Eq: hat omegas with Jtheta} and \eqref{Eq: hat omegas with Jzeta} into Eq. \eqref{Eq: G at GC}, we analytically determine $G(\bar{J}_{\theta}^{rc},\bar{J}_{\zeta}^{rc};\mu)$ in the case of LAR equilibrium.

\subsection{Computation of $F(\bar{J}_{\theta}^{rc},\bar{J}_{\zeta}^{rc};\mu)$}\label{Sec: Compute F}

To compute $F(\bar{J}_{\theta}^{rc},\bar{J}_{\zeta}^{rc};\mu)$, we need to express the first-order perturbation term of the perturbed GC Hamiltonian (see Eq. \eqref{Eq: H1}):
\begin{equation}\label{Eq: H_1 with orginal variables}
      H_1(\theta,\zeta,P_{\theta},P_{\zeta};\mu) = 2\alpha(\theta,\zeta,P_{\theta},P_{\zeta})\left(P_{\zeta} + \psi_{p}(P_{\zeta}, P_{\theta})\right) \frac{ B^2(P_{\zeta}, P_{\theta}, \theta)}{2 g(P_{\zeta}, P_{\theta})}  
\end{equation}
where the time-independent function $\alpha$ is given by Eqs. \eqref{alpha} and \eqref{canon moments perturbed} as
\begin{equation} \label{Eq: alpha with Ptheta Pzeta}
    \alpha(\theta,\zeta,P_{\theta},P_{\zeta})=\sum_{m,n} \alpha_{m,n}(P_{\theta},P_{\zeta})e^{i(n\zeta - m\theta)},
\end{equation}
as a multiple Fourier series of the angles $\hat{\theta}$ and $\hat{\zeta}$ in the form:
\begin{equation} \label{perturbation_in_AA}
    \hat{H}_{1}(\hat{\theta},\hat{\zeta},J_{\theta},J_{\zeta};\mu)= \sum_{m',n'}\hat{H}_{1(m',n')}(J_{\theta},J_{\zeta};\mu)e^{i(n'\hat{\zeta}-m'\hat{\theta})}.
\end{equation}
It can be written in this way because, as discussed in Sec. \ref{Action Angle Variables}, any function of the original unperturbed variables\footnote{Unperturbed because the action-angle transformation is performed in the unperturbed system. However, as a canonical transformation, it can be applied to any Hamiltonian system of these variables, including the perturbed GC Hamiltonian studied here.}—which, in the case of the GC system, are $(\theta,\zeta,P_{\theta},P_{\zeta})$—can be expressed as a multiple Fourier series expansion, being a periodic function of the angles with period $2\pi$.

To do it, we use the unperturbed GC Hamiltonian \eqref{GC H 1} to write $P_{\theta}$ as $P_{\theta} = P_{\theta}(\theta, E,\mu,P_{\zeta})$. In this way, $H_1(\theta,\zeta,P_{\theta},P_{\zeta};\mu)$ in Eq. \eqref{Eq: H_1 with orginal variables} can be written as a function of $(\theta,\zeta,E(J_{\zeta},J_{\theta};\mu),$ $ P_{\zeta}(J_{\zeta});\mu)$. Examining the unperturbed GC Hamiltonian \eqref{GC H 1}, we see that Eq. \eqref{Eq: H_1 with orginal variables} can be simplified. Substituting $\alpha$ from Eq. \eqref{Eq: alpha with Ptheta Pzeta} and $P_{\theta} = P_{\theta}(\theta, E,\mu,P_{\zeta})$, Eq. \eqref{Eq: H_1 with orginal variables} is written as
\begin{equation}\label{Eq: H1 with theta zeta}
   \begin{aligned}
        H_1(\theta,\zeta,E,P_{\zeta};\mu) &= \\
       &\hspace{-1.5cm} 2\frac{E-\mu B(P_{\zeta},P_{\theta}(\theta,E,P_{\zeta};\mu),\theta)}{P_{\zeta} + \psi_p(P_{\zeta}, P_{\theta}(\theta,E,P_{\zeta};\mu))}\sum_{m,n} \alpha_{m,n}(P_{\zeta}, P_{\theta}(\theta,E,P_{\zeta};\mu))e^{i(n\zeta - m\theta)}
    \end{aligned} 
\end{equation}

Substituting $\theta$ and $\zeta$ in terms of the angles $\hat{\theta}$ and $\hat{\zeta}$ using the first and the second of Eqs. \eqref{eq:Angles}, respectively, and substituting $E$ with $E(J_{\theta},J_{\zeta};\mu)$ and $P_{\zeta}$ with $P_{\zeta}(J_{\zeta})$, the above relation is written in terms of the action-angle variables as
\begin{equation}
    \begin{aligned}\label{Eq: hat H1 1}
    \hat{H}_1 &=
    &\hspace{-0.2cm}\sum_{m,n}\left[2\hat{\alpha}_{m,n}(\hat{\theta},J_{\theta},J_{\zeta}; \mu) \frac{E(J_{\theta},J_{\zeta};\mu)-\mu \hat{B}(\hat{\theta},J_{\theta},J_{\zeta}; \mu)}{P_{\zeta}(J_{\zeta}) + \hat{\psi}_p(\hat{\theta},J_{\theta},J_{\zeta}; \mu)}e^{-i\left(m\theta(\hat{\theta}) - n\frac{\partial f(J_{\theta},J_{\zeta},\theta(\hat{\theta});\mu)}{\partial J_{\zeta}}\right)}\right]\cdot e^{in\hat{\zeta}}
\end{aligned}
\end{equation}
where 
\begin{equation}\label{Eq: hat B}
    \hat{B}(\hat{\theta},J_{\theta},J_{\zeta}; \mu) = B(P_{\zeta},P_{\theta}(\theta(\hat{\theta}),E(J_{\theta},J_{\zeta};\mu),P_{\zeta}(J_{\zeta});\mu),\theta(\hat{\theta})),
\end{equation}
\begin{equation}\label{Eq: hat psip}
    \hat{\psi}_p(\hat{\theta},J_{\theta},J_{\zeta}; \mu) = \psi_p(P_{\zeta},P_{\theta}(\theta(\hat{\theta}),E(J_{\theta},J_{\zeta};\mu),P_{\zeta}(J_{\zeta});\mu))
\end{equation}
and
\begin{equation}\label{Eq: hat alpha}
    \hat{\alpha}_{m,n}(\hat{\theta},J_{\theta},J_{\zeta}; \mu) = \alpha_{m,n}(P_{\zeta},P_{\theta}(\theta(\hat{\theta}),E(J_{\theta},J_{\zeta};\mu),P_{\zeta}(J_{\zeta});\mu))
\end{equation}
The term in the bracket depends on $\theta(\hat{\theta})$ and, as a result, it is written as a Fourier series expansion of only $\hat{\theta}$ (not involving $\hat{\zeta}$) as:
\begin{equation}
    \begin{aligned}\label{Eq: f(hat theta)}
    2\frac{E(J_{\theta},J_{\zeta};\mu)-\mu \hat{B}(\hat{\theta},J_{\theta},J_{\zeta}; \mu)}{P_{\zeta}(J_{\zeta}) + \hat{\psi}_p(\hat{\theta},J_{\theta},J_{\zeta}; \mu)}\hat{\alpha}_{m,n}(\hat{\theta},J_{\theta},J_{\zeta}; \mu)e^{-i\left(m\theta(\hat{\theta}) - n\frac{\partial f(J_{\theta},J_{\zeta},\theta(\hat{\theta});\mu)}{\partial J_{\zeta}}\right)} &=\\
    &\hspace{-3.5cm} \sum_{m'} c_{m',n,m}(J_{\theta},J_{\zeta};\mu)e^{-i m'\theta}
\end{aligned}
\end{equation}
for each $m,n$. Thus, Eq. \eqref{Eq: hat H1 1}, changing the summation between $m'$ and $m$, is written as
\begin{align}\label{Eq: hat H1 2}
    \hat{H_1}(\hat{\theta},\hat{\zeta},J_{\theta},J_{\zeta};\mu) = \sum_{m',n}\sum_{m} c_{m',n,m}(J_{\theta},J_{\zeta};\mu)e^{-i m'\theta}e^{in\hat{\zeta}}
\end{align}
and setting
\begin{equation}\label{Eq: hat H1m',n sum}
    \hat{H}_{1(m',n)} = \sum_{m} c_{m',n,m}(J_{\theta},J_{\zeta};\mu)
\end{equation}
Eq. \eqref{Eq: hat H1 2} becomes
\begin{equation} \label{perturbation_in_AA 2}
    \hat{H}_{1}(\hat{\theta},\hat{\zeta},J_{\theta},J_{\zeta};\mu)= \sum_{m',n}\hat{H}_{1(m',n)}(J_{\theta},J_{\zeta};\mu)e^{i(n\hat{\zeta}-m'\hat{\theta})},
\end{equation}
which has been written in the form of \eqref{perturbation_in_AA}, where we see that $n=n'$ as we have already discussed in Sec. \ref{Sec: Number of islands}.

As a result, from Eqs. \eqref{Eq: hat hat H1m',n'}, \eqref{Eq: F bar J GC}, \eqref{Eq: hat H1m',n sum}, we find
\begin{equation}
  F(\bar{J}_{\theta}^{rc},\bar{J}_{\zeta}^{rc};\mu) =    \sum_{m} c_{m',n,m}(J_{\theta}^{rc}(\bar{J}_{\theta}^{rc},\bar{J}_{\zeta}^{rc}),J_{\zeta}^{rc}(\bar{J}_{\theta}^{rc},\bar{J}_{\zeta}^{rc});\mu)
\end{equation}
$c_{m',n,m}(J_{\theta}^{rc}(\bar{J}_{\theta}^{rc},\bar{J}_{\zeta}^{rc}),J_{\zeta}^{rc}(\bar{J}_{\theta}^{rc},\bar{J}_{\zeta}^{rc});\mu)$ can be computed with a Fourier series expansion in $\hat{\theta}$ of the left-hand side of Eq. \eqref{Eq: f(hat theta)} at $(J_{\theta},J_{\zeta})=(J_{\theta}^{rc},J_{\zeta}^{rc})$, which can be done numerically using the method of discrete Fourier transform (DFT). Then, applying Eq. \eqref{Eq: width of island in action angle}, we obtain the resonance width in $J_{\zeta}$ as:
\begin{equation}
    W_{J_{\zeta}} = 4n'\left(\frac{F}{G}\right)^{1/2}, \quad n'=n
\end{equation}

This width can directly give the width of islands of the resonance $(-m',n)$ in $P_{\zeta}$ in Poincar\'e surfaces, either of constant $\theta$ or constant $\zeta$, of the perturbed GC Hamiltonian, like the Poincar\'e surfaces in Figs. \ref{fig:Fig55} and \ref{fig:Fig56}(a), (c). Since, from the second of \eqref{gen Jzeta}, $J_{\zeta} = \sigma P_{\zeta}$, with $\sigma=\pm1$, the width of islands in $P_{\zeta}$ will be the same as the width in $J_{\zeta}$.

Note that to compute the width of islands in $J_{\theta}$, Eq. \eqref{Eq: Jzeta Jtheta with barred ones} has to be modified as 
\begin{equation}\label{Eq: Jzeta Jtheta with barred ones 2}
    J_{\theta} = -m'\bar{J}_{\theta}, \quad J_{\zeta} = \bar{J}_{\zeta} + n'\bar{J}_{\theta}
\end{equation}
However, with this modification, we observe that $F$ and $G$ take the same values as before, giving
\begin{equation}
    W_{J_{\theta}} = 4m'\left(\frac{F}{G}\right)^{1/2}.
\end{equation}
As expected, this relation can also be obtained by calculating $W_{J_{\zeta}}$ using the second of Eqs. \eqref{Eq: Jzeta Jtheta with barred ones 2} and considering that on the resonance, $\bar{J}_{\theta}$ remains constant, $\bar{J}_{\theta} = \bar{J}_{\theta}^{rc}$.

\section{Calculation of the kinetic $q$-factor in realistic equilibrium magnetic fields}\label{Sec: realistic equilibrium}

In this section, we extend the applicability of our Orbital Spectrum Analysis (OSA), methodology from LAR to numerically reconstructed equilibria. We employ a computationally efficient semi-analytical geometrical method applied on any given unperturbed equilibrium. 

\subsection{Semi-analytical geometric method}\label{Sec: geomtric method}
\begin{figure}[h!]
    \centering
    \includegraphics[width=0.5\textwidth]{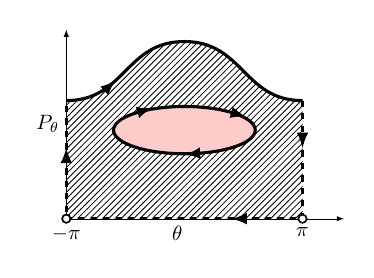}
    \caption[Shoelace formula]{A qualitative sketch of contour lines of the GC Hamiltonian \eqref{GC H 1} for passing and trapped orbits. The shaded areas enclosed by the respective curves are computed with the Shoelace formula, Eq. \eqref{eq: Shoelace formula}, and equal to the $J_{\theta}$, Eq. \eqref{gen Jtheta}, value for the respective $(E, \mu, P_{\zeta})$ set.}
    \label{fig:Fig510}
\end{figure}

In analogy to the analytical calculation of $q_{kin}$ in Ref. \cite{Antonenas2024}, this method directly computes the action integral in Eq. \eqref{gen Jtheta}. To achieve this, we exploit the fact that the particle guiding center Hamiltonian $H(\theta, P_{\theta}; P_{\zeta}, \mu)$, given by Eq. \eqref{GC H 1}, is integrable. Consequently, its phase space, represented by the poloidal plane $(\theta, P_{\theta})$, is filled with closed periodic curves $P_{\theta}(\theta; E, \mu, P_{\zeta})$ (see Fig. \ref{fig:Fig510}). Each phase-space orbit is characterized by a specific set of constants of motion $(E, P_{\zeta}, \mu)$, where $E$ is the energy of the system and is equal to the Hamiltonian. Two distinct types of closed orbits exist: \textit{Trapped orbits}, confined within a limited range of the poloidal angle $\theta$, have an elliptical shape encircling the elliptic fixed point (the $o$-point) of the Hamiltonian \eqref{GC H 1}. \textit{Passing orbits}, which periodically traverse the entire range of $\theta$ (from $\theta = -\pi$ to $\theta = \pi$), pass through the hyperbolic fixed points of Eq. \eqref{GC H 1}.

Obtaining the orbits in the form $P_{\theta}(\theta; E, \mu, P_{\zeta})$ requires solving Eq. \eqref{GC H 1} with respect to $P_{\theta}$. However, such a solution is not feasible analytically for experimental equilibrium magnetic fields, which are typically provided as numerical data, such as the equilibria used in this work. Even in the case of the large aspect ratio equilibrium considered in our previous work (Ref. \cite{Antonenas2024}), we had to approximate the magnetic field on a magnetic surface of reference $\psi_0$ to solve Eq. \eqref{GC H 1} with respect to $P_{\theta}$.

In this work, we use a contouring algorithm to compute the curves $P_{\theta}(\theta; E, \mu, P_{\zeta})$. For each set of $(E, \mu, P_{\zeta})$, we generate a contour curve in the $(\theta, P_{\theta})$ plane. This process requires a grid in the $(\theta, P_{\theta})$ plane and the corresponding values of $H$ on this grid. However, the numerical equilibrium data are provided in terms of the Boozer magnetic coordinates, $\psi$ (toroidal magnetic flux) and $\theta$ (poloidal angle), such that $B = B(\theta, \psi)$, with the poloidal and toroidal currents $g(\psi)$ and $I(\psi)$, and the safety factor $q(\psi)$. 

To express $H$ in terms of $(\theta, P_{\theta})$, we use Eq. \eqref{Eq: Ptheta with psi 1} obtained by the canonical coordinate definitions in Eqs. \eqref{canon moments}. This allows us to calculate $P_{\theta}$. In this equation $\psi_p(\psi)$ is determined from $q(\psi)$ as:
\begin{equation} \label{eq:psi_p_psi}
\psi_p(\psi) = \int_{0}^{\psi} \frac{1}{q(\psi')} d\psi'.
\end{equation}
The contours for specific $(E, \mu, P_{\zeta})$ values yield the orbits $P_{\theta}(\theta; E, \mu, P_{\zeta})$ which, depending on $(E, \mu, P_{\zeta})$, can be trapped, co-passing, or counter-passing (Fig. \ref{fig:Fig510}).

Subsequently, the action $J_{\theta}$ (Eq. \eqref{gen Jtheta}) is computed for each distinct curve in the $(\theta, P_{\theta})$ plane using the Shoelace formula, which calculates the area of a polygon based on the coordinates of its vertices. The formula is expressed as \cite{Lee2017}:
\begin{equation} \label{eq: Shoelace formula}
\text{Area} = \frac{1}{2}  \sum_i \left(y_i x_{i+1} - x_i y_{i+1}  \right),
\end{equation}
where $(x_i, y_i)$ are the vertices of the polygon. The Shoelace formula, as given in Eq. \eqref{eq: Shoelace formula}, provides the signed area enclosed by the curve. While trapped curves are closed due to the system's dynamics, passing curves are not. Thus, for Eq. \eqref{eq: Shoelace formula} to be applied, we technically close them with the $P_{\theta} = 0$ axis, as is dictated by the action integral, Eq. \eqref{gen Jtheta}, see Fig. \ref{fig:Fig510}. Any alternative numerical integration or area computation algorithm could also be used.

Following our analysis in Sec. \ref{Analytical}, for particular values of energy $(E, \mu)$, $q_{kin}$ is given by the relation
\begin{equation} \label{eq:qkin}
    q_{\text{kin}} = \frac{\hat{\omega}_{\zeta}}{\hat{\omega}_{\theta}} = -\frac{\partial J_{\theta}(E, \mu, P_{\zeta})}{\partial J_{\zeta}}
\end{equation}

\subsection{Calculation of kinetic $q$-factor}
\begin{figure}
    \centering
    \includegraphics[width=0.69\linewidth]{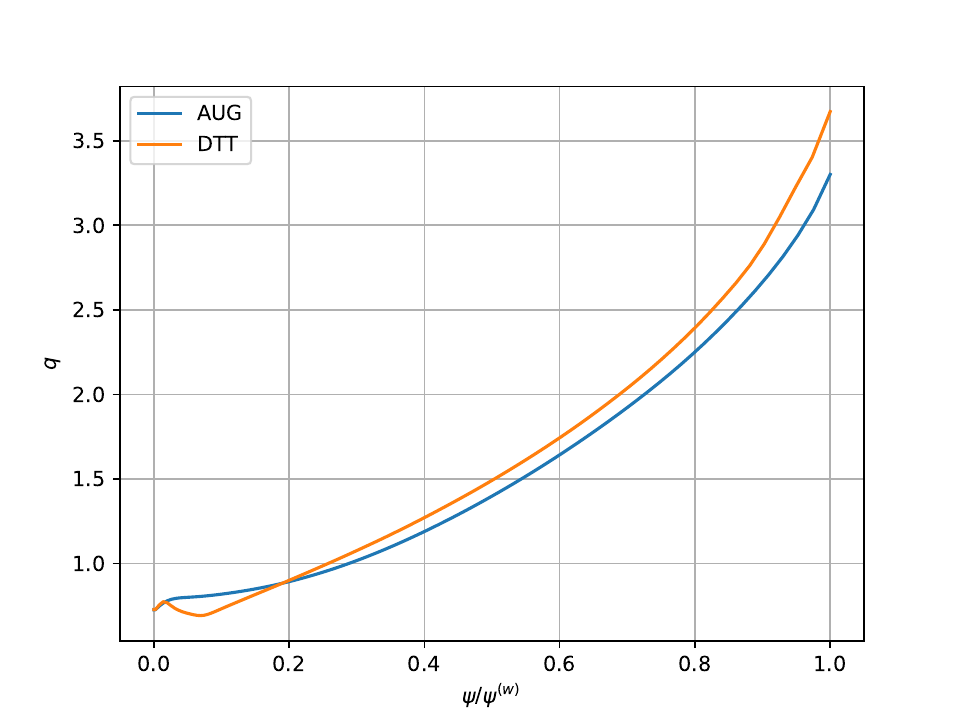}
    \caption[$q$ factor of DTT and AUG equilibria]{The safety factor of the two equilibria we study here (DTT and AUG). The profiles are largely similar across most of the plasma volume, except for a shallow minimum away from the magnetic axis, which is present only in the DTT equilibrium. This feature indicates that the DTT equilibrium is in a reverse shear configuration.}
    \label{fig:Fig511}
\end{figure}

The methodology described in Sec. \ref{Sec: geomtric method} is applied to two equilibria: one from the ASDEX Upgrade (AUG), with parameters $B_0 = 1.45\,\SI{}{\tesla}$ and $R_0 = 1.66\,\SI{}{\meter}$, and one from the Divertor Tokamak Test facility (DTT), with parameters $B_0 = 5.96\,\SI{}{\tesla}$ and $R_0 = 2.26\,\SI{}{\meter}$. Both have comparable safety factor profiles, except that the DTT exhibits a minimum in its $q$ profile, as shown in Fig. \ref{fig:Fig511}. 

The particle that we examine is deuterium (mass number: $A = 2$, atomic number: $Z = 1$). In both cases we consider $\mu = 2.66\cdot10^{-05}$ in normalized units which taking into account the normalizations (tables \ref{tab:Tab41} and \ref{tab:Tab42}) corresponds to $\mu B_0 = 230.8 \SI{}{\kilo\electronvolt}$ for DTT equilibrium and $\mu B_0 = 7.37 \SI{}{\kilo\electronvolt}$ for AUG equilibrium deuterium energy. For demonstration purposes, we will focus on two selected energy levels, one at $E/\mu B_0=1.73$, for passing orbits and one at $E/\mu B_0 = 1.1$ for trapped orbits,
in both AUG and DTT. The selected energy level for passing orbits corresponds to deuterium of approximately $\SI{25.6}{keV}$ in the case of AUG and $\SI{800}{keV}$ in the case of DTT. The energy of
the trapped orbits is approximately $\SI{16.2}{keV}$ for AUG and $\SI{500}{keV}$ for DTT.

\begin{figure}[h!]
    \centering
    \includegraphics[width=0.49\textwidth,keepaspectratio]{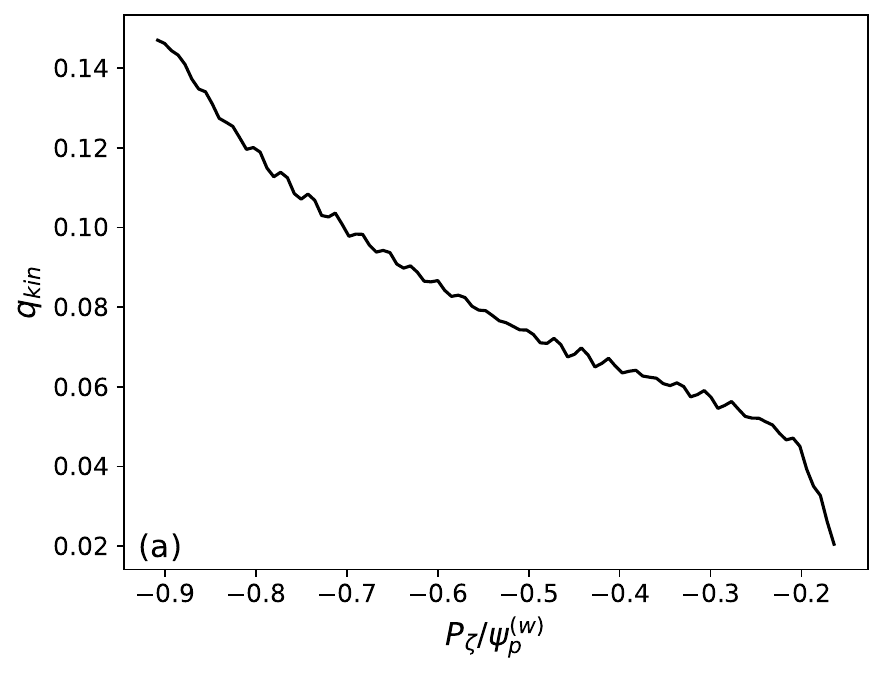}
    \includegraphics[width=0.49\textwidth,keepaspectratio]{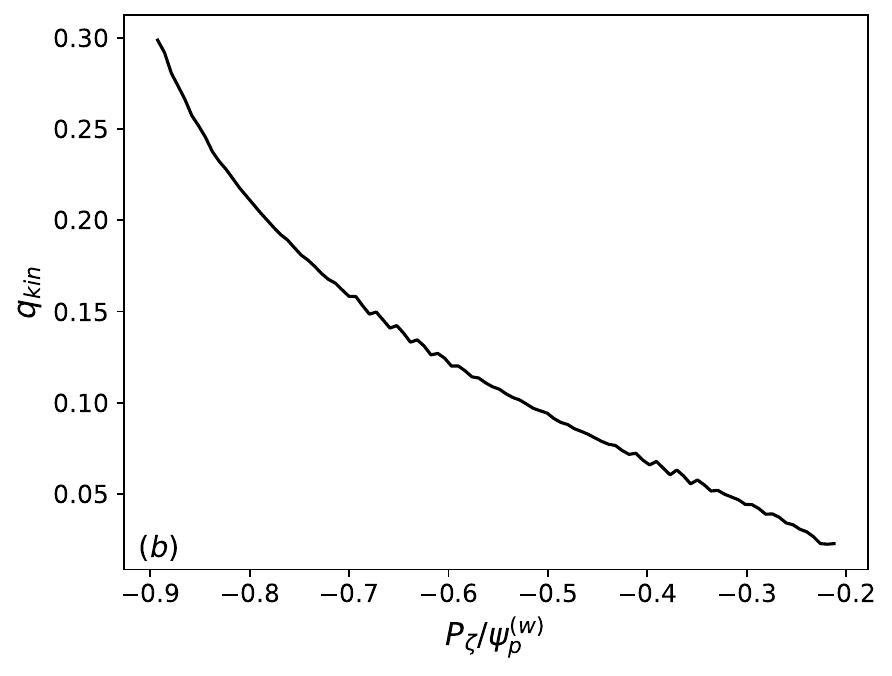}
    \includegraphics[width=0.49\textwidth,keepaspectratio]{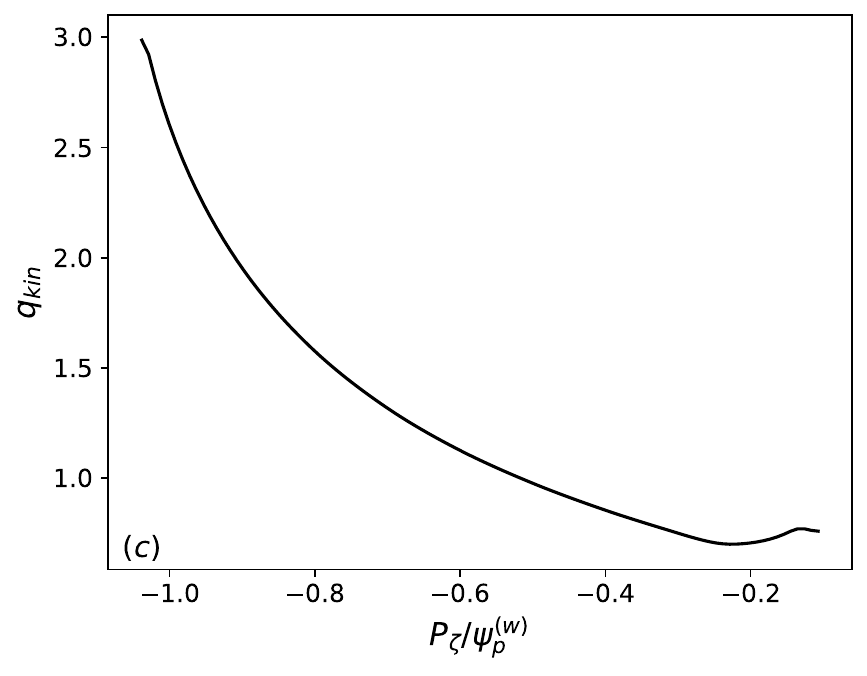}
    \includegraphics[width=0.49\textwidth,keepaspectratio]{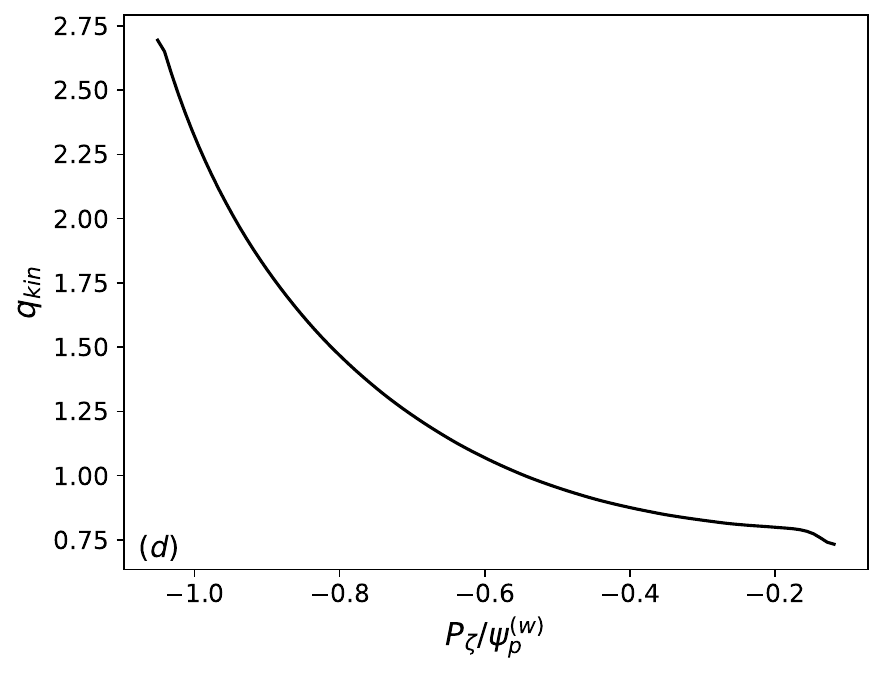}
    \includegraphics[width=0.49\textwidth,keepaspectratio]{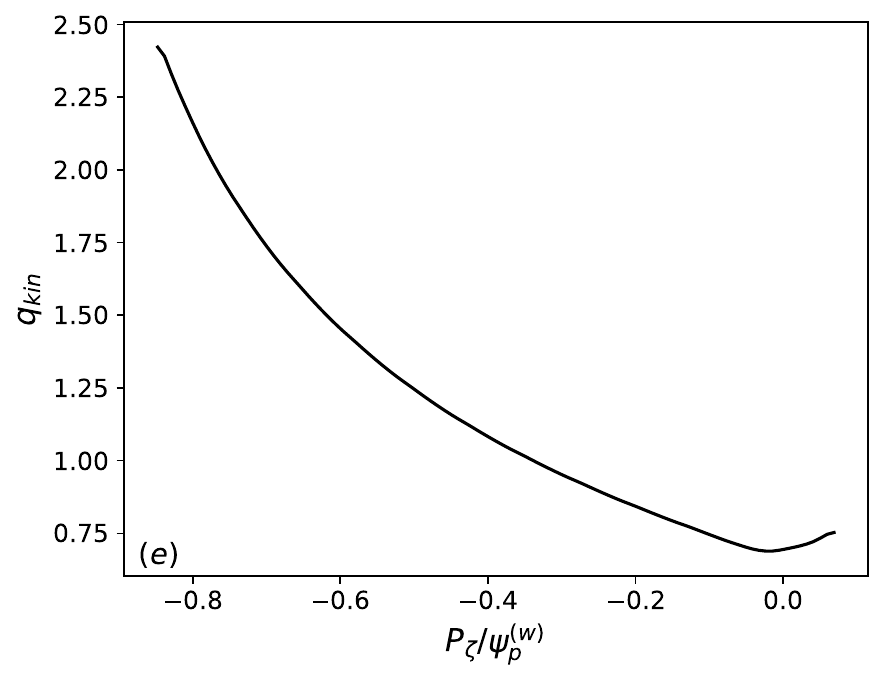}
    \includegraphics[width=0.49\textwidth,keepaspectratio]{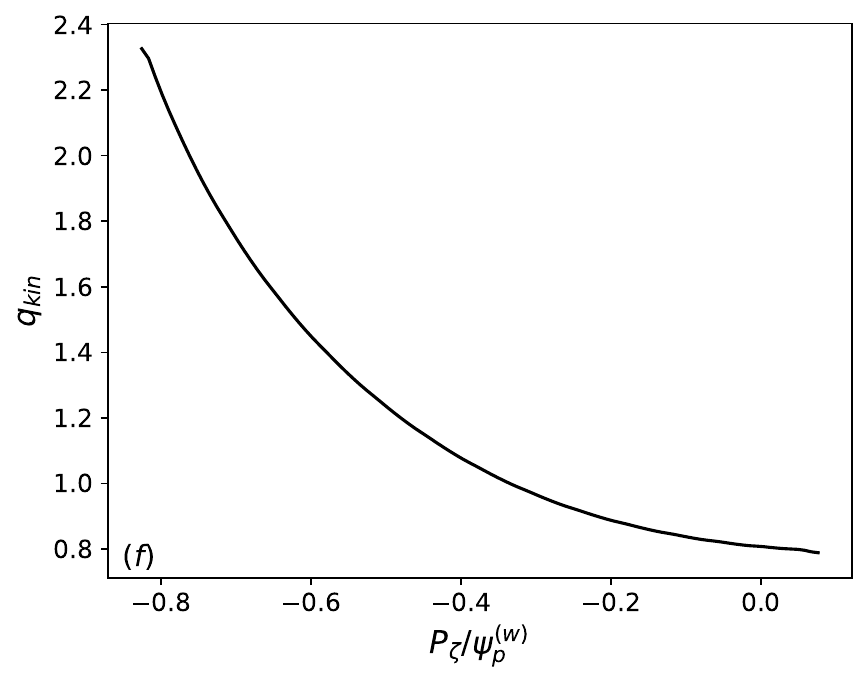}
    \caption[$q_{kin}$ as a function of the canonical momentum $P_{\zeta}$ for DTT and AUG equilibria]{$q_{kin}$ as a function of the canonical momentum $P_{\zeta}$ for a constant normalized $\mu = 2.66 \times 10^{-5}$. For trapped particles (panels (a),(b)) $E/\mu = 1.1$, while for passing orbits (panels (c), (d) for co-passing and panels (e),(f) for counter-passing particles) $E/\mu = 1.73$. The left column corresponds to the DTT equilibrium, where $\mu B_0 = 230.8~\SI{}{\kilo\electronvolt}$, and the right column corresponds to the AUG equilibrium, where $\mu B_0 = 7.37~\SI{}{\kilo\electronvolt}$. Both energy values refer to deuterium particles.}
    \label{fig:Fig512}
\end{figure}

The results for the calculation of $q_{kin}$ using the semi-analytical method (Sec.~\ref{Sec: geomtric method}), as a function of the canonical momentum $P_{\zeta}$, are presented in Fig. \ref{fig:Fig512} for trapped, counter-passing, and co-passing particles from top to bottom. It is important to note that the semi-analytical method does not involve the DC approximation used in Fig.~\ref{fig:Fig53}. Consequently, the semi-analytical calculation of $q_{kin}$ is exact and coincides with the $q_{kin}$ obtained numerically by solving the equations of motion for a single poloidal transit. For this reason, the numerical results for $q_{kin}$ are not shown in Fig. \ref{fig:Fig512}.  

In the $q_{kin}$ profiles of the DTT equilibrium for passing orbits (Fig. \ref{fig:Fig512}, panels (c) and (e)), we observe a local minimum, which arises due to the local minimum of the magnetic $q$-factor in the DTT equilibrium (Fig. \ref{fig:Fig511}). As we will see in the following discussion, this minimum corresponds to a transport barrier, as previously observed in the study of the LAR equilibrium (Figs. \ref{fig:Fig53}, \ref{fig:Fig57}).

\begin{figure}[h!]
    \centering
    \includegraphics[width=1.0\textwidth,keepaspectratio]{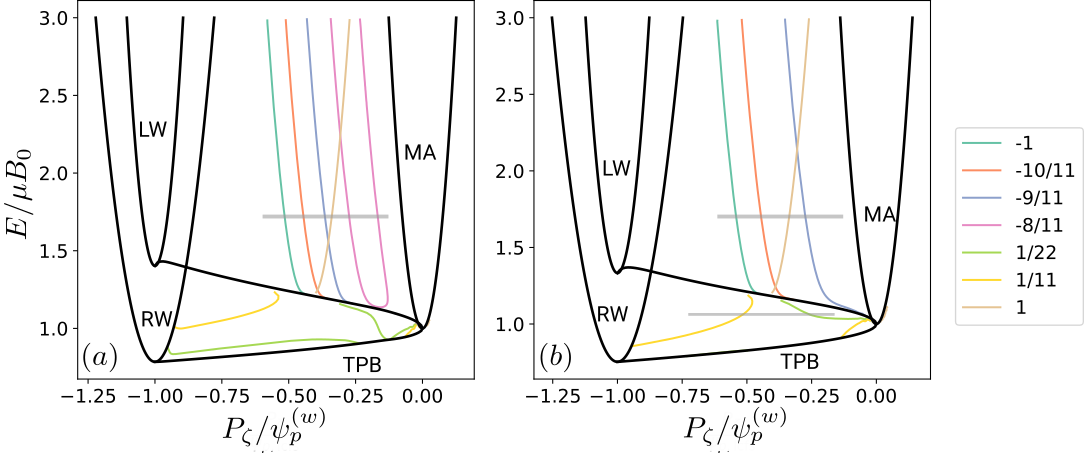}
    \caption[Resonances curves in COM space for AUG and DTT equilibria]{Analogous to Fig.~\ref{fig:Fig54}, this figure presents constant-$\mu = 2.66 \times 10^{-5}$ slices (plane cuts) of the three-dimensional COM space $(E, \mu, P_{\zeta})$. Panel (a) corresponds to the DTT equilibrium, while panel (b) corresponds to the AUG equilibrium. A constant pitch line, $\Lambda = E / (\mu B_{0})$, intersects a passing particle resonance twice only in the case of DTT, whereas twice-intersected trapped particle resonances are located very close to the magnetic axis in both DTT and AUG (resonance $1/11$).}

    \label{fig:Fig513}
\end{figure}

Analogous to Fig. \ref{fig:Fig54}, the resonant diagrams in the COM space $(E, \mu, P_\zeta)$ for $\mu = 2.66 \times 10^{-5}$ are presented in Fig. \ref{fig:Fig513} for both the DTT (panel (a)) and AUG (panel (b)) equilibria. Notable differences between AUG and DTT include the displacement of the $1/11$ and $1/22$ levels in the trapped domain and the appearance of two $-8/11$ branches (violet lines), which are present only in DTT. This reflects the reversed shear configuration of the DTT equilibrium (Fig.~\ref{fig:Fig511}).

\subsection{Resonant response to non-axisymmetric time-independent perturbations} \label{Sec: responce in realistic}

\begin{figure}[h!]
    \centering
    \includegraphics[width=0.49\textwidth,keepaspectratio]{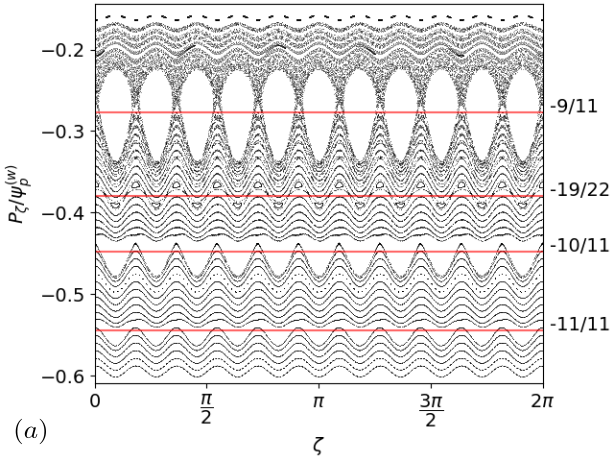}
    \includegraphics[width=0.49\textwidth,keepaspectratio]{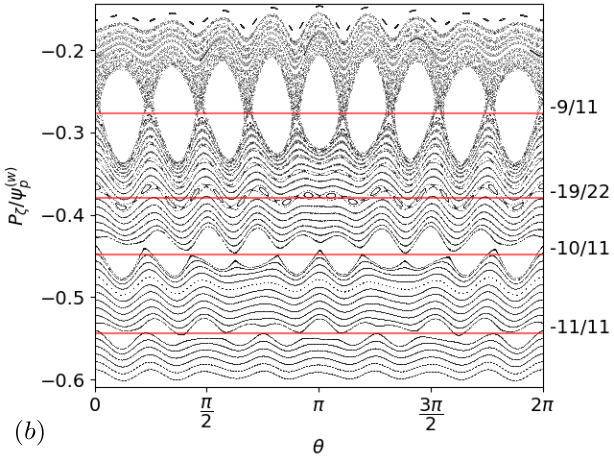}
    \includegraphics[width=0.49\textwidth,keepaspectratio]{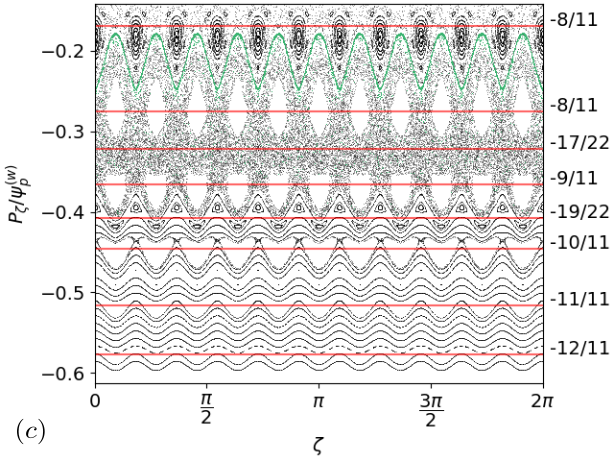}
    \includegraphics[width=0.49\textwidth,keepaspectratio]{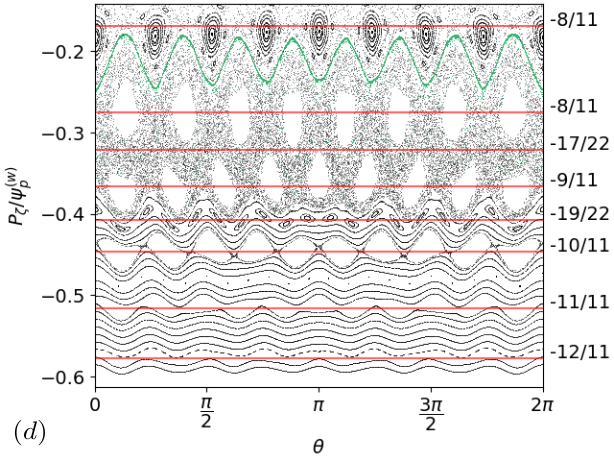}
    \caption[Poincaré plots for counter-passing orbits in the AUG equilibrium and DTT equilibria.]{Poincaré plots for counter-passing orbits in the AUG equilibrium (panels (a) and (b)) and DTT equilibrium (panels (c) and (d)) under the influence of a single static perturbative mode with toroidal mode number $n=11$ and poloidal mode number $m=-8$. Panels (a) and (c) correspond to a constant $\theta$ surface, while panels (b) and (d) correspond to a constant $\zeta$ surface of section. The red horizontal lines indicate the predicted locations of the corresponding resonant island chains, in agreement with either Fig. \ref{fig:Fig512}(c) (DTT) and Fig. \ref{fig:Fig512}(d) (AUG) or Fig. \ref{fig:Fig513}(a) (DTT) and Fig. \ref{fig:Fig513}(b) (AUG). The green orbit represents the transport barrier observed in the DTT equilibrium, whose location corresponds to the local minimum of $q_{kin}$ in Fig. \ref{fig:Fig512}(c).}
    \label{fig:Fig514}
\end{figure}

\begin{figure}[h!]
    \centering
    \includegraphics[width=0.49\textwidth,keepaspectratio]{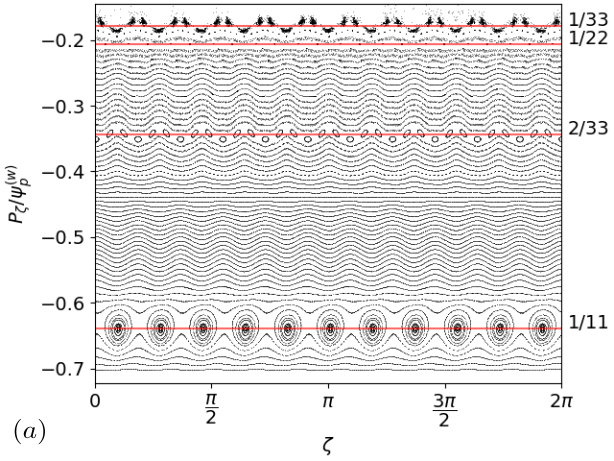}
    \includegraphics[width=0.49\textwidth,keepaspectratio]{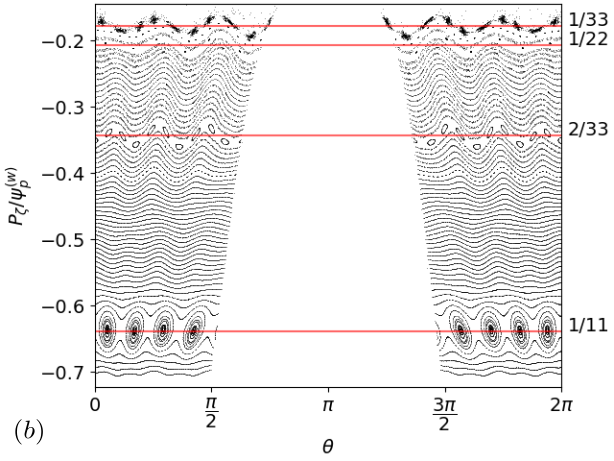}
    \caption[Poincaré plots for trapped orbits in the DTT equilibrium.]{Poincaré plots for trapped orbits in the DTT equilibrium under the influence of a single static perturbative mode with toroidal mode number $n=11$ and poloidal mode number $m=-8$. Panel (a) corresponds to a constant $\theta$ surface, while panel (b) corresponds to a constant $\zeta$ surface of section. The red horizontal lines indicate the predicted locations of the corresponding resonant island chains, in agreement with Fig.~\ref{fig:Fig512}(a) or Fig.~\ref{fig:Fig513}(a).}
    \label{fig:Fig515}
\end{figure}

For demonstration purposes, we construct Poincaré diagrams (Figs. \ref{fig:Fig514} and \ref{fig:Fig515}) at constant energy levels corresponding to those used in the calculation of $q_{kin}$ in Fig. \ref{fig:Fig512}, namely $E/\mu = 1.1$ for trapped particles and $E/\mu = 1.73$ for passing particles, with $\mu = 2.66 \times 10^{-5}$. For a passing deuterium particle, these energy levels correspond to $25.6~\SI{}{\kilo\electronvolt}$ in AUG and $800~\SI{}{\kilo\electronvolt}$ in DTT, while for a trapped deuterium particle, they correspond to $16.2~\SI{}{\kilo\electronvolt}$ in AUG and $500~\SI{}{\kilo\electronvolt}$ in DTT. The energy levels of the Poincaré diagrams are also indicated in Fig. \ref{fig:Fig513} by the gray horizontal line segments. To generate the Poincaré diagrams, we consider a static (time-independent) perturbation of the form given in Eq. \eqref{Eq: flute modes}, with single toroidal and poloidal mode numbers $n = 11$ and $m = -8$, expressed as $\alpha = \epsilon\sqrt{\psi} \cos(11\zeta - 8\theta)$. The radial dependence $\sim \psi^{1/2}$ is introduced to counteract the $1/r$ factor introduced by the $\nabla \times$ operator, ensuring a uniform $\delta B / B$ perturbation profile.  In Fig. \ref{fig:Fig514}, we present the Poincaré diagrams for counter-passing orbits at the same energy level, $E/\mu = 1.73$, and $\mu = 2.66 \times 10^{-5}$, for the AUG (panels (a) and (b)) and DTT (panels (c) and (d)) equilibria. Panels (a) and (c) correspond to a constant $\theta$ surface, while panels (b) and (d) correspond to a constant $\zeta$ surface of section. In Fig. \ref{fig:Fig515}, we present the Poincaré diagram for the trapped particle energy level $E/\mu = 1.1$ with $\mu = 2.66 \times 10^{-5}$ in the DTT equilibrium. Panel (a) corresponds to a constant $\theta$ surface, while panel (b) corresponds to a constant $\zeta$ surface.

Poincaré plots (Figs. \ref{fig:Fig514} and \ref{fig:Fig515}) confirm that the resonance island chains induced by the perturbation are located around the unperturbed resonant orbits, which are labeled by the constants of motion $(E, \mu, P_{\zeta})$ at which $q_{kin}$ takes rational values. These locations, as discussed in the case of the LAR equilibrium, can be predicted from the resonance diagrams in COM space by identifying the intersection points of the Poincaré plot energy levels (gray horizontal lines in Fig. \ref{fig:Fig513}) with the respective resonance lines. From the resonance diagrams in COM space (Fig. \ref{fig:Fig513}), the location of any resonance can be determined, regardless of whether it is a first-order, second-order, or higher-order resonance. The order of the resonance depends on the amplitude of the respective mode $\hat{a}_{m^{\prime},n^{\prime}}$ (see Eq.~\eqref{alpha hat}). In Figs. \ref{fig:Fig514} and \ref{fig:Fig515}, we see that we predict the locations (marked with red horizontal lines)  both of first and second order resonances. In the case of the DTT equilibrium, we also predict the location of the transport barrier, marked in green in Fig. \ref{fig:Fig514}(c) and (d), from the location of the minimum in $q_{kin}$ in Fig. \ref{fig:Fig512}(c) and (e), respectively.  

Similar to the LAR case (Sec. \ref{Resonse}), we observe that despite considering a single perturbative mode with $(m,n) = (-8,11)$, additional first and second order resonances appear. To fully understand mode-particle resonance interactions, even a single perturbative mode must be analyzed in action-angle variables (see Eq. \eqref{alpha hat}). Notably, in the trapped case (Fig.~\ref{fig:Fig515}), the number of resonance islands in the $\zeta = 0$ Poincaré surface does not correspond to the denominator of $q_{kin}$, as observed in the counter-passing cases (Fig.~\ref{fig:Fig514}). This behavior is attributed to the characteristics of unperturbed trapped orbits, which can leave multiple traces on the Poincaré surface within a single toroidal drift rotation due to their bounce motion. To determine the exact number of islands in such cases, the procedure described in Sec. \ref{Sec: Number of islands} must be followed.

\section{Particle energy and momentum transport on the Arnold web of the GC motion} \label{Sec: GC Arnold diffusion}

As discussed in the previous sections, the Orbital Spectrum (OS) of the Guiding Center (GC) motion in an axisymmetric toroidal equilibrium determines the resonance conditions for strong particle interactions with non-axisymmetric perturbations, leading to significant energy and momentum transport. The dependence of the OS on the kinetic characteristics of the GC motion has been analytically calculated for the case of a Large Aspect Ratio (LAR) equilibrium magnetic field (Ref. \cite{Antonenas2021}, \cite{Antonenas2024}), allowing for the pinpointing of the exact resonance locations in the three-dimensional Constants Of the Motion (COM) space, for the case of stationary non-axisymmetric perturbative modes (Secs. \ref{Action-Angle Transformation}-\ref{Sec: Number of islands}). For the case of time-varying perturbations, an additional degree of freedom is introduced to the GC Hamiltonian system, leading to significant excursions along the resonance curves. In this case, KAM surfaces do not isolate different resonances, and extended transport can take place even without resonance overlap in the phase space. Moreover, in contrast to the case of stationary perturbations resonance curves intersect in the COM space forming the so-called Arnold web (Sec. \ref{Sec: Arnold Diffusion}).

In this section, using the analytically calculated orbital frequencies for the LAR equilibrium (Sec. \ref{Analytical}), we construct the Arnold web in the COM space of GC motion, along which diffusive particle energy and momentum transport occurs. This calculation provides an a priori determination of the exact kinetic characteristics of particles that strongly interact with time-dependent non-axisymmetric modes. Numerical simulations confirm the Arnold diffusion mechanism (Sec. \ref{Sec: Arnold Diffusion}) and illustrate characteristic cases in the COM space.

Note that when the perturbation \eqref{alpha} is time-dependent, a parallel electric field along the equilibrium magnetic field $\bm{B}$ arises as $\bm{E} = \bm{B} \partial\alpha / \partial t$. The ideal MHD modes considered in this work require a zero parallel electric field (as dictated by Eq. \eqref{Ideal Ohm's law}). To eliminate the parallel component, an electric potential $\Phi$ is introduced in the GC Hamiltonian as a perturbation, expressed as 
\begin{equation}\label{Eq: Phi}
    \Phi(\psi,\theta,\zeta,t)=\sum_{m,n} \Phi_{m,n}(\psi)e^{i(n\zeta - m\theta - \omega t)}
\end{equation}
Ensuring the parallel electric field vanishes imposes the condition:
\begin{equation} \label{Eq: Phi_m,n}
    \left[g(\psi)q(\psi) + I(\psi)\right] k \omega \alpha_{m,n}(\psi) = \left[n q(\psi) - m\right] \Phi_{m,n}(\psi).
\end{equation}
When numerically solving the GC equations of motion in this section, we include both $\alpha$ and $\Phi$, with $\alpha$ given by Eq. \eqref{alpha}, $\Phi$ by Eq. \eqref{Eq: Phi}, and $\alpha$ and $\Phi$ related through Eq. \eqref{Eq: Phi_m,n}.

The perturbed GC Hamiltonian \eqref{Per GC H 1} can be expressed as:
\begin{align}
    H &= H_0(\theta, P_{\theta}, P_{\zeta}; \mu) + \epsilon H_1(\theta, \zeta, P_{\theta}, P_{\zeta}, t) + \epsilon^2 H_2(\theta, \zeta, P_{\theta}, P_{\zeta}, t) \label{Eq: GC H with time}\\
    &= \frac{\left[P_{\zeta} + \psi_{p}(P_{\zeta}, P_{\theta}) \right]^2}{2 g(P_{\zeta}, P_{\theta})} B^2(P_{\zeta}, P_{\theta}, \theta) + \mu B(P_{\zeta}, P_{\theta}, \theta) \\
    &\quad + 2\alpha\left(P_{\zeta} + \psi_{p}(P_{\zeta}, P_{\theta})\right)\frac{ B^2(P_{\zeta}, P_{\theta}, \theta)}{2 g(P_{\zeta}, P_{\theta})} + \Phi(\theta, \zeta, P_{\theta}, P_{\zeta}, t)  + \alpha^2 \frac{ B^2(P_{\zeta}, P_{\theta}, \theta)}{2 g(P_{\zeta}, P_{\theta})},
\end{align}
where the electric potential $\Phi$ is included. Here, $H_0$ is the unperturbed GC Hamiltonian \eqref{GC H 1}, for which we perform the action-angle (AA) transformation (Sec. \ref{Action-Angle Transformation}). The perturbation terms are defined as:
\begin{align}
    \epsilon H_1 &= 2\alpha\left(P_{\zeta} + \psi_{p}(P_{\zeta}, P_{\theta})\right) \frac{ B^2(P_{\zeta}, P_{\theta}, \theta)}{2 g(P_{\zeta}, P_{\theta})} + \Phi(\theta, \zeta, P_{\theta}, P_{\zeta}, t) , \label{Eq: H1}  \\
    \epsilon^2 H_2 &= \alpha^2 \frac{ B^2(P_{\zeta}, P_{\theta}, \theta)}{2 g(P_{\zeta}, P_{\theta})}.\label{Eq: H2}
\end{align}
The parameter $\epsilon$ represents the order of the coefficients $a_{m,n}$ and $\Phi_{m,n}$.   
    
According to the discussion in Sec. \ref{Sec: Number of islands}, $n = n'$ thus the perturbation terms in action-angle (AA) variables, in analogy to Eq. \eqref{alpha hat} will be expressed as:
\begin{equation} \label{perturbation_in_AA}
    \hat{H}_{1,2}(\hat{\theta},\hat{\zeta},J_{\theta},J_{\zeta}, J_{\xi})= \sum_{m',n}\hat{H}_{1,2(m',n)}(J_{\theta},J_{\zeta}, J_{\xi})e^{i(n\hat{\zeta}-m'\hat{\theta} - \omega t)},
\end{equation}
which leads to the resonance condition in the time-dependent case:
\begin{equation}\label{Eq:time_dependent_res_condition}
    n\hat{\omega}_{\zeta}(E,\mu,P_{\zeta}) - m'\hat{\omega}_{\theta}(E,\mu,P_{\zeta}) - \omega= 0.
\end{equation}
The frequencies $\hat{\omega}_{\theta}(E,\mu,P_{\zeta})$ and $\hat{\omega}_{\zeta}(E,\mu,P_{\zeta})$, defined in Eqs. \eqref{gen hat omega theta} and \eqref{gen hat omega zeta}, can be computed analytically for LAR equilibrium (Sec. \ref{Analytical}) or semi-analytically using the geometric method for realistic equilibria (Sec. \ref{Sec: geomtric method}).

\subsection{Single time-dependent perturbative mode}

When the perturbed GC Hamiltonian is time-dependent, the total Hamiltonian $H$ is not a constant of motion. Consequently, we cannot construct a Poincaré surface of section on a constant energy surface $H=E$, as we did in Secs. \ref{Resonse} and \ref{Sec: responce in realistic}. However, when the perturbation \eqref{alpha} consists of a single toroidal mode number $n$ and a single mode frequency $\omega$, even if multiple $m'$ modes arise, the combination $n\zeta - \omega t$ enables a canonical transformation that reveals a conserved quantity of the system. In the following, we construct Poincaré surfaces of section on this invariant surface. The generating function for the canonical transformation is \cite{White2012, White2018, Meng2018}.
\begin{equation}
    F = (n\zeta - \omega t)P'_{\phi} + P^{\prime}_{\theta}\theta,
\end{equation}
which reduces the system's degrees of freedom by one. In particular, applying this generating function, the new canonical variables in terms of the old ones are:
\begin{equation}\label{trasf_phi_1}
    \phi^{\prime} = n\zeta - \omega t, \quad \theta^{\prime} = \theta,
\end{equation}
\begin{equation} \label{trasf_phi_2}
      P_{\zeta} = nP_{\phi}^{\prime}, \qquad P_{\theta} = P_{\theta}^{\prime}.
\end{equation}
The new Hamiltonian, given by $H^{\prime} = H(\theta, \zeta, P_{\theta}, P_{\zeta}, t; \mu) + \partial F / \partial t$, is obtained by substituting the original variables $(\theta, \zeta, P_{\theta}, P_{\zeta}, t)$ with the transformed ones from Eqs. \eqref{trasf_phi_1} and \eqref{trasf_phi_2}. The resulting expression for the new Hamiltonian is:
\begin{align} \label{new_H}
    H^{\prime}(\theta, P_{\theta}, \phi, nP_{\phi}; \mu) &= H(\theta, P_{\theta}, \phi, P_{\phi}; \mu) - \omega P_{\phi}\\
    &= H_0(\theta, P_{\theta}, nP_{\phi}; \mu) - \omega P_{\phi} \\
    &+ \epsilon H_1(\theta,\phi,P_{\theta},nP_{\phi}) + \epsilon^2 H_2(\theta, \phi, P_{\theta}, nP_{\phi})
\end{align}
with $H_0$ given by Eq. \eqref{GC H 1}, $H_1$ and $H_2$ by Eqs. \eqref{Eq: H1} and \eqref{Eq: H2} respectively. For simplicity, the prime notation is omitted. Notably, the new Hamiltonian exhibits no explicit time dependence, making it a conserved quantity. The new unperturbed Hamiltonian is given by:
\begin{equation}\label{Eq:E_prime}
    E^{\prime} = E(\theta, P_{\theta}; nP_{\phi}, \mu) - \omega P_{\phi} = E(\theta, P_{\theta}, P_{\zeta}; \mu) - \frac{\omega}{n}P_{\zeta},
\end{equation}
where with $E$ is the unperturbed GC Hamiltonian $H_0$, and $E'$ is the new unperturbed Hamiltonian $H_0'$. In the unperturbed system, $P_{\phi}$ remains a constant of motion, as $E'$ is $\phi$ independent. Additionally, since the new unperturbed system is time-independent, $E^{\prime}$ is also a conserved quantity. As a result from Eq. \eqref{Eq:E_prime} we see that $E$ is conserved in the new unperturbed system.

\begin{figure}[h!]
    \centering
    \includegraphics[width=0.49\textwidth,keepaspectratio]{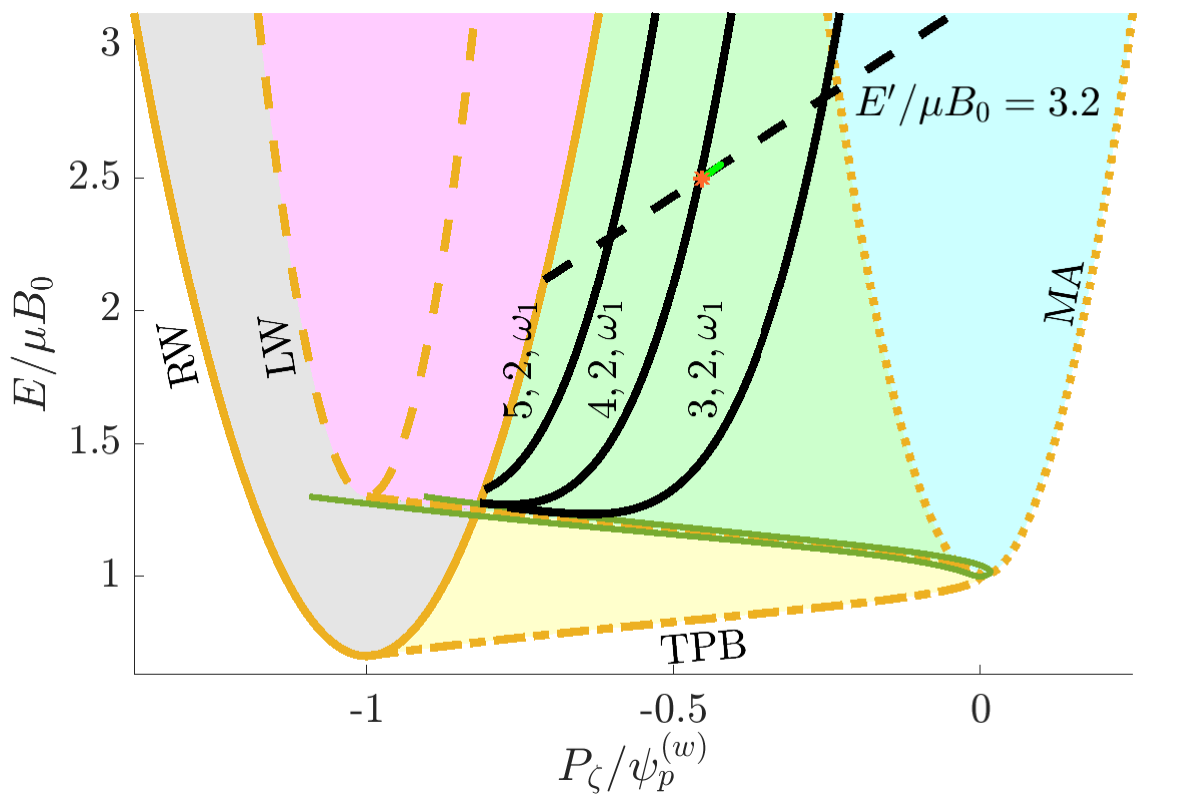}
    \includegraphics[width=0.49\textwidth,keepaspectratio]{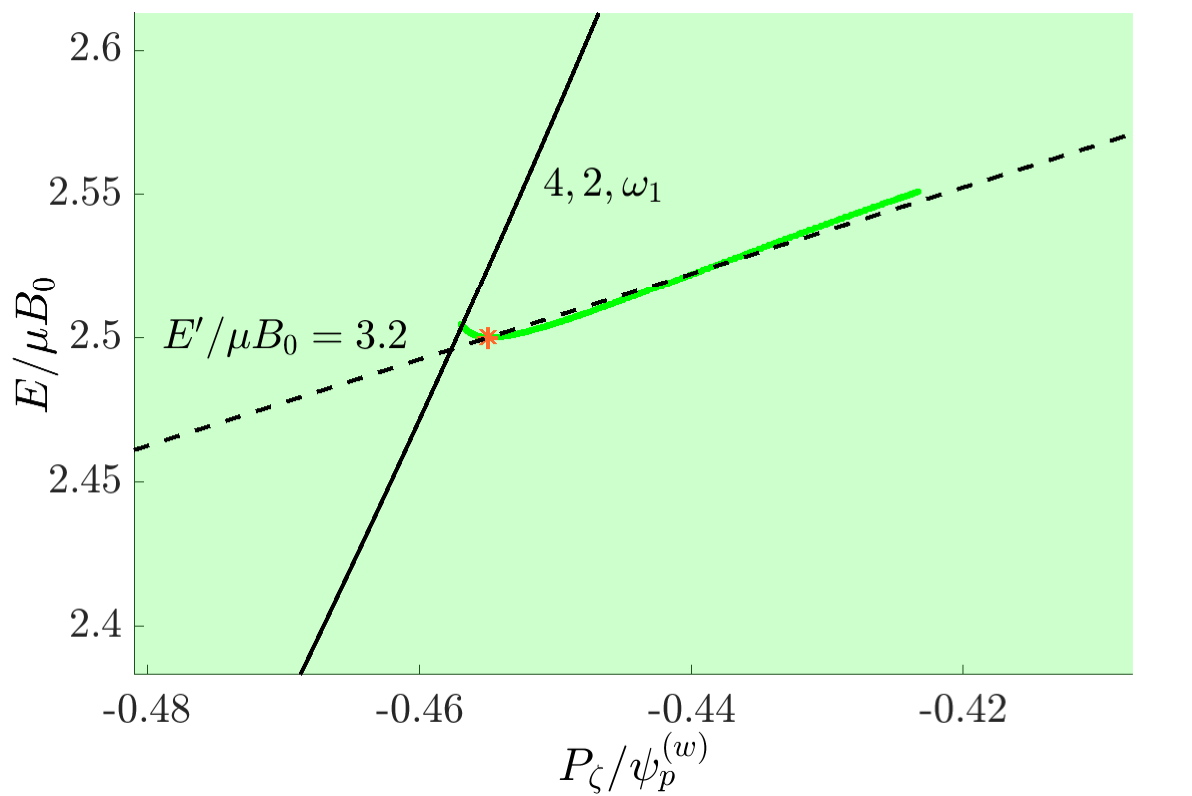}
    \caption[Time-dependent resonances curves in COM space in LAR equilibrium.]{Constant-$\mu$ slice (plane cut) of the three-dimensional COM space $(E, \mu, P_{\zeta})$ for Case \#1. Black lines correspond to resonances co-passing particles in the time-dependent case, fulfilling the resonance condition \eqref{Eq:time_dependent_res_condition}. In the time-dependent case with a single $(n, \omega)$ mode (with any $m$ or $m'$ that arises), the constant unperturbed energy of the system is the $E'$, Eq. \eqref{Eq:E_prime} (or $H'$ when perturbation is included, Eq. \eqref{new_H}). Here we depict with a black dashed line the $E'/\mu B_0 = 3.2$. This line in $(P_{\zeta}/\psi_p^{(w)}, E/\mu B_0)$ plane is obtained by setting in Eq. \eqref{Eq:E_prime} $E'/\mu B_0 = 3.2$. The intersection of this line with the resonance lines gives the location of the resonance island chains both in $P_{\zeta}/\psi_p^{(w)}$ (panels (a) and (b) in Fig. \ref{fig:Fig572}) and in $E/\mu B_0$ (panels (c) and (d) in Fig. \ref{fig:Fig572}). Right panel is an enlargement of the respective region in left panel, showing with the green line the variation with time $P_{\zeta}(t)/\psi_p^{(w)}, E(t)/\mu B_0$ of a particular orbit (green orbit in Fig. \ref{fig:Fig572}) of the perturbed system with an initial value in $(E, \mu, P_{\zeta})$ noted with a star. The initial value is on the $E'$ line, which is the total energy of the orbit (assuming that at the initial time the perturbation terms $H_1$ and $H_2$ are zero) and remains constant during the orbit's time evolution. As expected, the $E, P_{\zeta}$ of this orbit vary around the constant energy line $E'$.}
    \label{fig:Fig571}
\end{figure}

Since $n = n'$, the single $n$ mode will not produce multiply $n'$ in the perturbation when it is written in AA variables, thus, the single $n$ time-dependent mode-particle resonance location in terms of the COMs $(E,\mu,P_{\zeta})$ is determined by Eq. \eqref{Eq:time_dependent_res_condition}. However, the constant energy surface around which the perturbed motion occurs is not given by $E$, like in the case of the time-independent perturbations discussed in Secs. \ref{Resonse} and \ref{Sec: responce in realistic}, but rather by the conserved quantity $E'$ defined in Eq. \eqref{Eq:E_prime} (black dash line in Fig. \ref{fig:Fig571}).

For demonstration purposes, all the figures we present in this section correspond to the LAR equilibrium and use the same parameters as in \textbf{Case \#1}, discussed in Sec. \ref{Analytical}, i.e., low-energy particles with $\mu B_{0} = \SI{2}{\kilo\electronvolt}$ ($\mu = 7.7\times 10^{-6}$ in normalized units) in the $q_1$ profile. 

In Fig. \ref{fig:Fig571}, we show resonances for co-passing orbits in COM space, as determined by the resonance condition Eq. \eqref{Eq:time_dependent_res_condition}, for $m' = 3,4,5$, $n' = n = 2$, and $\omega = 10^{-3}$. The black dashed line denotes the constant unperturbed energy surface given by Eq. \eqref{Eq:E_prime} when $E'/\mu B_0 = 3.2$. 

\begin{figure}[h!]
    \centering
    \includegraphics[width=0.49\textwidth,keepaspectratio]{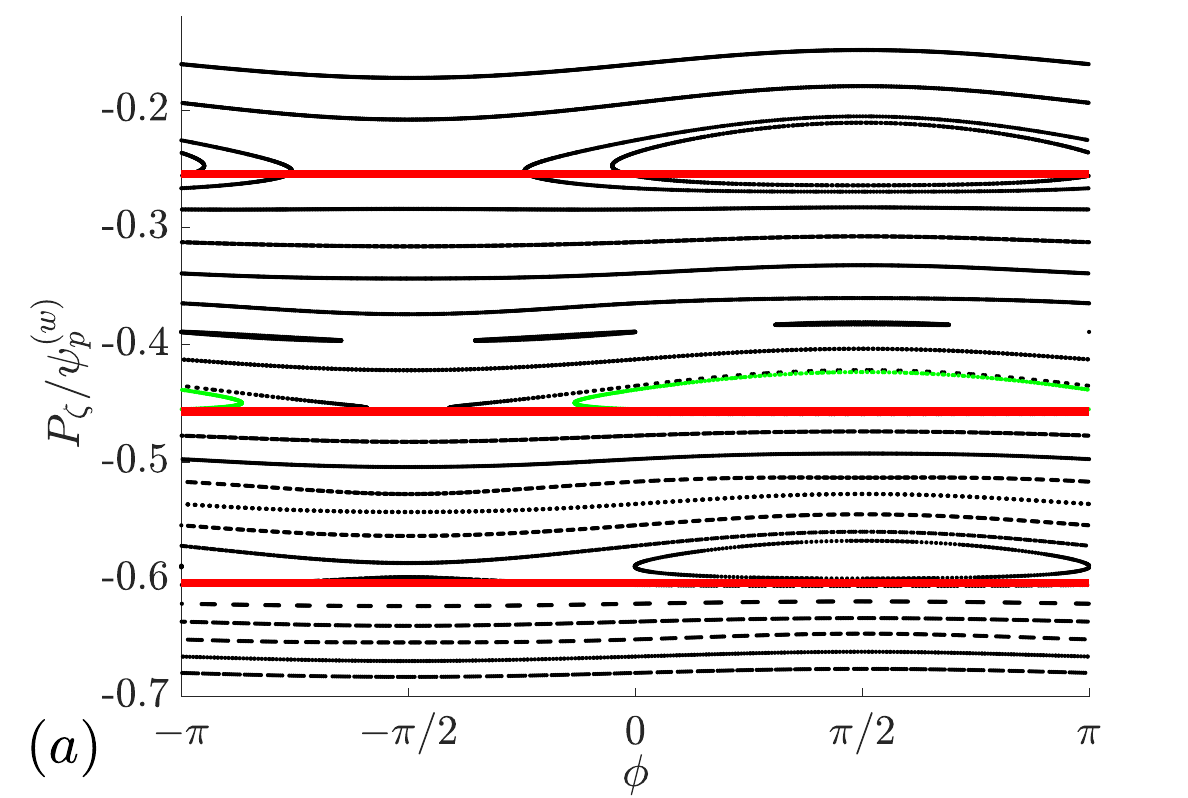}
    \includegraphics[width=0.49\textwidth,keepaspectratio]{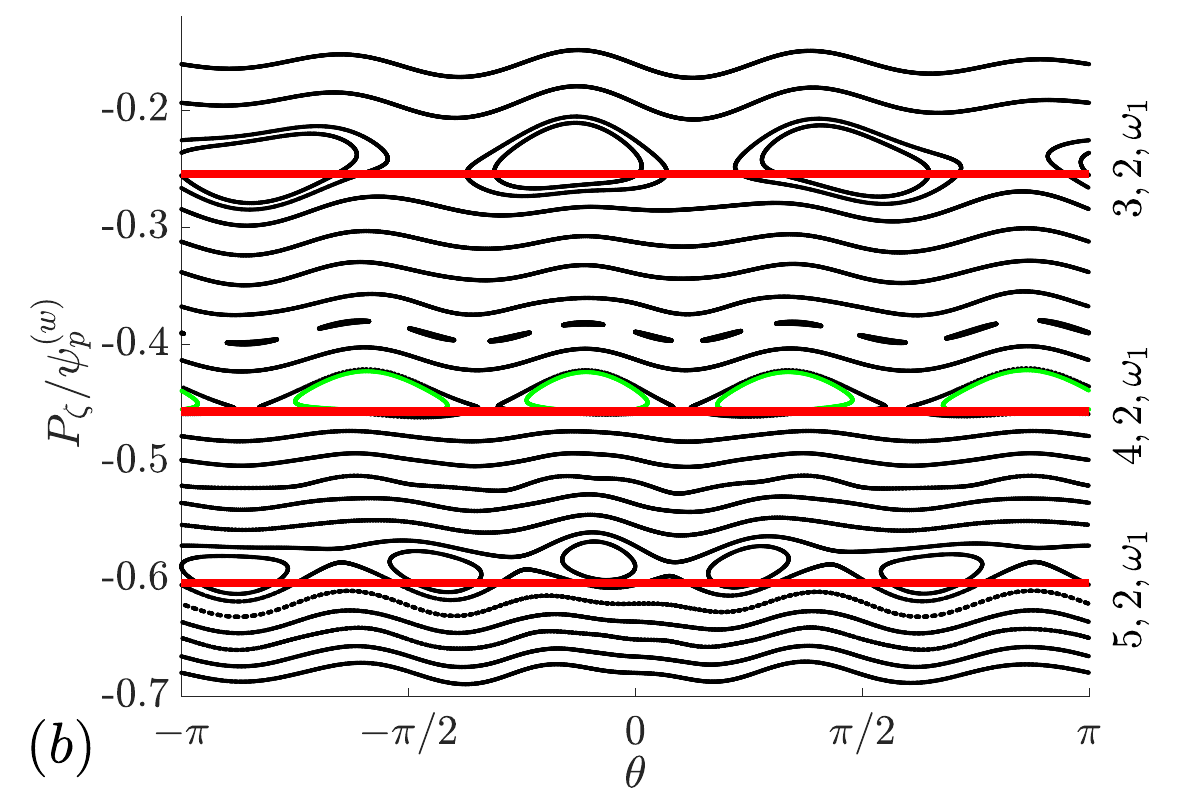}
    \includegraphics[width=0.49\textwidth,keepaspectratio]{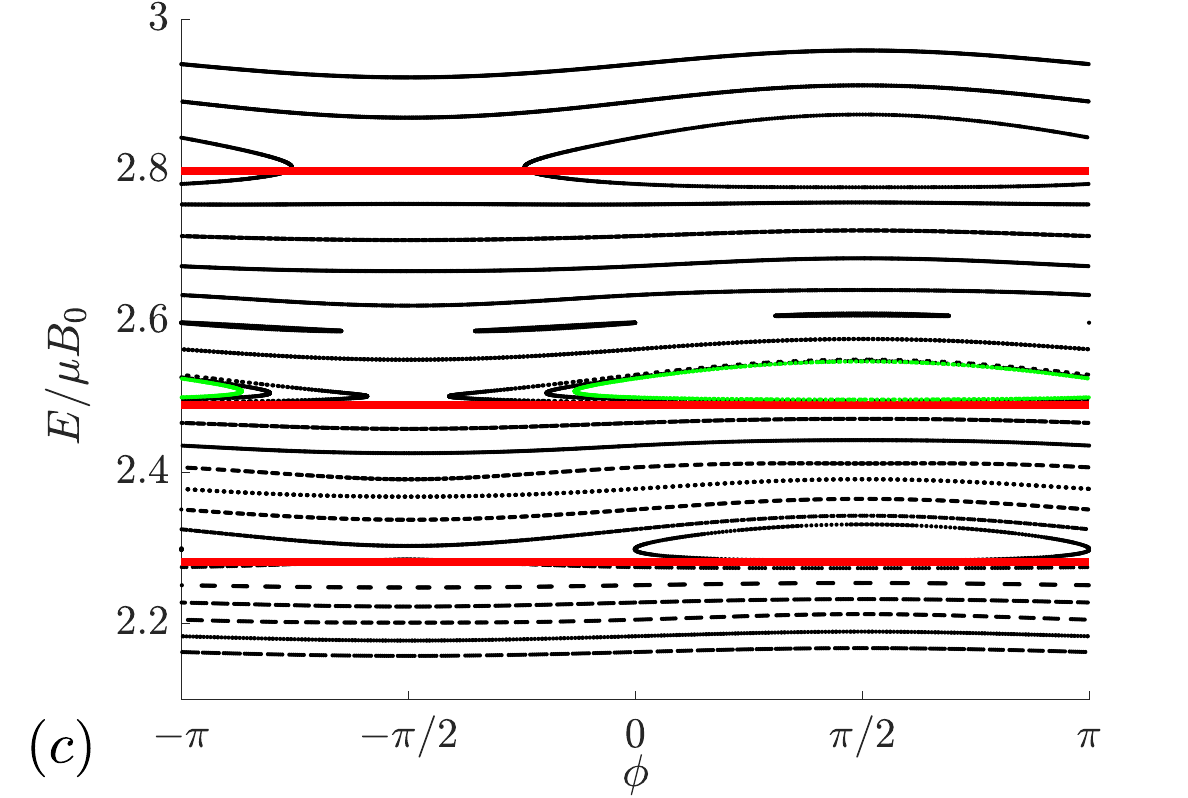}
    \includegraphics[width=0.49\textwidth,keepaspectratio]{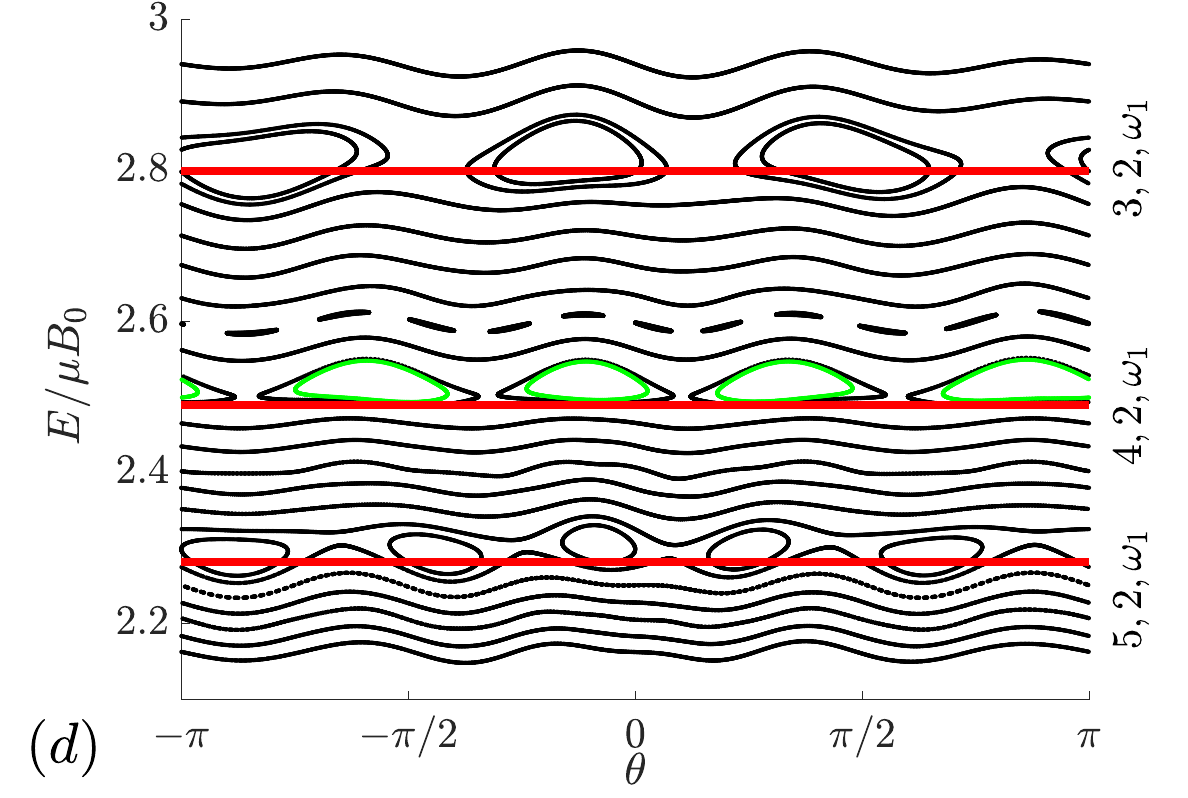}
    
    \caption[Poincaré surfaces of section for passing particles for single time-dependent perturbation in LAR equilibrium.]{Poincaré surfaces of section for passing particles of \textbf{Case \#1} ($\mu B_{0} = \SI{2}{\kilo\electronvolt}$ in the $q_{1}$ profile), under the presence of a single time-dependent perturbative mode $(m,n,\omega)=(4,2,10^{-3})$ with amplitude $\Phi_{4,2}=0.09\times10^{-6}$. Co-passing particles with $E'/\mu B_{0} = 1.20$, crossing the surfaces of section in the positive directions are depicted. Panels (a) and (c) display Poincaré surfaces of section at constant poloidal angle $\theta=0$, whereas panels (b) and (d) show Poincaré surfaces at a constant $\phi$ angle $\phi=0$. Panels (a) and (b) are in terms of $P_{\zeta}/\psi_{p}^{(w)}$ while panels (c) and (d) in terms of $E/\mu$. The red horizontal lines denote the predicted locations of the corresponding resonant island chains in accordance to Fig. \ref{fig:Fig571}, showing a remarkable accuracy.}
    \label{fig:Fig572}
\end{figure}

In Fig. \ref{fig:Fig572}, we present Poincaré surfaces of section for constant $\theta$ (panels (a), (c)) and constant $\phi$ (panels (b), (d)) for co-passing orbits on the constant energy surface $E'$. Panels (a) and (b) display one action, i.e., $J_{\zeta} = P_{\zeta}$ on the vertical axis, while panels (c) and (d) show the other constant of motion, $E$, which is related to the action $J_{\theta}$ through $E = E(J_{\theta}, J_{\zeta}; \mu)$ (Eq. \eqref{gen Jtheta}). Similar to the time-independent perturbation case discussed in Sec. \ref{Resonse}, the island locations here can also be predicted by the intersection points of the constant $E'$ surface (black dashed line in Fig. \ref{fig:Fig571}) with the respective resonance lines.

\subsection{Multiple Time-Dependent Perturbative Modes: Arnold Web $\&$ Diffusion}

\subsubsection{Arnold Web in the COM space}
When the perturbation includes multiple toroidal mode numbers $n$ and/or different mode frequencies $\omega$, the canonical transformation discussed in the previous section is no longer applicable. In this case, to study the system, we apply a canonical transformation in the extended phase space to render the system time-independent.  The time-dependent guiding center Hamiltonian is given by Eq. \ref{Eq: GC H with time}. The transformation from a time-dependent Hamiltonian to a time-independent one is achieved by introducing an additional degree of freedom and extending the phase space (Ref.~\cite{Lieberman1992}, Sec. 1.2). This is accomplished through the following generating function:
\begin{equation}
    \bar{F} = \bar{P}_{\theta}\theta + \bar{P}_{\zeta}\zeta - \bar{p}t
\end{equation}
where $\bar{p} = -H$. This generating function results in the following coordinate and momentum canonical transformation equations within the extended phase space:
\begin{equation}
    \bar{\theta} = \theta \qquad \bar{\zeta} = \zeta \qquad \bar{q} = t \label{time transform cor}
\end{equation} 
\begin{equation}
        \bar{P}_{\theta} = P_{\theta} \qquad \bar{P}_{\zeta} = P_{\zeta} \qquad \bar{p} = -H \label{time tranform moment}
\end{equation}

In this extended phase space, the Hamiltonian becomes time-independent, as the original time variable $t$ is transformed into a new coordinate $\bar{q}$, with its conjugate momentum $\bar{p} = -H$. In the new system, time is parametrized by any parameter independent of the variations implied by the variational principles (lets denote it here as $\bar{t}$). The new time independent Hamiltonian is given by $\bar{H} = H(\theta,\zeta, P_{\theta}, P_{\zeta},t) + \partial \bar{F}/\partial t$, where the old coordinates are expressed in terms of the new ones. Simplifying notation, we remove the bars from the canonical variables and obtain.
\begin{align}\label{time ind H}
    \bar{H}(\theta,\zeta,q,P_{\theta},P_{\zeta},p) &= H_{0}(\theta,P_{\theta},P_{\zeta};\mu) + p \\
    &+ \epsilon H_{1}(\theta,\zeta,q,P_{\theta},P_{\zeta}) + \epsilon^2 H_{1}(\theta,\zeta,q,P_{\theta},P_{\zeta}) 
\end{align}

In this time-independent form, the previous expression reveals that the new unperturbed Hamiltonian is given by:
\begin{equation}\label{time ind H0}
\bar{H}_{0} = H_{0}(\theta, P_{\theta}, P_{\zeta}; \mu) + p
\end{equation}
It is evident that in the Hamiltonian $\bar{H}_{0}$, the coordinate $q$ is cyclic, implying that its conjugate momentum $p$ is a constant of the motion. Therefore, from Eq. \eqref{time ind H0}, $\bar{H}_{0}$, being a constant of the motion, is expressed as the sum of two constants of the motion, $H_{0}$ and $p$. This indicates that this Hamiltonian is separable. To calculate the action integral, the momentum has to be expressed as a function of its conjugate coordinate and the constants of the motion of the system (see Secs. \ref{Separability and Integralbility} and \ref{Action Angle Variables}) leading to the following expressions:
\begin{equation} \label{time ind actions}
    J_{\zeta} = \frac{1}{2\pi}\oint_{\zeta} P_{\zeta} \, d\zeta \qquad J_{\theta} = \frac{1}{2\pi}\oint_{\theta} P_{\theta}(\theta, E, P_{\zeta}; \mu) \, d\theta \qquad J_{p} = \frac{1}{2\pi}\oint_{q} p \, dq
\end{equation}
where $E = H_0$. The first two integrals in Eq. \ref{time ind actions} are identical to those of the unperturbed, time-independent system with Hamiltonian $H_0$, which have been analytically computed and examined in Sec. \ref{Action-Angle Transformation}, whereas, the third integral evaluates to:
\begin{equation} \label{Jp}
    J_{p} = p
\end{equation}

Similarly to Eq. \eqref{genarating function}, the generating function for this action-angle transformation is given by  
\begin{equation} \label{genarating function extended}
   F(\theta, \zeta, \xi, J_{\theta}, J_{\zeta}, J_{\xi}) = J_{\zeta} \zeta + J_{\xi} \xi + J_p q + \int^{\theta} P_{\theta}(\theta^{\prime}, E, \mu, P_{\zeta}) \, d\theta^{\prime}.
\end{equation}  
From the generating function, it follows that the angles $\hat{\theta}$, $\hat{\zeta}$, and $\hat{\xi}$ are also given by Eqs. \eqref{eq:Angles},
\begin{equation}\label{eq:Angles extended}
    \hat{\theta}=\frac{\partial f(\mathbf{J,\theta})}{\partial J_\theta}, \qquad \hat{\zeta}=\zeta+\frac{\partial f(\mathbf{J,\theta})}{\partial J_\zeta}, \qquad \hat{\xi}=\xi+\frac{\partial f(\mathbf{J,\theta})}{\partial J_\xi}. 
\end{equation}
while,  
\begin{equation}\label{hat q}
    \hat{q} = \frac{\partial F}{\partial J_p} = q
\end{equation}  
since the integral in Eq. \eqref{genarating function extended} is independent of $J_p$.

Furthermore, from Eqs. \eqref{time ind actions}, it is evident that in terms of the actions, the $H_{0}$ in Eq. \eqref{time ind H0} is expressed as a function solely of $J_{\theta}$, $J_{\zeta}$, and $\mu$. This implies that Eq. \eqref{time ind H0} transforms into
\begin{equation}
    \bar{H}_{0} = H_{0}(J_{\theta},J_{\zeta};\mu) + J_{p}
\end{equation}
Thus, the canonical frequencies of the system are defined as follows:
\begin{equation} \label{hat omega zeta time}
    \hat{\omega}_{\zeta} = \frac{\partial \bar{H}_{0}}{\partial J_{\zeta}} = \frac{\partial H_{0}}{\partial J_{\zeta}} 
\end{equation}
\begin{equation} \label{hat omega theta time}
    \hat{\omega}_{\theta} = \frac{\partial \bar{H}_{0}}{\partial J_{\theta}} = \frac{\partial H_{0}}{\partial J_{\theta}} 
\end{equation}
\begin{equation} \label{hat omega q time}
    \hat{\omega}_{q} = \frac{\partial \bar{H}_{0}}{\partial J_{p}} = 1
\end{equation}
From these equations, it is observed that the frequencies corresponding to the $\theta$ and $\zeta$ coordinates are the canonical frequencies of the unperturbed time-independent system (Eqs. \eqref{gen hat omega theta}, \eqref{gen hat omega zeta}), while the frequency of the extra degree of freedom $q$, introduced due to the time dependence, is $\hat{\omega}_{q} = 1$. Additionally, it is noted that all canonical frequencies are independent of the action $J_{p} = p$.

The perturbation terms $H_1$ and $H_2$ in Eq. \eqref{Eq: GC H with time} are expressed in the extended phase space as  
\begin{equation}
    H_{1,2} = H_{1,2}(\theta,\zeta,P_{\theta},P_{\zeta},q;\mu).
\end{equation}  
Their transformation into action-angle (AA) variables takes the form  
\begin{equation} \label{perturbation_in_AA extended}
    \hat{H}_{1,2}(\hat{\theta},\hat{\zeta},J_{\theta},J_{\zeta}, J_{\xi}) = \sum_{m',n} \hat{H}_{1,2(m',n)}(J_{\theta},J_{\zeta}, J_{\xi}) e^{i(n\hat{\zeta}-m'\hat{\theta} - \omega \hat{q})},
\end{equation}  
which is identical to Eq. \eqref{perturbation_in_AA}, except that $t$ is directly replaced with $\hat{q}$. This substitution follows from the relation $t = q = \hat{q}$ and the fact that the perturbation depends on time through $e^{i\omega t}$. Additionally, the remaining variables on which the perturbation depends, namely $(\theta,\zeta,P_{\theta},P_{\zeta};\mu)$, are independent of $\hat{q}$ and $J_p$, as can be seen from the transformation equations \eqref{eq:Angles extended}, \eqref{hat q}, \eqref{time ind actions}, and \eqref{Jp}.  Thus, they are written is AA as functions of $\hat{\theta},\hat{\zeta},J_{\theta},J_{\zeta}$ and $\mu$ ($J_{\xi}$) only. Finally, the perturbed system in AA variables is given by  
\begin{align}\label{Eq: GC perturbed extended in AA}
    \bar{H} &= H_{0}(J_{\theta},J_{\zeta};\mu) + J_{p} + \epsilon \sum_{m',n} \hat{H}_{1 (m',n)}(J_{\theta},J_{\zeta};\mu) e^{i(n\hat{\zeta}-m'\hat{\theta} - \omega \hat{q})} \\
    &\quad + \epsilon^2 \sum_{m',n} \hat{H}_{2(m',n)}(J_{\theta},J_{\zeta};\mu) e^{i(n\hat{\zeta}-m'\hat{\theta} - \omega \hat{q})}.
\end{align}  

Taking into account that $\hat{q} = q = t$, which implies $\dot{\hat{q}} = 1$, the resonance condition remains the same as Eq. \eqref{Eq:time_dependent_res_condition}:  
\begin{equation} \label{res eq time}
    n\hat{\omega}_{\zeta}(J_{\theta},J_{\zeta};\mu) - m'\hat{\omega}_{\theta}(J_{\theta},J_{\zeta};\mu) - \omega = 0,
\end{equation} 
or
\begin{equation} \label{res eq time 2}
    n\hat{\omega}_{\zeta}(E,P_{\zeta};\mu) - m'\hat{\omega}_{\theta}(E,P_{\zeta};\mu) - \omega = 0,
\end{equation}  
where the frequencies are given by Eqs. \eqref{hat omega zeta time}, \eqref{hat omega theta time}, and \eqref{hat omega q time}.  

\begin{figure}[h!]
    \centering
    \includegraphics[width=0.69\textwidth,keepaspectratio]{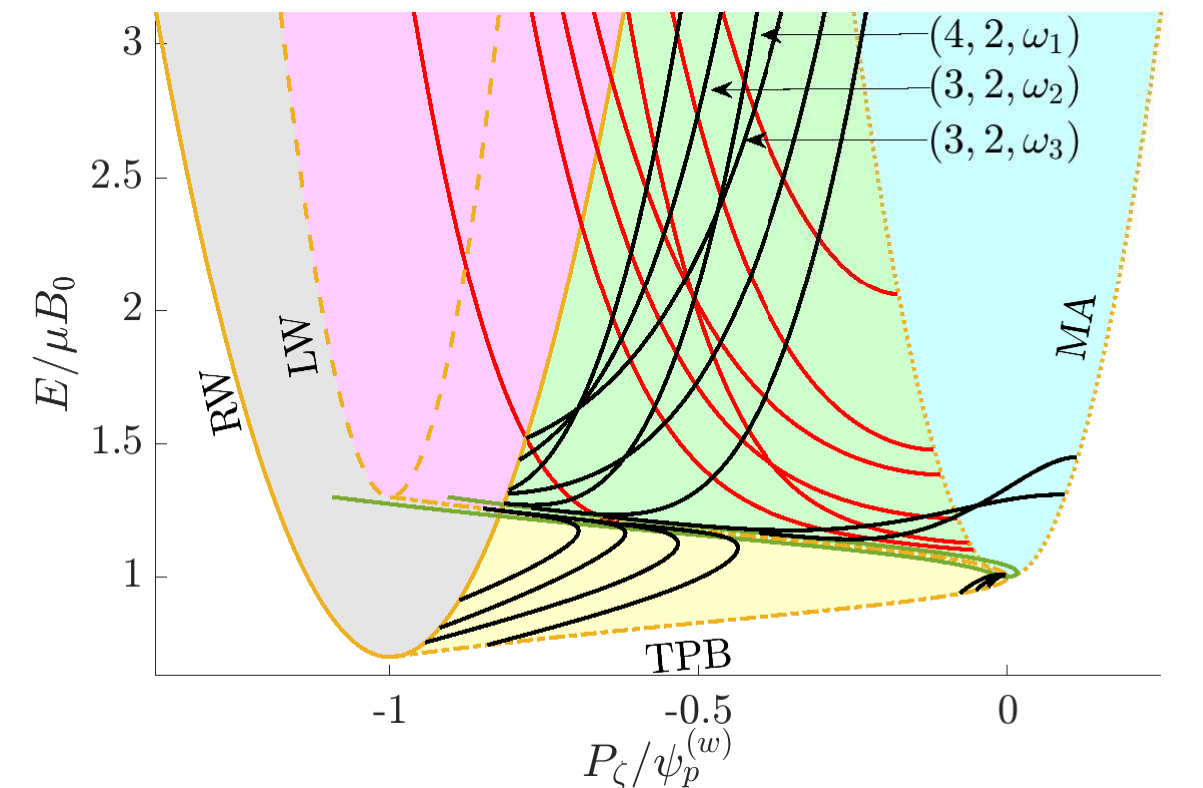}
    \caption[Arnold wed in COM space in LAR equilibrium]{Constant-$\mu$ slices (plane cuts) of the three-dimensional COM space $(E,\mu, P_{\zeta})$ for \textbf{Case \#1}. Black lines represent resonances (Eq. \eqref{res eq time 2}) associated with trapped and co-passing particles, while red lines correspond to resonances with counter-passing particles. Each resonance line corresponds to a different set of $(m',n,\omega)$ values. Unlike the time-independent case (Fig. \ref{fig:Fig54}) or the single-$(n,\omega)$ mode case (Fig. \ref{fig:Fig571}a), the resonance lines here intersect, forming the Arnold web. The intersections considered for the Arnold web are those between lines of the same color, i.e., between co-passing, counter-passing, or trapped resonance lines.}
    \label{fig:Fig573}
\end{figure}

As discussed in Sec. \ref{Sec: Arnold Diffusion}, in systems with three degrees of freedom, such as the system described by Eq. \eqref{Eq: GC perturbed extended in AA}, resonance surfaces in the action space $(J_{\theta},J_{\zeta},J_p) \leftrightarrow (E,P_{\zeta},J_p)$ can intersect, forming the Arnold web. In a constant-$\mu$ slice for \textbf{Case \#1}, the Arnold web for different values of $m'$, $n$, and $\omega$ is shown in Fig.~\ref{fig:Fig573}.  

We observe that the resonance lines obtained from Eq.~\eqref{res eq time} intersect at specific points. Since the system has three degrees of freedom, at a point two distinct resonance lines can intersect at most. Note that the resonance condition \eqref{res eq time} is independent of $J_p$, so Fig. \ref{fig:Fig573} presents a projection of the three-dimensional resonance surfaces onto a constant-$J_p$ plane.

\subsubsection{Example of Arnold diffusion}

Following the analysis in Sec. \ref{Sec: Arnold Diffusion}, Arnold diffusion occurs along the lines formed by the intersection of the resonance surfaces with a constant unperturbed energy surface. In the case of the GC system \eqref{Eq: GC perturbed extended in AA}, the unperturbed energy is given by  
\begin{equation}\label{Eq: bar H0}
    \bar{H}_0 = E(J_{\theta},P_{\zeta};\mu) + J_p,
\end{equation}  
where we have replaced $H_0$ with $E$. The specific energy surface depends on the initial conditions of the orbit under examination.  

We aim to investigate how unperturbed guiding center (GC) orbits, characterized by particular kinetic parameters—i.e., the constants of motion $(E,\mu,P_{\zeta})$—evolve under time-dependent perturbations. To achieve this, we study perturbed orbits by selecting specific initial conditions $(E^{init},P_{\zeta}^{init},\mu)$. For demonstration purposes, we consider the scenario of \textbf{Case \#1}. When numerically solving the GC equations of motion, we select $(P_{\theta}^{init}, \theta^{init})$ such that, given the chosen $(P_{\zeta}^{init},\mu)$, the unperturbed energy $E$ obtained from Eq. \eqref{GC H 1} equals $E^{\text{init}}$.  

Since $J_p = -H$, from Eq. \eqref{Eq: GC H with time}, we obtain  
\begin{equation}
    J_p^{\text{init}} = -E^{\text{init}} - \epsilon H_1^{\text{init}} - \epsilon^2 H_2^{\text{init}}.
\end{equation}  
As a result, from Eq. \eqref{Eq: bar H0}, the initial unperturbed energy is  
\begin{equation}\label{Eq: bar H0 init}
    \bar{H}_0^{\text{init}} = - \epsilon H_1^{\text{init}} - \epsilon^2 H_2^{\text{init}}.
\end{equation}  
Thus, from Eq. \eqref{Eq: bar H0}, the unperturbed orbit we examine lies on the constant energy surface in $(E,P_{\zeta},J_p)$ space, given by  
\begin{equation}\label{Eq: bar H0 plane}
    \bar{H}_0^{\text{init}} = E + J_p, \quad \forall P_{\zeta},
\end{equation}  
where $\bar{H}_0^{\text{init}}$ is given by Eq. \eqref{Eq: bar H0 init}.  

\begin{figure}[h!]
    \centering
    \includegraphics[width=0.79\textwidth,keepaspectratio]{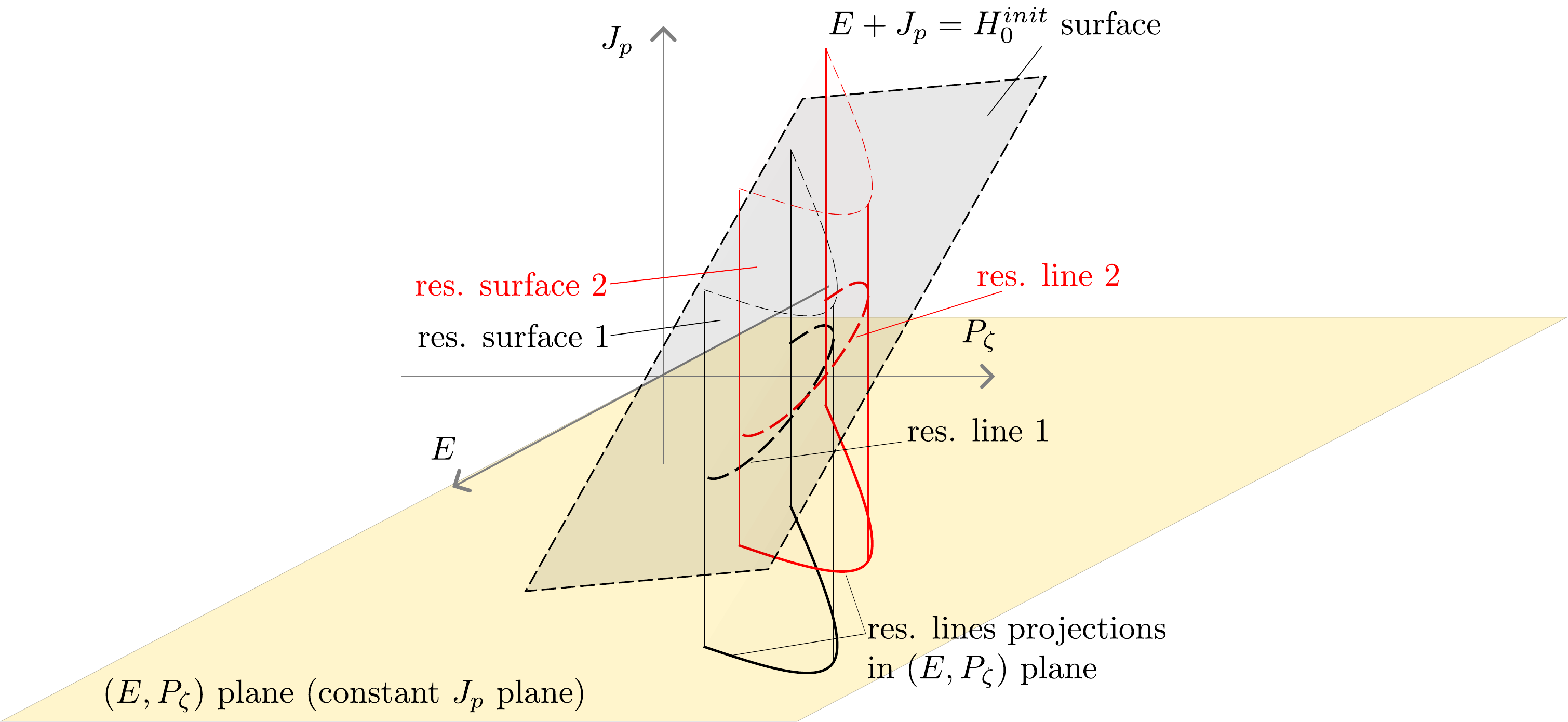}
    \caption[A qualitative depiction of the projection of resonance lines in COM space.]{The solid black and red resonance lines are the projections on the $(E,P_{\zeta})$ plane (constant $J_p$ plane) of the dashed black and red resonance lines formed by the intersection of the unperturbed energy surface—given by Eq. \eqref{Eq: bar H0 plane}, which is a plane in our system—with the resonance surfaces. Different values of $\bar{H}_0^{\text{init}}$ shift the energy (gray) plane up or down in the $J_p$ direction. However, since the resonance surfaces are independent of $J_p$, the resonance lines and their projections in the $(E,P_{\zeta})$ plane remain unchanged, regardless of $\bar{H}_0^{\text{init}}$. In the figures that follow we present projections on the $(E,P_{\zeta})$ plane.}
    \label{fig:Fig574}
\end{figure}

Equation \eqref{Eq: bar H0 plane} defines a plane surface. The intersections of this plane with the resonance surfaces generate resonance lines on the constant $\bar{H}_0(E,J_p;\mu)$ surface (depicted as dashed black and red lines in Fig. \ref{fig:Fig574}). However, our analysis thus far has been conducted in the $(E,P_{\zeta})$ plane, which generally differs from the constant $\bar{H}_0(E,J_p;\mu)$ plane (beige and gray planes in Fig. \ref{fig:Fig574}). For this reason we focus on the projections of these resonance lines—originally defined on the constant $\bar{H}_0(E,J_p;\mu)$ plane—onto the  
$(E,P_{\zeta})$(constant $J_p$) plane, where they appear as solid black and red lines in Fig. \ref{fig:Fig574}. As seen in Fig. \ref{fig:Fig574}, varying $\bar{H}_0^{\text{init}}$ shifts the energy plane up or down in the $J_p$ direction. However, since the resonance surfaces are independent of $J_p$, the resonance lines—and consequently their projections in the $(E,P_{\zeta})$ plane—remain unchanged. This implies that, despite introducing the additional degree of freedom $(\hat{q},J_p)$ due to time dependence, analyzing the system on a constant-$J_p$ plane, i.e., in the $(E,P_{\zeta})$ plane, suffices to capture the essential dynamical behavior of any orbit, regardless of its initial conditions (which correspond to an unperturbed energy surface characterized by $\bar{H}_0^{\text{init}}$) \cite{Wobig2001}.

\begin{figure}[h!]
    \centering
    \includegraphics[width=0.45\textwidth,keepaspectratio]{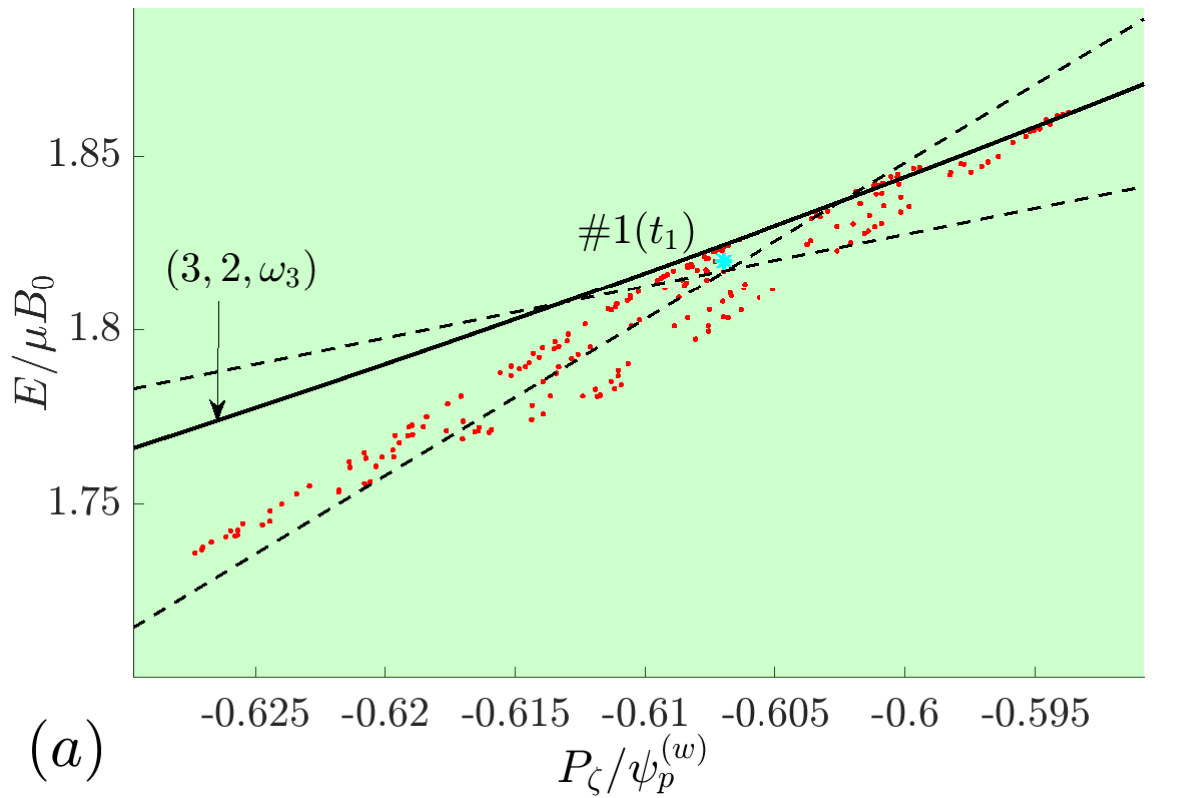}
    \includegraphics[width=0.45\textwidth,keepaspectratio]{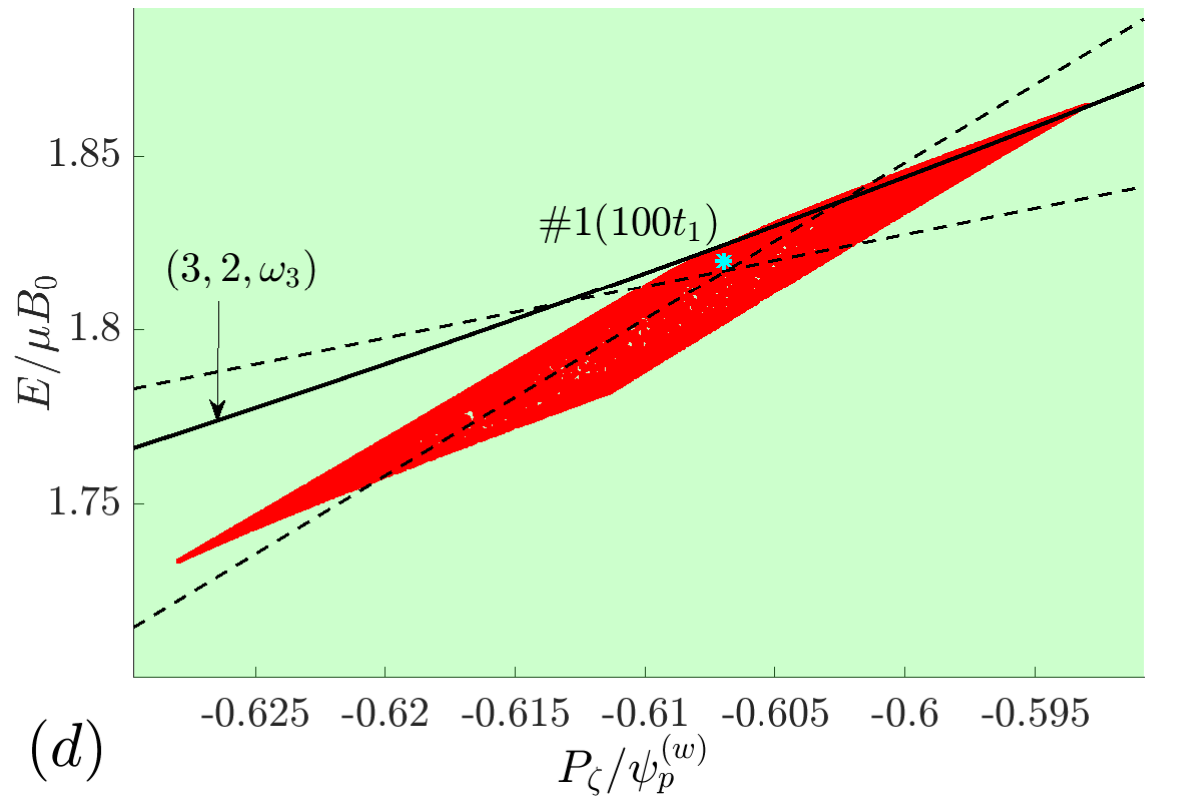}
    \includegraphics[width=0.45\textwidth,keepaspectratio]{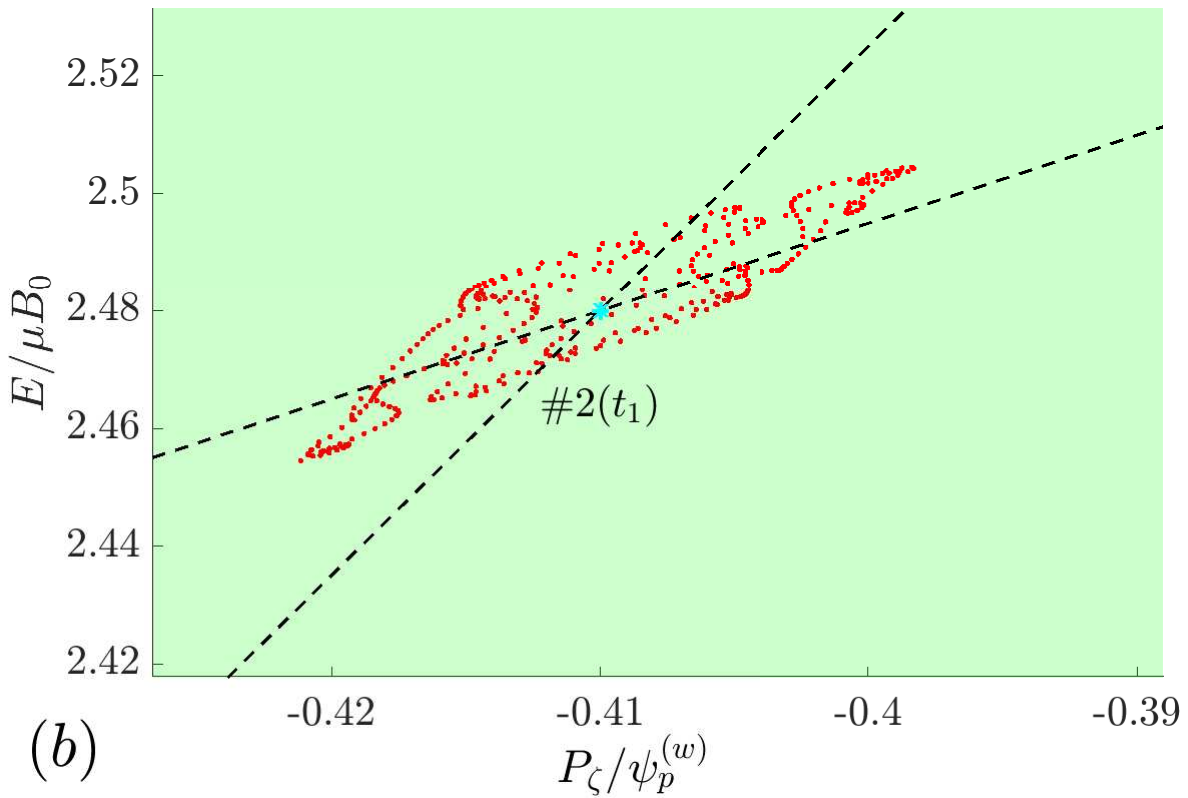}
    \includegraphics[width=0.45\textwidth,keepaspectratio]{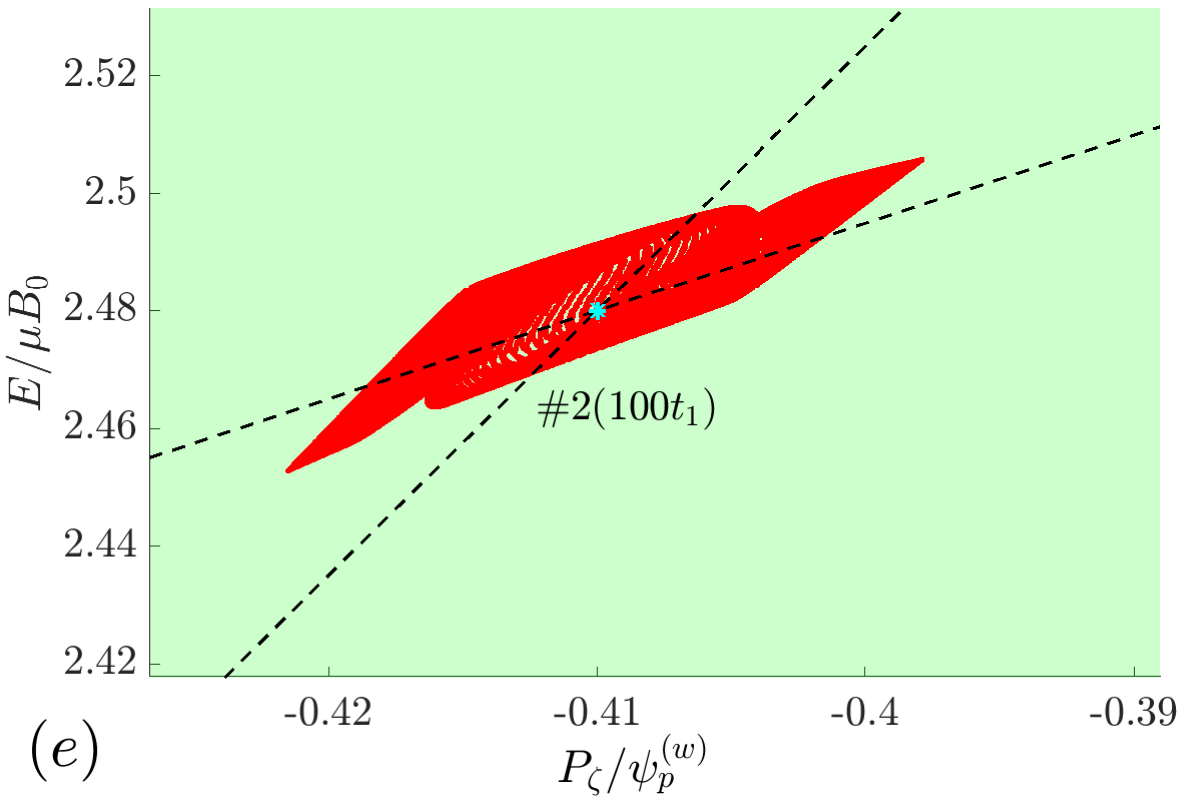}
    \includegraphics[width=0.45\textwidth,keepaspectratio]{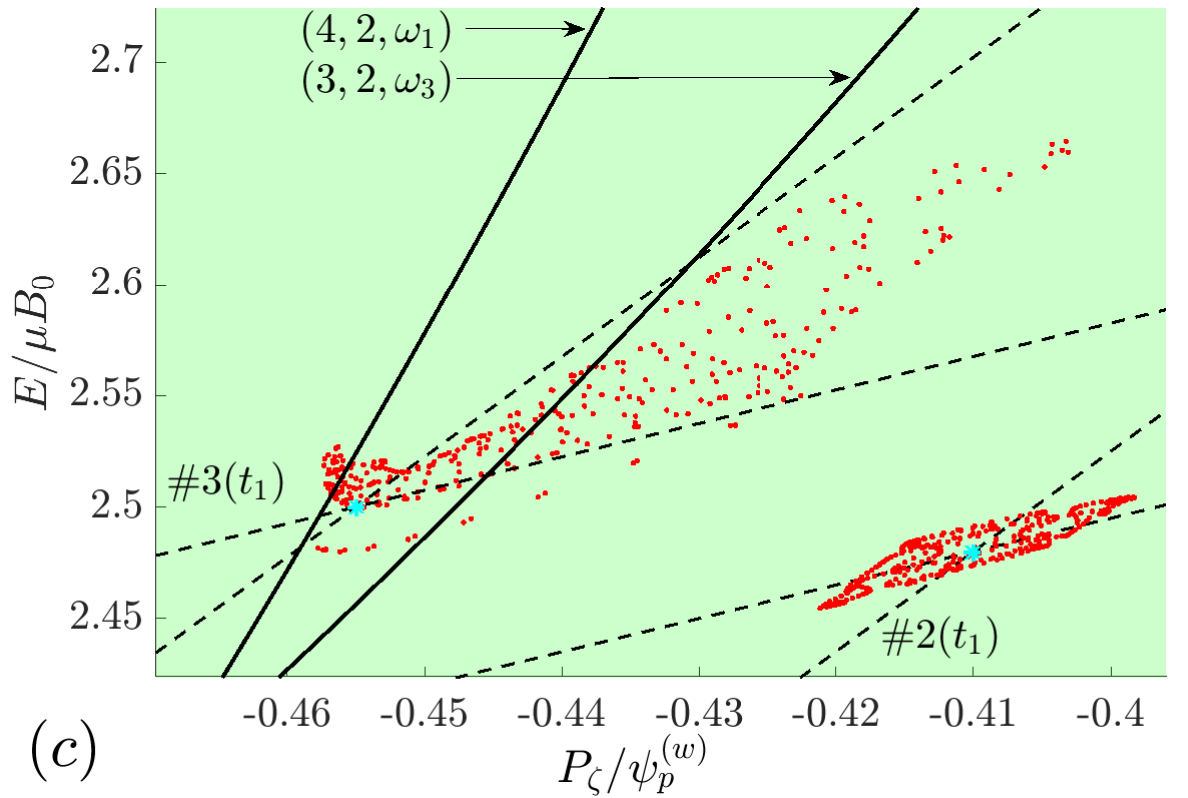}
    \includegraphics[width=0.45\textwidth,keepaspectratio]{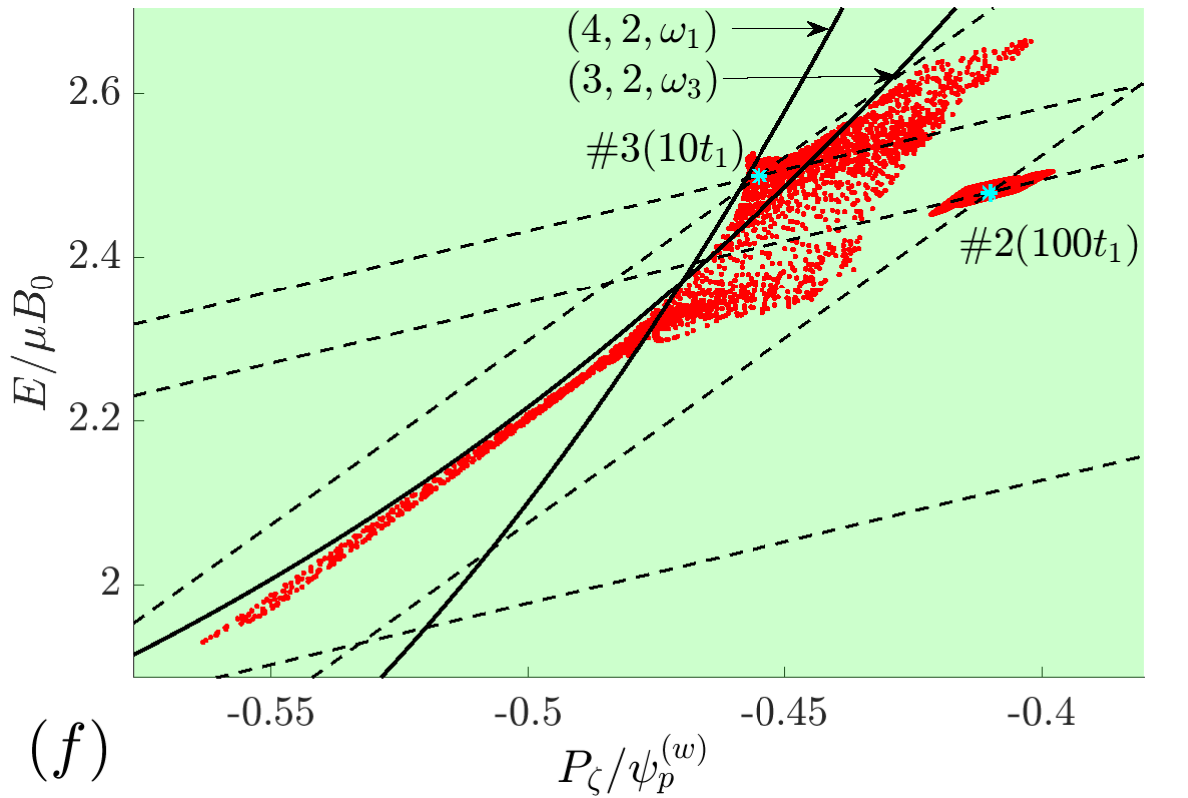}
    \caption[Arnold diffusion in the COM space in LAR equilibrium 1.]{Enlarged domains of Fig. \ref{fig:Fig573}, where we also display three distinct orbits \#1 (panels (a), (b)), \#2 (panels (b), (e)), and \#3 (panels (c), (f)), with initial conditions in $(E/\mu B_0, P_{\zeta}/\psi_p^{(w)})$ marked by cyan stars. Red points represent $(P_{\zeta}(t)/\psi_p^{(w)}, E(t)/\mu B_0)$ for each orbit. On the left column (panels (a)-(c)) orbits evolved up to time $t_1$, while on the right column (panels(d)-(f)) obits evolved up to a longer time. The runtime in each case is indicated in parentheses next to its corresponding number. Dashed lines represent the constant energy lines $E'_1$ and $E'_2$ for each orbit, given by Eq. \eqref{Eq:E_prime}, corresponding to the modes $(4,2,\omega_1)$ and $(3,2,\omega_3)$, respectively.}
    \label{fig:Fig578}
\end{figure}

In Fig. \ref{fig:Fig578}, we select three different co-passing initial conditions in the $(E,P_{\zeta})$ plane for Case \#1, marked by cyan stars (\#1, \#2, \#3). For each initial condition, we consider two perturbation modes with same amplitude for $\alpha$ (Eq. \eqref{alpha}) and $\Phi$ (Eq. \eqref{Eq: Phi}), corresponding to the resonance lines near the initial conditions. These modes are $(m=4, n=2, \omega_1 = 10^{-3})$ and $(m=3, n=2, \omega_3 = 3\cdot10^{-3})$ with amplitude $\epsilon_1$. We solve the perturbed GC equations of motion twice for each initial condition: first up to normalized time $t_1=10^6$, Fig. \ref{fig:Fig578}(a)-(c), and then up to $100 t_1$ for \#1 and \#2, Fig. \ref{fig:Fig578}(d)-(e), and $10\cdot t_1$ for initial condition \#3, Fig. \ref{fig:Fig578}(f), displaying the evolution of $E(t)$ and $P_{\zeta}(t)$ for the perturbed orbit (the integration time for each initial condition is indicated in parentheses next to its corresponding number in Fig. \ref{fig:Fig578}). As expected (see Sec. \ref{Sec: Arnold Diffusion}), initial condition \#3, which is very close to the intersection of two resonance lines, exhibits diffusion along these lines. Specifically, up to time $t_1$, it diffuses slightly, while at ten times longer integration time, it diffuses significantly. In contrast, \#1, located in a region where the $(m=3, n=2, \omega_3 = 3\cdot10^{-3})$ resonance is isolated (i.e., far from intersection points between the respective resonances lines), and \#2, situated in a KAM region, do not exhibit diffusion. Even at $100 t_1$, their trajectories remain confined to the same region as at time $t_1$.

Note that initial condition \#3 is the same as the initial condition of the green orbit in Fig. \ref{fig:Fig572} (red star in Fig. \ref{fig:Fig571}). In Figs. \ref{fig:Fig571} and \ref{fig:Fig572}, we considered a single perturbative mode with the same perturbation strength as in Fig. \ref{fig:Fig578}, the mode $(m=4, n=2, \omega_1 = 10^{-3})$. As a result, the resonance is isolated (see Sec. \ref{Sec: Arnold Diffusion}). In these figures, where resonance lines do not intersect near the initial condition, the orbit is trapped in an island (green orbit in Fig. \ref{fig:Fig572}) and does not diffuse. However, when an additional mode $(m=3, n=2, \omega_3 = 3\cdot10^{-3})$ is introduced, the same initial condition exhibits diffusion along the resonance lines of these two intersecting modes, as observed in Fig. \ref{fig:Fig578}.

In Fig. \ref{fig:Fig578}, for each orbit, we also depict with dashed black lines the constant unperturbed energy lines $E'_1$ and $E'_2$, given by Eq. \eqref{Eq:E_prime}, for the modes $(4,2,\omega_1=10^{-3})$ and $(m=3, n=2, \omega_3 = 3\cdot10^{-3})$, respectively. Unlike in Fig. \ref{fig:Fig572}, where only the first mode is present, here the second mode is also active. Consequently, instead of forming a single trajectory evolving around a constant $E'$ line (see right panel of Fig. \ref{fig:Fig572}), orbits \#1 and \#2, Fig. \ref{fig:Fig578}(a)-(b) and (d)-(e), form a parallelogram, with one side nearly parallel to the $E'_1$ line of the first mode and the other side nearly parallel to the $E'_2$ line of the second mode. Similarly, orbit \#3, up to time $t_1$, Fig. \ref{fig:Fig578}(c),  forms a parallelogram between $E/\mu B_0 = 2.48$ and $E/\mu B_0 = 2.66$ however at longer time, $10t_1$, the parallelogram extends along the resonance line $(3,2,\omega_3)$ form $E/\mu B_0 = 1.92$ to $E/\mu B_0 = 2.66$. This behavior resembles the findings in Ref. \cite{White2012}, Fig. 6, where two distinct time-dependent modes are considered simultaneously. The authors in \cite{White2012} argue that the resulting motion corresponds to diffusion produced by combined motion along the two constant $E'$ lines. We observe the same phenomenon in Fig. \ref{fig:Fig578}, where orbits evolve along directions dictated by $E'_1$ and $E'_2$ with the difference that orbit \#3 exhibits also Arnold diffusion. Inside these parallelograms, either rapid chaotic motion, orbits \#1 and \#3, or regular motion, orbit \#2, takes place. However, for orbit \#3, whose initial condition is very close to the intersection point of the two resonances, in addition to this fast chaotic motion inside the parallelogram, slow diffusion occurs along the $(3,2,\omega_3)$ resonance line. This represents Arnold diffusion: the center of the island in the $(E, P_{\zeta})$ plane diffuses slowly along the resonance line. The chaotic motion around the island, as depicted in $(E/\mu B_0,P_{\zeta}/\psi_p^{(w)})$ plane of Fig. \ref{fig:Fig578}, takes place within the parallelogram.
\clearpage

Next, we integrate orbit \#3 up to $10^3 t_1$. Even over this extended time, diffusion occurs only up to $E/\mu B_0 = 1.9$ (Fig. \ref{fig:Fig578}(f)) along the $(3,2,\omega_3)$ resonance line, indicating that the Arnold diffusion range due to the intersection of resonances $(4,2,\omega_1)$ and $(3,2,\omega_3)$ extends approximately to this point.

\begin{figure}[h!]
    \centering
    \includegraphics[width=0.45\textwidth,keepaspectratio]{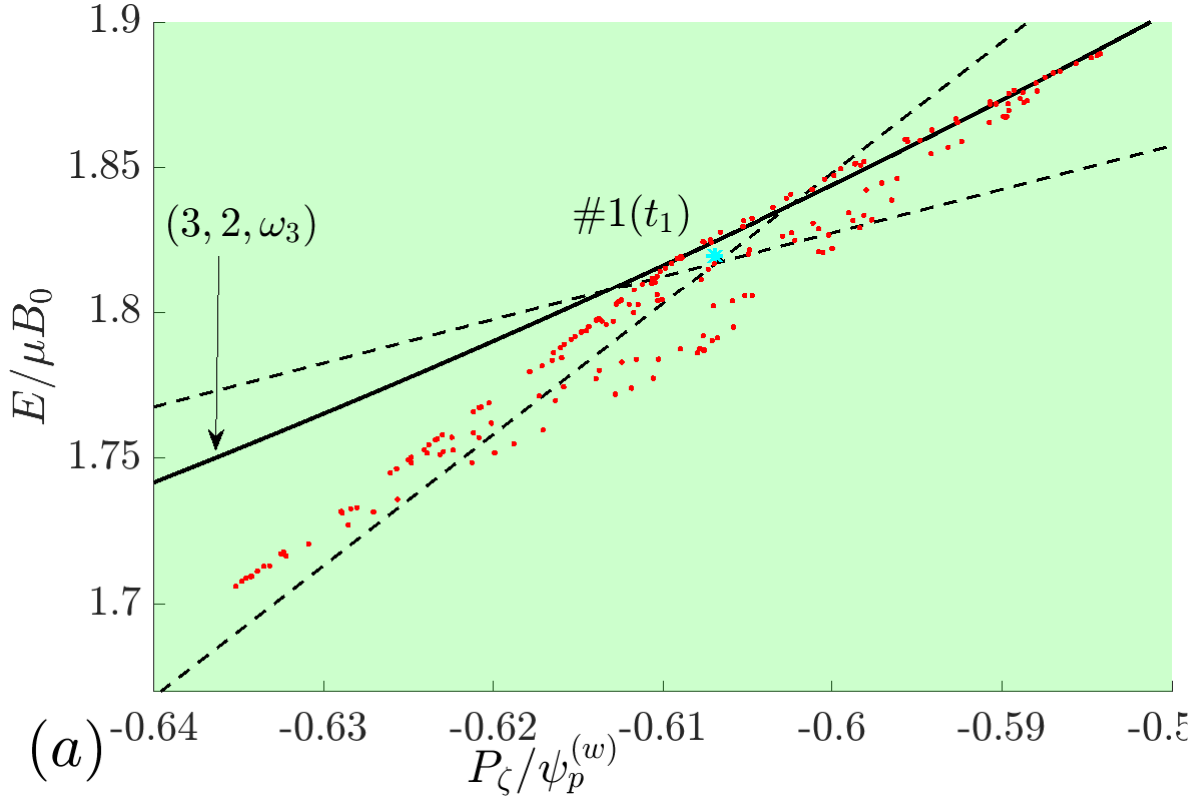}
    \includegraphics[width=0.45\textwidth,keepaspectratio]{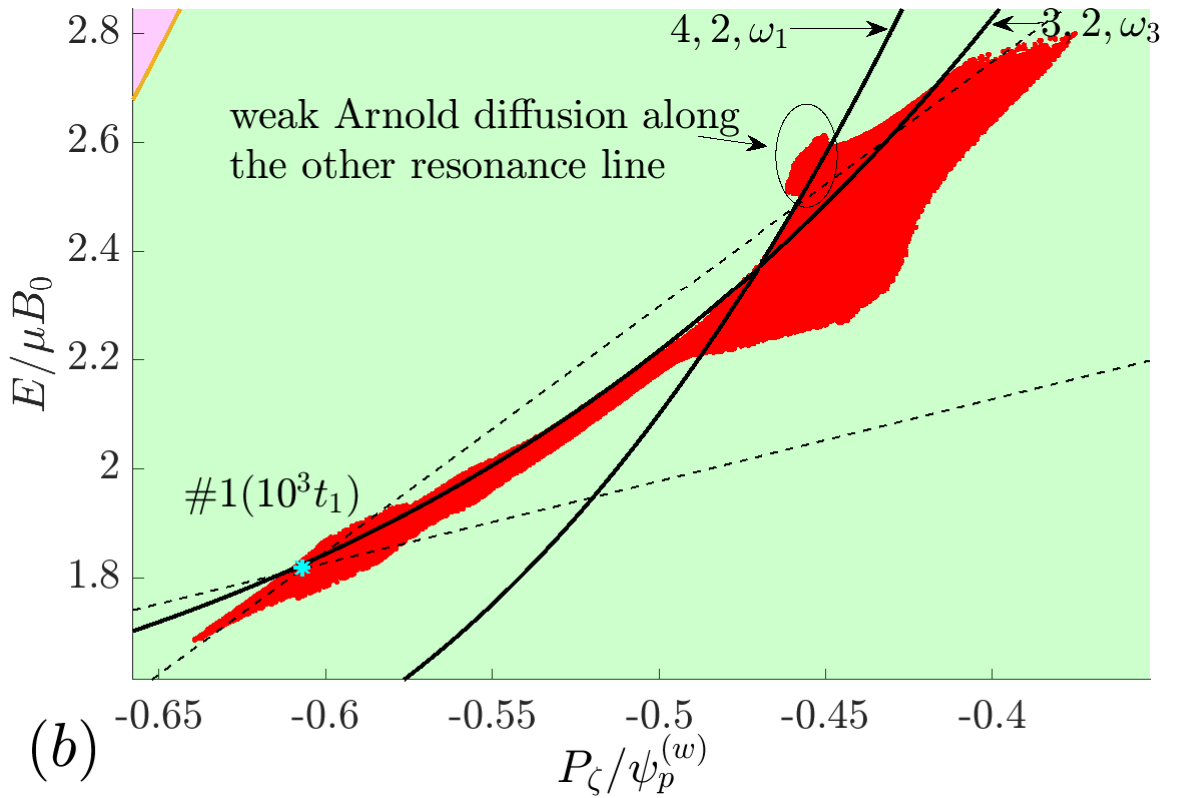}
    \caption[Arnold diffusion in the COM space in LAR equilibrium 2.] {Orbit \#1 is simulated with a perturbation amplitude 1.5 times that used for the orbits in Fig. \ref{fig:Fig578}.  In panel (a), the parallelogram's range is more extended than in Fig. \ref{fig:Fig578}(a) and (b), as expected due to the increased perturbation strength. As a result, the orbit reaches the Arnold diffusion range (around $E/\mu B_0 = 1.9$). In panel (b), the same orbit is evolved for a longer duration, $10^3 t_1$, demonstrating extensive diffusion along the resonance line $(3,2,\omega_3)$.}
    \label{fig:Fig579}
\end{figure}

We then repeat the integration of \#1 up to $t_1$, but now with a perturbation amplitude 1.5 times higher than that in Fig. \ref{fig:Fig578}. This increases the width of the island (i.e., the area of the parallelogram associated with orbit \#1) and enhances the chaotic motion within it, as shown in Fig. \ref{fig:Fig579}(a). For this reason, the parallelogram extends up to $E/\mu B_0 = 1.9$, meaning it now overlaps with the Arnold diffusion, sparked by the intersection point of the resonances $(4,2,\omega_1)$ and $(3,2,\omega_3)$, range. Running the same orbit up to $10^3 t_1$ reveals in Fig. \ref{fig:Fig579}(b) that the orbit diffuses from $E/\mu B_0=1.7$ to nearly $E/\mu B_0 = 2.8$. Furthermore, near the intersection point of the two resonances, the angle between the resonance curves decreases, bringing them closer together. There we see that the parallelogram is expanded and chaotic motion in it becomes stronger, possibly indicating resonance overlap between the islands of the two resonances.

It is worth noting that the intersection points of resonance lines analytically computed here (for the LAR equilibrium) in the Arnold web allow us to predict locations in COM space where Arnold diffusion may occur. Determining the full range of diffusion associated with each resonances intersection point, as well as computing specific diffusion coefficients, is left for future work.

\chapter{Summary and Perspectives}

\section{Conclusion}

A Hamiltonian action-angle formulation of the guiding center (GC) motion was utilized to analytically calculate the particle GC orbital frequencies for a large aspect ratio (LAR) equilibrium using the drift center (DC) approximation. The comparison between analytical and numerical results demonstrates a remarkable agreement, while the domain of validity of the analytical formulas is systematically investigated and explained. The calculation of orbital frequencies enables the representation of resonance curves in the three-dimensional space of the Constants of Motion (COM) of the particles, along with curves characterizing particles as trapped or passing, and confined or lost. These diagrams provide an overview of the plasma's kinetic response under perturbing modes and allow for the accurate identification of resonance island chains in phase space, precisely where the mode-particle resonance condition is satisfied for different $(m,n,\omega)$ modes.

For time-dependent perturbations, resonances in COM space form the Arnold web of the GC Hamiltonian system. In this case, we showed that particles characterized by given $(E, \mu, P_{\zeta})$ and distributed near the intersections of resonance lines in COM space experience Arnold diffusion along these resonances, even for weak perturbations, where strong chaotic motion is absent due to the lack of island overlap.

Moreover, by transforming to action-angle variables, we predict the exact number of islands within a specific resonance chain  on the basis of unperturbed particle orbits with its non-trivial correspondence to the perturbing mode numbers demonstrated for the case of trapped particles. Both the locations and the number of islands of resonance chains are systematically confirmed by numerically calculated Poincaré surfaces of section. Finally, conditions for the existence of transport barriers along with their locations, corresponding to local extrema of the kinetic $q$-factor are accurately predicted and their implications for enhanced confinement are discussed.

For realistic numerically obtained equilibria, the same calculations were performed using a semi-analytical geometrical method that computes actions without solving the unperturbed GC equations of motion. This method, which requires only contouring the unperturbed GC Hamiltonian in phase space, is exact and significantly reduces computational cost.

In conclusion, the application of the action-angle formulation and the analytical calculation of orbital frequencies for an LAR equilibrium, as performed in this work, enables the \textit{a priori} prediction of perturbative mode effects in terms of resonant island chains, transport barriers, and Arnold diffusion in phase space. Even for more general and realistic equilibria, this is a low-computational-cost semi-analytical approach, performed in the absence of perturbations, making it a valuable tool for predicting and understanding transport and confinement properties in toroidal fusion plasmas subject to multi-scale perturbations.

\section{Future Work}

The analytical tools developed in this work can be further applied to a variety of realistic equilibrium magnetic fields corresponding to different magnetic confinement fusion devices. A key direction is the extension of the drift center (DC) approximation to general equilibria, thereby eliminating the need for taking contours of the GC Hamiltonian. This will further reduce the computational cost, making the overall process significantly faster. While this approach will no longer yield exact results, it is expected to provide highly accurate approximations, especially in regimes where GC drift motion remains relatively weak, as demonstrated in the LAR equilibrium. Future studies will focus on identifying the parameter ranges where the DC approximation deviates significantly from the true behavior of mode-particle resonances, and on evaluating the practical implications of these deviations for modeling wave-particle interactions.

To advance the practical implementation of our framework, can also analyze a wide range of experimentally relevant modes that interact with energetic particles, including MHD modes, Resonant Magnetic Perturbations (RMPs), Edge-Localized Modes (ELMs), and toroidal field ripples. These perturbations will be expressed in action-angle variables via multi-dimensional Fourier expansions in the angle space, enabling a direct application of the island width analysis presented in Sec. \ref{Sec: Resonance islands width}.

Another important future direction is the extension of the current single-particle analysis to the study of particle ensembles interacting with multiple modes. By expressing the fast particle distribution function in action-angle variables, we aim to compute diffusion coefficients in regions of COM space where the mode-particle resonance condition is satisfied \cite{Kaufman1972}. This will help quantify transport processes in realistic scenarios involving overlapping resonances and complex mode spectra. Predictions for the evolution of the distribution function can also be used in the studies of the self-consistent nonlinear problem of mode–particle–mode interactions.

Moreover, the semi-analytical results developed in this work can be integrated to machine learning techniques. One approach is to generate a large dataset describing the location, width, number of islands, and transport barriers associated with resonances in phase space, across different equilibria and particle parameters. These data can be used to train neural networks that, given an equilibrium magnetic field and a specified region in COM space $(E, \mu, P_\zeta)$, predict the corresponding resonance structures. A key advantage of this approach is the compactness of neural network models, which store only trained weights and biases, thus avoiding the need to store extensive datasets for many scenarios. This could ultimately lead to the development of fast-access resonance maps for real-time use in experiments. A second approach involves the use of Physics-Informed Neural Networks (PINNs), which integrate physical models directly into the training process. The Hamiltonian and action-angle formalism, along with reduced models and analytical predictions developed in this work, can be embedded into the PINN framework by modifying the loss function to enforce physical constraints. This physics-based guidance can reduce the amount of required training data and make the behavior of the network more interpretable, especially in terms of identifying regions where the model diverges from known physics.


\begin{spacing}{0.9}


\printbibliography[title={References}]

@book{Chen2016,
   author = {Chen, F. F.},
   doi = {10.1007/978-3-319-22309-4},
   isbn = {978-3-319-22308-7},
   publisher = {Springer International Publishing},
   title = {Introduction to Plasma Physics and Controlled Fusion},
   year = {2016}
}

@book{Freidberg2007,
   author = {Freidberg, J. P.},
   publisher = {Cambridge University Press},
   title = {Plasma Physics and Fusion Energy},
   year = {2007}
}

@book{Wesson2005,
   author = {Wesson, J. A.},
   publisher = {Oxford University Press},
   title = {Tokamaks 3rd Edition},
   year = {2005}
}

@book{Goldston1995,
   author = {Goldston, R. and Rutherford, P.},
   doi = {10.1201/9781439822074},
   isbn = {978-0-7503-0183-1},
   publisher = {Taylor \& Francis},
   title = {Introduction to Plasma Physics},
   year = {1995}
}

@book{Piel2010,
   author = {Piel, A.},
   doi = {10.1007/978-3-642-10491-6},
   publisher = {Springer},
   title = {Plasma Physics: An Introduction to Laboratory, Space, and Fusion Plasmas},
   year = {2010}
}

@book{White2013,
   author = {White, R. B.},
   doi = {10.1142/P916},
   publisher = {World Scientific},
   title = {The Theory of Toroidally Confined Plasmas, Third Edition},
   year = {2013}
}

@book{Lieberman1992,
   author = {Lichtenberg, A. J. and Lieberman, M. A.},
   title = {Regular and Chaotic Dynamics},
   publisher = {Springer-Verlag},
   year = {1992}
}

@book{Goldstein2002,
    title = {Classical Mechanics},
    author = {Goldstein, H. and Poole, C. and Safko, J.},
    publisher = {Addison-Wesley},
    year = {2002}
}

@book{MEllo2007,
   author = {Ferraz-Mello, S.}, 
   publisher = {Springer Science+Business Media, LLC},
   title = {Canonical Perturbation Theories: Degenerate Systems and Resonance},
   year = {2007}
}

@book{Meletlidou2015,
    title = {Introduction to Non-linear Dynamical Systems},
    author = {Voyatzis, G. and Meletlidou, E.},
    publisher = {Kallipos},
    year = {2015}
}

@article{Antonenas2021,
  author = "Y. Antonenas and G. Anastasiou and Y. Kominis",
  title = "Analytical Calculation of the Orbital Spectrum of the Guiding Center Motion in Axisymmetric Magnetic Fields",
  journal = "J. Plasma Phys.",
  volume = "87",
  pages = "855870101",
  year = "2021"
}

@article{Bernard2016,
   author = {Bernard, P. and Kaloshin, V. and Zhang, K.},
   title = {Arnold diffusion in arbitrary degrees of freedom and normally hyperbolic invariant cylinders},
   journal = {Acta Math.},
   volume = {217},
   pages = {1-75},
   year = {2016}
}

@article{Anastassiou2024,
  author = "G. Anastassiou and P. Zestanakis and Y. Antonenas and E. Viezzer and Y. Kominis",
  title = "Role of the edge electric field in the resonant mode-particle interactions and the formation of transport barriers in toroidal plasmas",
  journal = "J. Plasma Phys.",
  volume = "90",
  number = "1",
  pages = "905900110",
  year = "2024",
  doi = "10.1017/S0022377824000047"
}

@article{Moges2024,
  author = "H. T. Moges and Y. Antonenas and G. Anastasiou and Ch. Skokos and Y. Kominis",
  title = "Kinetic vs. magnetic chaos in toroidal plasmas: A systematic quantitative comparison",
  journal = "Phys. Plasmas",
  volume = "31",
  number = "1",
  pages = "012302",
  year = "2024",
  doi = "10.1063/5.0173642"
}

@article{Antonenas2024,
  author = "Y. Antonenas and G. Anastasiou and Y. Kominis",
  title = "Analytical calculation of the kinetic q factor and resonant response of toroidally confined plasmas",
  journal = "Phys. Plasmas",
  volume = "31",
  pages = "102302",
  year = "2024",
  doi = "10.1063/5.0222886"
}

@article{Skokos2001,
  author = "Ch. Skokos",
  title = "Alignment indices: a new, simple method for determining the ordered or chaotic nature of orbits",
  journal = "J. Phys. A",
  volume = "34",
  number = "47",
  pages = "10029",
  year = "2001",
  publisher = "IOP Publishing"
}

@article{Benjamin2023,
  author = "S. Benjamin and H. Järleblad and M. Salewski and L. Stagner and M. Hole and D. Pfefferlé",
  title = "Distribution transforms for guiding center orbit coordinates in axisymmetric tokamak equilibria",
  journal = "Comput. Phys. Commun.",
  volume = "292",
  pages = "108893",
  year = "2023",
  doi = "10.1016/j.cpc.2023.108893"
}

@article{Bierwage2022b,
  author = "A. Bierwage and M. Fitzgerald and P. Lauber and M. Salewski and Y. Kazakov and Ž. Štancar",
  title = "Representation and modeling of charged particle distributions in tokamaks",
  journal = "Comput. Phys. Commun.",
  volume = "275",
  pages = "108305",
  year = "2022",
  doi = "10.1016/j.cpc.2022.108305"
}

@article{Meng2018,
  author = "G. Meng and N. N. Gorelenkov and V. N. Duarte and H. L. Berk and R. B. White and X. G. Wang",
  title = "Resonance frequency broadening of wave-particle interaction in tokamaks due to Alfvénic eigenmode",
  journal = "Nucl. Fusion",
  volume = "58",
  year = "2018",
  doi = "10.1088/1741-4326/aaa918"
}

@article{Wobig2001,
  author = "H. Wobig and D. Pfirsch",
  title = "On guiding centre orbits of particles in toroidal systems",
  journal = "Plasma Phys. Control. Fusion",
  volume = "43",
  year = "2001",
  doi = "10.1088/0741-3335/43/5/305"
}

@article{He2020,
  author = "K. He and Y. Sun and B. N. Wan and S. Gu and M. Jia",
  title = "Roles of primary and sideband resonances in the confinement of energetic passing ions under resonant magnetic perturbations",
  journal = "Nucl. Fusion",
  volume = "60",
  pages = "126027",
  year = "2020",
  doi = "10.1088/1741-4326/abb422"
}

@article{Levinton1995,
  title = {Improved Confinement with Reversed Magnetic Shear in {TFTR}},
  author = {Levinton, F. M. and Zarnstorff, M. C. and Batha, S. H. and Bell, M. and Bell, R. E. and Budny, R. V. and Bush, C. and Chang, Z. and Fredrickson, E. and Janos, A. and Manickam, J. and Ramsey, A. and Sabbagh, S. A. and Schmidt, G. L. and Synakowski, E. J. and Taylor, G.},
  journal = {Phys. Rev. Lett.},
  volume = {75},
  issue = {24},
  pages = {4417--4420},
  numpages = {0},
  year = {1995},
  month = {Dec},
  publisher = {American Physical Society},
  doi = {10.1103/PhysRevLett.75.4417},
}

@article{Lee2017,
   author = {Lee, Y. and Lim, W.},
   doi = {10.5951/mathteacher.110.8.0631},
   issn = {0025-5769},
   issue = {8},
   journal = {Math. Teach. Educ.},
   title = {Shoelace Formula: Connecting the Area of a Polygon and the Vector Cross Product},
   volume = {110},
   year = {2017},
}

@article{Strait1995,
  title = {Enhanced Confinement and Stability in {DIII-D} Discharges with Reversed Magnetic Shear},
  author = {Strait, E. J. and Lao, L. L. and Mauel, M. E. and Rice, B. W. and Taylor, T. S. and Burrell, K. H. and Chu, M. S. and Lazarus, E. A. and Osborne, T. H. and Thompson, S. J. and Turnbull, A. D.},
  journal = {Phys. Rev. Lett.},
  volume = {75},
  issue = {24},
  pages = {4421--4424},
  numpages = {0},
  year = {1995},
  month = {Dec},
  publisher = {American Physical Society},
  doi = {10.1103/PhysRevLett.75.4421},
}

@article{Nazikian2008,
    author = {Nazikian, R. and Fu, G. and Austin, M. and Berk, H. and Budny, R. and Gorelenkov, N. and Heidbrink, W. and Holcomb, C. and Kramer, G. and McKee, G. and Makowski, M. and Solomon, W. and Shafer, M. and Strait, E. and Van Zeeland, M.},
    title = {Intense geodesic acousticlike modes driven by suprathermal ions in a tokamak plasma},
    year = {2008},
    journal = {Phys. Rev. Lett.},
    volume = {101},
    number = {18},
    pages = {185001},
    type = {Article},
}

@article{Zhang2024,
   author = {Zhang, Y. N. and He, K. Y. and Sun, Y. W. and Wan, B. N. and Wu, X. M. and Xie, P. C. and Liu, Y. Q.},
   doi = {10.1088/1741-4326/ad249e},
   issn = {17414326},
   issue = {4},
   journal = {Nucl. Fusion},
   title = {Influence of the far non-resonant components of high-n resonant magnetic perturbations on energetic passing ions loss},
   volume = {64},
   year = {2024}
}

@article{Boozer1980,
   author = {Boozer, A. H.},
   doi = {10.1063/1.863080},
   issn = {10706631},
   issue = {5},
   journal = {Phys. Fluids},
   title = {Guiding center drift equations},
   volume = {23},
   year = {1980}
}

@article{Brizard2014,
   author = {Brizard, A. J. and Duthoit, F. X.},
   doi = {10.1063/1.4879811},
   issn = {10897674},
   issue = {5},
   journal = {Phys. Plasmas},
   title = {Canonical transformation for trapped/passing guiding-center orbits in axisymmetric tokamak geometry},
   volume = {21},
   year = {2014},
   pages = {052509},
}

@article{White2013c,
    author = {White, R. and Spizzo, G. and Gobbin, M.},
    title = {Guiding center equations of high accuracy},
    year = {2013},
    journal = {Plasma Phys. Controlled Fusion},
    volume = {55},
    number = {11},
    pages = {115002},
    type = {Article},
}

@article{Artsimovich1972,
   author = {Artsimovich, L. A.},
   doi = {10.1088/0029-5515/12/2/012},
   issn = {17414326},
   issue = {2},
   journal = {Nucl. Fusion},
   title = {Tokamak devices},
   volume = {12},
   year = {1972},
}

@article{Spitzeb1958,
   author = {Spitzer, L.},
   doi = {10.1063/1.1705883},
   issn = {10706631},
   issue = {4},
   journal = {Phys. Fluids},
   title = {The stellarator concept},
   volume = {1},
   year = {1958},
}

@article{Cabal2017,
   author = {Cabal, H. and Lechón, Y. and Bustreo, C. and Gracceva, F. and Biberacher, M. and Ward, D. and Dongiovanni, D. and Grohnheit, P. E.},
   doi = {10.1016/j.esr.2016.11.002},
   issn = {2211467X},
   journal = {Energy Strat. Rev.},
   title = {Fusion power in a future low carbon global electricity system},
   volume = {15},
   year = {2017},
}

@article{Gi2020,
   author = {Gi, K. and Sano, F. and Akimoto, K. and Hiwatari, R. and Tobita, K.},
   doi = {10.1016/j.esr.2019.100432},
   issn = {2211467X},
   journal = {Energy Strat. Rev.},
   title = {Potential contribution of fusion power generation to low-carbon development under the Paris Agreement and associated uncertainties},
   volume = {27},
   year = {2020},
}

@article{Sehila ,
   author = {Vicente, S. M. G. D. and Smith, N. A. and El-Guebaly, L. and Ciattaglia, S. and Pace, L. D. and Gilbert, M. and Mandoki, R. and Rosanvallon, S. and Someya, Y. and Tobita, K. and Torcy, D.},
   issue = {8},
   journal = {Nucl. Fusion},
   title = {Overview on the management of radioactive waste from fusion facilities: ITER, demonstration machines and power plants},
   volume = {62},
   year = {2022},
}

@article{Hawryluk2009,
   author = {Hawryluk, R. J. and Campbell, D. J. and Janeschitz, G. and Thomas, P. R. and al., e.},
   doi = {10.1088/0029-5515/49/6/065012},
   issn = {00295515},
   issue = {6},
   journal = {Nucl. Fusion},
   title = {Principal physics developments evaluated in the ITER design review},
   volume = {49},
   year = {2009},
}

@article{Rimini2024,
   author = {Rimini, F. G. and {JET} {C}ontributors},
   doi = {10.1109/TPS.2024.3352233},
   issn = {19399375},
   journal = {IEEE Trans. Plasma Sci.},
   title = {JET: 40 Successful Years of Fusion Research},
   year = {2024},
}

@article{Tomarchio2017,
   author = {Tomarchio, V. and Barabaschi, P. and Pietro, E. D. and Hanada, M. and Kamada, Y. and Sakasai, A. and Shirai, H.},
   doi = {10.1016/j.fusengdes.2017.05.041},
   issn = {09203796},
   journal = {Fusion Eng. Des.},
   title = {Status of the JT-60SA project: An overview on fabrication, assembly and future exploitation},
   volume = {123},
   year = {2017},
}

@article{Stroth2022,
   author = {Stroth, U. and Aguiam, D. and Alessi, E. and Angioni, C. and al., e.},
   doi = {10.1088/1741-4326/ac207f},
   issn = {17414326},
   issue = {4},
   journal = {Nucl. Fusion},
   title = {Progress from ASDEX Upgrade experiments in preparing the physics basis of ITER operation and DEMO scenario development},
   volume = {62},
   year = {2022},
}

@article{Sykes2001,
   author = {Sykes, A. and Ahn, J. W. and Akers, R. and Arends, E. and al., e.},
   doi = {10.1063/1.1352595},
   issn = {1070664X},
   issue = {5 II},
   journal = {Phys. Plasmas},
   title = {First physics results from the MAST Mega-Amp Spherical Tokamak},
   volume = {8},
   year = {2001},
}

@article{Coda2017,
   author = {Coda, S. and Ahn, J. and Albanese, R. and Alberti, S. and Alessi, E. and al., e.},
   doi = {10.1088/1741-4326/aa6412},
   issn = {17414326},
   issue = {10},
   journal = {Nucl. Fusion},
   title = {Overview of the TCV tokamak program: Scientific progress and facility upgrades},
   volume = {57},
   year = {2017},
}

@article{Bourdelle2015,
   author = {Bourdelle, C. and Artaud, J. F. and Basiuk, V. and Bécoulet, M. and al., e.},
   doi = {10.1088/0029-5515/55/6/063017},
   issn = {17414326},
   issue = {6},
   journal = {Nucl. Fusion},
   title = {WEST Physics Basis},
   volume = {55},
   year = {2015},
}

@article{Mailloux2022,
   author = {Mailloux, J. and Abid, N. and Abraham, K. and Abreu, P. and al., e.},
   doi = {10.1088/1741-4326/ac47b4},
   issn = {17414326},
   issue = {4},
   journal = {Nucl. Fusion},
   title = {Overview of JET results for optimising ITER operation},
   volume = {62},
   year = {2022},
}

@article{Newcomb1958,
   author = {Newcomb, W. A.},
   doi = {10.1016/0003-4916(58)90024-1},
   issn = {1096035X},
   issue = {4},
   journal = {Ann. Phys.},
   title = {Motion of magnetic lines of force},
   volume = {3},
   year = {1958},
}

@article{Lundquist1951,
   author = {Lundquist, S.},
   doi = {10.1103/PhysRev.83.307},
   issn = {0031899X},
   issue = {2},
   journal = {Phys. Rev.},
   title = {On the stability of magneto-hydrostatic fields},
   volume = {83},
   year = {1951},
}

@article{White2013b,
   author = {White, R. B.},
   doi = {https://doi.org/10.1063/1.4802094},
   number = {042116},
   journal = {Phys. Plasmas},
   title = {Guiding center equations for ideal magnetohydrodynamic modes},
   volume = {20},
   year = {2013},
}

@article{White2013a,
   author = {White, R. B.},
   doi ={https://doi.org/10.1063/1.4791661},
   number = {022105},
   journal = {Phys. Plasmas},
   title = {Representation of ideal magnetohydrodynamic modes},
   volume = {20},
   year = {2013},
}

@article{Cheng1986,
   author = {Cheng, C. Z. and Chance, M. S.},
   doi = {10.1063/1.865801},
   issn = {0031-9171},
   issue = {11},
   journal = {Phys. Fluids},
   title = { Low- n shear Alfvén spectra in axisymmetric toroidal plasmas },
   volume = {29},
   year = {1986},
   pages = {3695–3701}
}

@article{Sigmar1992,
   author = {Sigmar, D. J. and Hsu, C. T. and White, R. and Cheng, C. Z.},
   doi = {10.1063/1.860061},
   issn = {08998221},
   issue = {6},
   journal = {Phys. Fluids},
   title = {Alpha-particle losses from toroidicity-induced Alfvén eigenmodes. Part II: Monte Carlo simulations and anomalous alpha-loss processes},
   volume = {4},
   year = {1992},
   pages = {1506}
}

@article{White1984,
  title={Hamiltonian guiding center drift orbit calculation for plasmas of arbitrary cross section},
  author = {White, R. B. and Chance, M.},
  journal={Phys. Fluids},
  volume={27},
  number={10},
  pages={2455--2467},
  year={1984},
  publisher={American Institute of Physics}
}

@article{Heidbrink1991,
   author = {Heidbrink, W. W. and Strait, E. J. and Doyle, E. and Sager, G. and Snider, R. T.},
   doi = {10.1088/0029-5515/31/9/002},
   issn = {17414326},
   issue = {9},
   journal = {Nucl. Fusion},
   title = {An investigation of beam driven alfvén instabilities in the diii-d tokamak},
   volume = {31},
   year = {1991},
}

@article{Fasoli1997,
   author = {Fasoli, A. and Borba, D. and Gormezano, C. and Heeter, R. and Jaun, A. and Jacquinot, J. and Kerner, W. and King, Q. and Lister, J. B. and Sharapov, S. and Start, D. and Villard, L.},
   doi = {10.1088/0741-3335/39/12B/022},
   issn = {07413335},
   issue = {12B},
   journal = {Plasma Phys. Controlled Fusion},
   title = {Alfvén eigenmode experiments in tokamaks and stellarators},
   volume = {39},
   year = {1997},
}

@article{Fasoli1995,
   author = {Fasoli, A. and Lister, J. B. and Sharapov, S. E. and Ali-Arshad, S. and Bosia, G. and Borba, D. and Campbell, D. J. and Deliyanakis, N. and Dobbing, J. A. and Gormezano, C. and Holties, H. A. and Huysmans, G. T. and Jacquinot, J. and Jaun, A. and Kerner, W. and Lavanchy, P. and Moret, J. M. and Porte, L. and Santagiustina, A. and Villard, L.},
   doi = {10.1088/0029-5515/35/12/I09},
   issn = {00295515},
   issue = {12},
   journal = {Nucl. Fusion},
   title = {Overview of Alfven eigenmode experiments in JET},
   volume = {35},
   year = {1995},
}

@article{Fu1998,
   author = {Fu, G. Y. and Nazikian, R. and Budny, R. and Chang, Z.},
   doi = {10.1063/1.873165},
   issn = {1070664X},
   issue = {12},
   journal = {Phys. Plasmas},
   title = {Alpha particle-driven toroidal alfvén eigenmodes in tokamak fusion test reactor deuterium-tritium plasmas: Theory and experiments},
   volume = {5},
   year = {1998},
}

@article{Fasoli2007,
   author = {Fasoli, A. and Gormenzano, C. and Berk, H. L. and Breizman, B. and Briguglio, S. and Darrow, D. S. and Gorelenkov, N. and Heidbrink, W. W. and Jaun, A. and Konovalov, S. V. and Nazikian, R. and Noterdaeme, J. M. and Sharapov, S. and Shinohara, K. and Testa, D. and Tobita, K. and Todo, Y. and Vlad, G. and Zonca, F.},
   doi = {10.1088/0029-5515/47/6/S05},
   issn = {00295515},
   issue = {6},
   journal = {Nucl. Fusion},
   title = {Chapter 5: Physics of energetic ions},
   volume = {47},
   year = {2007},
}

@article{Duong1993,
   author = {Duong, H. H. and Heidbrink, W. W. and Strait, E. J. and Petrie, T. W. and Lee, R. and Moyer, R. A. and Watkins, J. G.},
   doi = {10.1088/0029-5515/33/5/I06},
   issn = {00295515},
   issue = {5},
   journal = {Nucl. Fusion},
   title = {Loss of energetic beam ions during TAE instabilities},
   volume = {33},
   year = {1993},
}

@article{Kim2022,
   author = {Kim, J. and Kang, J. and Rhee, T. and Jo, J. and Han, H. and Podest, M. and Lee, J. H. and Lee, S. and Bak, J. G. and Choi, M. J. and Nazikian, R. and Jhang, H. and Ko, J. and Joung, M. and Jeon, Y. M. and Na, Y. S. and Shinohara, K. and Cheng, C. Z.},
   doi = {10.1088/1741-4326/ac3e39},
   issn = {17414326},
   issue = {2},
   journal = {Nucl. Fusion},
   title = {Suppression of toroidal Alfvén eigenmodes by the electron cyclotron current drive in KSTAR plasmas},
   volume = {62},
   year = {2022},
}

@misc{Heidbrink1994,
   author = {Heidbrink, W. W. and Sadler, G. J.},
   doi = {10.1088/0029-5515/34/4/I07},
   issn = {00295515},
   issue = {4},
   journal = {Nucl. Fusion},
   title = {The behaviour of fast ions in tokamak experiments},
   volume = {34},
   year = {1994},
}

@article{Collins2016,
   author = {Collins, C. S. and Heidbrink, W. W. and Austin, M. E. and Kramer, G. J. and Pace, D. C. and Petty, C. C. and Stagner, L. and Zeeland, M. A. V. and White, R. B. and Zhu, Y. B.},
   doi = {10.1103/PhysRevLett.116.095001},
   issn = {10797114},
   issue = {9},
   journal = {Phys. Rev. Lett.},
   title = {Observation of Critical-Gradient Behavior in Alfvén-Eigenmode-Induced Fast-Ion Transport},
   volume = {116},
   year = {2016},
}

@article{White1983,
   author = {White, R. B. and Goldston, R. J. and McGuire, K. and Boozer, A. H. and Monticello, D. A. and Park, W.},
   doi = {10.1063/1.864060},
   issn = {10706631},
   issue = {10},
   journal = {Phys. Fluids},
   title = {Theory of mode-induced beam particle loss in tokamaks},
   volume = {26},
   year = {1983},
}

@article{White1982,
   author = {White, R. B. and Boozer, A. H. and Hay, R.},
   doi = {10.1063/1.863773},
   issn = {10706631},
   issue = {3},
   journal = {Phys. Fluids},
   title = {Drift Hamiltonian in magnetic coordinates},
   volume = {25},
   year = {1982},
}

@article{Hsu1992,
   author = {Hsu, C. T. and Sigmar, D. J.},
   doi = {10.1063/1.860060},
   issn = {08998221},
   issue = {6},
   journal = {Phys. Fluids},
   title = {Alpha-particle losses from toroidicity-induced Alfvén eigenmodes. Part I: Phase-space topology of energetic particle orbits in tokamak plasma},
   volume = {4},
   year = {1992},
}

@article{Heidbrink2008,
   author = {Heidbrink, W. W.},
   doi = {10.1063/1.2838239},
   issn = {1070664X},
   issue = {5},
   journal = {Phys. Plasmas},
   title = {Basic physics of Alfvn instabilities driven by energetic particles in toroidally confined plasmas},
   volume = {15},
   year = {2008},
}

@article{Pinches1998,
   author = {Pinches, S. D. and Appel, L. C. and Candy, J. and Sharapov, S. E. and Berk, H. L. and Borba, D. and Breizman, B. N. and Hender, T. C. and Hopcraft, K. I. and Huysmans, G. T. and Kerner, W.},
   doi = {10.1016/s0010-4655(98)00034-4},
   issn = {00104655},
   issue = {1-3},
   journal = {Comput. Phys. Commun.},
   title = {The HAGIS self-consistent nonlinear wave-particle interaction model},
   volume = {111},
   year = {1998},
}

@article{White2012,
   author = {White, R. B.},
   doi = {10.1016/j.cnsns.2011.02.013},
   issn = {10075704},
   issue = {5},
   journal = {Commun. Nonlinear Sci. Numer.},
   title = {Modification of particle distributions by MHD instabilities I},
   volume = {17},
   year = {2012},
}

@article{Kramer2012,
   author = {Kramer, G. J. and Chen, L. and Fisher, R. K. and Heidbrink, W. W. and Nazikian, R. and Pace, D. C. and Zeeland, M. A. V.},
   doi = {10.1103/PhysRevLett.109.035003},
   issn = {00319007},
   issue = {3},
   journal = {Phys. Rev. Lett.},
   title = {Fractional resonances between waves and energetic particles in tokamak plasmas},
   volume = {109},
   year = {2012},
}

@article{Gobbin2008,
   author = {Gobbin, M. and White, R. B. and Marrelli, L. and Martin, P.},
   doi = {10.1088/0029-5515/48/7/075002},
   issn = {00295515},
   issue = {7},
   journal = {Nucl. Fusion},
   title = {Resonance between passing fast ions and MHD instabilities both in the tokamak and the RFP configurations},
   volume = {48},
   year = {2008},
}

@article{Fiksel2005,
   author = {Fiksel, G. and Hudson, B. and Hartog, D. J. D. and Magee, R. M. and O'Connell, R. and Prager, S. C. and Beklemishev, A. D. and Davydenko, V. I. and Ivanov, A. A. and Tsidulko, Y. A.},
   doi = {10.1103/PhysRevLett.95.125001},
   issn = {00319007},
   issue = {12},
   journal = {Phys. Rev. Lett.},
   title = {Observation of weak impact of a stochastic magnetic field on fast-ion confinement},
   volume = {95},
   year = {2005},
}

@article{Arnold1963,
   author = {Arnol'd, V. I.},
   doi = {10.1070/rm1963v018n05abeh004130},
   issn = {0036-0279},
   issue = {5},
   journal = {Russ. Math. Surv.},
   title = {Proof of a theorem of A. N. Kolmogorov on the invariance of quasi-periodic motions under small perturbations of the Hamiltonian},
   volume = {18},
   year = {1963},
}

@article{White2018,
   author = {White, R. B. and Gorelenkov, N. N. and Duarte, V. N. and Berk, H. L.},
   doi = {10.1063/1.5046655},
   issn = {10897674},
   issue = {10},
   journal = {Phys. Plasmas},
   title = {Resonances between high energy particles and ideal magnetohydrodynamic modes in tokamaks},
   volume = {25},
   year = {2018},
}

@article{Littlejohn1983,
   author = {Littlejohn, R. G.},
   doi = {10.1017/S002237780000060X},
   issn = {14697807},
   issue = {1},
   journal = {J. Plasma Phys.},
   title = {Variational Principles of Guiding Centre Motion},
   volume = {29},
   year = {1983},
}

@article{Boozer1981,
   author = {Boozer, A. H.},
   doi = {10.1063/1.863297},
   issn = {10706631},
   issue = {11},
   journal = {Phys. Fluids},
   title = {Plasma equilibrium with rational magnetic surfaces},
   volume = {24},
   year = {1981},
}

@article{Evans2006,
   author = {Evans, T. E. and Moyer, R. A. and Burrell, K. H. and Fenstermacher, M. E. and Joseph, I. and Leonard, A. W. and Osborne, T. H. and Porter, G. D. and Schaffer, M. J. and Snyder, P. B. and Thomas, P. R. and Watkins, J. G. and West, W. P.},
   doi = {10.1038/nphys312},
   issn = {17452481},
   issue = {6},
   journal = {Nat. Phys},
   title = {Edge stability and transport control with resonant magnetic perturbations in collisionless tokamak plasmas},
   volume = {2},
   year = {2006},
}

@article{Zohm1996,
   author = {Zohm, H.},
   doi = {10.1088/0741-3335/38/2/001},
   issn = {07413335},
   issue = {2},
   journal = {Plasma Phys. Controlled Fusion},
   title = {Edge localized modes (ELMs)},
   volume = {38},
   year = {1996},
}

@article{Liang2007,
   author = {Liang, Y. and Koslowski, H. R. and Thomas, P. R. and Nardon, E. and Alper, B. and Andrew, P. and Andrew, Y. and Arnoux, G. and Baranov, Y. and Bécoulet, M. and Beurskens, M. and Biewer, T. and Bigi, M. and Crombe, K. and Luna, E. D. L. and Vries, P. D. and Fundamenski, W. and Gerasimov, S. and Giroud, C. and Gryaznevich, M. P. and Hawkes, N. and Hotchin, S. and Howell, D. and Jachmich, S. and Kiptily, V. and Moreira, L. and Parail, V. and Pinches, S. D. and Rachlew, E. and Zimmermann, O.},
   doi = {10.1103/PhysRevLett.98.265004},
   issn = {00319007},
   issue = {26},
   journal = {Phys. Rev. Lett.},
   title = {Active control of type-I edge-localized modes with n=1 perturbation fields in the JET tokamak},
   volume = {98},
   year = {2007},
}

@article{Suttrop2011,
   author = {Suttrop, W. and Eich, T. and Fuchs, J. C. and Günter, S. and Janzer, A. and Herrmann, A. and Kallenbach, A. and Lang, P. T. and Lunt, T. and Maraschek, M. and McDermott, R. M. and Mlynek, A. and Pütterich, T. and Rott, M. and Vierle, T. and Wolfrum, E. and Yu, Q. and Zammuto, I. and Zohm, H.},
   doi = {10.1103/PhysRevLett.106.225004},
   issn = {00319007},
   issue = {22},
   journal = {Phys. Rev. Lett.},
   title = {First observation of edge localized modes mitigation with resonant and nonresonant magnetic perturbations in ASDEX upgrade},
   volume = {106},
   year = {2011},
}

@article{Shinohara2018,
   author = {Shinohara, K. and Bierwage, A. and Suzuki, Y. and Kim, J. and Matsunaga, G. and Honda, M. and Rhee, T.},
   doi = {10.1088/1741-4326/aab170},
   issn = {17414326},
   issue = {8},
   journal = {Nucl. Fusion},
   title = {Estimation of orbit island width from static magnetic island width, using safety factor and orbit pitch},
   volume = {58},
   year = {2018},
}

@article{Shinohara2020,
   author = {Shinohara, K. and Bierwage, A. and Matsuyama, A. and Suzuki, Y. and Matsunaga, G. and Honda, M. and Sumida, S. and Kim, J.},
   doi = {10.1088/1741-4326/aba0c8},
   issn = {17414326},
   issue = {9},
   journal = {Nucl. Fusion},
   title = {Efficient estimation of drift orbit island width for passing ions in a shaped tokamak plasma with a static magnetic perturbation},
   volume = {60},
   year = {2020},
}

@article{Munoz2019,
   author = {Garcia-Munoz, M. and Sharapov, S. E. and Zeeland, M. A. V. and Ascasibar, E. and Cappa, A. and Chen, L. and Ferreira, J. and Galdon-Quiroga, J. and Geiger, B. and Gonzalez-Martin, J. and Heidbrink, W. W. and Johnson, T. and Lauber, P. and Mantsinen, M. and Melnikov, A. V. and Nabais, F. and Rivero-Rodriguez, J. F. and Sanchis-Sanchez, L. and Schneider, P. and Stober, J. and Suttrop, W. and Todo, Y. and Vallejos, P. and Zonca, F.},
   doi = {10.1088/1361-6587/aaef08},
   issn = {13616587},
   issue = {5},
   journal = {Plasma Phys. Controlled Fusion},
   title = {Active control of Alfvén eigenmodes in magnetically confined toroidal plasmas},
   volume = {61},
   year = {2019},
}

@article{Bortolon2013,
   author = {Bortolon, A. and Heidbrink, W. W. and Kramer, G. J. and Park, J. K. and Fredrickson, E. D. and Lore, J. D. and Podestà, M.},
   doi = {10.1103/PhysRevLett.110.265008},
   issn = {00319007},
   issue = {26},
   journal = {Phys. Rev. Lett.},
   title = {Mitigation of Alfvén activity in a tokamak by externally applied static 3d fields},
   volume = {110},
   year = {2013},
}

@article{Caldas2012,
   author = {Caldas, I. L. and Viana, R. L. and Abud, C. V. and Fonseca, J. C. and Filho, Z. O. G. and Kroetz, T. and Marcus, F. A. and Schelin, A. B. and Szezech, J. D. and Toufen, D. L. and Benkadda, S. and Lopes, S. R. and Morrison, P. J. and Roberto, M. and Gentle, K. and Kuznetsov, Y. and Nascimento, I. C.},
   doi = {10.1088/0741-3335/54/12/124035},
   issn = {07413335},
   issue = {12},
   journal = {Plasma Phys. Controlled Fusion},
   title = {Shearless transport barriers in magnetically confined plasmas},
   volume = {54},
   year = {2012},
}

@article{Littlejohn1985,
   author = {Littlejohn, R. G.},
   doi = {10.1063/1.865379},
   issn = {10706631},
   issue = {6},
   journal = {Phys. Fluids},
   title = {Differential forms and canonical variables for drift motion in toroidal geometry},
   volume = {28},
   year = {1985},
}

@article{Evans2008,
   author = {Evans, T. E. and Fenstermacher, M. E. and Moyer, R. A. and Osborne, T. H. and Watkins, J. G. and Gohil, P. and Joseph, I. and Schaffer, M. J. and Baylor, L. R. and Bécoulet, M. and Boedo, J. A. and Burrell, K. H. and Degrassie, J. S. and Finken, K. H. and Jernigan, T. and Jakubowski, M. W. and Lasnier, C. J. and Lehnen, M. and Leonard, A. W. and Lonnroth, J. and Nardon, E. and Parail, V. and Schmitz, O. and Unterberg, B. and West, W. P.},
   doi = {10.1088/0029-5515/48/2/024002},
   issn = {00295515},
   issue = {2},
   journal = {Nucl. Fusion},
   title = {RMP ELM suppression in DIII-D plasmas with ITER similar shapes and collisionalities},
   volume = {48},
   year = {2008},
}

@article{Kaufman1972,
   author = {Kaufman, A. N.},
   doi = {10.1063/1.1694031},
   issn = {10706631},
   issue = {6},
   journal = {Phys. Fluids},
   title = {Quasilinear diffusion of an axisymmetric toroidal plasma},
   volume = {15},
   year = {1972},
}

@article{Abdullaev2008,
   author = {Abdullaev, S. S. and Jakubowski, M. and Lehnen, M. and Schmitz, O. and Unterberg, B.},
   doi = {10.1063/1.2907163},
   issn = {1070664X},
   issue = {4},
   journal = {Phys. Plasmas},
   title = {On description of magnetic stochasticity in poloidal divertor tokamaks},
   volume = {15},
   year = {2008},
}

@article{Zestanakis2016,
    title={Orbital spectrum analysis of non-axisymmetric perturbations of the guiding-center particle motion in axisymmetric equilibria},
    author = {Zestanakis, P. and Kominis, Y. and Anastassiou, G. and Hizanidis, K.},
    journal={Phys. Plasmas},
    volume={23},
    number={3},
    pages={032507},
    year={2016},
    publisher={AIP Publishing LLC}
}

@article{White2023,
    author = {White, R. B. and Duarte, V. N.},
    title = "{Assessment of radial transport induced by Alfvénic resonances in tokamaks and stellarators}",
    journal = {Phys. Plasmas},
    volume = {30},
    number = {1},
    pages = {012502},
    year = {2023},
    month = {01},
    issn = {1070-664X},
    doi = {10.1063/5.0100215},
    url = {https://doi.org/10.1063/5.0100215},    
}

@article{Taylor1964,
   author = {Taylor, J. B.},
   doi = {10.1063/1.1711283},
   issn = {10706631},
   issue = {6},
   journal = {Phys. Fluids},
   title = {Equilibrium and stability of plasma in arbitrary mirror fields},
   volume = {7},
   year = {1964},
}

@article{Berk1995,
   author = {Berk, H. I. and Breizman, B. N. and Pekker, M. S.},
   doi = {10.1088/0029-5515/35/12/I36},
   issn = {00295515},
   issue = {12},
   journal = {Nucl. Fusion},
   title = {Simulation of Alfven-wave-resonant-particle interaction},
   volume = {35},
   year = {1995},
}

@article{Heidbrink2020,
   author = {Heidbrink, W. W. and White, R. B.},
   doi = {10.1063/1.5136237},
   issn = {10897674},
   issue = {3},
   journal = {Phys. Plasmas},
   title = {Mechanisms of energetic-particle transport in magnetically confined plasmas},
   volume = {27},
   year = {2020},
}

@article{Brizard2011,
   author = {Brizard, A. J.},
   doi = {10.1063/1.3554696},
   issn = {1070664X},
   issue = {2},
   journal = {Phys. Plasmas},
   title = {Compact formulas for guiding-center orbits in axisymmetric tokamak geometry},
   volume = {18},
   year = {2011},
}

@article{Brizard2013,
   author = {Brizard, A. J.},
   doi = {10.1016/j.cnsns.2012.08.023},
   issn = {10075704},
   issue = {3},
   journal = {Commun. Nonlinear Sci. Numer.},
   title = {Jacobi zeta function and action-angle coordinates for the pendulum},
   volume = {18},
   year = {2013},
}

@article{Cary2009,
   author = {Cary, J. R. and Brizard, A. J.},
   doi = {10.1103/RevModPhys.81.693},
   issn = {00346861},
   issue = {2},
   journal = {RMP},
   title = {Hamiltonian theory of guiding-center motion},
   volume = {81},
   year = {2009},
}

@article{Shinohara2016,
   author = {Shinohara, K. and Suzuki, Y. and Kim, J. and Kim, J. Y. and Jeon, Y. M. and Bierwage, A. and Rhee, T.},
   doi = {10.1088/0029-5515/56/11/112018},
   issn = {17414326},
   issue = {11},
   journal = {Nucl. Fusion},
   title = {Investigation of fast ion behavior using orbit following Monte-Carlo code in magnetic perturbed field in KSTAR},
   volume = {56},
   year = {2016},
}

@article{Chiricov1979,
author = {Chirikov, B. V.},
title = {A universal instability of many-dimensional oscillator systems},
journal = {Phys. Rep.},
volume = {52},
number = {5},
pages = {263-379},
year = {1979},
issn = {0370-1573},
doi = {https://doi.org/10.1016/0370-1573(79)90023-1},
}

@article{Spizzo2021,
   author = {Zonca, F.},
   doi = {https://doi.org/10.1088/1741-4326/ac1e08},
   journal = {Nucl. Fusion},
   title = {Collisionless losses of fast ions in the divertor tokamak test due to toroidal field ripple},
   volume = {61},
   year = {2021},
   number = {116016},
}

@article{Arnold1964,
   author = {Arnol’d, VI},
   journal = {Sov. Math. Dokl.},
   title = {Instabilities in dynamical systems with several degrees of freedom},
   volume = {5},
   pages = {581-585},
   year = {1964},
}

@article{Bierwage2022,
  author =       "A. Bierwage and R. B. White and A. Matsuyama",
  title =        "Testing the conservative character of particle simulations: {I}. Canonical and noncanonical guiding center model in {B}oozer coordinates",
  journal =      "Phys. Plasmas",
  volume =       "29",
  number =        "113905",
  year =         "2022",
  pages = {113905}
}



\end{spacing}


\begin{appendices} 

\end{appendices}

\printthesisindex 

\end{document}